%% file: paper.tex
\documentclass[11pt]{article}
\usepackage[utf8]{inputenc}
\usepackage[T1]{fontenc}
\usepackage[sc]{mathpazo}
\usepackage{microtype}
\usepackage{amsmath,xcolor}
\usepackage{amssymb}
\usepackage{amsthm}
\usepackage{bm,float}
\usepackage{dsfont}
\usepackage{authblk}
\usepackage{fullpage,caption,wrapfig}
\usepackage{comment}
\usepackage{mathtools}
\usepackage[shortlabels]{enumitem}
\usepackage{complexity}
\usepackage[backend=bibtex, style=alphabetic, backref=true,maxbibnames=99, url=false]{biblatex} 
\usepackage{bbm}
\usepackage{hyperref}
\hypersetup{
    colorlinks=true,
    linkcolor=violet,
    filecolor=magenta,      
    urlcolor=cyan,
    citecolor=blue,
    pdffitwindow=true,
}\usepackage[capitalize, nameinlink]{cleveref}
\usepackage{nag,tikz}
\usetikzlibrary{calc}
\usetikzlibrary{decorations.pathreplacing}

\usepackage{algorithm,algpseudocode}
\usepackage{multicol}
\usepackage[scaled]{helvet} 
\usepackage{thmtools}

\theoremstyle{plain}
\newtheorem{theorem}{Theorem}[section]
\newtheorem{lemma}[theorem]{Lemma}
\newtheorem{corollary}[theorem]{Corollary}
\newtheorem{proposition}[theorem]{Proposition}
\newtheorem{fact}[theorem]{Fact}

\newtheorem{observation}[theorem]{Observation}
\newtheorem{claim}[theorem]{Claim}

\newtheorem{definition}[theorem]{Definition}

\theoremstyle{remark}
\newtheorem{remark}[theorem]{Remark}
\newtheorem{example}[theorem]{Example}

\newcommand{\old}[1]{}
\def\cR{{\cal R}}

\newcommand{\st}{\text{s.t.}}

\renewcommand{\R}{\ensuremath{\mathbb R}}

\renewcommand{\P}[1]{{\mathbb{P}}\left[#1\right]}
\renewcommand{\PP}[2]{{\mathbb{P}}_{#1}\left[#2\right]}

\newcommand{\I}[1]{ {\mathbb{I}}\left\{#1\right\} }

\renewcommand{\E}[1]{{\mathbb{E}}\left[#1\right]}
\renewcommand{\EE}[2]{{\mathbb{E}}_{#1}\left[#2\right]}

\renewcommand{\path}[2]{{ S_{#1}, \ldots, S_{#2} }}

\def\argmax{\textup{argmax}}

\def\decrease{\beta}

\def\bone{{\bf 1}}

\def\b1{{\bf 1}}
\def\1{{\bf 1}}

\def\cB{{\cal B}}
\def\oS{{\overline{S}}}

\def\eps{{\epsilon}}
\def\cD{{\cal D}}
\def\cI{{\cal I}}
\def\cH{{\cal H}}

\def\cA{{\cal A}}

\def\cE{{\cal E}}
\def\cost{c}
\def\cL{{\cal L}}

\def\setminus{-}
\def\R{\mathbb{R}}
\def\cR{{\mathcal{R}}}

\def\cC{{\cal C}}
\newcommand{\norm}[1]{\|#1\|}
\def\bbe{{\bf e}}
\def\bbf{{\bf f}}
\def\bbg{{\bf g}}

\newcommand{\declareperson}[1]{\expandafter\newcommand\csname#1\endcsname[1]{\textcolor{orange}{#1: ##1}}}

\declareperson{Anna}
\declareperson{Shayan}
\declareperson{Nathan}
\title{A (Slightly) Improved Approximation Algorithm\\ for Metric TSP}
\author{Anna R. Karlin\thanks{\href{mailto:karlin@cs.washington.edu}{karlin@cs.washington.edu}. Research supported by Air Force Office of Scientific Research grant FA9550-20-1-0212 and NSF grant CCF-1813135.}}
\author{Nathan Klein\thanks{\href{mailto:nwklein@cs.washington.edu}{nwklein@cs.washington.edu}. Research supported in part by NSF grants CCF-1813135 and CCF-1552097.}}
\author{Shayan Oveis Gharan\thanks{\href{mailto:shayan@cs.washington.edu}{shayan@cs.washington.edu}. Research supported by Air Force Office of Scientific Research grant FA9550-20-1-0212, NSF grants  CCF-1552097, CCF-1907845,  ONR YIP grant N00014-17-1-2429, and a Sloan fellowship.}} 
\affil{University of Washington}

\begin{document}
\maketitle 
\begin{abstract}
For some $\eps > 10^{-36}$ we give a randomized $3/2-\eps$ approximation algorithm for metric TSP.
\end{abstract}
\thispagestyle{empty} 

\newpage
\tableofcontents
\thispagestyle{empty}

\newpage 
\setcounter{page}{1}
\input{introduction}

\input{preliminaries}

\input{overview}

\input{polygons}

\input{probabilistic}

\input{matching}

\input{payment}

\printbibliography

\appendix
\input{notation}
\input{probabilistic-app}

\end{document}

%% file: introduction.tex

\section{Introduction}
One of the most fundamental problems in combinatorial optimization is the traveling salesperson problem (TSP), formalized as early as 1832 (c.f. \cite[Ch 1]{ABCC07}).
In an instance of  TSP we are given a set of $n$ cities $V$ along with their pairwise symmetric distances, $c:V\times V \to\R_{\geq 0}$. The goal is to find a Hamiltonian cycle of minimum cost. It is well known that for a general distance function, it is NP-Hard to approximate TSP within any polynomial factor. Therefore it is natural to study metric TSP, in which the distances satisfy the triangle inequality, i.e. $$c(u,w) \le c(u,v) + c(v,w) \quad\quad \forall u,v,w \in V.$$ In this case, the problem is equivalent to finding a closed Eulerian connected walk of minimum cost.\footnote{Given such an Eulerian cycle, we can use the triangle inequality to shortcut vertices visited more than once to get a Hamiltonian cycle.}

It is NP-hard to approximate metric TSP within a factor of $\frac{123}{122}$ \cite{KLS15}. An algorithm of Christofides-Serdyukov~\cite{Chr76,Ser78} from four decades ago gives a $\frac32$-approximation for TSP (see \cite{VS20} for a historical note about TSP). This remains the best known approximation algorithm for the general case of the problem despite significant work, e.g., ~\cite{Wol80,SW90,BP91,Goe95,CV00,GLS05,BM10,BC11,SWV12, HNR17,HN19, KKO20}. 

In contrast, there have been major improvements to this algorithm for a number of special cases of TSP. For example, polynomial-time approximation schemes (PTAS) have been found for Euclidean \cite{Aro96,Mitchell99}, planar \cite{GKP95, AGKKW98, Kle05},  and low-genus metric \cite{DHM07} instances. 
In addition, the case of graph metrics has received significant attention. In 2011, the third author, Saberi, and Singh~\cite{OSS11} found a $\frac{3}{2} - \epsilon_0$ approximation for this case. M\"omke and Svensson \cite{MS11} then
obtained a combinatorial algorithm for graphic TSP with an approximation ratio of 1.461. This ratio was later improved by Mucha \cite{Muc12} to $\frac{13}{9} \approx 1.444$, and then by Seb\"o and Vygen \cite{SV12} to $1.4$.

In this paper we prove the following theorem:
\begin{theorem}\label{thm:main}
	For some absolute constant $\eps > 10^{-36}$, there is a randomized algorithm that outputs a tour with expected cost at most $\frac{3}{2}-\eps$ times the cost of the optimum solution. 
\end{theorem}
We note that while the algorithm makes use of the \ref{eq:tsplp}, we do not prove that the integrality gap of this polytope is bounded away from 3/2. 
We also remark that although our approximation factor is only slightly better than Christofides-Serdyukov, we are not aware of any example where the approximation ratio of the algorithm we analyze exceeds $4/3$ in expectation. 

Following a new exciting result of Traub, Vygen, Zenklusen \cite{TVZ20} we also get the following theorem.
\begin{theorem}
	For some absolute constant $\eps>0$ there is a randomized algorithm that outputs a TSP path with expected cost at most $\frac{3}{2}-\eps$ times the cost of the optimum solution.
\end{theorem}


\subsection{Algorithm}
First, we recall the classical Christofides-Serdyukov algorithm: Given an instance of TSP, choose a minimum spanning tree and then add the minimum cost matching on the odd degree vertices of the tree. The algorithm we study is very similar, except we choose a random spanning tree based on the standard linear programming relaxation of TSP. 

Let $x^0$ be an optimum   solution of the following TSP   linear program relaxation \cite{DFJ59,HK70}:
\begin{equation}\label{eq:tsplp}
\begin{aligned}
	\min \quad& \sum_{u,v} x_{(u,v)} c(u,v)& \\
	\text{s.t.,} \quad &  \sum_{u} x_{(u,v)} = 2&\forall v\in V,\\
	& \sum_{u\in S, v\notin S} x_{(u,v)}\geq 2,&\forall S\subsetneq V,\\
	& x_{(u,v)}\geq 0 &\forall u,v\in V.
\end{aligned}\tag{Held-Karp relaxation}	
\end{equation}
Given  $x^0$, we pick an arbitrary node, $u$, split it into two nodes $u_0,v_0$ and set $x_{(u_0,v_0)}=1, c(u_0,v_0)=0$ and we assign half of every edge incident to $u$ to $u_0$ and the other half to $v_0$.  This allows us to assume without loss of generality that $x^0$ has an edge $e_0=(u_0,v_0)$  such that  $x_{e_0}=1, c(e_0)=0$. 

Let $E_0=E\cup\{e_0\}$ be the support of $x^0$ and let $x$ be $x^0$ restricted to $E$ and $G=(V,E)$.
$x^0$ restricted to $E$ is in the spanning tree polytope \eqref{eq:spanningtreelp}.

For a vector $\lambda:E\to\R_{\geq 0}$, a $\lambda$-uniform distribution $\mu_\lambda$ over spanning trees of $G=(V,E)$ is a distribution where for every spanning tree $T\subseteq E$, $\PP{\mu}{T}=\frac{\prod_{e\in T} \lambda_e}{\sum_{T'} \prod_{e\in T'} \lambda_e}$.
Now, find a vector $\lambda$ such that for every edge $e\in E$, $\PP{\mu_\lambda}{e\in T}=x_e(1\pm\eps)$, for some $\eps<2^{-n}$. Such a vector $\lambda$ can be found using the multiplicative weight update algorithm \cite{AGMOS10} or by applying interior point methods \cite{SV19} or the ellipsoid method \cite{AGMOS10}. (We note that the multiplicative weight update method can only guarantee $\eps<1/\text{poly}(n)$ in polynomial time.) We will sometimes call such a distribution the \textit{maximum entropy} distribution because a $\lambda$-uniform distribution has maximal entropy over all distributions with marginals $x$.\footnote{Although, note that since we do not always find a distribution which preserves marginals exactly (as one does not necessarily exist), for precision we generally refer to the distribution used by the algorithm as $\lambda$-uniform instead.} 
\begin{theorem}[{\cite[Theorem 5.2]{AGMOS10}}]
\label{thm:maxentropycomp}
Let $z$ be a point in the spanning tree polytope (see \eqref{eq:spanningtreelp}) of a graph $ G=(V, E)$.
For any $\eps>0$, a vector $\lambda:E\to\R_{\geq 0}$ can be found such that the corresponding $\lambda$-uniform spanning tree distribution, $\mu_\lambda$, satisfies
$$
\sum_{T\in {\cal T}: T \ni e} \PP{\mu_\lambda}{T}  \leq (1+\varepsilon)z_e,\hspace{3ex}\forall e\in E,$$
i.e., the marginals are approximately preserved.  In the above ${\cal T}$ is the set of all spanning trees of $(V,E)$. The running
time is polynomial in $n=|V|$, $- \log \min_{e\in E} z_e$ and $\log(1/\eps)$.
\end{theorem}

Finally, we sample a tree $T\sim\mu_{\lambda}$ and then add the minimum cost matching on the odd degree vertices of $T$.
\begin{algorithm}[htb]
\begin{algorithmic}
	\State Find an optimum solution $x^0$ of  \ref{eq:tsplp}, and let $e_0=(u_0,v_0)$ be an edge with $x^0_{e_0}=1,c(e_0)=0$.
	\State Let $E_0=E\cup \{e_0\}$ be the support of $x^0$ and $x$ be $x^0$ restricted to $E$ and $G=(V,E)$.
	\State Find a vector $\lambda:E\to\R_{\geq 0}$ such that for any $e\in E$, $\PP{\mu_\lambda}{e}=x_e(1\pm 2^{-n})$.
	\State Sample a tree $T\sim\mu_\lambda$.
	\State Let $M$ be the minimum cost matching on odd degree vertices of $T$.
	\State Output $T \cup M$.
\end{algorithmic}
\label{alg:tsp}
\caption{An Improved Approximation Algorithm for TSP}
\end{algorithm}
The above algorithm is a slight modification of the algorithm proposed in  \cite{OSS11}. 
We refer the interested reader to exciting work of Genova and Williamson \cite{GW17} on the empirical performance of the max-entropy rounding algorithm. We also remark that although the algorithm implemented in \cite{GW17} is slightly different from the above algorithm, we expect the performance to be similar. 

\subsection{New Techniques}
Here we discuss new machinery and technical tools that we developed for this result which could be of independent interest.

\subsubsection{Polygon Structure for Near Minimum Cuts Crossed on one Side.}
Let \hyperlink{tar:G=(V,E,x)}{$G=(V,E,x)$} be an undirected graph equipped with a weight function $x:E\to\R_{\geq 0}$ such that for any cut $(S,\overline{S})$ such that $u_0,v_0\not\in S$, $x(\delta(S))\geq 2$.

For some (small) $\eta\geq 0$, consider the family of \hyperlink{tar:nearmincut}{$\eta$-near min cuts} of $G$. Let ${\cal C}$ be a connected component of \hyperlink{tar:crossing}{crossing} $\eta$-near min cuts. Given ${\cal C}$ we can partition vertices of $G$ into sets $a_0,\dots,a_{m-1}$ (called atoms); this is the coarsest partition such that for each $a_i$, and each $(S,\overline{S})\in {\cal C}$, we have $a_i\subseteq S$ or $a_i\subseteq \overline{S}$. 
Here $a_0$ is the atom that contains $u_0,v_0$.

There have been  several works studying the structure of edges between these atoms and the structure of cuts in a connected component of cuts ${\cal C}$ w.r.t. the $a_i$'s. The {\em cactus structure} (see \cite{DKL76}) shows that if $\eta=0$, then we can arrange the $a_i$'s of a connected component around a cycle, say $a_1,\dots,a_m$ (after renaming), such that $x(E(a_i,a_{i+1}))=1$ for all $i$.

Bencz\'ur and Goemans \cite{Ben95,BG08} studied the case when $\eta\leq 6/5$ and introduced the notion of {\em polygon representation}, in which case atoms can be placed on the sides of an equilateral polygon and some atoms  placed inside  the polygon, such that every cut in ${\cal C}$ can be represented by a diagonal of this polygon. Later, \cite{OSS11} studied the structure of edges of $G$ in this polygon when $\eta<1/100$.

In this paper, we show it suffices to study the structure of edges in a special family of polygon representations: Suppose we have a polygon representation for a connected component $\cC$ of $\eta$-near min cuts of $G$ such that 
\begin{itemize}
\item No atom is mapped inside,
\item If we identify each cut $(S,\overline{S})\in \cC$ with the interval along the polygon that does not contain $a_0$, then any interval is only crossed on one side (only on the left or only on the right).
\end{itemize}
Then, we have (i) For any atom $a_i$, $x(\delta(a_i))\leq 2+O(\eta)$ and (ii) For any pair of atoms $a_i,a_{i+1}$, $x(E(a_i,a_{i+1})\geq 1-\Omega(\eta)$ (see \cref{thm:poly-structure} for details).

We expect to see further applications of our theorem in studying variants of TSP.

\subsubsection{Generalized Gurvits' Lemma}
Given a real stable polynomial $p\in\R_{\geq 0}[z_1,\dots,z_n]$ (with non-negative coefficients), Gurvits proved the following inequality \cite{Gur06,Gur08}
\begin{equation}
	e^{-n}\inf_{z>0} \frac{p(z_1,\dots,z_n)}{z_1\dots z_n}\leq \partial_{z_1}\dots\partial_{z_n} p|_{z=0} \leq \inf_{z>0} \frac{p(z_1,\dots,z_n)}{z_1\dots z_n}.
\end{equation}

As an immediate consequence, one can prove the following theorem about \hyperlink{tar:SR}{strongly Rayleigh} (SR) distributions.
\begin{theorem}\label{thm:gurvits}
Let $\mu:2^{[n]}\to\R_{\geq 0}$ be SR and $A_1,\dots,A_m$ be random variables corresponding to the number of elements sampled in $m$ disjoint subsets of $[n]$ such that $\E{A_i}=n_i$ for all $i$. If $n_i=1$ for all $1\leq i\leq n$, then $\P{\forall i, A_i=1}\geq e^{-m}$.
\end{theorem}
One can ask what happens if the vector $(n_1,\dots,n_m)$  in the above theorem is not equal but close to the all ones vector, $\bone$. 

A related theorem was proved in \cite{OSS11}.
\begin{theorem}
Let $\mu:2^{[n]}\to\R_{\geq 0}$ be SR and $A,B$ be random variables corresponding to the number of elements sampled in two disjoint sets. If $\P{A+B=2}\geq \eps$, $\P{A\leq 1},\P{B\leq 1}\geq \alpha$ and $\P{A\geq 1},\P{B\geq 1}\geq \beta$ then $\P{A=B=1}\geq \eps\alpha\beta/3$.
\end{theorem}

We prove a generalization of both of the above statements; roughly speaking, we show that as long as $\sum_{i=1}^m |n_i-1| <1-\eps$ then $\P{\forall i, A_i=1}\geq f(\eps,m)$ where $f(\eps,m)$ has no dependence on $n$, the number of underlying elements in the support of $\mu$. 
\begin{theorem}[Informal version of \cref{lem:427gen}]\label{thm:gurinf}
Let $\mu:2^{[n]}\to\R_{\geq 0}$ be SR and let $A_1,\dots,A_m$ be random variables corresponding to the number of elements sampled in $m$ disjoint subsets of $[n]$. Suppose that there are integers $n_1,\dots,n_m$ such that for any set $S\subseteq [m]$, $\P{\sum_{i\in S} A_i=\sum_{i\in S} n_i}\geq \eps$. Then, 
$$\P{\forall i, A_i=n_i}\geq f(\eps,m).$$
\end{theorem}
The above statement is even  stronger than \cref{thm:gurvits} as we only require $\P{\sum_{i\in S} A_i=\sum_{i\in S} n_i}$  to be bounded away from $0$ for any set $S\subseteq [m]$ and we don't need a bound on the expectation.
Our proof of the above theorem has double exponential dependence on $\eps$. We leave it an open problem to find the optimum dependency on $\eps$. Furthermore, our proof of the above theorem is probabilistic in nature; we expect that an algebraic proof based on the theory of real stable polynomials will provide a significantly improved lower bound. Unlike the above theorem, such a proof may  possibly extend to the more general class of completely log-concave distributions \cite{AOV18}. In an independent work, Gurvits and Leake \cite{GL21} proved a variant of the above theorem with a much better dependence on $\epsilon$ and $m$ for a homogeneous strong Rayleigh distribution.

\subsubsection{Conditioning while Preserving Marginals}
\label{sec:marginals}
Consider a \hyperlink{tar:SR}{SR} distribution $\mu:2^{[n]}\to\R_{\geq 0}$ and let $x:[n]\to\R_{\geq 0}$, where for all $i$, $x_i=\PP{T\sim\mu}{i\in T}$, be the marginals. 

Let $A,B\subseteq [n]$ be two disjoint sets such that $\E{\hyperlink{tar:AT}{A_T}},\E{B_T}\approx 1$. It follows from \cref{thm:gurinf} that $\P{A_T=B_T=1}\geq \Omega(1)$. Here, however, we are interested in a stronger event; let $\nu = \mu | A_T=B_T=1$ and let $y_i=\PP{T\sim\mu}{i\in T}$. It turns out that the $y$ vector can be very different from the $x$ vector, in particular, for some $i$'s we can have $|y_i-x_i|$ bounded away from $0$. We show that there is an event of non-negligible probability that is a subset of $A_T=B_T=1$ under which the marginals of elements in $A,B$ are almost preserved. 
\begin{theorem}[Informal version of \cref{lem:maxflow}]
Let $\mu:2^{[n]}\to\R_{\geq 0}$ be a SR distribution and let $A,B\subseteq [n]$ be two disjoint subsets such that $\E{A_T},\E{B_T}\approx 1$. For any $\alpha\ll 1$ there is an event $\cE_{A,B}$ such that $\P{\cE_{A,B}}\geq \Omega(\alpha^2)$ and
\begin{itemize}
\item $\P{A_T=B_T=1 | \cE_{A,B}}=1$,
\item $\sum_{i\in A} |\P{i} - \P{i | \cE_{A,B}}| \leq \alpha$,
\item $\sum_{i\in B} |\P{i} - \P{i | \cE_{A,B}}| \leq \alpha$.
\end{itemize}
\end{theorem}
We remark that the quadratic lower bound on $\alpha$ is necessary in the above theorem for a sufficiently small $\alpha>0$.
The above theorem can be seen as a generalization of \cref{thm:gurvits} in the special case of two sets.

We leave it an open problem to extend the above theorem to arbitrary $k$ disjoint sets. We suspect that in such a case the ideal event $\cE_{A_1,\dots,A_k}$ occurs with probability $\Omega(\alpha)^k$ and preserves all marginals of elements in each of the sets $A_1,\dots,A_k$ up to a total variation distance of $\alpha$.

%% file: preliminaries.tex

\section{Preliminaries}
\subsection{Notation}
We write $[n]:=\{1,\dots,n\}$ to denote the set of integers from $1$ to $n$.

For a set $S\subseteq V$, we write 
$$E(S)=\{(u,v)\in E: u,v\in S\}$$ to denote the set of edges in $S$ and we write 
$$\delta(S)=\{(u,v)\in E: |\{u,v\}\cap S|=1\}$$ 
to denote the  set of edges that leave $S$. 

For two {\em disjoint} sets of vertices $A,B\subseteq V$, we write
$$ E(A,B)=\{(u,v)\in E: u\in A, v\in B\}.$$

For a set $A\subseteq E$ and a function $x:E\to\R$ we write
$$ x(A):=\sum_{e\in A} x_e.$$

\hypertarget{tar:crossing}{For two sets $A,B\subseteq V$, we say $A$ {\em crosses} $B$ if all of the following sets are non-empty:
$$ A\cap B, A\smallsetminus B, B\smallsetminus A, \overline{A\cup B}.$$}

\hypertarget{tar:G=(V,E,x)}{We write $G=(V,E,x)$ to denote an (undirected) graph $G$ together with special vertices $u_0,v_0$ and a weight function $x:E\to\R_{\geq 0}$ such that 
$$x(\delta(S))\geq 2, \quad\quad \forall S\subsetneq V: u_0,v_0\notin S.$$}
\hypertarget{tar:nearmincut}{For such a graph, we say a cut $S\subseteq V$ is an {\em $\eta$-near min cut w.r.t., $x$} (or simply $\eta$-near min cut when $x$ is understood) if $x(\delta(S))\leq 2+\eta$.}
Unless otherwise specified, in any statement about a cut $(S,\overline{S})$ in $G$,  we assume $u_0,v_0\not\in S$. 

\subsection{Polyhedral Background}
For any graph $G=(V,E)$,
Edmonds \cite{Edm70} gave the following description for the convex hull of spanning trees of a graph $G=(V,E)$, known as the {\em spanning tree polytope}.
\begin{equation}
\begin{aligned}
& z(E) = |V|-1 & \\
& z(E(S)) \leq |S|-1 &  \forall S\subseteq V\\
& z_e \geq 0 & \hspace{6ex} \forall e\in E.
\end{aligned}
\label{eq:spanningtreelp}
\end{equation}
Edmonds \cite{Edm70} proved that the extreme point solutions of this polytope are the characteristic vectors of the spanning trees of $G$.

\begin{fact} \label{fact:sptreepolytope}
Let $x^0$ be a feasible solution of the \ref{eq:tsplp} such that $x^0_{e_0}=1$ with support $E_0=E\cup \{e_0\}$. 
Let $x$ be $x^0$ restricted to $E$; then $x$ is in the spanning tree polytope of $G=(V,E)$. 
\end{fact}
\begin{proof}
For any set $S\subseteq V$ such that $u_0,v_0\notin S$, $x(E(S))=\frac{2|S|-x^0(\delta(S))}{2}\leq |S|-1$.
If $u_0\in S, v_0\notin S$, then
$x(E(S)) = \frac{2|S|-1 - (x^0(\delta(S)) -1 )}{2}\leq |S|-1$.
Finally, if $u_0,v_0\in S$, then 
$x(E(S)) = \frac{2|S|-2 - x^0(\delta(S))}{2} \leq |S|-2$.
The claim follows because $x(E)=x^0(E_0)-1=n-1$.
 \end{proof}

Since $c(e_0)=0$, the following fact is immediate.
\begin{fact} \label{fact:expcostT}Let $G=(V,E,x)$ where  $x$ is in the spanning tree polytope. Let $\mu$ be any distribution of spanning trees with marginals $x$, then $\EE{T\sim\mu}{c(T \cup e_{0})}=c(x)$.
 \end{fact}
 
To bound the cost of the min-cost matching on the set $O$ of odd degree vertices of the tree $T$, we use the following characterization of the $O$-join polytope\footnote{The standard name for this is the $T$-join polytope. Because we reserve $T$ to represent our tree, we call this the $O$-join polytope, where $O$ represents the set of odd vertices in the tree.} due to Edmonds and Johnson \cite{EJ73}.
\begin{proposition}
\label{prop:tjoin}
For any graph $G=(V,E)$, cost function $c: E \to \R_+$, and a set $O\subseteq V$ with an even number of vertices,  the minimum weight of an $O$-join equals the optimum value of the following integral linear program.
\begin{equation}
\begin{aligned}
\min \hspace{4ex} & \cost(y) \\
\st \hspace{3ex} & y(\delta(S)) \geq 1 & \forall S \subseteq V, |S\cap  O| \text{ odd}\\
& y_e \geq 0 & \forall e\in E
\end{aligned}
\label{eq:tjoinlp}
\end{equation}
\end{proposition}

\begin{definition}[Satisfied cuts]\label{def:satisfiedcuts}
\hypertarget{tar:satisfy}{For a set $S\subseteq V$ such that $u_0,v_0\notin S$ and a spanning tree $T\subseteq E$ we say a vector $y:E\to\R_{\geq 0}$ 	satisfies $S$ if one of the following holds:
\begin{itemize}
\item $\delta(S)_T$ is even, or
\item $y(\delta(S))\geq 1$.	
\end{itemize}}
\end{definition}
To analyze our algorithm, we will see that the main challenge is to construct a (random) vector $y$ that satisfies all  cuts and $\E{c(y)}\leq (1/2-\eps)OPT$.

\subsection{Structure of Near Minimum Cuts}\label{sec:structure-NMC}


\begin{lemma}[\cite{OSS11}]\label{lem:cutdecrement}
For $G=(V,E,x)$, let $A,B\subsetneq V$ be two crossing $\eps_A, \eps_B$ near min cuts respectively. Then,
$A\cap B, A\cup B, A\smallsetminus B, B\smallsetminus A$ are $\eps_A+\eps_B$ near min cuts.
\end{lemma}
\begin{proof}
We prove the lemma only for $A\cap B$; the rest of the cases can be proved similarly.
By submodularity,
$$ x(\delta(A\cap B)) + x(\delta(A\cup B)) \leq x(\delta(A)) + x(\delta(B)) \leq 4+\eps_A+\eps_B.$$
Since $x(\delta(A\cup B))\geq 2$, we have $x(\delta(A\cap B))\leq 2+\epsilon_A+\eps_B$, as desired.
\end{proof}

The following lemma is proved in \cite{Ben97}:
\begin{lemma}[{\cite[Lem 5.3.5]{Ben97}}]
\label{lem:nmcuts_largeedges}
For $G=(V,E,x)$, let $A,B\subsetneq V$ be two crossing $\eps$-near minimum cuts. 
Then, $$x(E(A\cap B, A\setminus B)),x(E(A\cap B, B\setminus A)), x(E(\overline{A\cup B}, A\setminus B)), x(E( \overline{A\cup B}, B\setminus A)) \geq (1-\epsilon/2).$$
\end{lemma}
%
\begin{lemma}
\label{lem:shared-edges}
For $G=(V,E,x)$, let $A,B\subsetneq V$ be two $\eps$ near min cuts  such that $A \subsetneq B$. Then 
$$x(\delta(A) \cap \delta(B)) = x(E(A,\overline{B}))\le 1 + \eps, \text{ and }$$
$$x(\delta(A)\smallsetminus \delta(B))\geq 1-\eps/2. $$
\end{lemma}
\begin{proof}
Notice
\begin{align*}&2+\epsilon \ge x(\delta(A)) = x(E(A,B \smallsetminus A)) + x(E(A,\overline{B}))\\
&2+\epsilon \ge x(\delta(B)) = x(E(B \smallsetminus A,\overline{B})) + x(E(A,\overline{B}))
\end{align*}
Summing these up, we get
$$2x(E(A,\overline{B})) + x(E(A,B \smallsetminus A)) + x(E(B \smallsetminus A, \overline{B})) = 2x(E(A,\overline{B}))+x(\delta(B\smallsetminus A)) \le 4+2\eps.$$
Since $B \smallsetminus A$ is non-empty,
	$x(\delta(B\smallsetminus A)) \ge 2$,
which implies the first inequality.
To see the second one, let $C=B\smallsetminus A$ and note
$$ 4\leq x(\delta(A))+x(\delta(C)) = 2 x(E(A,C)) + x(\delta(B))\leq 2 x(E(A,C))+ 2+\eps$$
which implies $x(E(A,C))\geq 1-\eps/2$.
\end{proof}




\def\ber{\textup{Ber}}
\def\bin{\textup{Bin}}
\def\Poi{\textup{Poi}}

\def\ber{\textup{Ber}}
\def\bin{\textup{Bin}}
\def\Poi{\textup{Poi}}

\subsection{Strongly Rayleigh Distributions and $\lambda$-uniform Spanning Tree Distributions}
\label{sec:SRdistns}
Let $\cB_E$ be the set of all probability measures on the Boolean algebra $2^E$. 
Let $\mu\in\cB_E$. The generating polynomial $g_\mu: \R[\{z_{e}\}_{e\in E}]$ of $\mu$ is defined as follows:
$$ g_\mu(z):=\sum_S \mu(S) \prod_{e\in S} z_e.$$
\hypertarget{tar:SR}{We say $\mu$ is a strongly Rayleigh distribution if $g_\mu\neq 0$ over all $\{y_e\}_{e\in E} \in \mathbb{C}^E$ where $\text{Im}(z_e)>0$ for all $e\in E$. We say $\mu$ is $d$-homogenous if for any $\lambda \in \R$, $g_\mu(\lambda {\bf z}) = \lambda^d g_\mu({\bf z})$.}
Strongly Rayleigh (SR) distributions were defined in \cite{BBL09} where it was shown any $\lambda$-uniform spanning tree distribution is strongly Rayleigh. In this subsection we recall several properties of SR distributions proved in \cite{BBL09,OSS11} which will be useful to us.

\paragraph{Closure Operations of SR Distributions.}
SR distributions are closed under the following operations.

\begin{itemize}
	\item {\bf Projection.} For any $\mu\in \cB_E$, and any $F\subseteq E$, the projection of $\mu$ onto $F$ is the measure $\mu_F$
	where for any $A\subseteq F$,
	$$ \mu_F(A)=\sum_{S: S\cap F=A} \mu(S).$$
	\item {\bf Conditioning.} For any $e\in E$, $\{\mu | e \text{ out}\}$ and $\{\mu | e \text{ in}\}$.
	\item {\bf Truncation.} For any integer $k\geq 0$ and $\mu\in \cB_E$, truncation of $\mu$ to $k$, is the measure $\mu_k$ where for any $A\subseteq E$,
$$ \mu_k(A)=\begin{cases} \frac{\mu(A)}{\sum_{S:|S|=k} \mu(S)} & \text{if } |A|=k\\ 0 & \text{otherwise.}\end{cases}$$
	\item {\bf Product.} For any two disjoint sets $E,F$, and $\mu_E\in \cB_E,\mu_F\in\cB_F$ the product measure $\mu_{E\times F}$ is the measure where for any $A\subseteq E,B\subseteq F$, $\mu_{E\times F}(A\cup B)=\mu_E(A)\mu_F(B)$.
\end{itemize}
Throughout this paper we will repeatedly apply the above operations. 
We  remark that SR distributions are {\em not} necessarily closed under truncation of a subset, i.e., if we require exactly $k$ elements from $F\subsetneq E$.

Since $\lambda$-uniform spanning tree distributions are special classes of SR distributions, if we perform any of the above operations on a $\lambda$-uniform spanning tree distribution $\mu$ we get another SR distribution. Below, we see that by performing  the following particular operations we still have a $\lambda$-uniform spanning tree distribution (perhaps with a different $\lambda$). 
\paragraph{Closure Operations of $\lambda$-uniform Spanning Tree Distributions}
For $G=(V,E)$, a spanning tree distribution $\mu\in \cB_E$, and $T \sim \mu$, we have:
\begin{itemize}
	\item {\bf Conditioning}. For any $e\in E$, $\{\mu \mid e \not\in T\}, \{\mu \mid e \in T\}$. 
	\item {\bf Tree Conditioning}. For $S\subseteq V$, $\{\mu \mid |E(S) \cap T| = |S|-1\}$, i.e. $T$ restricted to $S$ is a tree. We will often just write $S\text{ is a tree}$ to denote such an event.
\end{itemize}
Note that arbitrary spanning tree distributions are not necessarily closed under truncation and projection. 
We remark that SR measures are also closed under an analogue of tree conditioning, i.e., for a set $F\subseteq E$, let $k=\max_{S\in\text{supp }\mu} |S\cap F|$. Then, $\{\mu \mid |S\cap F|=k\}$ is SR. But if $\mu$ is a spanning tree distribution we get an extra {\em independence} property. The following independence is crucial to several of our proofs.
\begin{fact}
\label{fact:treeIndep}
For a graph $G = (V,E)$, and a vector $\lambda (G): E\rightarrow \R _{\ge 0}$, let $\mu_{\lambda(G)}$ be the corresponding $\lambda$-uniform spanning tree distribution.
Then for any $S\subsetneq V$, 
	$$\{\mu_{\lambda(G)} \mid S\text{ is a tree}\} = \mu_{\lambda(G[S])} \times  \mu_{\lambda(G/S)}.$$ 
\end{fact}
\begin{proof}
	Intuitively, this holds because in the max entropy distribution (recall a $\lambda$-uniform distribution maximizes entropy subject to matching the marginals of $x$), conditioned on $S$ being a tree, any tree chosen inside $S$ can be composed with any tree chosen on $G/S$ to obtain a spanning tree on $G$. So, to maximize the entropy these trees should be chosen independently. 
	More formally for any $T_1 \in G[S]$ and $T_2 \in G/S$, 
	\begin{align*}
	\P{T=T_1\cup T_2 \mid S\text{ is a tree}} &= \frac{\lambda^{T_1}\lambda^{T_2}}{\sum_{T'_1 \in G[S],T'_2 \in G/S} \lambda^{T'_1}\lambda^{T'_2}} \\ 
	&= \frac{\lambda^{T_1}}{\sum_{T'_1 \in G[S]} \lambda^{T'_1}} \cdot \frac{\lambda^{T_2}}{\sum_{T'_2 \in G/S} \lambda^{T'_2}} \\
	&= \PP{T'_1 \sim G[S]}{T'_1 = T_1}\PP{T'_2 \sim G/S}{T'_2 = T_2},
	\end{align*}
giving independence.
\end{proof}

\paragraph{Negative Dependence Properties.}
An {\em upward event}, ${\cal A}$, on $2^E$ is a collection of subsets of $E$ that is closed  under upward containment, i.e. if $A\in{\cal A}$ and $A\subseteq B\subseteq E$, then $B\in{\cal A}$. Similarly, a {\em downward event} is closed  under downward  containment.
An {\em increasing function} $f:2^E\rightarrow \R$, is a function where for any $A\subseteq B\subseteq E$, we have $f(A)\leq f(B)$. We also say $f:2^E\to\R$ is a {\em decreasing function} if $-f$ is an increasing function.
So, an indicator of an upward event is an increasing function.
For example, if $E$ is the set of edges of a graph $G$, then the existence of a Hamiltonian cycle is an increasing function, and the $3$-colorability of $G$ is a decreasing function.

\begin{definition}[Negative Association]
\label{def:negativeassociation}
A measure $\mu \in {\cal B}_E$ is {\em negatively associated} if 
for any increasing functions $f,g: 2^E\to \R$, that depend on {\em disjoint} sets of edges,
$$ \EE{\mu}{f}\cdot \EE{\mu}{g} \geq  \EE{\mu}{f\cdot g} $$
\end{definition}
It is shown in \cite{BBL09,FM92}  that strongly Rayleigh measures
are negatively associated. 

\paragraph{Stochastic Dominance.}
For two measures $\mu,\nu:2^{E}\to\R_{\geq 0}$, we say $\mu \preceq \nu$ if there exists a {\em coupling} $\rho: 2^{E}\times 2^{E} \to\R_{\geq 0}$ such that 
\begin{eqnarray*}
	\sum_B \rho(A,B)&=&\mu(A), \forall A\in 2^{E},\\
	\sum_A \rho(A,B)&=& \nu(B), \forall B\in 2^{E},
\end{eqnarray*} 
and for all $A,B$ such that $\rho(A,B)>0$ we have $A\subseteq B$ (coordinate-wise).

\begin{theorem}[\cite{BBL09}]
\label{thm:stochDom}
	If $\mu$ is strongly Rayleigh and $\mu_k,\mu_{k+1}$ are well-defined, then $\mu_k \preceq \mu_{k+1}$.
\end{theorem}
Note that in the above particular case the coupling $\rho$ satisfies the following: For any $A,B\subseteq E$ where $\rho(A,B)>0$, $B\supseteq A$ and $|B\smallsetminus A|=1$, i.e., $B$ has exactly one more element.

Let $\mu$ be a strongly Rayleigh measure on edges of $G$. Recall that for a set $A\subseteq E$, we write $A_T=|A\cap T|$ to denote the random variable indicating the number of edges in $A$ chosen in a random sample $T$ of $\mu$. The following facts immediately follow from the negative association and stochastic dominance properties. We will use these facts repeatedly in this paper. 

\begin{fact}[{\cite[Theorems 4.8, 4.19]{BBL09}}]
\label{fact:updown}
Let $\mu$ be any SR distribution on $E$, then
for any $F\subset E$, and any integer $k$
\begin{enumerate}
\item (Negative Association) If $e\notin F$, then $\PP{\mu}{e \big| F_T \geq k} \leq \PP{\mu}{e}$ and 
$\PP{\mu}{e | F_T \leq k} \geq \PP{\mu}{e} $
\item (Stochastic Dominance) If $e\in F$, then $\PP{\mu}{e | F_T \geq k} \geq \PP{\mu}{e}$ and $\PP{\mu}{e | F_T\leq k} \leq \PP{\mu}{e}$.
\end{enumerate}
\end{fact}

The following fact is a direct consequence of the above, see e.g. Corollary 6.10 of \cite{OSS11}.
\begin{fact}
\label{fact:e1}
Let $\mu$ be a homogenous SR distribution on $E$. Then,
\begin{itemize}
\item (Negative association with homogeneity) For any $A\subseteq E$, and  any $B\subseteq \overline{A}$ 
\begin{equation} \EE{\mu}{ B_T | A_T = 0} \leq \EE{\mu}{B_T} + \EE{\mu}{A_T} 
\label{fact:e0}
\end{equation}	
\item Suppose that $\mu$ is a spanning tree distribution. For  $S\subseteq V$, let $q:=|S|-1-\EE{\mu}{E(S)_T}$.  For any $A\subseteq E(S), B\subseteq \overline{E(S)}$,
\begin{align}
&\EE{\mu}{B_T}  - q \le \EE{\mu}{ B_T | S\text{ is a tree}} \le \EE{\mu}{B_T}\tag{Negative association and homogeneity}\\
 &\EE{\mu}{A_T}
 \leq \EE{\mu}{A_T | S\text{ is a tree}} \leq \EE{\mu}{A_T}+q
\tag{Stochastic dominance and tree}
\end{align}
\end{itemize}
\end{fact}
%

\paragraph{Rank Sequence.}

The {\em rank sequence} of $\mu$ is the sequence
$$\P{|S|=0}, \P{|S|=1}, \ldots,\P{|S|=m},$$
where $S\sim\mu$.
Let $g_\mu({\bf z})$ be the generating polynomial of $\mu$.
The {\em diagonal specialization} of $\mu$ is the univariate polynomial 
$$\bar{g}_\mu(z):=g_\mu(z,z,\dots,z). $$ Observe that $\bar{g}(.)$ is the generating polynomial of the rank sequence of $\mu$. It follows that if $\mu$ is SR then $\bar{g_\mu}$ is real rooted.

It is not hard to see that the rank sequence of $\mu$ corresponds to sum of independent Bernoullis iff $\bar{g_\mu}$ is real rooted. It follows that the rank sequence of an SR distributions has the law of a sum of independent Bernoullis. As a consequence, 
it follows  (see \cite{HLP52,Dor64,BBL09})
that the rank sequence of any strongly Rayleigh measure is  log concave (see below for the definition), unimodal, and its mode differs from the mean by less than 1.
\begin{definition}[Log-concavity {\cite[Definition 2.8]{BBL09}}]
A real sequence $\{a_k\}^m_{k=0}$ is log-concave if $a^2_k \geq a_{k-1}\cdot a_{k+1}$ for all $1 \leq k\leq m-1$, and it is said to have no internal zeros if the indices of its non-zero terms form an interval (of non-negative integers). 
\end{definition}

\subsection{Sum of Bernoullis}
In this section, we collect a number of properties of sums of Bernoulli random variables.

\begin{definition}[Bernoulli Sum Random Variable]\label{def:BS}\hypertarget{tar:BS}{
We say $BS(q)$ is a {\em Bernoulli-Sum} random variable if it has the law of a sum  of independent Bernoulli random variables, say $B_1 + B_2 + \ldots +B_n$  for some $n\ge 1$, with $\E{B_1+ \dots +B_n} = q$. }
\end{definition}

We start with the following theorem of Hoeffding.
\begin{theorem}[{\cite[Corollary 2.1]{Hoe56}}]\label{thm:hoeffding}
Let $g:\{0,1,\dots,n\}\to \R$ and $0\leq q\leq n$ for some integer $n\geq 0$.  Let $B_1,\dots,B_n$ be $n$ independent Bernoulli random variables with success probabilities $p_1,\dots,p_n$, where $\sum_{i=1}^n p_n = q$ that minimizes (or maximizes)
$$ \E{g(B_1+\dots+B_n)}$$
over all such distributions. Then,  $p_1,\dots,p_n\in\{0,x,1\}$ for some $0<x<1$. In particular, if only $m$ of $p_i$'s are nonzero and $\ell$ of $p_i$'s are 1, then the remaining $m-\ell$ are $\frac{q-\ell}{m-\ell}$.
\end{theorem}

\begin{fact} \label{fact:evensum} Let $B_1, \ldots, B_n$ be independent Bernoulli random variables each with expectation $0\le p\le 1$.
Then $$\P{\sum_i B_i \text{ even}} = \frac{1}{2} (1+ (1-2p)^n)$$
\end{fact}
\begin{proof}
Note that
$$
 (p+(1-p))^n = \sum_{k=0}^n p^k (1-p)^{n-k} {n\choose k}\quad\text{ and }\quad
 ((1-p)-p)^n = \sum_{k=0}^n (-p)^k (1-p)^{n-k} {n\choose k}	
$$
Summing them up we get,
$$ 1+ (1-2p)^n = \sum_{0\leq k\leq n, k\text{ even}} 2p^k (1-p)^{n-k} {n\choose k}.$$
\end{proof}

\begin{corollary}\label{cor:bernoullisumeven}
Given a $BS(q)$ random variable with $0 < q \le 1.2$,  then
$$ \P{BS(q) \text{ even}} \leq \frac12(1+e^{-2q})$$
\end{corollary}
\begin{proof}
First, if $q\leq 1$, then by Hoeffding's theorem we can write $BS(q)$ as sum of $n$ Bernoullis with success probability $p=q/n$. If $n=1$, then the statement obviously holds. Otherwise, by the previous fact, we have (for some $n$),
 $$\P{BS(q) \text{ even} } \leq \frac12 (1+(1-2p)^n)) \leq \frac12(1+e^{-2q})$$
 where we used that $|1-2p|\leq e^{-2p}$ for $p\leq 1/2$.
 
So, now assume $q>1$.
Write $BS(q)$ as the sum of $n$ Bernoullis, each with success probabilities $1$ or $p$. First assume we 0have no ones. 
Then, either we only have two non-zero Bernoullis with success probability $q/2$ in which case $\P{BS(q)\text{ even}}\leq 0.6^2 + 0.4^2$ and we are done.
Otherwise, $n\geq 3$ so $p\leq 1/2$ and similar to the previous case we get $\P{BS(q) \text{ even}} \leq \frac12(1+e^{-2q})$.

Finally, if $q>1$ and one of the Bernoullis is always $1$, i.e. $BS(q) = BS(q-1)+1$, then we get
$$\P{BS(q)\text{ even}} = \P{BS(q-1) \text{ odd}} = \frac12(1-(1-2p)^{n-1}) \le 1/2$$
where we used that $p \le 0.5$ (since $q \le 1.2$).
\end{proof}


\begin{lemma}\label{lem:logconcaveexpecation}
Let $p_0,\dots,p_n$ be a log-concave sequence. If for some $i$, $\gamma p_i \geq p_{i+1}$ for some $\gamma<1$, then,
\begin{align*} 
& \sum_{j=k}^n p_j \leq \frac{p_k}{1-\gamma}, \quad \forall k\geq i\\
& \sum_{j=i+1}^n  p_j\cdot j  \leq \frac{p_{i+1}}{1-\gamma} \left(i+1 + \frac\gamma{1-\gamma}\right).	
\end{align*}
\end{lemma}
\begin{proof}
Since we have a log-concave sequence we can write
\begin{equation}\label{eq:logconcaveseq} \frac{1}{\gamma} \leq \frac{p_i}{p_{i+1}}  \leq \frac{p_{i+1}}{p_{i+2}} \leq \dots 	
\end{equation}
Since all of the above ratios are at least $1/\gamma$, for all $l\geq 1$ we can write
$$  p_{i+l} \leq \gamma^{l-1} p_{i+1}\leq \gamma^l p_i.$$
Therefore, the first statement is immediate and the second one follows,
$$ \sum_{j=i+1}^ n p_j j \leq \sum_{l=0}^\infty \gamma^l p_{i+1} (i+l+1) = p_{i+1} \left(\frac{i+1}{1-\gamma} + \frac\gamma{(1-\gamma)^2}\right)$$
\end{proof}

\begin{corollary}\label{cor:logconcaveexpecation}
	Let $X$ be a $BS(q)$ random variable such that $\P{X=k}\geq 1-\eps$ for some integer $k\geq 1$, $\eps<1/10$.
	Then, $k(1-\eps)\leq q\leq k(1+\eps)+3\eps$.
\end{corollary}
\begin{proof}
	The left inequality simply follows since $X\geq 0$. 	Since $\P{X=k+1}\leq \eps$, we can apply \cref{lem:logconcaveexpecation}  with $\gamma=\eps/(1-\eps)$ to get 
	$$\E{X | X\geq k+1}\P{X\geq k+1}\leq \frac{\eps(1-\eps)}{1-2\eps}\left(k+1+\frac{\eps}{1-2\eps}\right)$$
	Therefore,
	$$ q=\E{X}\leq k(1-\eps)+\frac{\eps(1-\eps)}{1-2\eps}(k+1+\frac{\eps}{1-2\eps})\leq k(1+\eps)+3\eps$$
	as desired.
\end{proof}

\begin{fact}\label{fact:1-pm^m}
For  integers $k<t$ and $k-1\leq p\leq k$, 
$$\prod_{i=1}^{k-1} (1-i/t) (1-p/t)^{t-k}\geq e^{-p}.$$
\end{fact}
\begin{proof}
We show that the LHS is a decreasing function of $t$. Since $\ln$ is monotone, it is enough to show 
\begin{align*} 0\geq \partial_t  \ln (\text{LHS}) &= \partial_t \left(\sum_{i=1}^{k-1} \ln(1-i/t) + (t-k)\ln(1-p/t)\right)\\
	&= \frac{1}{t^2}\sum_{i=1}^{k-1} \frac{1}{\frac1{i} - \frac{1}{t}} + \ln(1-p/t) + \frac{(t-k)p}{t(t-p)} 
\end{align*}
Using $\sum_{i=1}^{k-2} \frac{1}{t^2/i-t}\leq \int_{0}^{k-1} \frac{dx}{t^2/x - t}=-(k-1)/t -\ln(1-(k-1)/t)$
it is enough to show
\begin{align*}
0&\geq 	-\frac{k-1}{t} - \ln (1-\frac{k-1}{t}) + \ln(1-p/t) + \frac{(t-k)p}{t(t-p)} + \frac{1}{t^2(\frac1{k-1} - \frac1t)}\\
&= \ln\frac{t-p}{t-k+1} + \frac{p-k}{t-p}+\frac{1}{t}+\frac{k-1}{t(t-k+1)}
\end{align*}
Rearranging, it is equivalent to show
\begin{align*}
\ln (1+\frac{p-k+1}{t-p})	\geq \frac{p-k}{t-p} + \frac{1}{t-k+1}
\end{align*}
Since $p>k-1$, using taylor series of $\ln$, to prove the above it is enough to show
$$ \frac{p-k+1}{t-p} - \frac{(p-k+1)^2}{2(t-p)^2} \geq \frac{p-k}{t-p} + \frac{1}{t-k+1}.$$
This is equivalent to show
$$ \frac{p-k+1}{(t-p)(t-k+1)} \geq \frac{(p-k+1)^2}{2(t-p)^2} \Leftrightarrow \frac{1}{t-k+1} \geq \frac{p-k+1}{2(t-p)} $$
Finally the latter holds because $(t-k+1)(p-k+1)\leq (t-k+1)\leq 2(t-p)$ where we use $t\geq k+1$ and $p\leq k$.
\end{proof}

Let $\Poi(p,k) = e^{-p}p^k/k!$  be the probability that a Poisson random variable with rate $p$ is exactly $k$; similarly, define $\Poi(p,\leq k),\Poi(p,\geq k)$ as the probability that a Poisson with rate $p$ is at most $k$ or at least $k$.
\begin{lemma}
\label{thm:rayleigh_expectconstprob}
Let $X$ be a Bernoulli sum $BS(p)$ for some $n$. 
 For any integer $k\geq 0$ such that $k-1<p<k+1$, the following holds true
$$\P{X=k} \geq \min_{0\leq\ell \leq p,k} \Poi(p-\ell,k-\ell) \left(1-\frac{p-\ell}{k-\ell+1}\right)^{(p-k)_+}$$
where the minimum is over all nonnegative integers $\ell\leq p,k$, and for $z\in \R$, $z_+=\max\{z,0\}$. 
\end{lemma}
\begin{proof}
Let $X = B_1 + \dots + B_n$ where $B_i$ is a Bernoulli. 
Applying Hoeffding's theorem, if $\ell$ of them have success probability 1, it suffices to prove a lower bound of  $\Poi(p-\ell, k-\ell)(1-\frac{p-\ell}{k-\ell+1})^{(p-k)_+}$. Since without loss of generality none have success probability 1, it follows that each has success probability $p/n$. If $k\geq p$,
$$\P{X=k} = {n\choose k}\left(\frac{p}{n}\right)^k (1-p/n)^{n-k} = \prod_{i=1}^{k-1} (1-i/n)\frac{p^k}{k!} (1-p/n)^{n-k}\geq \frac{p^k}{k!}e^{-p} = \Poi(p,k),$$
where in the  inequality we used \cref{fact:1-pm^m} (also note if $n=k$ the inequality follows from Stirling's formula and that $p\geq k-1$).
If $k<p<k+1$, then as above
\begin{align*}
\P{X=k} &= \prod_{i=1}^{k-1} (1-i/n)\frac{p^k}{k!} (1-p/n)^{n-p} (1-p/n)^{p-k} \\ &\underset{p \ge k}{\ge} \prod_{i=1}^{k-1} (1-i/n)\frac{p^k}{k!} (1-p/n)^{n-k} (1-p/n)^{p-k} \ge  \Poi(p,k)(1-p/n)^{p-k},
\end{align*}
where we used \cref{fact:1-pm^m} in the last inequality.
\end{proof}
Note that if we further know $X\geq a$ with probability 1 we can restrict $\ell$ in the statement to be in the interval $[a,\min(p,k)]$.

\begin{lemma}\label{lem:SR>=}
Let $X$ be a Bernoulli sum $BS(p)$, where for some integer $k=\lceil p\rceil$,
Then,
$$\P{X\geq k} \geq \min_{0\leq \ell\leq p} \Poi(p-\ell,\geq k-\ell)$$
where the minimum is over all non-negative integers $\ell\leq p$.
\end{lemma}
\begin{proof}
%
Suppose that  $X$ is a $BS(p)$ with $n$ Bernoullis with probabilities $p_1, \dots, p_n$. If $p-1 < k-1 < p$, by  \cite[Thm 4, (25)]{Hoe56},
\begin{equation}
\label{eq:Hoeff2}	
 \P{X\leq k-1} \leq  \max_{0\leq\ell<p} \sum_{i=0}^{k-1-\ell} {{n-\ell} \choose i} q^i(1-q)^{n-\ell-i}
\end{equation}
where $q=\frac{p-\ell}{n-\ell}$.

If $Y$ is a $BS(p)$ with $m > n$ Bernoullis
with probabilities $q_1, \ldots, q_m$, the same upper bound applies of course, with $m$ replacing $n$.
Also, note that
 $$\max_{p_1\ldots p_n} \P{ X \le k-1} \le \max_{q_1, \ldots, q_m} \P{ Y \le k-1}$$ since it is always possible to set $q_i = p_i$ for $i \le n$ and $q_j = 0$ for $j > n$.

Therefore, the upper bound in \eqref{eq:Hoeff2} obtained by taking the limit as $n$ goes to infinity applies, from which it follows that  
$$ \P{X\leq k-1} \leq  \max_{0\leq\ell<p} \sum_{i=0}^{k-1-\ell} \Poi(p-\ell, i)$$
and therefore
$$\P{X\geq k} \geq  \min_{0\leq\ell<p}  \Poi(p-\ell, \ge k-\ell). $$ 
\end{proof}

\subsection{Random Spanning Trees}
\begin{lemma}\label{lem:treeconditioning}
Let $G=(V,E,x)$, and let $\mu$ be any distribution over spanning trees with marginals $x$. 
For any  $\eps$-near min cut $S\subseteq V$ 
(such that none of the endpoints of $e_0=(u_0,v_0)$ are in $S$), we have
$$\PP{T\sim\mu}{T\cap E(S)\text{ is tree}} \ge 1- \eps/2.$$ 
Moreover, if $\mu$ is a max-entropy distribution with marginals $x$, then for any set of edges $A\subseteq E(S)$ and $B\subseteq E\smallsetminus E(S)$, 
$$ \E{A_T}\leq \E{A_T|S\text{ is tree}}\leq \E{A_T}+\eps/2, \E{B_T}-\eps/2 \leq \E{B_T|S\text{ is tree}}\leq \E{B_T}.$$
\end{lemma}
\begin{proof}
First, observe that
$$\E{E(S)_T}= x(E(S)) \geq  \frac{2|S|-x(\delta(S))}{2} \geq |S|-1 -\eps/2,$$ 
where we used that since $u_0,v_0\notin S$, and that for any $v\in S$, $\E{\delta(v)_T)} = x(\delta(v))=2$.

Let $p_S=\P{S\text{ is tree}}$.
Then, we must have
$$ |S|-1 - (1-p_S) = p_S(|S|-1) + (1-p_S)(|S|-2)\ge \E{E(S)_T} \geq |S|-1 - \eps/2.$$
Therefore, $p_S\geq 1-\eps/2$.

The second part of the claim follows from \cref{fact:e1}.
\end{proof}
\begin{corollary}\label{lem:treeoneedge}
Let  $A,B\subseteq V$ be disjoint sets such that  $A,B,A\cup B$ are $\eps_A,\eps_B,\eps_{A\cup B}$-near minimum cuts w.r.t., $x$ respectively, where none of them  contain endpoints of $e_0$.  Then for any distribution $\mu$ of spanning trees on $E$ with marginals $x$,
$$\PP{T\sim \mu}{E(A,B)_T=1}\geq 1-(\eps_A+\eps_B+\eps_{A\cup B})/2.$$	
\end{corollary}
\begin{proof}
By the union bound, with probability at least $1-(\eps_A+\eps_B+\eps_{A\cup B})/2$, $A,B,$ and $A\cup B$ are trees. 
But this implies that we must have exactly one edge between $A,B$.
\end{proof}

The following simple fact also holds by the union bound.
\begin{fact}\label{fact:0edgerandomspanningtree}
Let $G=(V,E,x)$ and let $\mu$ be a distribution over spanning trees with marginals $x$. For any set $A\subseteq E$	, we have
$$ \PP{T\sim\mu}{T\cap A=\emptyset} \geq 1-x(A).$$
\end{fact}


\begin{lemma}\label{lem:wpath}
Let $G=(V,E,x)$, and let $\mu$ be a $\lambda$-uniform random spanning tree distribution with marginals $x$.
For any edge $e=(u,v)$ and any vertex $w\neq u,v$ we have
 $$ \E{W_T | e\not\in T} \leq \E{W_T} + \P{w\in P_{u,v} | e \not\in T}\cdot\P{e\in T} ,$$
 where $W_T=|T\cap \delta(w)|$ and for a spanning tree $T$ and vertices $u,v\in V$, $P_{u,v}(T)$ is the set of vertices on the path from $u$ to $v$ in $T$.
\end{lemma}
\begin{proof}
Define $E'=E\smallsetminus \{e\}$.
Let $\mu' = \mu|_{E'}$ be $\mu$ projected on all edges except $e$. Define $\mu_{in} = \mu'_{n-2}$ (corresponding to $e$ in the tree) and $\mu_{out}=\mu'_{n-1}$ (corresponding to $e$ out of the tree). Observe that any tree $T$ has positive measure  in exactly one of these distributions.

By \cref{thm:stochDom}, $\mu_{in} \preceq \mu_{out}$ so there exists a coupling $\rho:2^{E'}\times 2^{E'}$ between them such that for any $T_{in}, T_{out}$ such that $\rho(T_{in}, T_{out})>0$, the tree $T_{out}$ has exactly one more edge than $T_{in}$. Also, observe that $T_{out}$ is always a spanning tree whereas $T_{in}\cup\{e\}$ is a spanning tree. The added edge (i.e., the edge in $T_{out}\smallsetminus T_{in}$) is always along the unique path from $u$ to $v$ in $T_{out}$.

For intuition for the rest of the proof, observe that if $w$ is not on the path from $u$ to $v$ in $T_{out}$, then the same set of edges is incident to $w$ in both $T_{in}$ and $T_{out}$. So, if $w$ is almost never on the path from $u$ to $v$, the distribution of $W_T$ is almost independent of $e$. On the other hand, whenever $w$ is on the path from $u$ to $v$, then in the worst case, we may replace $e$ with one of the edges incident to $w$, so conditioned on $e$ out, $W_T$ increases by at most the probability that $e$ is in the tree.

Say $x_e$ is the marginal of $e$. Then, 
\begin{eqnarray}
	\E{W_T} &=& \E{W_T | e\notin T} (1-x_e) + \E{W_T | e\in T} x_e\nonumber \\
	&=& \sum_{T_{in},T_{out}} \rho(T_{in},T_{out}) W_{o} (1-x_e) + \sum_{T_{in},T_{out}} \rho(T_{in},T_{out}) W_{i} x_e\nonumber \\
	&=& \sum_{T_{in},T_{out}} \rho(T_{in},T_{out}) ((1-x_e)W_o + x_e W_i),\label{eq:EZT}
\end{eqnarray}
where we write $W_i$/$W_o$ instead of $W_{T_{in}}$/$W_{T_{out}}$

\begin{eqnarray*}
\E{W_T | e\notin T} &=& \sum_{T_{in},T_{out}} \rho(T_{in},T_{out}) W_o	\\
&=& \sum_{T_{in},T_{out}: w\in P_{u,v}(T_{out})} \rho(T_{in},T_{out}) W_o + \sum_{T_{in},T_{out}: w\notin P_{u,v}(T_{out})} \rho(T_{in},T_{out}) W_o\\
&\leq& \sum_{T_{in},T_{out}: w\in P_{u,v}(T_{out})} \rho(T_{in},T_{out}) (x_e(W_i+1)+ (1-x_e) W_o)  \\
&&\quad\quad\quad +\sum_{T_{in},T_{out}: w\notin P_{u,v}(T_{out})} \rho(T_{in},T_{out})(x_e W_i+(1-x_e)W_o)\\
&=& \E{W_T} + \sum_{T_{in},T_{out}: w\in P_{u,v}(T_{out})}\rho(T_{in},T_{out}) x_e\\
&=& \E{W_T} + \sum_{T_{out}: w\in P_{u,v}(T_{out})} \mu_{out}(T_{out}) x_e\\
&=& \E{W_T} + \P{w\in P_{u,v} | e\text{ out}} \cdot \P{e\text{ in}}
\end{eqnarray*}
where in the inequality we used the following: When $w\notin P_{u,v}(T_{out})$ we have $W_i=W_o$ and when $w\in P_{u,v}(T_{out})$ we have $W_o\leq W_i+1$. Finally, in the third to last equality we used \eqref{eq:EZT}.
\end{proof}

\begin{figure}[htb]\centering
\begin{tikzpicture}[scale=0.8]
\tikzstyle{every node}=[draw,circle]
\foreach \a/\x/\l in {u/-3/U, w/3/W}{
\path  (\x,0) node  (\a) {$\a$};
\path (\a)+(-0.7,0.5) node [color=red,draw=none] (){$\l$};
\foreach \xx in {0,...,2}{
\draw  (\a) -- ++(60+\xx*30: 1.5);}
\draw [dashed,color=red,line width=1.2] (\a)+(45:1) arc (45:135:1);
}
\path  (0,0) node  (v) {$v$};
\draw  (v) -- +(60:1.5) (v) -- +(90:1.5) (v) -- +(120:1.5);
\path (u) edge node [draw=none,below] {$e$} (v);
\path  (w) edge node [draw=none,below] {$f$} (v);
\end{tikzpicture}
\caption{Setting of \cref{lem:crossingcorrelation}} 
\label{fig:crossingcorrelation}
\end{figure}

\begin{lemma}\label{lem:crossingcorrelation}
Let \hyperlink{tar:G=(V,E,x)}{$G=(V,E,x)$}, and let $\mu$ be a $\lambda$-uniform spanning tree distribution with marginals $x$. 
For any pair of edges  $e=(u,v), f=(v,w)$ such that $|\P{e}-1/2|,|\P{f}-1/2|<\eps$ (see \cref{fig:crossingcorrelation}), if $\eps<1/1000$, then
$$\E{W_T | e\not\in T}+\E{U_T | f\not\in T}\leq \E{W_T+U_T}+0.81,$$
where $U=\delta(u)_{-e}$ and $W=\delta(w)_{-f}$.
\end{lemma}
\begin{proof}
All probabilistic statements are with respect to $\nu$ so we drop the subscript.
First, by \cref{lem:wpath}, and negative association we can write,
	\begin{align*}
		\E{W_T |e \not\in T} &\le \E{W_T} + \P{w \in P_{u,v}|e \not\in T}\P{e\in T}\\
		&\le \E{W_T} + \P{w\in P_{u,v}\wedge e\notin T} + 2\eps
	\end{align*}
	Note that the lemma only implies  $\E{\delta(w)_T|e\notin T}\leq \E{\delta(w)_T}+\P{w\in P_{u,v}| e\notin T }\P{e\in T}$. To derive the first inequality we also exploit negative association which asserts that the marginal of every edge only goes up under $e\notin T$, so any subset of $\delta(w)$ (in particular $W$) also goes up by at most $\P{e\notin T\wedge w\in P_{u,v}}$.
	Also, the second inequality uses $\P{e\in T} \leq \P{e\notin T} + 2\eps$.
	Using a similar inequality for $U_T$, to prove the lemma it is enough to show that
	\begin{align*}
		\P{w \in P_{u,v} \wedge e \not\in T}
		+\P{u \in P_{v,w} \wedge f \not\in T} \leq 0.806
	\end{align*}
	or that when this inequality fails, a different argument yields the lemma. 
	
	The main observation is that  in any tree it cannot be that both $u$ is on the $v-w$ path and $w$ is on the $u-v$ path. Therefore
	$$\P{u \in P_{v,w} \mid e,f \not\in T} + \P{w \in P_{u,v} \mid e,f \not\in T} \le 1$$
	So, we have
	\begin{align*}
		&\P{e \not\in T \wedge w \in P_{u,v}}+\P{f \not\in T \wedge u \in P_{v,w}} \\
		&\leq 	\P{e,f \not\in T \wedge w \in P_{u,v}}+\P{e\notin T,f\in T}+\PP{\nu}{e,f \not\in T \wedge u \in P_{v,w}}+\P{f\notin T,e\in T}\\
		&\leq  \P{e,f\notin T} + \P{e\notin T,f\in T} + \P{f\notin T, e\in T}\\
		&= 1-\P{e,f\in T}.
	\end{align*}
	It remains to upper bound the RHS. Let $\alpha=\P{f \in T| e\notin T}$. Observe that
	$$ \P{e,f\in T} = \P{f\in T} - \P{f\in T,e\notin T} \geq 1/2-\eps - (1/2+\eps)\alpha.$$
	If $\alpha\leq 0.6$, then $\P{e,f\in T} \geq 0.198$ (using $\eps<0.001$) and the claim follows.
	Otherwise, $\P{f | e\notin T}\geq 0.6$. Similarly, $\P{e | f\notin T}\geq 0.6$. But, by negative association, 
	$$\E{W_T | e\notin T}\leq\E{W_T}+ \P{e} - (\P{f |e\notin T}-\P{f}) \leq  \E{W_T}+2\eps+0.4\leq \E{W_T}+0.405$$ 
	and similarly,   $\E{U_T | f\notin T}\leq  \E{U_T}+0.405$, so the claim follows.
\end{proof}

%% file: overview.tex

\section{Overview of Proof}


As alluded to earlier, the crux of the proof of \cref{thm:main} is to show that the expected cost of the minimum cost matching on the odd degree vertices of the sampled tree is at most $OPT(1/2-\eps)$. We do this by showing the existence of a cheap feasible $O$-join solution to \eqref{eq:tjoinlp}.

First, recall that if we only wanted to get an $O$-join solution of value at most $OPT/2$, to \hyperlink{tar:satisfy}{satisfy} all cuts, it is enough to set $y_e := x_e/2$ for each edge \cite{Wol80}.
To do better, we want to take advantage of the fact that we only need to satisfy a constraint in the $O$-join for $S$ when $\delta(S)_T$ is odd. Here, we are aided by the fact that the sampled tree is likely to have many even cuts because it is drawn from  a \hyperlink{tar:SR}{Strong Rayleigh distribution}. 

If an edge $e$  is exclusively on even cuts then $y_e$ can be reduced below $x_e/2$. This, more or less, was the approach in \cite{OSS11} for graphic TSP, where it was shown that a constant fraction of LP edges will be exclusively on \textit{even} near min cuts with constant probability.
The difficulty in implementing this approach in the metric case comes from the fact that a high cost edge can be on many cuts and it may be exceedingly unlikely that {\em all} of these cuts will be even simultaneously. Overall, our approach to addressing this is  to start with $y_e:= x_e/2$ and then modify it with a random\footnote{where the randomness comes from the random sampling of the tree}  {\em slack vector} $s:E\to\R$: When  certain special (few) cuts that $e$ is on are even we let $s_e=-x_e \decrease$ (for a carefully chosen constant $\decrease>0$); for other cuts that contain $e$, whenever they are odd, we will increase the slack of other edges on that cut to satisfy them. The bulk of our effort is to show that we can do this while guaranteeing that $\E{s_e}<-\eps \decrease x_e$ for some $\eps>0$.

By carefully choosing $\decrease$ smaller than $\eta$, we do not need to worry about the reduction breaking a constraint for any cut $S$ such that $x(\delta(S)) > 2(1+ \eta)$. In particular, if we choose $\decrease \le \eta/4.1$, any such cut is always satisfied, even if every edge in $\delta(S)$ is decreased and no edge is increased.

Let OPT be the optimum TSP tour, i.e., a Hamiltonian cycle, with set of edges $E^*$; throughout the paper, we write $e^*$ to denote an edge in $E^*$. 
To bound the expected cost of the $O$-join for a random spanning tree $T\sim\mu_\lambda$, we also construct a random slack vector $s^*:E^*\to \R_{\geq 0}$ such that $(x+OPT)/4 + s + s^*$ is a feasible for \cref{eq:tjoinlp} with probability $1$. In \cref{sec:maintechnical} we explain how to use  $s^*$ to satisfy all but a linear number of near mincuts.
\begin{restatable}[Main Technical Theorem]{theorem}{maintechnical}\label{thm:maintechnical}
Let $x^0$ be a solution of the \ref{eq:tsplp} with support $E_0=E\cup \{e_0\}$, and $x$ be $x^0$ restricted to $E$.
Let $z:= (x+ OPT)/2$, $\eta\leq 10^{-12}$, $\decrease >0$, and let $\mu$ be  the max-entropy distribution with marginals  $x$. 
Also, let $E^*$ denote the support of OPT. 
There are two functions $s: E_0\rightarrow \R$ and $s^*: E^* \rightarrow \R _{\ge 0}$ (as functions of $T\sim\mu$), , such that
\begin{enumerate}[i)]
\item 	For each edge $e \in E$, $s_e \ge -x_e \decrease$. 
\item
For each $\eta$-near-min-cut $S$ of $z$, if $\delta(S)_T$ is odd, then
$  s(\delta(S)) + s^*(\delta(S)) \ge  0.$
\item For every $OPT$ edge $e^*$, $\E{s^*_{e^*}}\leq 218\eta\decrease$ and for every LP edge $e\neq e_0$, $\E{s_e}\leq -\frac{1}{3}x_e \eps_P\decrease$ for $\eps_P = 3.12 \cdot 10^{-16}$ (defined in \eqref{eq:epsP}).
\end{enumerate}
\end{restatable}
 In the next subsection, we explain the main ideas needed to prove this technical theorem. But first, we show how our main theorem follows readily from \cref{thm:maintechnical}.
\begin{proof}[Proof of \cref{thm:main}]
Let $x^0$ be an extreme point solution of the \ref{eq:tsplp}, with support $E_0$ and let $x$ be $x^0$ restricted to $E$. By \cref{fact:sptreepolytope} $x$ is in the spanning tree polytope. 
For $\mu=\mu_{\lambda^*}$ the max entropy distribution with marginals $x$ and $\beta > 0$ a parameter we choose below, let $s,s^*$ be as defined in \cref{thm:maintechnical}.
We will define $y:E_0\to\R_{\geq 0}$ and $y^*:E^*\to\R_{\geq 0}$. 
Let  
$$y_e=\begin{cases}
x_e/4+s_e & \text{if } e\in E\\
\infty & \text{if } e=e_0
\end{cases}
$$	
we also let $y^*_{e^*}=1/4+s^*_{e^*}$ for any edge $e^*\in E^*$.
We will show that $y + y^*$ is a feasible solution\footnote{Recall that we merely need to prove the {\em existence} of  a cheap O-join solution. The actual optimal O-join solution can be found in polynomial time.}
to \eqref{eq:tjoinlp}.
First, observe that for any $S$ where $e_0\in\delta(S)$, we have $y(\delta(S))+y^*(\delta(S))\geq 1$. Otherwise, we assume $u_0,v_0\notin S$.
If $S$ is an $\eta$-near min cut  w.r.t., $z$ and $\delta(S)_T$ is odd, then by property (ii) of \cref{thm:maintechnical}, we have
$$ y(\delta(S))+y^*(\delta(S)) = \frac{z(\delta(S))}{2}+ s(\delta(S))+s^*(\delta(S))\geq 1.$$
On the other hand, if $S$  is not an $\eta$-near min cut (w.r.t., $z$). 
\begin{align*} y(\delta(S)) + y^*(\delta(S)) &\geq  \frac{z(\delta(S))}{2} - \decrease x(\delta(S))\\
&\geq \frac{z(\delta(S))}{2} - \decrease 2(z(\delta(S))-1)	\\
&\geq z(\delta(S)) (1/2-2\decrease)+ 2\decrease\geq (2+\eta) (1/2-2\beta)+2\beta
\end{align*}
where in the first inequality we used property (i) of \cref{thm:maintechnical} which says that $s_e\geq x_e \beta$ with probability 1 for all LP edges and that $s^*_{e^*}\geq 0$ with probability 1.
In the second inequality we used that $z=(x+OPT)/2$, so, since $OPT\geq 2$ across any cut, $x(\delta(S))\leq 2(z(\delta(S))-1)$. Finally, if we choose \begin{align}\label{def:decrease-param}\decrease = \eta/4.1\end{align} then the righthand side is at least 1, so $y+y^*$ is a feasible $O$-join solution.

Finally, using $c(e_0)=0$ and part (iii) of \cref{thm:maintechnical},
\begin{align*}
\E{c(y)+c(y^*)}&= OPT/4 + c(x)/4 +  	\E{c(s)+c(s^*)}\\
&\leq OPT/4+c(x)/4 +218\eta \decrease OPT - \frac{1}{3}\eps_P \decrease c(x) \leq (1/2-\frac{1}{6}\eps_P\decrease) OPT
\end{align*}
choosing $\eta$ such that 
\begin{equation}\label{eq:whatiseta}
	218\eta= \frac{1}{6}\eps_P
\end{equation}
and using $c(x) \leq OPT$.

Now, we are ready to bound the approximation factor of our algorithm.
First, since $x^0$ is an  extreme point solution of the \ref{eq:tsplp}, $\min_{e\in E_0} x^0_e\geq \frac{1}{n!}$. So, by \cref{thm:maxentropycomp}, in polynomial time we can find $\lambda:E\to\R_{\geq 0}$ such that for any $e\in E$, $\PP{\mu_\lambda}{e}\leq x_e(1+\delta)$ for some $\delta$ that we fix later. It follows that
$$ \sum_{e\in E} |\PP{\mu}{e} - \PP{\mu_{\lambda}}{e}| \leq n\delta.$$
By stability of maximum entropy distributions (see  \cite[Thm 4]{SV19} and references therein), we have that $\norm{\mu-\mu_\lambda}_1\leq O(n^4\delta)=:q$. Therefore, {for some $\delta \ll n^{-4}$ we get} $\norm{\mu-\mu_{\lambda}}_1=q\leq \frac{\eps_P\decrease}{100}$. That means that 
$$ \EE{T\sim\mu_{\lambda}}{\text{min cost matching}} \leq \EE{T\sim\mu}{c(y)+c(y^*)}+  q (OPT/2) \leq \left(\frac12-\frac{1}{6}\eps_P\decrease
 + \frac{\eps_P \decrease }{100}\right)OPT, $$
where we used that for any spanning tree the cost of the minimum cost matching on odd degree vertices is at most $OPT/2$.
Finally, since $\EE{T\sim\mu_\lambda}{c(T)}\leq OPT(1+\delta)$, $\eps_P=3.12\cdot 10^{-16}$, and $\beta = \eta/4.1 = \eps_p/5362.8$ (from \eqref{eq:whatiseta}) we get a $3/2- 3\cdot 10^{-36}$ approximation algorithm for TSP.
\end{proof}

\subsection{Ideas underlying proof of \cref{thm:maintechnical}}\label{sec:maintechnical}
\begin{figure}[htb]
	\centering
	\begin{tikzpicture}
		\node [draw,circle] at (0,0) (a) {$a$};
		\node [draw,circle] at (0,2) (b) {$b$} edge (a);
		\node [draw,circle] at (2,0) (c) {$c$} edge [dashed] (a) edge [color=green,line width=6pt,opacity=0.2] (a) ;
		\node [draw,circle] at (2,2) (d) {$d$} edge (c) edge [dashed] (b) edge [line width=6pt,opacity=0.2,color=blue] (b);
		\draw [dashed] (a) -- +(-1,3)  (b) -- +(0,1.5) (d) -- +(0,1.5) (c) --  +(1,3);
		\draw [color=red,line width=6pt,opacity=0.2] (a) -- +(-1,3);
		\draw [color=orange,line width=6pt,opacity=0.2] (c) --  +(1,3);
		\draw [dotted,line width=1pt, color=red] (0,1) ellipse (0.75 and 1.75)
			(2,1) ellipse (0.75 and 1.75);
		\draw[line width=1pt,dotted,color=blue]	(1,1) ellipse (2.5 and 2.1);
		\node [color=red]at (-0.95,0.25) () {$u_1$};
		\node [color=red] at (2.95,0.25) () {$u_2$};
		\node [color=blue] at (4,1) () {$u_3$};
		
	\end{tikzpicture}
	\quad \quad \quad
	\begin{tikzpicture}
		\node [circle,draw] at (0,0) (a) {$a$};
		\node [circle,draw] at (2,0) (b) {$b$} edge (a);
		\node[circle,draw]  at (1,1) (u1) {$u_1$} edge (a) edge [color=green,opacity=0.2,line width=6pt,bend left=10] (a) edge [opacity=0.2,line width=6pt,color=red,bend right=10] (a) edge [line width=1pt](b) edge [color=blue,opacity=0.2,line width=6pt](b);
		\node[circle,draw]  at (4,0) (c) {$c$};
		\node [circle,draw] at (6,0) (d) {$d$} edge (c);
		\node [circle,draw] at (5,1) (u2) {$u_2$} edge (c) edge [color=orange,line width=6pt,opacity=0.2,bend left=10] (c) edge [color=green,opacity=0.2,line width=6pt,bend right=10] (c) edge (d) edge [color=blue,opacity=0.2,line width=6pt] (d) edge (u1) edge [color=blue,opacity=0.2,line width=6pt,bend right=6] (u1) edge [color=green,opacity=0.2,line width=6pt,bend left=6] (u1);
		\node[circle,draw]  at (3,2) (u3) {$u_3$} edge (u1) edge [color=red,opacity=0.2,line width=6pt] (u1) edge (u2) edge [color=orange,line width=6pt,opacity=0.2] (u2) ;
	\end{tikzpicture}
	\caption{An example of part of a hierarchy with three triangles. The graph on the left shows part of a feasible LP solution where dashed (and sometimes colored) edges have fraction $1/2$ and solid edges have fraction 1. The dotted ellipses on the left show the min-cuts $u_1,u_2,u_3$ in the graph. (Each vertex is also a min-cut). On the right is a representation of the corresponding hierarchy. Triangle $u_1$ corresponds to the cut $\{a,b\}$, $u_2$ corresponds to $\{c,d\}$ and $u_3$ corresponds to $\{a,b,c,d\}$. Note that, for example, the edge $(a,c)$, represented in green,  is in $\delta(u_1)$,  $\delta(u_3)$, and inside $u_3$.  For triangle $u_1$, we have $A=\delta(a)\smallsetminus (a,b)$ and $B = \delta(b)\smallsetminus (b,d)$.}
	\label{fig:hierarchywithtriangle}
\end{figure}
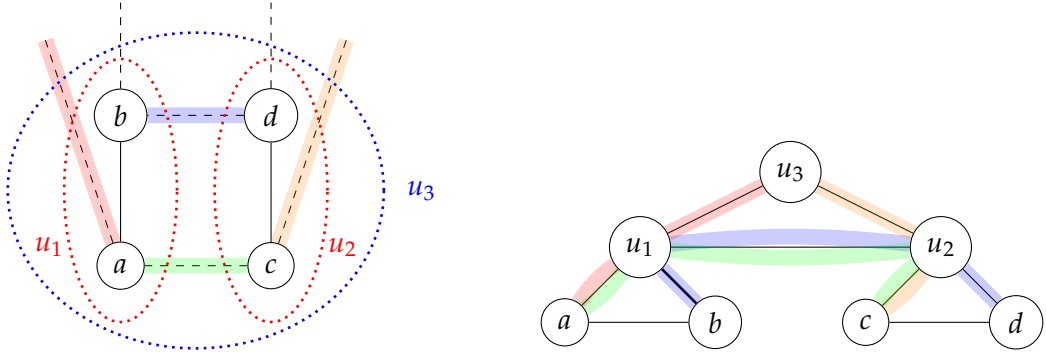

The first step of the proof is to show that it suffices to construct a slack vector $s$ for a ``cactus-like'' structure of near min-cuts that we call a {\em hierarchy}.
Informally, a hierarchy $\cH$ is a laminar family of mincuts\footnote{This is really a family of near-min-cuts, but for the purpose of this overview, assume $\eta =0$}, consisting of two types of cuts: {\em triangle cuts} and {\em degree cuts}. A triangle  $S$ is the union of two min-cuts $X$ and $Y$ in $\cH$ such that $x(E(X,Y)) = 1$. 
See \cref{fig:hierarchywithtriangle} for an example of a hierarchy with three triangles.

We will refer to the set of edges $E(X, \overline{S})$  (resp. $E(Y, \overline{S})$) as $A$ (respectively $B$) for a triangle cut $S$.  In addition, we say a triangle cut $S$ is {\em happy} if $A_T$ and $B_T$ are both odd. All non-triangle cuts are called degree cuts. A degree cut $S$ is {\em happy} if $\delta(S)_T$ is even.

\begin{theorem}[Main Payment Theorem (informal)]\label{thm:paymentinformal}
Let \hyperlink{tar:G=(V,E,x)}{$G=(V,E,x)$} for LP solution $x$ and let $\mu$ be the max-entropy distribution with marginals $x$ and $\decrease > 0$.
Given a hierarchy $\cH$, there is a slack vector $s: E\rightarrow \R$  such that
\begin{enumerate}[i)]
\item 	For each edge $e \in E$, $s_e \ge -x_e \decrease$. 
\item
For each cut $S\in \cH$  if $S$ is not happy, then $s(\delta(S))  \ge  0.$
\item For every LP edge $e\neq e_0$, $\E{s_e}\leq -\beta \eps_P x_e$ for  $\eps_P>0$.
\end{enumerate}

\end{theorem}

%
%
%
In the following subsection, we discuss how to prove this theorem. Here we explain at a high level how to define the hierarchy and reduce \cref{thm:maintechnical} to this theorem. The details are in \cref{sec:polygons}.

First, observe that, given \cref{thm:paymentinformal}, cuts in $\cH$ will automatically satisfy (ii) of \cref{thm:maintechnical}. The approach we take to satisfying 
all other cuts is to introduce additional slack, the vector $s^*$, on $OPT$ edges. 

Consider the set of all near-min-cuts of $z$, where $z := (x + OPT)/2$.  Starting with $z$ rather than $x$ allows us to restrict attention to a significantly {\em more structured} collection of near-min-cuts. The key observation here is that in $OPT$, all min-cuts have value 2, and any non-min-cut has value {\em at least} 4. Therefore averaging  $x$ with $OPT$ guarantees that every $\eta$-near min-cut of $z$  must consist of a {\em contiguous sequence of vertices (an interval) along the OPT cycle}. Moreover, each of these cuts is a $2\eta$-near min-cut of $x$.  Arranging the vertices in the $OPT$ cycle around a circle, we identify every such cut with the interval of vertices that does not contain $(u_0, v_0)$. Also, we say that a cut is crossed on both sides if it is crossed on the left and on the right.

To ensure that any cut $S$ that is {\em crossed on both sides} is satisfied, we first observe that $S$ is odd with probability  $O(\eta)$.  To see this, let $S_L$ and $S_R$ be the cuts crossing $S$ on the left and right with minimum intersection with $S$ and consider
the two (bad) events $\{E(S\cap S_L, S_L \smallsetminus S))_T\ne 1\}$
and $\{E(S\cap S_R, S_R \smallsetminus S))_T\ne 1\}$. Recall that if $A,B$ and $A\cup B$ are all near-min-cuts, then $\P{ E(A,B)_T \ne  1}=  O(\eta)$ (see \cref{lem:treeoneedge}).
Applying this fact to the two aforementioned bad events implies that each of them has probability $O(\eta)$. Therefore, we will let the two $OPT$ edges in $\delta(S)$ be responsible for these two events, i.e.,   we will increase the slack $s^*$ on these two $OPT$ edges by $O(\eta)$ when the respective bad events happens. This gives  $\E{s^*(e^*)} = O(\eta^2)$ for each OPT edge $e^*$. As we will see, this  simple step will reduce the number of near-min-cuts of $z$ that we need to worry about satisfying to $O(n)$.

Next, we consider the set of near-min-cuts of $z$ that are crossed on at most one side. Partition these into maximal connected components of crossing cuts. Each such component corresponds to an interval along the OPT cycle and, by definition, these intervals form a laminar family. 

A single connected component $\cC$ of at least two crossing cuts  is called a {\em polygon}. 
We prove the following structural theorem about the polygons induced by $z$:

\begin{theorem}[Polygons look like cycles (Informal version of \cref{thm:poly-structure})]\label{thm:approxpoly}
Given a connected component $\cC$ of near-min-cuts of $z$ that are crossed on one side, consider the coarsest partition of vertices of the OPT cycle into 
a sequence  $a_1, \ldots, a_{m-1}$ of sets called atoms (together with $a_0$ which is the set of vertices not contained in any cut of $\cC$). Then 
\begin{itemize}
\item Every cut in $\cC$ is the union of some number of consecutive atoms in $a_1, \ldots, a_{m-1}$.
\item 	For each $i$ such that  $0\le i < m-1$, $x(E(a_i, a_{i+1})) \approx 1$ and similarly $x(E(a_{m-1}, a_{0})) \approx 1$.
\item For each $i>0$, $x(\delta(a_i))\approx 2$.
\end{itemize}
 
\end{theorem}

The main observation used to prove \cref{thm:approxpoly} is that the cuts in $\cC$ crossed on one side can be partitioned into two laminar families $\cL$ and $\cR$, where $\cL$ (resp. $\cR$) is the set of cuts crossed on the left (resp. right).
This immediately implies that $|\cC|$ is linear in $m$. Since cuts in $\cL$ cannot cross each other (and similarly for $\cR$), the proof boils down to understanding the interaction between $\cL$ and $\cR$.

The approximations in \cref{thm:approxpoly} are correct up to $O(\eta)$. Using additional slack in $OPT$, at the cost of an additional $O(\eta^2)$ for edge, we can treat these  approximate equations as if they are exact. Observe that if $x(E(a_i, a_{i+1})) =1$, and $x(\delta(a_i))=x(\delta(a_{i+1}))=2$
for $1 \le i \le m-2$, then with probability 1, $E(a_i, a_{i+1})_T = 1$. Therefore, any cut in $\cC$ which doesn't include $a_1$ or $a_{m-1}$ is even with probability 1. The  cuts in $\cC$ that contain $a_1$ are even precisely\footnote{Roughly, this corresponds to the definition of the polygon being left-happy.} when $E(a_0, a_1)_T$ is odd and similarly the cuts in $\cC$ that contain $a_{m-1}$ are even when $E(a_0, a_{m-1})_T$ is odd. These observations are what allow us to imagine that each polygon is a triangle, i.e., assume $m=3$. (Note that often it is convenient to look at the event in which $E(a_0, a_1)_T = 1$ and $E(a_0, a_{m-1})_T = 1$ since this is a simple criteria which implies that all cuts in $\cC$ are even.)

The hierarchy $\cH$ is the set of all $\eta$-near mincuts of $z$ that are not crossed at all (these will be the degree cuts), together with a triangle for every polygon. In particular, for a connected component $\cC$ of size more than 1, the corresponding triangle cut is $a_1\cup \ldots \cup a_{m-1}$, with $A = E(a_0, a_1)$ and $B = E(a_0, a_{m-1})$. Observe that from the discussion above, when a triangle cut is happy, then all of the cuts in the corresponding polygon $\cC$ are even.

 Summarizing, we show that if we can construct a good slack vector $s$ for a hierarchy of degree cuts and triangles, then there is a nonnegative slack vector $s^*$, that satisfies all near-minimum cuts of $z$ not represented in the hierarchy, while maintaining slack for each OPT edge $e^*$ such that $\E{s^*(e^*)} = O(\eta^2)$.

 \paragraph{Remarks:}
 
 The reduction that we sketched above only uses the fact that $\mu$ is an arbitrary distribution of spanning trees with marginals $x$ and not necessarily a maximum-entropy distribution.
 
 We also observe that to prove \cref{thm:main}, we crucially used that $28 \eta \ll \eps$.  This forces us to take $\eta$ very small, which is why we get only a ``very slightly'' improved approximation algorithm for TSP. Furthermore, since we use OPT edges in our construction, we don't get a new upper bound on the integrality gap.  We leave it as an open problem to find a reduction to the ``cactus'' case that doesn't involve using a slack vector for OPT (or a completely different approach).

 \subsection{Proof ideas for \cref{thm:paymentinformal}}
 
 We now address the problem of constructing a good slack vector $s$ for a hierarchy of degree cuts and triangle cuts.
 For each LP edge $f$, consider the lowest cut in the hierarchy, that contains both endpoints of $f$. We call this cut $\p(f)$.
 If $\p(f)$ is a degree cut, then we call $f$ a {\em top edge} and otherwise, it is a {\em bottom edge}\footnote{For example, in \cref{fig:hierarchywithtriangle}, $\p(a,c) = u_3$, and $(a,c)$ is a bottom edge.}. We will see that bottom edges are easier to deal with, so we start by discussing the slack vector $s$ for top edges.
  
 Let $S$ be a degree cut and let $\bbe = (u,v)$ (where $u$ and $v$ are children of $S$ in $\cH$) be the set of all top edges $f = (u', v')$ such that $u' \in u$ and $v' \in v$. We call $\bbe$ a {\em top edge bundle} and say that $u$ and $v$ are the {\em top cuts} of each $f \in \bbe$. We will also sometimes say that $\bbe \in S$.
 
 Ideally, our plan is to reduce the slack of every edge $f \in \bbe$ when it is {\em happy}, that is, both of its top cuts are even in $T$. Specifically, we will set  $s_f := -\eta x_f$ when  $\delta(u)_T$ and $ \delta(v)_T $ are even. When this happens, we say that $f$ is {\em reduced}, and refer to the event $\{\delta(u)_T,\delta(v)_T\text{ even}\}$ as the {\em reduction event} for $f$.
 Since this latter event doesn't depend on the actual endpoints of $f$, we view this as a simultaneous reduction of $s_\bbe$. 
 
Now consider the situation from the perspective of the degree cut $u$ (where $\p(u)=S$) and consider any incident edge bundle in $S$, e.g., $\bbe = (u,v)$. Either its top cuts are both even and $s_\bbe := - \eta x_\bbe$, or they aren't even, because, for example, $\delta(u)_T$ is odd. In this latter situation,  edges in   $\delta^\uparrow(u):= \delta(u)\cap \delta(S)$ might have been reduced (because {\em their} top two cuts are even), which a priori could leave $\delta(u)$ \hyperlink{tar:satisfy}{unsatisfied}. In such a case, we {\em increase} $s_\bbe$ for edge bundles in $ \delta^\rightarrow(u):= \delta (u) \smallsetminus \delta(S)$ to compensate for this reduction.  Our main goal is then to prove is that for any edge bundle its expected reduction is greater than its expected increase.
  The next example shows this analysis in an ideal setting.

\begin{example}[Simple case] \label{ex:simple}
Fix a top edge bundle $\bbe = (u,v)$ with $\p(\bbe) = S$.
Let $x_u := x(\delta^\uparrow (u))$ and let $x_v := x(\delta^\uparrow (v))$. Suppose we have constructed a (fractional) {\em matching} between edges whose top two cuts are children of $S$ in $\cH$ and the edges in $\delta(S)$, and this matching satisfies the following three conditions: (a) $\bbe=(u,v)\in S$ is matched (only) to edges going higher from its top two cuts (i.e., to edges in $\delta^\uparrow (u)$ and $\delta^\uparrow (v)$),  (b) $\bbe$ is matched to  
an $m_{\bbe,u}$ fraction of every edge in $\delta^\uparrow(u)$ and to an $m_{\bbe,v}$ fraction of each edge in $\delta^\uparrow(v)$, where
$$m_{\bbe,u}  + m_{\bbe, v}  = x_\bbe,$$ and (c) the fractional value of edges in $\delta^\rightarrow(u):= \delta(u) \smallsetminus \delta^\uparrow(u)$ matched to edges in $\delta^\uparrow(u)$  is equal to $x_u$. That is, for each $u\in S$, $\sum_{\bbf \in \delta^\rightarrow(u)} m_{\bbf,u} = x_u$.

\begin{figure}[htb]
\centering	
\begin{tikzpicture}
	\node [circle,draw] at (0,0) (u) {$u$};
	\node [circle,draw] at (2,0) (v) {$v$} edge node [above] {$\bbe$} (u);
	\draw (u) -- node [left] {\color{blue}{$x_u$}} +(0,1.5) (v) -- node [right] {\color{blue}{$x_v$}} +(0,1.5);
	\draw [dotted,color=red,line width=1pt] (1,-0.5) ellipse (2 and 1.2) ;
	\draw (u) -- +(-0.5,-0.5) (v) -- +(-0.5,-0.5) (u) -- +(0.5,-0.5) (v) -- +(0.5,-0.5) (u) -- +(0,-0.75) (v) -- +(0,-0.75);
	\node [color=red] at (-1,-1) () {$S$};
\end{tikzpicture}
\end{figure}

The plan is for $\bbe\in S$ to be tasked with part of the responsibility for {\em fixing the cuts} $\delta(u)$ and $\delta(v)$ when they are odd and edges going higher are reduced. Specifically, $s_\bbe$ is increased to compensate for an $m_{\bbe,u}$  fraction of the reductions in edges in $\delta^\uparrow(u)$ when $\delta(u)_T$ is odd. (And similarly for reductions in $v$.) 
Thus,
\begin{align}
	\E{s_\bbe} &= -\P{\bbe\text{ reduced}} \eta x_\bbe + m_{\bbe,u}\sum_{g\in \delta^\uparrow(u)} \P{\delta(u)_T\text{ odd}|g \text{ reduced} }\P{g \text{ reduced}}\eta \frac{x_g}{x(\delta^\uparrow(u))}\notag\\
	&\quad\quad +  m_{\bbe,v}\sum_{g\in \delta^\uparrow(v)} \P{\delta(v)_T\text{ odd}|g \text{ reduced} }\P{g \text{ reduced}}\eta \frac{x_g}{x(\delta^\uparrow(v))} \label{eq:sfUB}
\end{align}
We will lower bound  $\P{\delta(u)_T\text{ even}|g\text{ reduced}}$. We can write this as
$$\P{\delta^\rightarrow (u)_T\text{ and }\delta^\uparrow (u)_T\text{ have same parity } | g \text{ reduced}}.$$
Unfortunately, we do not currently have a  good handle on the parity of $\delta^\uparrow (u)_T$ conditioned on $g$ reduced. 
However,  we can use 
the following simple but crucial property: Since $x(\delta(S)) = 2$, by \cref{lem:treeconditioning}, $T$ consists of two independent trees, one on $S$ and one on $V \smallsetminus S$, each with the corresponding marginals of $x$.
Therefore, we can write$$\P{\delta(u)_T\text{ even}|g\text{ reduced}} \ge \min (\P{(\delta^\rightarrow (u))_T \text{ even}}, \P{(\delta^\rightarrow (u))_T \text{ odd}}).$$
This gives us a reasonable bound when $\eps \le x_u, x_v \le 1- \eps$ since, because
$x(\delta(u)) = x(\delta(v))=2$, 
by the SR property, $(\delta^\rightarrow (u))_T$ (and similarly $(\delta^\rightarrow (v))_T$) is the sum of Bernoulis with expectation in $[1+\eps, 2-\eps]$. From this it follows that 
$$ \min (\P{(\delta^\rightarrow (u))_T \text{ even}}, \P{(\delta^\rightarrow (u))_T \text{ odd}}) =\Omega (\epsilon). $$We can therefore conclude that 
$\P{\delta(u)_T\text{ odd}|g\text{ reduced}}  \le 1- O(\eps).$

The rest of the analysis of this special case follows  from (a) the fact that our construction will guarantee that for {\em all} edges $g$, the probability that $g$ is reduced is {\em exactly} $p$, i.e., it is the same for all edges, and 
(b) the fact that $m_{\bbe,u}x_u + m_{\bbe,v}x_v = x_\bbe$.
Plugging these facts back into \eqref{eq:sfUB}, gives
\begin{align}
\E{s_\bbe} &\le  -p \eta x_\bbe + m_{\bbe,u} (1-\eps) p \eta  + 
m_{\bbe,v} (1-\eps) p \eta \notag\\
& \le -p\eta x_\bbe + (1-\eps)p\eta x_\bbe =
 -\eps p\eta x_\bbe.	\label{eq:sbbe}
\end{align}
If we could prove \eqref{eq:sbbe} for {\em every} edge $f$ in the support of $x$, that would complete the proof that the expected cost of the min $O$-join for a random spanning tree $T\sim\mu$ is at most $(1/2 - \eps) OPT$.
\end{example}

\paragraph{Remark:} 
Throughout this paper, we repeatedly use a mild generalization of the above "independent trees fact": that if $S$ is a cut with $x(\delta(S)) \le 2 + \eps$, then $S_T$ is very likely to be a tree. Conditioned on this fact, marginals inside $S$ and outside $S$ are nearly preserved and the trees inside $S$ and outside $S$ are sampled independently (see \cref{lem:treeconditioning}).

\paragraph{Ideal reduction:}
In the example, we were able to show that $\P{\delta(u)_T \text{ odd} \mid g\text{ reduced}}$ was bounded away from 1 for every edge $g\in \delta^\uparrow(u)$, and this is how we proved that the expected reduction for each edge was greater than the expected increase on each edge, yielding negative expected slack. 

This motivates the following definition: A reduction for an edge $g$ is $k$-{\em ideal}
 if, conditioned on $g$ reduced, every cut $S$ that is in the top $k$ levels of cuts containing $g$  is odd with probability that is bounded away from 1.

\paragraph{Moving away from an idealized setting:}
In \cref{ex:simple}, we oversimplified in four ways:
\begin{enumerate}[(a)]
\item We assumed that it would be possible to show that each top edge is {\em good}. That is, that its top two cuts are even {\em simultaneously} with constant probability.

\item We considered only top edge bundles (i.e., edges whose top cuts were inside a degree cut).

\item 	We assumed that $x_u, x_v \in [\epsilon, 1-\epsilon]$.
\item We assumed the existence of a nice matching between edges  whose top two cuts were children of $S$ and the edges in $\delta(S)$.
\end{enumerate}
Our proof needs to address all four anomalies that result from deviating from these assumptions. 

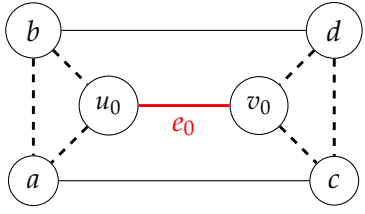
\begin{figure}[htb]\centering
\begin{tikzpicture}
	\node [draw,circle,inner sep=4] at (0,0) (a)  {$a$};
	\node [draw,circle] at (1,1) (b)  {{\small$u_0$}}  edge [dashed, line width=1.1pt] (a);
	\node [draw,circle,inner sep=4] at (0,2) (c)  {$b$}  edge [dashed, line width=1.1pt] (b) edge [dashed, line width=1.1pt] (a);
	
	\node [draw,circle,inner sep=4] at (4,0) (d)  {$c$} edge  (a);
	\node [draw,circle] at (3,1) (e)  {{\small$v_0$}}  edge [dashed, line width=1.1pt] (d) edge  [color=red,line width=1.1pt]  node [below] {$e_0$} (b);
	\node [draw,circle,inner sep=4] at (4,2) (f)  {$d$}  edge [dashed, line width=1.1pt] (d) edge [dashed, line width=1.1pt] (e) edge(c);
\end{tikzpicture}
\caption{An Example with Bad Edges. A feasible solution of the \ref{eq:tsplp} is shown; dashed edges have fraction 1/2 and solid edges have fraction 1. Writing $E=E_0\smallsetminus \{e_0\}$ as a maximum entropy distribution $\mu$ we get the following: Edges $(a,b),(c,d)$ must be completely negatively correlated (and independent of all other edges). So, $(b,u_0), (a,u_0)$ are also completely negatively correlated. This implies $(a,b)$ is a bad edge. }
\label{fig:badedgeexistence}
\end{figure}

\paragraph{Bad edges.} Consider first (a). Unfortunately, it is not the case that all top edges are good. Indeed, some are {\em bad}.  However, it turns out that bad edges are rare in the following senses: First, for an edge  to be bad, it must be a half edge, where we say that an edge $\bbe$ is a half edge if  $x_\bbe \in 1/2 \pm \eps_{1/2}$ for a suitably chosen constant $\eps_{1/2}$. Second, of any two half edge bundles sharing a common endpoint in the hierarchy, at least one is good. For example, in \cref{fig:badedgeexistence}, $(a,u_0)$ and $(b,u_0)$ are good half-edge bundles. We advise the reader to ignore half edges in the first reading of the paper. Correspondingly, we note that our proofs would be much simpler if half-edge bundles never showed up in the hierarchy. It may not be a coincidence that half edges are hard to deal with, as it is conjectured that TSP instances with half-integral LP solutions are the hardest to round~\cite{SWV12,SWvZ13}.

Our solution is to {\em never} reduce bad edges. But this in turn poses two problems. First, it means that we need to address the possibility that the bad edges constitute most of the cost of the LP solution. Second, our objective is to get negative expected slack on each good edge and non-positive expected slack on bad edges. Therefore, if we never reduce bad edges, we can't increase them either, which means that the responsibility for fixing an odd cut with reduced edges going higher will have to be split amongst fewer edges (the incident good ones).

We deal with the first problem by showing that 
in every cut $u$ in the hierarchy at least 3/4 of the fractional mass in $\delta(u)$ is good and these edges suffice to compensate for reductions on the edges going higher. Moreover, because  there are sufficiently many good edges incident to each cut, we can show that either using the slack vector $\{s_e\}$ gives us a low-cost O-join, or we can average it out with another O-join solution concentrated on bad edges to obtain a reduced cost matching of odd degree vertices.

We deal with the second problem by proving \cref{lem:matching}, which guarantees a matching between  {\em good} edge bundles $\bbe = (u,v)$ and fractions $m_{\bbe,u}, m_{\bbe,v}$ of edges in $\delta^\uparrow(u), \delta^\uparrow(v)$ such that, roughly, $m_{\bbe,u} + m_{\bbe,v} = (1 + O(\eps_{1/2}))x_\bbe$.

\begin{figure}\centering
\begin{tikzpicture}
	\node [draw,circle] at (0,0) (a1) {$a_1$};
	\node [draw,circle] at (2,0) (a2) {$a_2$} edge node [below] {$\bbf$} (a1) ;
	\node [draw,circle,inner sep=4] at (1,2) (u) {$u$} edge node [left] {$A$} (a1) edge node [right] {$B$} (a2) ;
	\node [draw,circle] at (5,0) (a3) {$a_3$};
	\node [draw,circle] at (7,0) (a4) {$a_4$} edge node [below] {${\bbg}$} (a3);
	\node [draw,circle,inner sep=4] at (6,2) (v) {$v$} edge node [above] {$\bbe$} (u) edge node [left] {$A'$} (a3) edge node [right] {$B'$} (a4);
	\draw [color=red, line width=1.1pt](0.5,-0.4) arc (-30:80:0.8)
	(1.5,-0.4) arc (210:110:0.8);
	\draw (u) -- +(60:1) (u) -- +(90:1) (u) -- +(120:1) (v) -- +(60:1) (v) -- +(90:1) (v) -- +(120:1);;
\end{tikzpicture}
\caption{In this representation of the cut hierarchy (as in \cref{fig:hierarchywithtriangle}), for the triangle $u$ corresponding to the cut $\delta(a_1\cup a_2)$, when $A_T$ and $B_T$ are odd, all 3 cuts ($\delta(a_1)_T, \delta(a_2)_T$ and $\delta(a_1 \cup a_2)_T = \delta(u)_T$ are odd (since $\bbf_T$ is always 1). (Recall also that the edges in the bundle $\bbe$ must have one endpoint in $\{a_1 \cup a_2\}$ and one endpoint in $\{a_3 \cup a_4\}$, as was the case, e.g., for the edge $(a,c)$ in \cref{fig:hierarchywithtriangle}.)}
\label{fig:twotriangles}
\end{figure}
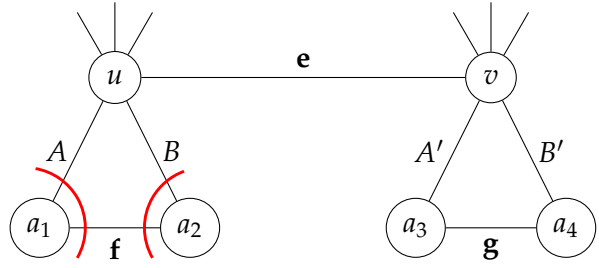

\paragraph{Dealing with triangles.} \hypertarget{211-explanation}{Turning to (b)}, consider  a triangle cut $S$, for example $\delta(a_1 \cup a_2)$ in \cref{fig:twotriangles}.
Recall that in a triangle, we can assume that there is an edge of fractional value 1 connecting $a_1$ and $a_2$ in the tree, and this is why we defined the cut to be happy when $A_T$ and $B_T$ are odd: this guarantees that all 3 cuts  defined by the triangle ($\delta(a_1),\delta(a_2), \delta(a_1 \cup a_2)$ are even. 

Now suppose that $\bbe = (u,v)$ is a top edge bundle, where $u$ and $v$ are both triangles, as shown in \cref{fig:twotriangles}. Then we'd like to reduce $s_\bbe$ when both cuts $u$ and $v$ are happy. But this would require more than simply both cuts being even. This would require  {\em all} of 
$A_T, B_T, A'_T, B'_T$ to  be odd. Note that if, for whatever reason,  $\bbe$ is reduced only when $\delta(u)_T$ and $\delta(v)_T$ are both even, then it could be, for example, that this only happens when $A_T$ and $B_T$ are both even. In this case, both $\delta(a_1)_T$ and $\delta(a_2)_T$ will be odd with probability 1 (recalling that $\bbf_T=1$), which would then necessitate an increase in $s_\bbf$ whenever $\bbe$ is reduced. In other words, the reduction will not even be 1-ideal.

It turns out to be easier for us to get a 1-ideal reduction rule for $\bbe$ as follows: Say that
$\bbe$ is {\em 2-1-1 happy with respect to $u$} if  $\delta(u)_T$ is even and both $A'_T, B'_T$ are odd. We reduce $\bbe$ with probability $p/2$ when it is 2-1-1 happy with respect to $u$ and with probability $p/2$ when it is 2-1-1 happy with respect to $v$. This means that when $\bbe$ is reduced, half of the time no increase in $s_\bbf$ is needed since $u$ is happy. Similarly for $v$.

The 2-1-1 criterion for reduction introduces a new kind of bad edge: a half edge that is good, but not 2-1-1 good. We are able to show that non-half-edge bundles are 2-1-1 good (\cref{lem:x_e<=1/2-eps1,lem:x_e>=1/2+eps_1/2}), and that if there are two half edges which are both in $A$ or are both in $B$, then at least one of them is 2-1-1 good (\cref{lem:one-of-two-211}). Finally, we show that if there are two half edges, where one is in  $A$ and the other is in $B$, and neither is 2-1-1 good, then we can apply a different reduction criterion that we call {\em 2-2-2 good}. When the latter applies, we are guaranteed to decrease both of the half edge bundles simultaneously.
All together, the various considerations discussed in this paragraph force us to come up with a relatively more complicated set of rules under which we reduce $s_\bbe$ for a top edge bundle $\bbe$ whose children are triangle cuts. \cref{sec:probabilistic} focuses on developing the relevant probabilistic statements.

\paragraph{Bottom edge reduction.} Next, consider a bottom edge bundle $\bbf = (a_1,a_2)$ where $\p(a_1) = \p(a_2)$ is a triangle. Our plan is to reduce $s_\bbf$ (i.e., set it to $-\eta x_\bbf$) when the triangle is happy, that is, $A_T = B_T = 1$. The good news here is that every triangle is happy with constant probability. However, when a triangle is {\em not} happy, $s_\bbf$ may need to increase to make sure that the O-join constraint for $\delta(a_1)$ and $\delta(a_2)$  are satisfied, if
 edges in $A$ and $B$ going higher are reduced. Since  $x_\bbf = x(A) = x(B) = 1$, this means that $\bbf$ may need to compensate at {\em twice} the rate at which it is getting reduced. This would result in $\E{s_\bbf}>0$, which is the opposite of what we seek.
 
 We use two key ideas to address this problem.  First, we reduce top edges and bottom edges by different amounts: Specifically, when the relevant reduction event occurs, we reduce a bottom edge $\bbf$  by $\beta x_\bbf$ and top edges $\bbe$ by $\tau x_\bbe$, where $\beta >   \tau$ (and $\tau$ is a multiple of $\eta$). 

Thus, the expected reduction in $s_\bbf$ is $p\beta x_\bbf = p\beta$, whereas the expected increase (due to compensation of, say, top edges going higher) is $p\tau (x(A) + x(B))q = p \tau 2 q$,
where $$q = \P{\text{ triangle not happy} \mid \text{reductions in $A$ and $B$}}.$$ Thus, so long as $2 \tau q < \beta - \epsilon$, we get the expected reduction in $s_\bbf$ that we seek.

The discussion so far suggests that we need to take $\tau$ smaller than $\beta/2q$, which is $\beta/2$ if $q$ is 1, for example. On the other hand, if $\tau = \beta/2$, then when a top edge needs to fix a cut due to reductions on bottom edges, we have the opposite problem -- their expected increase will be greater than their expected reduction, and we are back to square one.

Coming to our aid is the second key idea, already discussed in \cref{sec:marginals}.  We reduce bottom edges only when $A_T = B_T = 1$ {\em and} the marginals of edges in $A,B$ are approximately preserved (conditioned on $A_T = B_T = 1$). This allows us to get much stronger upper bounds on the probability that a lower cut a bottom edge is on is odd, given that the bottom edge is reduced, and enables us to show that bottom edge reduction is $\infty$-ideal.

It turns out that the combined effects of (a) choosing $\tau =0.571 \beta$, and (b) getting better bounds on the probability that a lower cut is even given that a bottom edge is reduced, suffice to deal with the interaction between the reductions and the increases in slack for top and bottom edges.

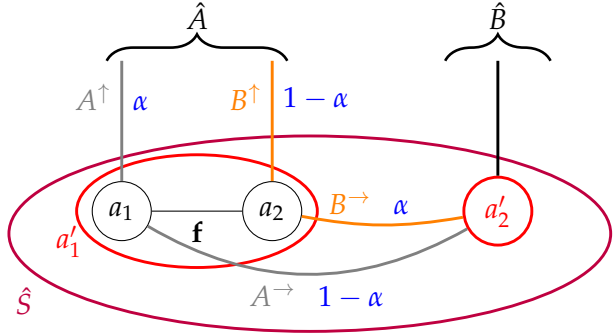
\begin{figure}[htb]\centering
\begin{tikzpicture}
\node [draw,circle]at (0,0) (a1) {$a_1$} ;
\node [draw,circle] at (2,0) (a2) {$a_2$} edge node [below] {$\bbf$} (a1);
\draw [color=red,line width=1.1pt] (1,0) ellipse  (1.6 and 0.75);
\draw [line width=1.1pt,color=purple] (2.5,-0.3) ellipse (4 and 1.3);
\node [color=purple] at (-1.3,-1.2) () {$\hat{S}$};
\draw [line width=1pt, decorate,decoration={brace,amplitude=10pt}]
(-0.2,2) -- node [above=8] {$\hat{A}$} (2.2,2);
\draw [line width=1pt, decorate,decoration={brace,amplitude=10pt}]
(4.3,2) -- node [above=8] {$\hat{B}$} (5.7,2);

\node [color=red,line width=1.1pt] at (-0.7,-0.4)  () {$a_1'$};
\node [color=red,draw,circle,line width=1.1pt] at (5,0) (a3) {$a_2'$} edge [color=gray,bend left=30,line width=1.1pt] node [below left] {$A^\rightarrow$} node [below right,color=blue] {\textcolor{blue}{$1-\alpha$}} (a1) edge [color=orange, line width=1.1pt,bend left=10] node [above left] {$B^\rightarrow$} node [above right,color=blue] {\textcolor{blue}{$\alpha$}} (a2);
\draw [line width=1.1pt,color=gray] (a1) -- node [above left]{$A^\uparrow$} node [above right, color=blue ] {\textcolor{blue}{$\alpha$}} +(0,2);
\draw [line width=1.1pt,color=orange] (a2) -- node [above left]{$B^\uparrow$} node [above right,color=blue] {\textcolor{blue}{$1-\alpha$}} +(0,2);
\draw [line width=1.1pt] (a3) --  +(0,2);
\end{tikzpicture}	
\caption{Setting of \cref{ex:bottombottom}. Note that the set $A = \delta(a_1) \cap \delta(a_1')$  decomposes into two sets of edges, $A^\uparrow$, those that are also in $\delta(S)$, and the rest, which we call $A^\rightarrow$. Similarly for $B$.}
\label{fig:bottombottom}
\end{figure}

\begin{example}\label{ex:bottombottom} [Bottom-bottom case]
To see how preserving marginals helps us handle the interaction between bottom edges at consecutive levels, consider a triangle cut $a_1' = \{a_1, a_2\}$ whose parent cut $\hat{S} = \{a_1', a_2'\}$ is also a triangle cut (as shown in \cref{fig:bottombottom}). 
Let's analyze $\E{s_\bbf}$ where $\bbf = (a_1, a_2)$. Observe first that
$A^\rightarrow\cup B^\rightarrow$ is a bottom edge bundle in the triangle $\hat S$ and all edges in this bundle are reduced simultaneously when $\hat{A}_T = \hat{B}_T = 1$
and marginals of all edges in $\hat{A}\cup\hat{B}$ are approximately preserved. (For the purposes of this overview, we'll assume they are preserved exactly). Furthermore, since the tree inside $\hat{S}$ is picked independently of the tree on $G/\hat{S}$ (using \cref{lem:treeconditioning} and assuming $\epsilon = 0$ for this overview), exactly one edge in $A^\rightarrow\cup B^\rightarrow$ is selected independently of the reduction event $\hat{A}_T = \hat{B}_T = 1$. Let $x(A^\uparrow) = \alpha$. Then since $A = A^\uparrow \cup A ^\rightarrow$ and $x(A)=1$, we have $x(A ^\rightarrow) = 1-\alpha$.
Moreover, since $\hat A = A^\uparrow \cup B^\uparrow$ and $x(\hat A)=1$, we also have
$x(B ^\uparrow) =1- \alpha$ and  $x(B ^\rightarrow) = \alpha$.

Therefore, using the fact that when $A^\rightarrow \cup B^\rightarrow$ is reduced, exactly one edge in $A^\uparrow \cup B^\uparrow$ is selected (and also exactly one edge in $A^\rightarrow \cup B^\rightarrow$ is selected independently since it is a bottom edge bundle, as mentioned above),  and marginals are preserved given the reduction,
 we conclude that
$$\P{a_1'\text{ happy} \mid A^\rightarrow \cup B^\rightarrow\text{ reduced}}  =\P{A_T= B_T = 1 \mid A^\rightarrow \cup B^\rightarrow\text{ reduced}}  =\alpha^2 + (1-\alpha)^2. $$
Now, we calculate $\E{s_\bbf}$. First, note that $\bbf$ may have to increase to compensate either for reduced edges in $A^\uparrow\cup B\uparrow$ or in $A^\rightarrow\cup B
^\rightarrow$.
For the sake of this discussion, suppose that $A^\uparrow \cup B^\uparrow$ is a set of top edges. Then, in the worst case we need to increase $\bbf$ by $p\tau$ in expectation to fix the cuts $a_1,a_2$ due to the reduction in $A^\uparrow \cup B^\uparrow$.
Now, we calculate the expected increase due to the reduction in $A^\rightarrow\cup B^\rightarrow$.
The crucial observation is that edges in $A^\rightarrow \cup B^\rightarrow$ are reduced simultaneously, so both cuts $\delta(a_1)$ and $\delta(a_2)$ can be fixed simultaneously by an increase in $s_\bbf$. Therefore, when they are both odd,  it suffices for $\bbf$ to increase by 
$$\max\{x(A^\rightarrow), x(B^\rightarrow)\}\beta = \max \{\alpha, 1-\alpha\}\beta,$$ 
to fix cuts $a_1,a_2$. Putting this together, we get
\begin{align*}
	\E {s_\bbf} & = -p\beta + \E{\text{increase due to }A^\rightarrow \cup B^\rightarrow} + \E{\text{increase due to }A^\uparrow \cup B^\uparrow}\\
	&\le -p\beta  +p\beta \max_{\alpha \in [1/2, 1] }\alpha[1-\alpha^2 - (1-\alpha)^2] + p\tau\\
	\intertext{which, since  $\max_{\alpha \in [1/2, 1] }\alpha[1-\alpha^2 - (1-\alpha)^2]= 8/27$ and  $\tau = 0.571 \beta$ is}
	&= p\beta (-1 + \frac{8}{27} + 0.571) = -0.13 p \beta.
\end{align*}
\end{example}

\paragraph{Dealing with $x_u$ close to $1$.}\footnote{Some portions of this discussion might be easier to understand after reading the rest of the paper.} 
\hypertarget{ABC-explanation}{Now}, suppose that $\bbe=(u,v)$ is a top edge bundle with $x_u:=x(\delta^\uparrow(u))$ is close to  $1$.
 Then, the  analysis in \cref{ex:simple}, bounding $r:= \P{\delta(u)_T \text{ odd} | g\text{ reduced}}$ away from 1 for an edge $g\in \delta^\uparrow(u)$ doesn't hold.
To address this, we consider two cases:
The first case, is that the edges in $\delta^\uparrow(u)$ break up into many groups that end at different levels in the hierarchy.
In this case, we can analyze $r$ separately for the edges that end at any given level, taking advantage of the independence between the trees chosen at different levels of the hierarchy.

The second case is when nearly all of the edges in   $\delta^\uparrow(u)$ end at the same level, for example, they are all in $\delta^\rightarrow(u')$ where $\p(u')$ is a degree cut. In this case,
we introduce a more complex (2-1-1) reduction rule for these edges.
The observation is that from the perspective of these edges $u'$ is a "pseudo-triangle". That is, it looks like a triangle cut, with atoms $u$ and $u'\smallsetminus u$ where $\delta(u)\cap\delta(u')$ corresponds to the ``$A$''-side of the triangle. 

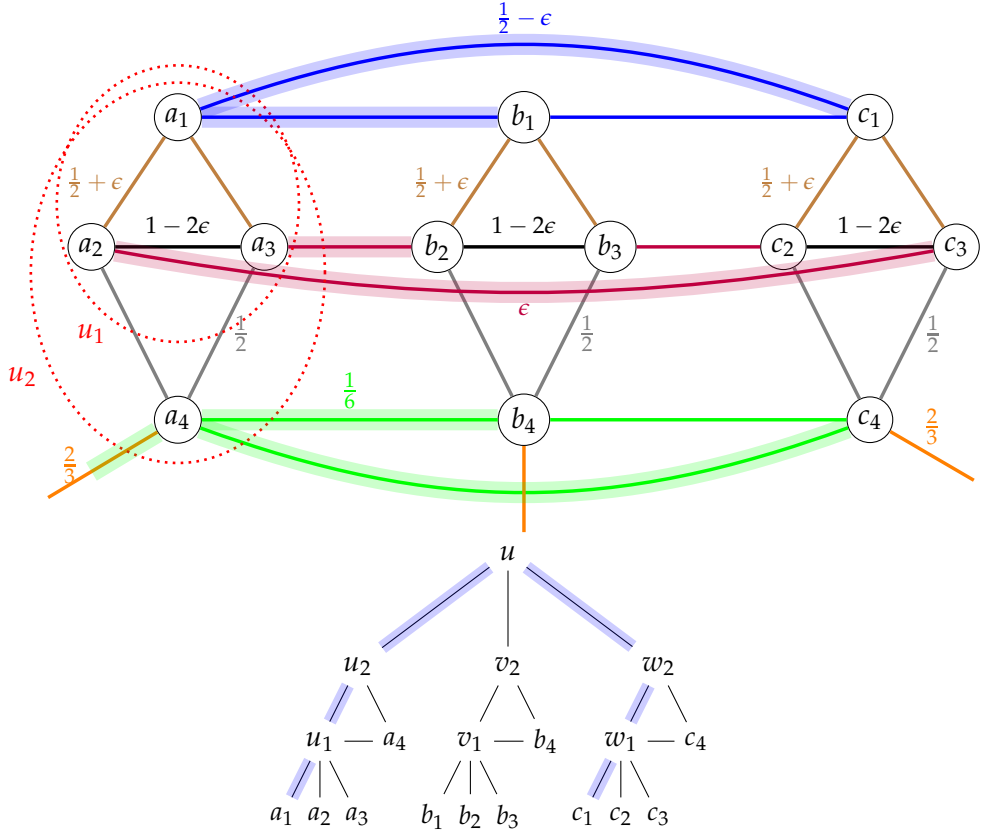
\begin{figure}[htb]
\centering
\begin{tikzpicture}[scale=1.15]
\foreach \i/\a in {0/a, 1/b, 2/c}{
	\node [draw,circle,inner sep=2] at (\i*4,2) (\a1) {$\a_1$};
	\node [draw,circle,inner sep=2] at (\i*4-1,0.5) (\a2) {$\a_2$} edge [color=brown,line width=1.3pt] node [left] {\footnotesize$\frac12+\eps$} (\a1);
	\node [draw,circle, inner sep=2] at (\i*4+1,0.5) (\a3) {$\a_3$} edge [color=brown,line width=1.3pt] (\a1) edge [line width=1.3pt] node [above] {\footnotesize{$1-2\eps$}} (\a2);
	\node [draw,circle, inner sep=2] at (\i*4, -1.5) (\a4) {$\a_4$} edge [line width=1.3pt,color=gray]  (\a2) edge [color=gray,line width=1.3pt] node [right] {$\frac12$} (\a3);
}
	\draw [color=green,line width=1.3pt] (a4) edge node [above] {$\frac16$} (b4) (b4) edge (c4) (c4) edge [bend left=20] (a4);
	\draw [color=orange,line width=1.3pt] (a4) -- node [left=5] {$\frac23$} +(-1.5,-0.9) (b4) --  +(0,-1.3)  (c4) -- node [above] {$\frac23$} +(1.2,-0.7);
	\draw [color=blue,line width=1.3pt] (a1) edge  (b1) (b1) edge (c1) (c1) edge [bend right=20] node [above] {\footnotesize$\frac12-\eps$} (a1);
	\draw [color=purple,line width=1.3pt] (a3) edge (b2) (b3) edge (c2) (c3) edge [bend left=10] node [below] {\footnotesize$\eps$} (a2);
	\draw [dotted,color=red,line width=1pt](0,1) ellipse (1.4 and 1.6)
	(0,0.2) ellipse (1.7 and 2.2);
	\node [color=red]at (-1.8,-1) () {$u_2$};
	\node [color=red] at (-1,-0.5) () {$u_1$};
	\draw [color=blue,line width=8pt,opacity=0.2] (a1) edge (b1)
	(a1) edge [bend left=20] (c1);
	\draw [color=purple,line width=8pt,opacity=0.2] (a3) edge (b2) (a2) edge [bend right=10] (c3);
	\draw [color=green,line width=8pt,opacity=0.2] (a4) edge (b4) (a4) edge [bend right=20] (c4) (a4) -- +(-1,-0.6);
\end{tikzpicture}	  
\begin{tikzpicture}
	\foreach \i/\a/\b in {0/a/u, 1/b/v, 2/c/w}{
		\node at (\i*2,0) (\a1) {\small$\a_1$};
		\node at (\i*2+0.5,0) (\a2) {\small$\a_2$};
		\node at (\i*2+1,0) (\a3) {\small$\a_3$};
		\node at (\i*2+0.5,1) (\b1) {$\b_1$} edge (\a1) edge (\a2) edge (\a3);
		\node at (\i*2+1.5,1) (\a4) {\small$\a_4$} edge (\b1);
		\node at (\i*2+1,2) (\b2) {$\b_2$} edge (\a4) edge (\b1);
	}
	\node at (3,3.5) (u) {$u$} edge (u2) edge (v2) edge (w2);
	\draw[color=blue,line width=5pt,opacity=0.2] (u1) edge (a1) (u2) edge (u1) (u) edge (u2) (u) edge (w2) (w2) edge (w1) (w1) edge (c1);
\end{tikzpicture}
\caption{\small Part of the hierarchy of the graph is shown on top. Edges of the same color have the same fraction and $\eps\gg \eta$ is a small constant. $u_1$ corresponds to the degree cut 
$\{a_1,a_2,a_3\}$, $u_2$ corresponds to the triangle cut $\{u_1,a_4\}$ and $u$ corresponds to the degree cut containing all of the vertices shown. Observe that edges in $\delta^\uparrow(a_1)$ are top edges in the degree cut $u$. 
If $\eps<\frac12 \eps_{1/1}$ then the $(A,B,C)$-degree partitioning
of edges in $\delta(u_2)$ is as follows: 
$A=\delta(a_1)\cap\delta(u_2)$ are the blue highlighted edges each of fractional value $1/2-\eps$, $B=\delta(a_4)\cap\delta(u_2)$ are the green highlighted edges of total fractional value 1, and $C$ are the red highlighted edges each of fractional value $\eps$. The cuts that contain edge $(a_1,c_1)$ are highlighted in the  hierarchy at the bottom.
}
\label{fig:topedgeABCmotivation}
\end{figure}

Now, we define this more complex 2-1-1 reduction rule: Consider a top edge $\bbf=(u',v')\in\delta^\rightarrow(u')$. So far, we only considered the following reduction rule for $\bbf$: If both $u',v'$ are degree cuts, $\bbf$ reduces when they are both even in the tree; otherwise if say $u'$ is a triangle cut, $\bbf$ reduces when it is 2-1-1 good w.r.t., $u'$ (and similarly for $v'$). 
But clearly these rules ignore the pseudo triangle.  The simplest adjustment is, if $u'$ is a pseudo triangle with partition $(u,u'\smallsetminus u)$, to require $\bbf$ to reduce when $A_T=B_T=1$ and $v'$ is happy. However, as stated, it is not clear that the sets $A$ and $B$ are well-defined. For example, $u'$ could be an actual triangle or there  could be multiple ways to see $u'$ as a pseudo triangle only one of which is $(u, u'\smallsetminus u)$. Our solution is to find the {\em smallest} disjoint pair of cuts $a,b\subset u'$ in the hierarchy such that $x(\delta(a)\cap\delta(u')),x(\delta(b)\cap\delta(u'))\geq 1-\eps_{1/1}$, where $\eps_{1/1}$ is a fixed universal constant, and then let $A=\delta(a)\cap\delta(u'), B=\delta(b)\cap\delta(u')$ and $C=\delta(u')\smallsetminus A\smallsetminus B$ (see \cref{fig:topedgeABCmotivation} for an example). Then, we say $\bbf$ is 2-1-1 happy w.r.t., $u'$ if $A_T=B_T=1$ and $C_T=0$.

A few observations are  in order:
\begin{itemize}
\item Since $u$ is a candidate for, say $a$, it must be that $a$ is a descendent of $u$  in the hierarchy (or equal to $u$). In addition, $b$ cannot simultaneously be in $u$, since $a\cap b=\emptyset$ and $x(\delta(u) \cap\delta(u')) \le 1$ by \cref{lem:shared-edges}.  So, when $\bbf$ is 2-1-1 happy w.r.t.   $u'$ we get $(\delta(u)\cap\delta(u'))_T=1$.
\item If $u'=(X,Y)$ is a actual triangle cut, then we must have $a\subseteq X, b\subseteq Y$. So, when $\bbf$ is 2-1-1 happy w.r.t. $u'$, we know that $u'$ is a happy triangle, i.e., $(\delta(X)\cap\delta(u'))_T=1$ and $(\delta(Y)\cap\delta(u'))_T=1$.
\end{itemize}

Now, suppose for simplicity that all top edges in $\delta(u')$ are 2-1-1 good w.r.t. $u'$. Then, when an edge $g\in\delta(u)\cap\delta(u')$ is reduced, $(\delta(u)\cap\delta(u'))_T=1$, so 
$$ \P{\delta(u)_T\text{ odd} | g\text{ reduced}} \leq \P{E(u,u'\smallsetminus u)_T\text{ even}| g\text{ reduced}}\leq 0.57,$$
since edges in $E(u,u'\smallsetminus u)$ are in the tree independent of the reduction and $\E{E(u,u'\smallsetminus u)_T}\approx 1$.


\paragraph{Dealing with $x_u$ close to 0 and the matching.}
\hypertarget{xu-close-to-zero}{We already} discussed how the matching is modified to handle the existence of bad edges. We now observe that we can handle the case $x_u \approx 0$ by further modifying the matching. 
The key observation is that  in this case, $x(\delta ^\rightarrow(u))\gg x(\delta ^\uparrow(u))$. Roughly speaking, this enables us to find a matching in which 
each edge in $\delta^\rightarrow(u)$ has to increase about half as much as would normally be expected to fix the cut of $u$. This eliminates the need to prove a nontrivial bound on $\P{\delta(u)_T\text{ odd} | g\text{ reduced}}$. The details of the matching are in \cref{sec:matching}.



%% file: polygons.tex

\def\es{e^*}
\def\ss{s^*}
\section{Polygons and the Hierarchy of Near Minimum Cuts}
\label{sec:polygons}
Let OPT be a minimum TSP solution, i.e., minimum cost Hamiltonian cycle and without loss of generality assume it visits $u_0$ and $v_0$ consecutively (recall that $c(u_0,v_0)=0$). We write  $E^*$ to denote the edges of OPT and we write $\es$ to denote an edge of OPT. Analogously, we use $\ss:E^*\to\R_{\geq 0}$ to denote the slack vector that we will construct for OPT edges.

Throughout this section we study $\eta$-near minimum cuts of $G=(V,E,z)$ 
Note that these cuts are $2\eta$-near minimum cuts w.r.t., $x$. For every such near minimum cut, $(S,\overline{S})$, we identify the cut with the side, say $S$, such that $u_0,v_0\notin S$. Equivalently, we can identify these cuts with an interval along the optimum cycle, OPT, that does not contain $u_0,v_0$. 

We will use ``left" synonymously with ``clockwise" and ``right" synonymously with ``counterclockwise." We say a vertex is to the left of another vertex if it is to the left of that vertex and to the right of edge $e_0=(u_0,v_0)$. Otherwise, we say it is to the right (including the root itself in this case).

\begin{definition}[Crossed on the Left/Right, Crossed on Both Sides]
	For two crossing near minimum cuts $S,S'$, we say $S$ {\em crosses $S'$ on the left} if the leftmost endpoint of $S$ on the optimal cycle is to the left of the leftmost endpoint of $S$. Otherwise, we say {\em $S$ crosses $S'$ on the right}. 
	
	
	A near minimum cut is {\em crossed on both sides} if it is crossed on both the left and the right.
	We also say a a near minimum cut is {\em crossed on one side} if it is either crossed on the left or on the right, but not both.
\end{definition}

\subsection{Cuts Crossed on Both Sides}\label{sec:crossedBothSides}

The following theorem is the main result of this section:
\begin{theorem}\label{thm:cutsbothsides}
Given OPT TSP tour with set of edges $E^*$, and  a feasible LP solution $x^0$ of the \ref{eq:tsplp} with support $E_0=E\cup \{e_0\}$ and let $x$ be $x^0$ restricted to $E$.  For {\em any} distribution $\mu$ of spanning trees with marginals $x$ and $\decrease > 0$, if $\eta <1/100$, then there is a random vector $s^*:E^*\to \R_{\geq 0}$ (the randomness in $s^*$ depends exclusively on $T\sim\mu$) such that 
\begin{itemize}
\item For any vector $s:E\to\R$ where $s_e\geq -x_e \decrease$ for all $e$ and for any $\eta$-near minimum cut $S$ w.r.t., $z=(x+OPT)/2$ crossed on both sides where $\delta(S)_T$ is odd, we have
$s(\delta(S))+s^*(\delta(S)) \geq 0$;
\item For any $\es\in E^*$, $\E{s^*_{\es}} \leq 37\eta \decrease$.
\end{itemize}
\end{theorem}


\begin{figure}[htb]
\begin{center}
\begin{tikzpicture}[inner sep=1.7pt,scale=.7,pre/.style={<-,shorten <=2pt,>=stealth,thick}, post/.style={->,shorten >=1pt,>=stealth,thick}]

\node at (-3,-2.5) () {$L(e^*)$};
\node at (3,-2.5) () {$R(e^*)$};
\node at (0,-3) () {$e^*$};

\node at (-1.2,-3.2) () {$u$};
\node at (1.2,-3.2) () {$v$};


\tikzstyle{every node} = [draw, circle,color=red];
\foreach \i in {1,...,8}{
\path (22.5+\i*45:3) node  (a_\i) {};
}

\draw [color=black,rotate around={45*4.5:(4.3*45:1.6)},line width=1.2] (4*45:0.8) ellipse (0.9 and 2.7);
\draw [color=black,rotate around={45*7.5:(4.3*45:1.6)},line width=1.2] (4.25*45:-2.1) ellipse (0.9 and 2.7);


\foreach \a/\b in {1/2, 2/3, 3/4, 4/5, 5/6, 6/7, 7/8, 8/1}{
\path (a_\a) edge  (a_\b);
}

\end{tikzpicture}
\end{center}
\caption{$L$ and $R$ for an OPT edge $e^*$.}
\label{fig:opt-augmentation}
\end{figure}
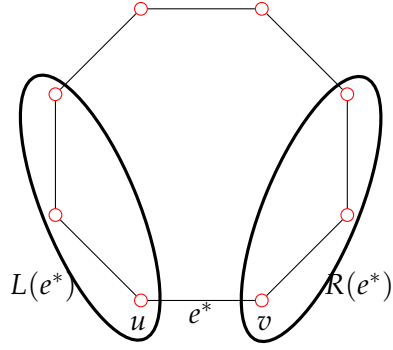


For an OPT edge $\es=(u,v)$, let $L(\es)$  be the largest $\eta$-near minimum cut  (w.r.t. $z$) containing $u$ and not $v$ which is crossed on both sides. Let $R(\es)$ be the largest near minimum cut containing $v$ and not $u$ which is crossed on both sides. (Note that $L(\es),R(\es)$ do not necessarily exist). For example, see \cref{fig:opt-augmentation}.



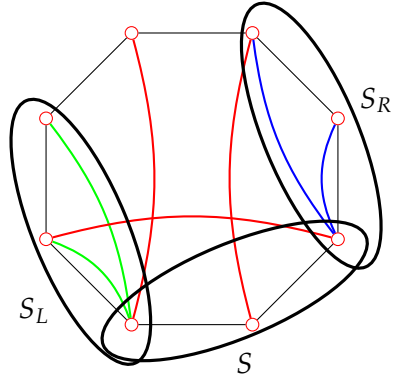
\begin{figure}[htb]
\begin{center}
\begin{tikzpicture}[inner sep=1.7pt,scale=.7,pre/.style={<-,shorten <=2pt,>=stealth,thick}, post/.style={->,shorten >=1pt,>=stealth,thick}]

\node at (-3,-2.5) () {$S_L$};
\node at (3.5,1.5) () {$S_R$};

\node at (1,-3.5) () {$S$};

\tikzstyle{every node} = [draw, circle,color=red];
\foreach \i in {1,...,8}{
\path (22.5+\i*45:3) node  (a_\i) {};
}

\path (a_5) edge [thick, bend right=25, green] (a_4);
\path (a_5) edge [thick, bend right=15, green] (a_3);

\path (a_7) edge [thick, bend left=25, blue] (a_8);
\path (a_7) edge [thick, bend left=15, blue] (a_1);

\path (a_7) edge [thick, bend right=15, red] (a_4);
\path (a_6) edge [thick, bend left=15, red] (a_1);
\path (a_5) edge [thick, bend right=15, red] (a_2);

\draw [color=black,rotate around={45*4.5:(4.3*45:1.6)},line width=1.2] (4*45:0.8) ellipse (0.9 and 2.7);
\draw [color=black,rotate around={45*8.5:(1*45:1.6)},line width=1.2] (4.25*45:-2.1) ellipse (0.9 and 2.7);

\draw [color=black,rotate around={45*6.5:(4.3*45:1.6)},line width=1.2] (1.1*45:1.5) ellipse (0.9 and 2.7);

\foreach \a/\b in {1/2, 2/3, 3/4, 4/5, 5/6, 6/7, 7/8, 8/1}{
\path (a_\a) edge  (a_\b);
}

\end{tikzpicture}
\end{center}
\caption{$S$ is crossed on the left by $S_L$ and on the right by $S_R$. In green are edges in $\delta(S)_L$, in blue edges in $\delta(S)_R$, and in red are edges in $\delta(S)_O$. }
\label{fig:crossed-both-sides}
\end{figure}

 \begin{definition}\label{def:SLR}
 For a near minimum cut $S$ that is crossed on both sides let $S_L$ be the near minimum cut crossing $S$ on the left which minimizes the intersection with $S$, and similarly for $S_R$; if there are multiple sets crossing $S$ on the left with the same minimum intersection, choose the smallest one to be $S_L$ (and similar do for $S_R$).

We partition $\delta(S)$ into three sets $\delta(S)_L, \delta(S)_R$ and $\delta(S)_O$ as in \cref{fig:crossed-both-sides} such that
 \begin{align*}
 	\delta(S)_L &= E(S \cap S_L, S_L \smallsetminus S) \\
 	 \delta(S)_R &= E(S \cap S_R, S_R \smallsetminus S) \\
 	 \delta(S)_O &= \delta(S) \smallsetminus (\delta(S)_L \cup \delta(S)_R)
 \end{align*}

 \end{definition}
For an OPT edge $e^*$ define an (increase) event (of second type) $\cI_2(e^*)$ as the event that at least one of the following {\em does not} hold. (If $L(e^*)$ does not exist, assume the first and third events always hold; similarly if $R(e^*)$ does not exist, assume the second and fourth events always hold.)
\begin{equation}\label{opt-increase-event}
	|T\cap \delta(L(e^*))_R| = 1, |T\cap \delta(R(e^*))_L| = 1, T\cap \delta(L(e^*))_O = \emptyset, \text{ and }T\cap \delta(R(e^*))_O = \emptyset. 
\end{equation}
In the proof of \cref{thm:cutsbothsides} we will increase an OPT edge $e^*$ whenever $\cI_2(e^*)$ occurs.

\begin{lemma}\label{lem:OPT-edge-increase-prob-xed-both-sides}
For any OPT edge $e^*$, $\P{\cI_2(e^*)}\leq 18\eta$.
\end{lemma}
\begin{proof}	Fix $e^*$.
To simplify notation we abbreviate $L(e^*),R(e^*)$ to $L,R$.
Since $L$ is crossed on both sides, $L_L, L_R$ are well defined. Since by \cref{lem:cutdecrement} $L_L\cap L, L_L\smallsetminus L$ are $4\eta$-near min cuts and $L$ is $2\eta$-near mincut with respect to $x$, by \cref{lem:treeoneedge}, $\P{|T\cap \delta(L)_L)|=1}\geq 1-5\eta$. Similarly, $\P{|T\cap \delta(R)_L| = 1}\geq 1-5\eta$. 
On the other hand, since $L, L_L, L_R$ are $2\eta$-near min cuts, by \cref{lem:nmcuts_largeedges}, $x(E(L \cap L_R, L_R)), x(E(L \cap L_L, L_L)) \ge 1-\eta$. Therefore 
$$x(\delta(L)_O) \le 2+2\eta - x(E(L \cap L_R, L_R)) - x(E(L \cap L_L, L_L)) \le 4\eta.$$ 
It follows that $\P{T\cap \delta(L)_O = \emptyset}\geq  1-4\eta$. Similarly, $\P{T\cap \delta(R)_O = \emptyset}\geq  1-4\eta$.
	Finally,	by the union bound, all events occur simultaneously with probability at least $1-18\eta$. So, $\P{\cI_2(e^*)}\leq 18\eta$ as desired.
%
	\end{proof}

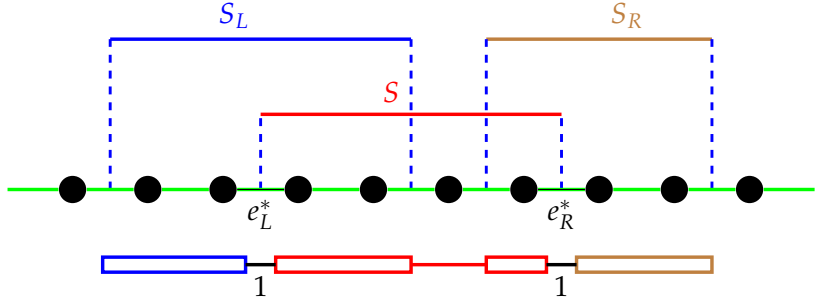
\begin{figure}[htb]\centering
\begin{tikzpicture}
	\foreach \i in {1,...,10}{
		\node [circle,fill=black] at (\i,0) (\i) {};
	}
	\node at (0,0) (0) {}; \node at (11,0) (11) {};
	\draw [line width=1.3pt,color=red] (3.5,1) -- node [above left] {$S$} (7.5,1);
	\draw [line width=1.3pt,color=blue] (1.5,2) -- node [above left] {$S_L$} (5.5,2);
	\draw [line width=1.3pt,color=brown] (6.5,2) -- node [above right] {$S_R$} (9.5,2);
	\draw [dashed,line width=1.1pt,color=blue] (1.5,0) -- (1.5,2) (3.5,0) -- (3.5,1) (6.5,0) -- (6.5,2) (7.5,0) -- (7.5,1) (9.5,0) -- (9.5,2) (5.5,0) -- (5.5,2);
	\foreach \i/\j in {0/1,1/2,2/3,3/4,4/5,5/6,6/7,7/8,8/9,9/10,10/11}{
		\draw [color=green,line width=1.3pt] (\i) edge (\j);
		}
	\draw (3) edge node [below] {$e_L^*$} (4) (7) edge node [below] {$e_R^*$} (8);
	\draw [color=blue,line width=1.3pt] (1.4,-1.1) rectangle (3.3,-0.9);
	\draw [color=red,line width=1.3pt] (3.7,-1.1) rectangle (5.5,-0.9)
		(6.5,-1.1) rectangle (7.3,-0.9);
	\draw[color=brown,line width=1.3pt] (7.7,-1.1) rectangle (9.5,-0.9);
	\draw [line width=1.3pt] (3.3,-1)-- node [below] {$1$} (3.7,-1)
		(7.3,-1) -- node [below] {$1$} (7.7,-1);
	\draw [color=red,line width=1.3pt] (5.5,-1) -- (6.5,-1);
\end{tikzpicture}
%
%
%
%
%
%
%
\caption{Setting of \cref{lem:xed-both-sides-one-increases}. Here we zoom in on a portion of the optimal cycle and assume the root is not shown. If $\cI_2(e_L^*)$ does not occur then $E(S\cap S_L, S_L\smallsetminus S)_T=1$.}
\label{fig:xed-both-sides-proof}
\end{figure}

\begin{lemma}\label{lem:xed-both-sides-one-increases} Let $S$ be a cut which is crossed on both sides and let $e^*_L,e^*_R$ be the OPT edges on its interval where $e^*_L$ is the edge further clockwise. Then, if $\delta(S)_T \not=2$, at least one of $\cI_2(e^*_L),\cI_2(e^*_R)$  occurs.
\end{lemma}
\begin{proof}
We prove by contradiction. Suppose none of $\cI_2(e^*_L),\cI_2(e^*_R)$ occur; we will show that this implies $\delta(S)_T = 2$. 

Let $R=R(e^*_L)$; note that $S$ is a candidate for $R(e^*_L)$, so $S\subseteq R$. Therefore, $S_L = R_L$ and we have
$$\delta(R)_L = E(R \cap R_L, R_L \smallsetminus R) = E(R \cap S_L, S_L \smallsetminus R) = \delta(S)_L.$$
where we used $S\cap S_L=R\cap S_L$ and that $S_L \smallsetminus S = S_L \smallsetminus R$.
Similarly let $L=L(e^*_R)$, and, we have $\delta(L)_R=\delta(S)_R$.

Now, since $\cI_2(e^*_L)$ has not occurred,
$1 = |T\cap \delta(R)_L| =|T\cap \delta(S)_L|,$
and since $\cI_2(e^*_R)$ has not occurred,
$1 = |T\cap \delta(L)_R|= |T\cap \delta(S)_R|,$
where  $L = L(e_R^*)$.
So, to get $\delta(S)_T=2$, it remains to show that  $T\cap \delta(S)_O = \emptyset$.
Consider any edge $e=(u,v)\in \delta(S)_O$ where  $u \in S$. We need to show $e\notin T$. Assume that $v$ is to the left of $S$ (the other case can be proven similarly). 
Then $e \in \delta(R)$. So, since $e$ goes to the left of $R$, either $e \in E(R \cap R_L, R_L \smallsetminus R)$ or $e \in \delta(R)_O$. But since $e\notin \delta(S)_L = \delta(R)_L$, we must have $e\in \delta(R)_O$. So, since $\cI_2(e^*_L)$ has not occurred, $e\notin T$ as desired.
%
\end{proof}

\begin{proof}[Proof of \cref{thm:cutsbothsides}]
For any OPT edge $e^*$  whenever $\cI_2(e^*)$ occurs, define $s^*_{e^*}=2.02\decrease $. Then, 
by \cref{lem:OPT-edge-increase-prob-xed-both-sides}, 
$\E{s_{e^*}} \leq 18 \cdot 2.02\beta$ and for any $2\eta$-near min cut $S$ (w.r.t., $x$) that is crossed on both sides if $\delta(S)_T$ is odd, then at least one of $\cI_2(e^*_L),\cI_w(e^*_R)$ occurs, so
$$ s(\delta(S)) + s^*(\delta(S)) \geq  - x(\delta(S))\beta + s^*_{e^*_L}+s^*_{e^*_R} \geq -(2+2\eta)\decrease + 2.02\decrease \geq 0$$
for $\eta<1/100$ as desired.
\end{proof}


\subsection{Proof of the Main Technical Theorem, \cref{thm:maintechnical}}
The following theorem is the main result of this section.
\begin{restatable}{theorem}{beforetechnical}\label{lem:beforetechnicalthm}
Let $x^0$ be a feasible solution of the \ref{eq:tsplp} with support $E_0=E\cup\{e_0\}$ and $x$ be $x^0$ restricted to $E$.
Let $\mu$ be the max entropy distribution with marginals $x$.
For $\eta\leq 10^{-12}$, $\decrease > 0$, there is a set $E_g\subset E\smallsetminus \delta(\{u_0,v_0\})$  of {\em good} edges
and two functions $s: E_0\rightarrow \R$ and $s^*: E^* \rightarrow \R _{\ge 0}$ (as functions of $T\sim\mu$) such that
	\begin{itemize}
\item[(i)] 	For each edge $e \in E_g$, $s_e \ge -x_e \decrease$ and for any $e\in E\smallsetminus E_g$, $s_e=0$.
\item[(ii)] For each $\eta$-near-min-cut $S$ w.r.t. $z$, if $\delta(S)_T$ is odd, then $  s(\delta(S)) + s^*(\delta(S)) \ge  0.$
\item[(iii)] We have $\E{s_e} \le -\epsilon_P \decrease x_e$  for all edges $e \in E_g$  and $\E{s^*_{e^*}} \le 218 \eta \decrease$ for all OPT edges $e^* \in E^*.$
for $\epsilon_P$ defined in \eqref{eq:epsP}.
\item[(iv)] For every $\eta$-near minimum cut $S$ of $z$ crossed on (at most) one side such that $S\neq V\smallsetminus \{u_0,v_0\}$, $x(\delta(S)\cap E_g) \ge 3/4.$
\end{itemize}
\end{restatable}
Before proving this theorem we use it to prove the main technical theorem from the previous section.
\maintechnical*
\begin{proof}[Proof of \cref{thm:maintechnical}]
Let $E_g$ be the good edges defined in \cref{lem:beforetechnicalthm} and let $E_b:=E\smallsetminus E_g$ be the set of bad edges; in particular, note all edges in $\delta(\{u_0,v_0\})$ are bad edges. We define a new vector $\tilde{s}:E\cup \{e_0\}\to\R$ as follows: 
\begin{equation}\label{eq:tildesdef}\tilde{s}(e)\gets \begin{cases}\infty & \text{if } e=e_0\\
-x_e(4\beta/5)(1-2\eta) & \text{if } e\in E_b,\\
x_e(4\beta/3) & \text{otherwise.}
\end{cases}
\end{equation}
Let $\tilde{s}^*$ be the vector $s^*$ from \cref{thm:cutsbothsides}.
We claim that for any  $\eta$-near minimum cut $S$ such that $\delta(S)_T$ is odd, we have 
$$ \tilde{s}(\delta(S))+\tilde{s}^*(\delta(S))\geq 0.$$
To check this note by (iv) of \cref{lem:beforetechnicalthm} for every set $S\neq V\smallsetminus \{u_0,v_0\}$ crossed on at most one side, we have $x(E_g \cap \delta(S)) \ge \frac{3}{4}$, so
\begin{equation}\label{eq:tildeScutsok}\tilde{s}(\delta(S))+\tilde{s}^*(\delta(S))\geq \tilde{s}(\delta(S)) = \frac{4\decrease}{3} x(E_g\cap\delta(S)) - \frac{4\decrease}{5}(1-2\eta)x(E_b\cap\delta(S))\geq 0.	
\end{equation}

For $S=V\smallsetminus \{u_0,v_0\}$, we have $\delta(S)_T=\delta(u_0)_T + \delta(v_0)_T=2$ with probability 1, so condition ii) is satisfied for these cuts as well.
Finally, consider cuts  $S$ which are crossed on both sides. By \cref{thm:cutsbothsides},
\begin{equation}\label{eq:tildeScutstwo}\tilde{s}(\delta(S)) + \tilde{s}^*(\delta(S)) \geq 0	
\end{equation}
since $\tilde{s}_e\geq -\frac{4}{5}\decrease x_e \ge -\decrease x_e$ for all $e$.

Now, we are ready to define $s,s^*$. Let $\hat{s},\hat{s}^*$ be the $s,s^*$ of \cref{lem:beforetechnicalthm} respectively.
Define $s=\gamma \tilde{s} + (1-\gamma) \hat{s}$ and similarly define $s^*=\gamma\tilde{s}^*+(1-\gamma)\hat{s}^*$ for some $\gamma$ that we choose later. 
We prove all three conclusions for $s,s^*$. (i) follows by (i) of \cref{lem:beforetechnicalthm} and  \cref{eq:tildesdef}. (ii) follows by (ii) of \cref{lem:beforetechnicalthm} and \cref{eq:tildeScutsok} above. 
It remains to verify (iii). For any OPT edge $e^*$, $\E{s^*_{e^*}}\leq 218\eta\decrease$ by (iii) of \cref{lem:beforetechnicalthm} and the construction of $\tilde{s}^*$. On the other hand, by (iii) of \cref{lem:beforetechnicalthm} and \cref{eq:tildesdef},
\begin{align*} 
\E{s_e} \begin{cases}\leq  x_e  (\gamma \frac{4}{3}\decrease - (1-\gamma)\eps_P\decrease ) & \forall e\in E_g,\\	
= -x_e\gamma\cdot (\frac{4}{5}\decrease)(1-2\eta)& \forall e\in E_b.
\end{cases}
\end{align*}
Setting $\gamma=\frac{15}{32}\eps_P$ we get $\E{s_e}\leq -\frac{1}{3}\eps_P\decrease x_e$ for $e\in E_g$ and $\E{s_e}\leq -\frac{1}{3}x_e\decrease\epsilon_P$ for $e\in E_b$ as desired.
\end{proof}

\subsection{Structure of Polygons of Cuts Crossed on One Side}\label{sec:crossedOneSide}

\begin{definition}[Connected Component of Crossing Cuts]
	Given a family of cuts crossed on at most one side, construct a graph where two cuts are connected by an edge  if they cross. Partition this graph into {\em maximal connected components}.
	We call a path in this graph, a {\em path of crossing cuts}.
\end{definition}

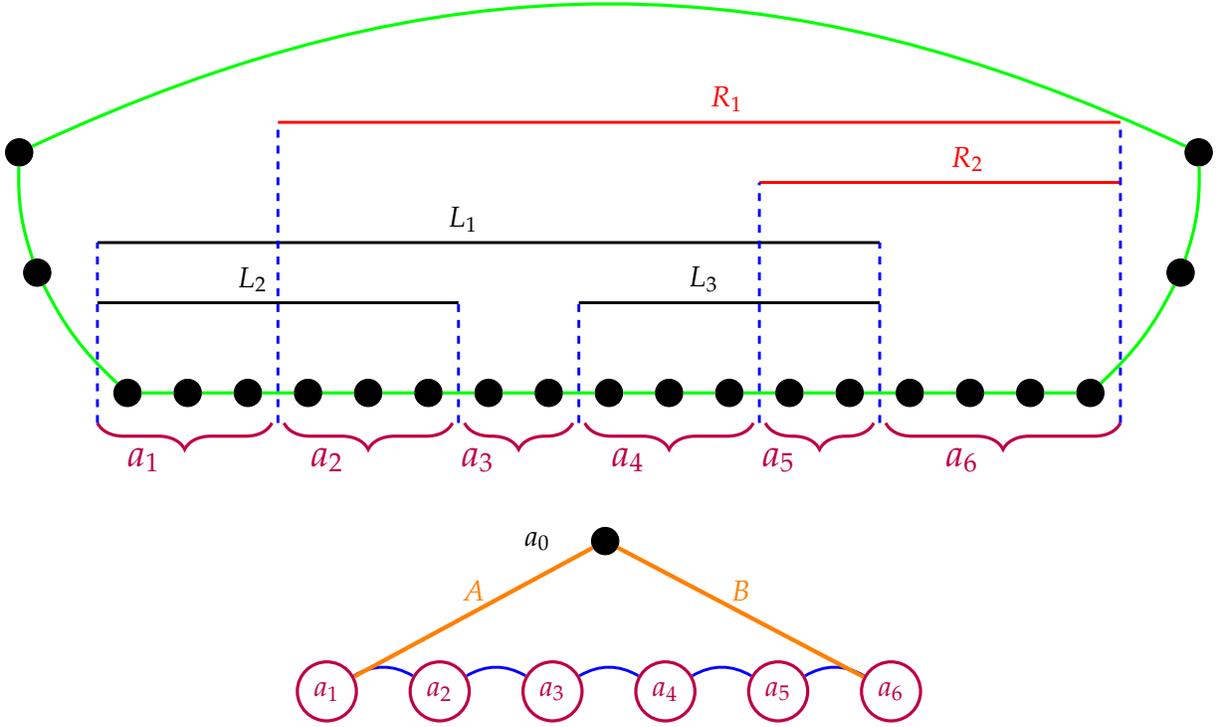
\begin{figure}[htb]
	\centering
	\begin{tikzpicture}[scale=0.8]
		\foreach \i in {3,...,19}{
			\node [draw,fill=black,circle] at (\i,0) (\i) {};
		}
		\node [draw,fill=black,circle] at (1.2,4) (1) {};
		\node [draw,fill=black,circle] at (1.5,2) (2) {};
		\node [draw,fill=black,circle] at (20.5,2) (20) {};
		\node [draw,fill=black,circle] at (20.8,4) (21) {};

	   \foreach \i/\j in {3/4,4/5,5/6,6/7,7/8, 8/9, 9/10, 10/11, 11/12, 12/13, 13/14, 14/15, 15/16, 16/17, 17/18, 18/19}{
	   	\draw [color=green, line width=1.2pt] (\i) edge (\j);
	   }
	   \draw [color=green, line width=1.2pt] (1) edge [bend left=25] (21)
	   (1) edge [bend right=10] (2) (2) edge [bend right=10] (3) (19) edge [bend right=10] (20) (20) edge [bend right=10] (21);
		\draw [line width=1.1pt] (2.5,2.5) -- node [above left] {$L_1$} (15.5,2.5) 
		(2.5,1.5) -- node [above left] {$L_2$} (8.5,1.5)
		(10.5,1.5) -- node [above left] {$L_3$} (15.5,1.5)
		;
		\draw[ line width=1.1pt, color=red] (5.5,4.5) -- node [above right] {$R_1$} (19.5,4.5)
		(13.5,3.5) -- node [above right] {$R_2$} (19.5,3.5);
		\foreach \i/\a/\b in {1/2.5/5.4, 2/5.6/8.4, 3/8.6/10.4, 4/10.6/13.4, 5/13.6/15.4, 6/15.6/19.5}{
			\draw [line width=1.2pt, color=purple,decorate,decoration={brace,amplitude=10pt}]
(\b,-0.5) -- node [below left=5] {\Large$a_\i$} (\a,-0.5);
}		
		\draw [dashed,color=blue,line width=1.1pt] (2.5,-0.5) -- (2.5,2.5)
		(5.5,-0.5) -- (5.5,4.5) (8.5,-0.5) -- (8.5,1.5) (10.5,-0.5) -- (10.5,1.5)  (19.5,-0.5) -- (19.5,4.5) (13.5,-0.5) -- (13.5,3.5) (15.5,-0.5) -- (15.5,2.5);
	\end{tikzpicture}\vspace{0.5cm} 
	\begin{tikzpicture}
	  \node [draw,fill=black,circle] at (3.7,2) (a0) {};
	  \foreach \i/\j in {0/1, 1/2, 2/3, 3/4, 4/5, 5/6}{
	  	\node [draw,circle,color=purple,line width=1.2pt] at (1.5*\i,0) (a\j) {$a_{\j}$};
	  }
	 \foreach \i/\j in {1/2, 2/3, 3/4, 4/5, 5/6}{
	  	\draw [line width=1.1pt, color=blue] (a\i) edge [bend left=30] (a\j);
	  }
	  \node at (2.8,2) () {$a_0$};
	\draw [line width=1.6pt, color=orange] (a1) edge node [above] {$A$} (a0) (a6) edge node [above] {$B$} (a0); 
	\end{tikzpicture}
	\caption{An example of a polygon with contracted atoms. In black are the cuts in the left polygon hierarchy, in red the cuts in the right polygon hierarchy. OPT edges around the cycle are shown in green.
	Here $R_1$ is an ancestor of $R_2$, however it is not a strict ancestor of $R_2$ since they have the same right endpoint. $L_1$ is a strict ancestor and the strict parent of $L_3$. By \cref{thm:poly-structure}, every edge in the bottom picture represents a set of LP edges of total fraction at least $1-\eps_\eta$.  }
\label{fig:polygon-example-detail}
\end{figure}

In the rest of this section we will focus on a single connected component ${\cal C}$ of cuts crossed on (at most) one side.
\begin{definition}[Polygon]\label{def:polygon}
	For a connected component ${\cal C}$ of crossing near min cuts that are crossed on one side, let 
$a_0,\dots,a_{m-1}$  be the coarsest partition of the vertices $V$	, such that for all $0\leq i\leq m-1$ and for any $A\in {\cal C}$ either $a_i\subseteq A$ or $a_i\cap A=\emptyset$. These are called atoms. We assume $a_0$ is the atom that contains the special edge $e_0$, and we call it the {\em root}. 
	Note that for any $A\in {\cal C}, a_0\cap A=\emptyset$. 
	
	Since every cut $A\in {\cal C}$ corresponds to an interval of vertices in $V$ in the optimum Hamiltonian cycle, we can arrange $a_0,\dots,a_{m-1}$ around a cycle (in the counter clockwise order). We label the arcs in this cycle from 1 to m, where $i+1$ is the arc connecting $a_i$ and $a_{i+1}$  (and $m$ is the name of the arc connecting $a_{m-1}$ and $a_0$). Then every cut $A\in {\cal C}$ can be identified by the two arcs surrounding its atoms. 
	 Specifically, $A$ is identified with arcs $i,j$ (where $i<j$) if $A$ contains  atoms $a_{i},\dots, a_{j-1}$, and we write  $\ell(A)=i, r(A)=j$. Note that  $A$ does not contain the root $a_0$. 
	
	By construction for every arc $1\leq i\leq m$, there exists a cut $A$ such that $\ell(A)=i$ or $r(A)=i$. Furthermore, $A,B\in {\cal C}$ (with $\ell(A)\leq \ell(B)$) cross iff $\ell(A) < \ell(B)<r(A)<r(B)$.
	
	See \cref{fig:polygon-example-detail} for a visual example.
\end{definition}

Notice that every atom of a polygon is an interval of the optimal cycle.
In this section, we prove the following structural theorem about polygons of near minimum cuts crossed on one side. 

\begin{theorem}[Polygon Structure]\label{thm:poly-structure}
For  $\eps_{\eta}\geq 14\eta$ and any polygon with atoms $a_0...a_{m-1}$ (where $a_0$ is the root) the following holds:
	\begin{itemize}
	\item For all adjacent atoms $a_i,a_{i+1}$ (also including $a_0,a_{m-1}$), we have $x(E(a_i,a_{i+1})) \ge 1-\eps_\eta$. 
	\item All atoms $a_i$ (including the root) have $x(\delta(a_i)) \le 2+\eps_\eta$. 
	\item $x(E(a_0, \{a_2,\dots,a_{m-2}\}))\leq \eps_{\eta}$.
	\end{itemize}
\end{theorem}
The interpretation of this theorem is that the structure of a polygon converges to the structure of an actual integral cycle as $\eta\to 0$. 
The proof of the theorem follows from the lemmas in the rest of this subsection.



\begin{definition}[Left and Right  Hierarchies]
For a polygon $u$ corresponding to a connected component ${\cal C}$ of cuts crossed on one side, let ${\cal L}$ (the {\em left  hierarchy}) be the set of all  cuts $A\in {\cal C}$ that are not crossed on the left.
We call any cut in ${\cal L}$ {\em open} on the left. Similarly, we let ${\cal R}$ be the set of  cuts that are open on the right.
So, ${\cal L},{\cal R}$ is a partitioning of all cuts in ${\cal C}$. 

For two distinct cuts $A,B\in {\cal L}$ we say $A$ is an {\em ancestor} of $B$ in the left polygon hierarchy if $A\supseteq B$. We say $A$ is a {\em strict} ancestor of $B$ if, in addition, $\ell(A) \not= \ell(B)$. We define the right  hierarchy similarly: $A$ is a strict ancestor of $B$ if $A \supseteq B$ and $r(A) \not= r(B)$. 

We say $B$ is a {\em strict parent} of $A$ if among all strict ancestors of $A$ in the (left or right) hierarchy, $B$ is the one closest to $A$.

See \cref{fig:polygon-example-detail} for examples of sets and their parent/ancestor relationships. 
\end{definition}

\begin{fact}\label{lem:same-hieararchy-no-cross}
If $A,B$ are in the same hierarchy and they are not ancestors of each other, then $A\cap B=\emptyset$.	
\end{fact}
\begin{proof}
If $A\cap B\neq \emptyset$ then they cross. So, they cannot be open on the same side.
\end{proof}
This lemma immediately implies that the cuts in each of the left (and right) hierarchies form a laminar family.

\begin{lemma}\label{lem:ABcommoncross}
For $A, B\in {\cal R}$  where $B$ is a strict parent of $A$, there exists a cut $C\in {\cal L}$ that crosses both $A,B$. Similarly, if $A,B \in {\cal L}$ and $B$ is a strict parent of $A$, there exists a cut $C \in {\cal R}$ that crosses $A,B$.
\end{lemma}
\begin{proof}
Since we have a connected component of near min cuts, there exists a path of crossing cuts from $A$ to $B$. Let $P=(A=C_0, C_1,\dots, C_k=B)$ be the shortest such path. We need to show that $k=2$. 

First, since $C_1$ crosses $C_0$ and $C_0$ is open on right, we have
$$ \ell(C_1) < \ell(C_0) < r(C_1) < r(C_0).$$
Let $I$ be the {\em closed} interval $[\ell(C_1),r(C_0)]$. Note that $C_k=B$ has an endpoint that does not belong to $I$. Let $C_i$ be the {\em first} cut in the path with an endpoint not in $I$ (definitely $i>1$).
This means $C_{i-1}\subseteq I$; so, since $C_{i-1}$ crosses $C_i$,  exactly one of the endpoints of $C_i$ is strictly inside $I$. We consider two cases:

{\bf Case 1: $r(C_{i}) > r(C_0)$.} In this case, $C_i$ must be crossed on the left (by $C_{i-1}$) and $C_{i}\in {\cal R}$ and it does not cross $C_0$. So, $C_0\subsetneq C_i$ and
 $$\ell(C_1) < \ell(C_i)\leq \ell(C_0)$$
 where the first inequality uses that the left endpoint of $C_i$ is strictly inside $I$. Therefore, $C_1$ crosses both of $C_0,C_i$, and $C_i$ is a strict ancestor of $A=C_0$. If $C_i=B$ we are done, otherwise, $A\subseteq B \subseteq C_i$, but since $C_1$ crosses both  $A$ and $C_i$, it also crosses $B$ and we are done.

{\bf Case 2: $\ell(C_i)<\ell(C_1)$.} In this case, $C_i$ must be crossed on the right (by $C_{i-1}$) and $C_i\in {\cal L}$ and it does not cross $C_1$. So, we must have
$$ r(C_1) \leq r(C_i) < r(C_0), $$
where the second inequality uses that the right endpoint of $C_i$ is strictly inside $I$. But, this implies that $C_i$ also crosses $C_0$. So, we can obtain a shorter path by excluding all cuts $C_1,\dots, C_{i-1}$ and that is a contradiction.
\end{proof}

\begin{lemma}\label{lem:ABcommonancestor}
Let $A,B\in {\cal R}$ such that $A\cap B=\emptyset$, i.e., they are not ancestors of each other. Then, they have a common ancestor, i.e., there exists a set $C\in {\cal R}$ such that $A,B\subseteq C$.	
\end{lemma}
\begin{proof}
WLOG assume $r(A)\leq \ell(B)$.
Let $C$ be the highest ancestor of $A$ in the hierarchy, i.e., $C$ has no ancestor. For the sake of contradiction suppose $B\cap C=\emptyset$ (otherwise, $C$ is an ancestor of $B$ and we are done). So, $r(C)\leq \ell(B)$. Consider the path of crossing cuts from $C$ to $B$, say $C=C_0,\dots,C_k=B$. 

Let $C_i$ be the first cut in this path such that $r(C_i)>r(C_0)$. Note that such a cut always exists as $r(B)>r(C)$. 
Since $C_{i-1}$ crosses $C_i$ and $r(C_{i-1})\leq r(C_0)$, $C_{i-1}$ crosses $C_i$ on the left and $C_i$ is open on the  right. We show that $C_i$ is an ancestor of $C=C_0$ and we get a contradiction to $C_0$ having no ancestors (in ${\cal R}$).
If $\ell(C_0)<\ell(C_i)$, then $C_i$ crosses $C_0$ on the right and that is a contradiction. So, we must have $C_0\subseteq C_i$, i.e., $C_i$ is an ancestor of $C_0$.
\end{proof}
It follows from the above lemma that each of the left and right hierarchies have a unique cut  with no ancestors.

\begin{lemma}\label{lem:strictparentexists}
If $A$ is a cut in ${\cal R}$ such that $r(A)<m$, then $A$ has a strict ancestor. And, similarly, if $A\in{\cal L}$ satisfies $\ell(A)>1$, then it has a strict ancestor.
\end{lemma}
\begin{proof}
Fix a cut $A\in {\cal R}$. If there is a cut in $B\in {\cal R}$ such that $r(B)>r(A)$, then either $B$ is a strict ancestor of $A$ in which case we are done, or $A\cap B=\emptyset$, but then by \cref{lem:ABcommonancestor} $A,B$ have a common ancestor $C$, and $C$ must be a strict ancestor of $A$ and we are done.

Now, suppose for any $R\in {\cal R}$, $r(R)\leq r(A)$. So, there must be a cut $B\in {\cal L}$ such that $r(B)>r(A)$ (otherwise we should have less than $m$ atoms in our polygon). The cut $B$ must be crossed on the right by a cut $C\in {\cal R}$. But then, we must have $r(C)>r(B)>r(A)$ which is a contradiction. 
\end{proof}

\begin{corollary}\label{cor:leftmostrightmostancestor}
	If $A\in {\cal C}$ has no strict ancestor, then $r(A)=m$ if $A\in {\cal R}$ and $\ell(A)=1$ otherwise.
\end{corollary}

\begin{restatable}[Polygons are Near Minimum Cuts]{lemma}{lemNMC}\label{lem:NMC}
$x(\delta(a_1\cup\dots\cup a_{m-1}))\leq 2+4\eta.$ 
\end{restatable}
\begin{proof}
	Let $A\in {\cal L}$ and $B\in {\cal R}$ be the unique cuts in the left/right  hierarchy with no ancestors. Note that $A$ and $B$ are crossing (because there is a cut $C$ that crosses $A$ on the right, and $B$ is an ancestor of $C$). Therefore, since $A,B$ are both $2\eta$ near min cuts (with respect to $x$), by \cref{lem:cutdecrement}, $A\cup B$ is a $4\eta$ near min cut.
\end{proof}

\begin{restatable}[Root Neighbors]{lemma}{lemrootededges}\label{lem:rootedgesarebig}
$x(E(a_0, a_1)), x(E(a_0, a_{m-1})) \geq 1-2\eta$.	
\end{restatable}
\begin{proof}
Here we prove $x(E(a_0,a_1))\geq 1-2\eta$. One can prove $x(E(a_0, a_{m-1})) \ge 1-2\eta$   similarly. Let $A\in {\cal L}$ and $B\in {\cal R}$ be the unique cuts in the left/right hierarchy with no ancestors.
First, observe that if $\ell(B)=2$, then since $A,B$ are crossing, by \cref{lem:nmcuts_largeedges} we have
$$x(E(A\smallsetminus B, \overline{A\cup B})) = x(E(a_1, a_0)) \geq 1-\eta.$$
as desired.

By definition of atoms, there exists a cut $C\in{\cal C}$ such that either $\ell(C)=2$ or $r(C)=2$; but if $r(C)=2$ we must have $\ell(C)=1$ in which case $C$ cannot be crossed, so this does not happen. So, we must have  $\ell(C)=2$. 	
If $C\in {\cal R}$, then since $C$ is a descendent of $B$, we must have $\ell(B)=2$, and we are done by the previous paragraph.

Otherwise, suppose $C\in {\cal L}$. 
We claim that $B$ crosses $C$. This is because, $C$ is crossed on the right by some cut $B'$ and $B$ is an ancestor of $B'$, so $B\cap C\neq \emptyset$ and $C\not\subseteq B$ since $\ell(B)>2$. 
Therefore, by \cref{lem:cutdecrement} $B\cup C$ is a $4\eta$ near min cut. Since $A$ crosses $B\cup C$, by \cref{lem:nmcuts_largeedges} we have
$$x(E(A\smallsetminus (B\cup C), \overline{A\cup B\cup C})) = x(E(a_1, a_0)) \geq 1-2\eta$$ 
as desired.
\end{proof}

\begin{lemma}\label{lem:polygonedgesarebig}
For any pair of atoms $a_i,a_{i+1}$ where $1\leq i\leq m-2$ we have $x(\delta(\{a_i,a_{i+1}\}))\leq 2+12\eta$, so $x(E(a_i,a_{i+1}))\geq 1-6\eta$.	
\end{lemma}
\begin{proof}
We prove the following claim: There exists $j\leq i$ such that $x(\delta(\{a_j,\dots,a_{i+1}\}))\leq 2+6\eta$. 
Then, by a similar argument we can find $j'\geq i+1$ such that $x(\delta(\{a_i,\dots,a_{j'}\}))\leq 2+6\eta$. By \cref{lem:cutdecrement} it follows that $x(\delta(\{a_i,a_{i+1}\}))\leq 2+12\eta$. Since $x(\delta(a_i)),x(\delta(a_{i+1}))\geq 2$, we have
$$x(\delta(\{a_i,a_{i+1}\})) + 2x(E(a_i,a_{i+1})) \ge 4.$$
But due to the bound on $x(\delta(\{a_i,a_{i+1}\}))$ we must have $x(E(a_i,a_{i+1}))\geq 1-6\eta$ as desired.

It remains to prove the claim. 	
First, observe that there is a cut $A$ separating $a_{i+1},a_{i+2}$ (Note that if $i+1=m-1$ then  $a_{i+2}=a_0$); so, either $\ell(A)=i+2$ or $r(A)=i+2$. If $r(A)=i+2$ then, $A$ is the cut we are looking for and we are done. So, assume $\ell(A)=i+2$. 

{\bf Case 1: $A\in {\cal L}.$}
Let $L\in {\cal L}$ be the strict parent of $A$. If $\ell(L)\leq i$ then we are done (since there is a cut $R\in\cR$ crossing $A,L$ on the right so $L\smallsetminus (A\cup R)$ is the cut that we want. If $\ell(L)=i+1$, then let $L'$ be the strict parent of $L$).
Then, there is a cut $R\in\cR$ crossing $A,L$ and a cut 
$R'$ crossing $L,L'$. 
First, since both $R,R'$ cross $L$ (on the right) they have a non-empty intersection, so one of them say $R'$ is an ancestor of the other ($R$) and therefore $R'$ must intersect $A$.
On the other hand, since $R'$ crosses $L$ and $\ell(L)=i+1$, $\ell(R')\geq i+2=\ell(A)$. Since $R'$ intersect $A$, either they cross, or $A\subseteq R'$, so we must have $x(\delta(A\cup R))\leq 2+4\eta$.
Finally, since $R'$ crosses $L'$ (on the right) we have $x(\delta (L' \smallsetminus (A \cup R))) \le 2 + 6 \eta$ and $L'\smallsetminus (A\cup R)$ is our desired set.

{\bf Case 2: $A\in {\cal R}.$}
We know that $A$ is crossed on the left by, say,  $L\in {\cal L}$. 
If $\ell(L) \le i$, we are done, since then $L\smallsetminus A$ is the cut that we seek and we get $x(\delta(L\smallsetminus A))\leq 2+4\eta$.

Suppose then that $\ell(L) = i+1$. Let $L'$ be the strict parent of $L$, which must have $\ell(L') \le i$. 
If $L'$ crosses $A$, then $L' \smallsetminus A$ is the cut we seek and we get $x(\delta(L\smallsetminus A))\leq 2+4\eta$.

Finally, if $L'$ doesn't cross $A$, i.e., $r(A) \le r(L')$,
then consider the cut $R \in \cR$ that crosses $L$ and $L'$ on the right. Since $r(L) < r(A)$, and $A$ is not crossed on the right, it must be that $\ell(R) = i+2$.
In this case, $L' \smallsetminus R$ is the cut we want, and we get
$x(\delta(L'\smallsetminus R))\leq 2+4\eta$.
\end{proof}

\begin{restatable}[Atoms are Near Minimum Cuts]{lemma}{lematomsNMC}\label{lem:smallatoms}
For any $1\leq i\leq m-1$, we have $x(\delta(a_i))\leq 2+14\eta.$ 
\end{restatable}
\begin{proof}
By \cref{lem:polygonedgesarebig}, $x(\delta(\{a_i,a_{i+1}\}))\leq 2+12\eta$ (note that in the special case $i=m-1$ we take the pair $a_{i-1},a_i$). There must be a $2\eta$-near minimum cut $C$ (w.r.t., $x$) separating $a_{i}$ from $a_{i+1}$. Then either $a_i=C\cap \{a_i,a_{i+1}\}$ or $a_i=\{a_i,a_{i+1}\}\smallsetminus C$. In either case, we get $x(\delta(a_i))\leq 2+14\eta$ by \cref{lem:cutdecrement}.
\end{proof}

\def\cH{\mathcal{H}}

\subsection{Happy Polygons}
\label{sec:happypolygons}
\begin{definition}[$A,B,C$-Polygon Partition]\label{def:abcpolygonpartitioning}
Let $u$ be a polygon with atoms $a_0,\dots,a_{m-1}$ with root  $a_0$ where $a_1,a_{m-1}$ are the atoms left and right of the root.
The $A,B,C$-polygon partition of $u$ is a partition of edges of $\delta(u)$ into sets $A=E(a_1,a_0)$ and $B=E(a_{m-1},a_0)$, $C=\delta(u)\smallsetminus A\smallsetminus B$. 
\end{definition}
Note that by \cref{thm:poly-structure}, $x(A),x(B)\geq 1-\eps_\eta$ and $x(C)\leq \eps_\eta$ where we set 
\begin{align}\label{def:eps-eta}
	\eps_\eta= 14\eta
\end{align}
as needed for \cref{thm:poly-structure}.
\begin{definition}[Leftmost and Rightmost cuts]
	Let $u$ be a polygon with atoms $a_0,\dots,a_{m-1}$ and arcs labelled $1,\dots,m$ corresponding to a connected component ${\cal C}$ of $\eta$-near minimum cuts (w.r.t., $z$).  We call any cut $C\in {\cal C}$ with $\ell(C)=1$ a {\em leftmost} cut of $u$ and any cut $C\in {\cal C}$ with $r(C)=m$ a {\em rightmost} cut of $u$. 
		We also call $a_1$ the leftmost atom of $u$ (resp.  $a_{m-1}$ the rightmost atom).
\end{definition}
Observe that by \cref{cor:leftmostrightmostancestor}, any cut that is not a leftmost or a rightmost cut has a strict ancestor. 

\begin{definition}[Happy Polygon]
Let $u$ be a polygon with polygon partition $A,B,C$.
For a spanning tree $T$, we say that $u$ is {\em happy} if 
$$A_T \text{ and } B_T \text{ odd}, C_T=0.$$

We say that $u$ is {\em left-happy} (respectively {\em right-happy}) if 
$$A_T \text{ odd}, C_T=0,$$
(respectively $B_T \text{ odd}, C_T=0$).
\end{definition}

\begin{definition}[Relevant Cuts]
Given a polygon $u$ corresponding to a connected component ${\cal C}$ of cuts crossed on one side with atoms $a_0,\dots,a_{m-1}$, define a family of relevant cuts
$${\cal C}'={\cal C}\cup \{a_i:  1\leq i\leq m-1, z(\delta(a_i))\leq 2+\eta\}.$$
\end{definition}
Note that atoms of $u$ are always $\eps_\eta/2$-near minimum cuts w.r.t., $z$ but not necessarily $\eta$-near minimum cuts.
The following theorem is the main result of this section.
\begin{restatable}[Happy Polygons and Cuts Crossed on One Side]{theorem}{thmcrossedoneside}\label{thm:crossed-one-side}
	Let $G=(V,E,x)$ for $x$ be an LP solution and  $z=(x+OPT)/2$.
	For a connected component ${\cal C}$ of near minimum cuts of $z$, let $u$ be the polygon with atoms $a_0,a_1...a_{m-1}$ with polygon partition $A,B,C$. 
	For $\mu$ an arbitrary distribution of spanning trees with marginals $x$, $\decrease > 0$, there is a random vector $s^*:E^*\to\R_{\geq 0}$ (as a function of $T\sim\mu$) such that for any vector $s:E\to\R$ where $s_e\geq -\decrease x_e$  for all $e\in E$ the following holds:
	
	\begin{itemize}
		\item If $u$ is happy then, for any cut $S\in {\cal C}'$ if  $\delta(S)_T$ is odd then we have $s(\delta(S))+s^*(\delta(S))\geq 0$,
		\item For any $S\in {\cal C}'$ that is not a rightmost/leftmost cut or rightmost/leftmost atom, if $\delta(S)_T$ is odd, then we have 
		$s(\delta(S))+s^*(\delta(S))\geq 0$. 
		\item For all OPT edges $e^*_2,\dots,e^*_{m-1}$ with respect to the above polygon, $\E{s^*_{e^*_i}}\leq 181\eta \decrease$. $\E{s^*_{e^*}} = 0$ for all other OPT edges.
	\end{itemize}
\end{restatable}

Before proving the above theorem, we study a special case.
\begin{lemma}[Triangles as Degenerate Polygons]\label{lem:trianglereduction}
	Let $S=X\cup Y$ where $X,Y,S$ are $\eps_{\eta}$-near min cuts (w.r.t., $x$) and each of these sets is a contiguous interval around the OPT cycle. Then, viewing $X$ as $a_1$ and $Y$ as $a_2$ (and $a_0=\overline{X\cup Y}$) the above theorem holds viewing $S$ as a degenerate polygon. 
\end{lemma}
\begin{proof}
In this case $A=E(a_1,a_0), B=E(a_2,a_0), C=\emptyset$.
For the OPT edge $e^*$ between $X,Y$ we define $\cI_1(e^*)$ to be the event that at least one of $T\cap E(X)$, $T\cap E(Y)$, $T\cap E(S)$ is not a tree.
Whenever this happens we define $s^*_{e^*}=2.05 \cdot \decrease$.
If $S$ is left-happy we need to show when $\delta(X)_T$ is odd, then $s(\delta(X))+s^*(\delta(X))\geq 0$. This is because when $S$ is left-happy we have $A_T$ is odd (and $C_T=0$), so either $\cI_1(e^*)$ does not happen and $\delta(X)_T$ is even, or it happens in which case $s(\delta(X))+s^*(\delta(X))\geq 0$ as $s(\delta(X))\ge -(2+2\eta)\decrease$ and $s^*_{e^*}=2.05\decrease$.
Finally, observe that by \cref{lem:treeoneedge}, 
$\P{ {\cI_1(e^*)}}\le 3 \eps_\eta$, so
$\E{s^*_{e^*}}=3\eps_\eta \cdot 2.05\decrease \leq 87\eta \decrease$ using $\eta < 1/100$ and $\eps_\eta$ as defined in \cref{def:eps-eta}.
\end{proof}




\begin{lemma}\label{lem:Astrictpeven}
For every cut $A\in {\cal C}$ that is not a leftmost or a rightmost cut, $\P{\delta(A)_T=2} \geq 1-22\eta$.
\end{lemma}
\begin{proof}
Assume $A\in \cR$; the other case can be proven similarly. 
Let $B$ be the strict parent of $A$. By \cref{lem:ABcommoncross} there is a cut $C\in{\cal L}$ which crosses $A,B$ on their left. It follows by \cref{lem:cutdecrement} that $C \smallsetminus A, C\cap A$ are $4\eta$ near minimum cuts (w.r.t., $x$). So, by \cref{lem:treeoneedge}, $\P{E(A\cap C, C\smallsetminus A)_T=1}\geq 1-5\eta$. 
On the other hand, $B\smallsetminus (A\cup C)$ is a $6\eta$ near minimum cut and $A\smallsetminus C, B\smallsetminus C$ are $4\eta$ near min cuts (w.r.t., $x$). So, by \cref{lem:treeoneedge} $\P{E(A\smallsetminus C, B\smallsetminus (A\cup C))_T=1}\geq 1-7\eta$.

Finally,  by  \cref{lem:nmcuts_largeedges}, $x(E(A\cap C, C\smallsetminus A)), x(E(A\smallsetminus C, B\smallsetminus (A\cup C))) \geq 1-3\eta$. Since $A$ is a $2\eta$ near min cut (w.r.t., $x$), all remaining edges have fractional value at most $8\eta$, so with probability $1-8\eta$, $T$ does not choose any of them. Taking a union bound over all of these events, $\P{\delta(A)_T=2}\geq 1-22\eta$.
\end{proof}

\begin{lemma}\label{lem:atoms-even-whp}
	For any atom $a_i\in {\cal C}'$ that is not the leftmost or the rightmost atom we have
	$$\P{\delta(a_i)_T = 2} \ge 1-42\eta.$$
\end{lemma}
\begin{proof}
	By \cref{lem:polygonedgesarebig}, $x(\delta(\{a_i,a_{i+1}\}))\leq 2+12\eta$, 
	and by \cref{lem:smallatoms}, $x(\delta(a_{i+1}))\leq 2+14\eta$ (also recall by the assumption of lemma $x(\delta(a_i))\leq 2+2\eta$,
	 Therefore, by \cref{lem:treeoneedge}, 
	 $$\P{E(a_{i},a_{i+1})_T=1},\P{E(a_{i-1},a_i)_T=1}\geq 1-14\eta,$$
	 where the second inequality holds similarly.
Also, by \cref{lem:polygonedgesarebig}, $x(E(a_{i-1},a_{i})),x(E(a_{i},a_{i+1}))	 \geq 1-6\eta$.
Since $x(\delta(a_i))\leq 2+2\eta$, $x(E(a_i, \overline{a_{i-1}\cup a_i\cup a_{i+1}}))\leq 14\eta$.
So, 
$$\P{T\cap E(a_i, \overline{a_{i-1}\cup a_i\cup a_{i+1}})=\emptyset}\geq 1-14\eta.$$
Finally,  by the union bound all events occur with probability at least $1-42\eta$. 
\end{proof}

Let $e^*_1,\dots,e^*_m$ be the OPT edges mapped to the arcs $1,\dots,m$ of the component $\cal C$ respectively.
\begin{lemma}\label{lem:4cutmapping}
There is a {\em mapping}\footnote{Each cut will be mapped to one or two OPT edges.} of cuts in ${\cal C}'$  to OPT edges $e^*_2,\dots e^*_{m-1}$ such that each OPT edge has at most 4 cuts mapped to it, an OPT edge $e^*$ is mapped to a cut $S$ only if $e^* \in \delta(S)$, and every atom of the polygon in ${\cal C}'$ gets mapped to two (not necessarily distinct) OPT edges. 
\end{lemma}
\begin{proof}
Consider first the set of cuts in $ {\cal C}'_\cR:= \cR \cup \{a_i:  1\leq i\leq m-1, z(\delta(a_i))\leq 2+\eta\}$ and similarly $ {\cal C}'_\cL:= \cL \cup \{a_i:  1\leq i\leq m-1, z(\delta(a_i))\leq 2+\eta\}$.  Observe that this is also a laminar family. Note that atoms are in both $\cC'_\cR$ and $\cC'_\cL$.
We  define a map from cuts in ${\cal C}'_R$  to OPT edges such that every OPT edge $e^*_2,\dots,e^*_{m-1}$ gets at most $2$ cuts mapped to it. A similar argument works for cuts in $\cC'_\cL$. 

For any  $2\leq i\leq m-1$, we map 
$$\argmax_{A\in \cC'_\cR: \ell(A)=i} |A|\text{ and }\argmax_{A\in \cC'_\cR: r(A)=i} |A|$$ 
to $e^*_i$, where recall $\ell(A)$ is the OPT edge leaving $A$ on the left side and $r(A)$ the OPT edge leaving on the right. 
By construction, each OPT edge gets at most two cuts mapped to it. 

Furthermore, we claim every cut $A\in {\cal C}'_\cR$ gets mapped to at least one OPT edge.
For the sake of contradiction let $A\in {\cal C}'_\cR$ be a cut that is not mapped to any OPT edge. First note that $a_1$ is mapped to edge $e_2^*$ (in both hierarchies) and $a_{m-1}$ is mapped to edge $e_{m-1}^*$. Otherwise, if $A\in \cR$, $\ell(A)\neq 1$. Furthermore, if $A \in \cR$ and $r(A)=m$, then $A$ is definitely the largest cut with left endpoint $\ell(A)$.
So assume, $1<\ell(A) < r(A)<m$.
Let $B=\argmax_{B\in \cC'_\cR: \ell(B)=\ell(A)} |B|$ and let $C=\argmax_{B\in \cC'_\cR: r(C)=r(A)} |C|$. Since $A$ is not mapped to any OPT edge but $B,C$ are mapped by above definition, we must have $B,C\neq A$. But that implies $A\subsetneq B,C$. And this means $B,C$ cross; but this is a contradiction with $\cR$ being a laminar family.
\end{proof}


\begin{definition}[Happy Cut]We say a leftmost cut $L\in \cL$  is {\em happy} if
$$E(L,\overline{a_0 \cup L})_T = 1$$
Similarly, the leftmost atom $a_1$ is {\em happy} if $E(a_1,\overline{a_0\cup a_1})_T=1$. Define  rightmost cuts in $u$ or the rightmost atom in $u$ to be happy, similarly.
\end{definition}
Note that, by definition, if leftmost cut $L$ is happy and $u$ is left happy then $L$ is even, i.e., $\delta(L)_T=2$. Similarly, $a_1$ is even if it is happy and $u$ is left-happy.

\begin{lemma}\label{lem:even-cuts-cond-on-happy}
For every leftmost or rightmost cut $A$  in $u$ that is an $\eta$-near min cut w.r.t. $z$, $\P{A\text{ happy}}\geq 1-10\eta$, and for the leftmost atom $a_1$ (resp. rightmost atom $a_{m-1}$), if it is an $\eta$-near min cut then  $\P{a_1\text{ happy}}\geq 1-24\eta$ (resp. $\P{a_{m-1}\text{ happy}}\geq 1-24\eta$).
\end{lemma}
\begin{proof}
Recall that if $A$ is a $\eta$-near min cut w.r.t. $z$ then it is a $2\eta$-near min cut w.r.t. $x$.
Also, recall for a cut $L\in\cL$, $L_R$ is the near minimum cut crossing $L$ on the right that minimizes the intersection (see \cref{def:SLR}).
	We prove this for the leftmost cuts and the leftmost atom; the other case can be proven similarly. Consider a cut $L \in {\cal L}$. 
	Since by \cref{lem:cutdecrement} $L_R\cap L, L_R \smallsetminus L$ are $4\eta$ near min cuts (w.r.t., $x$) and $L_R$ is a $2\eta$ near min cut, by \cref{lem:treeoneedge}, $\P{E(L_R\cap L, L_R\smallsetminus L)_T=1}\geq 1-5\eta$.
	 On the other hand, by \cref{lem:nmcuts_largeedges}, $x(E(L_R\cap L, L_R\smallsetminus L)) \ge 1-\eta$, and by \cref{lem:rootedgesarebig}, $x(E(L,a_0)) \ge 1-2\eta$. It follows that 
	 $$x(\delta(L) \smallsetminus E(L_R \cap L, L_R \smallsetminus L) \smallsetminus E(L,a_0)) \le 5\eta$$
	 Therefore, by the union bound, $\P{L\text{ happy}}\geq 1-10\eta$, since if ($\delta(L) \smallsetminus E(L_R \cap L, L_R \smallsetminus L) \smallsetminus E(L,a_0))_T = 0$ and $E(L_R\cap L, L_R\smallsetminus L)_T=1$ then $E(L,\overline{a_0 \cup L})_T = 1$	 and therefore $L$ is happy.
	 
	Now consider the atom $a_1$, and suppose it is an $\eta$ near min cut. 
By \cref{lem:polygonedgesarebig}, $x(\delta(\{a_1,a_2\})) \le 2+12\eta$
  and by \cref{lem:smallatoms}, $x(\delta(a_2))\leq 2+14\eta$.
	Therefore, by \cref{lem:treeoneedge}, $\P{E(a_1,a_2)_T=1}\geq 1-14\eta$.
	On the other hand, by  \cref{lem:polygonedgesarebig},	$x(E(a_1,a_2)) \ge 1-6\eta$ and by \cref{lem:rootedgesarebig}, $x(E(a_1,a_0)) \ge 1-2\eta$.
	Therefore,
$$x(E(a_1,a_3\cup\dots\cup a_{m-1}))\leq 2+2\eta - (1-6\eta)-(1-2\eta))\leq 10\eta.$$
Observe, $a_1$ is happy when both of these events occur; so, by the union bound, $\P{a_1\text{ happy}}\geq 1-24\eta$ as desired.
\end{proof}


\begin{proof}[Proof of \cref{thm:crossed-one-side}]
Consider an OPT edge $e^*_i$ for $1<i<m$.
For the at most four cuts mapped to $e^*_i$ in \cref{lem:4cutmapping}, we define the following three events:
\begin{enumerate}[i)]
\item A leftmost cut assigned to $e^*_i$ is not happy. (Equivalently, a leftmost cut $L \in \cL \cap \cC'$ with $r(L)=i$ is not happy.)
\item A rightmost cut assigned to $e^*_i$ is not happy. (Equivalently, a rightmost cut $R \in \cR \cap \cC'$ with $l(R)=i$ is not happy.\footnote{Note in the special case that $i=2$, $L$ in (i) will be the leftmost atom if it is a near min cut, and similarly in (ii) when $i=m-1$, $R$ will be the rightmost atom if it is a near min cut.})
\item A cut which is not leftmost or rightmost assigned to $e^*_i$ is odd.
\end{enumerate}
Observe that the cuts in (i) and (ii) are assigned to $e^*_i$ in \cref{lem:4cutmapping}. We say an atom $a$ is singly-mapped to $e^*_i$ if in the matching $a$ is only mapped to $e^*_i$ once, otherwise we say it is doubly-mapped to $e^*_i$. 

We say an event $\cI_1(e^*_i)$ occurs if either (i), (ii), or (iii) occurs. If $\cI_1(e^*_i)$ occurs then we set:
\begin{align*}
	s^*_{e^*_i} = \begin{cases}2.05\decrease  & \text{If (i),(ii), or (iii) occurred for at least one non-atom cut in $\cC'$, or for an atom}\\ &\text{which is doubly-mapped to $e^*_i$} \\
		2.05\decrease/2 & \text{Otherwise.}
	 \end{cases}
\end{align*}
If $\cI_1(e^*_i)$ does not occur we set $s^*_{e^*_i} = 0$. 
First, observe that for any non-atom cut $S\in {\cal C}'$  that is not a leftmost or a rightmost cut, 
if $\delta(S)_T$ is odd, then if $e^*_i$ is the OPT edge that $S$ is  mapped to, it satisfies $s^*_{e^*_i}=2.05\decrease$, so
$$ s(\delta(S))+s^*(\delta(S))\geq -x(\delta(S))\decrease + s^*(e^*_i)\geq -(2+2\eta)\decrease+2.05\decrease\geq 0,$$
for $\eta<1/100$.
The same inequality holds for non-leftmost/rightmost atom cuts $a\in {\cal C}'$ which are doubly-mapped to $e^*_i$. For non-leftmost/rightmost atom cuts $a\in {\cal C}'$ which are singly-mapped to $e^*_i$, $a$ is mapped (possibly even twice) to another edge $e^*_j$ (note $j=i-1$ or $i+1$), and in this case $s^*(e^*_i) + s^*(e^*_j) \ge 2.05\decrease$, and again the above inequality holds.

Now, suppose for a leftmost cut $S\in\cL \cap \cC'$ with $r(S)=i$ has $\delta(S)_T$ odd. If $u$ is not left-happy there is nothing to prove. If $u$ is left-happy, then we must have $S$ is not happy (as otherwise $\delta(S)_T$ would be even), so $\cI_1(e^*_i)$ occurs, so similar to the above inequality $s(\delta(S))+s^*(\delta(S))\geq 0$.
The same holds for rightmost cuts and the leftmost/rightmost atoms in $\cC'$ (note leftmost/rightmost atoms are always doubly-mapped: $a_1$ to $e^*_2$ and $a_{m-1}$ to $e^*_{m-1}$).

It remains to upper bound $\E{s^*(e^*_i)}$ for $1<i<m$. By \cref{lem:4cutmapping} at most four cuts are mapped to $e^*_i$. Then, either there is an atom which is doubly-mapped to $e^*_i$ or there is not.

First suppose exactly one atom is doubly-mapped to $e^*_i$. Then there are at most three cuts mapped to $e^*_i$, including that atom. The probability of an event of type (i) or (ii) occurring for the leftmost or rightmost atom is at most $1-24\eta$ by \cref{lem:even-cuts-cond-on-happy}. Atoms which are not leftmost or rightmost are even with probability at least $1-42\eta$ by \cref{lem:atoms-even-whp}. Therefore, in the worst case, the doubly-mapped atom is not leftmost or rightmost. For the remaining two cuts, leftmost and rightmost cuts are happy with probability at least $1-10\eta$ by \cref{lem:even-cuts-cond-on-happy}, and (non-atom) non leftmost/rightmost cuts are even with probability at least $1-22\eta$ by \cref{lem:Astrictpeven}. Therefore in the worst case the remaining two (non-atom) cuts mapped to $e^*_i$ are not leftmost/rightmost. Therefore, if an atom is doubly-mapped to $e^*_i$,
$$\E{s^*(e^*_i)}\leq 42\eta \cdot 2.05 \decrease + 2\cdot 22\eta \cdot 2.05\decrease \le 177\eta \decrease$$
Note if two atoms are doubly-mapped to $e^*_i$, 
$$\E{s^*(e^*_i)}\leq 2\cdot 42\eta \cdot 2.05\decrease \le 173\eta\decrease$$

Otherwise, any atoms mapped to $e^*_i$ are singly-mapped. In this case, if only an atom cut is odd/unhappy, we set $s^*(e_i^*) = 2.05\decrease/2$. The probability of an event of type (i) or (ii) occurring for the leftmost or rightmost atom is at most $1-24\eta$ by \cref{lem:even-cuts-cond-on-happy}, so we can bound the contribution of this event to $\E{s^*(e_i^*)}$ by $24\eta \cdot 2.05\decrease/2$. Atoms which are not leftmost or rightmost are even with probability at least $1-42\eta$ by \cref{lem:atoms-even-whp}, and so we can bound their contribution by $42\eta \cdot 2.05\decrease/2$. Therefore, in the worst case four non-leftmost/rightmost \textit{non}-atom cuts are mapped to $e^*_i$, in which case,
$$\E{s^*(e^*_i)}\leq 4\cdot 22\eta \cdot 2.05\decrease = 181\eta\decrease$$ as desired.
\end{proof}

\subsection{Hierarchy of Cuts and Proof of \cref{lem:beforetechnicalthm}}\label{subsec:poly-summary}

\begin{definition}[Hierarchy]\label{def:hierarchy}
\hypertarget{tar:hierarchy}{For an LP solution $x^0$ with support $E_0=E\cup \{e_0\}$ and $x$ be $x^0$ restricted to $E$,  a hierarchy ${\cal H}$ is a {\em laminar} family of $\eps_\eta$-near min cuts of $G=(V,E,x)$ with root $V\smallsetminus \{u_0,v_0\}$, where every cut $S\in \cH$  is either a polygon cut (including triangles) or a degree cut and $u_0,v_0\notin S$. Furthermore, every cut $S$ is a union of its children.
For any (non-root) cut $S\in \cH$, define the parent of $S$, $\p(S)$, to be the smallest cut $S'\in\cH$ such that $S\subsetneq S'$.}

\hypertarget{tar:AS}{For a cut $S\in \cH$, 
let $\cA(S):=\{u\in \cH: \p(u)=S\}$. If $S$ is a polygon cut, then we can order cuts in $\cA(S)$, $u_1,\dots,u_{m-1}$ such that 
\begin{itemize}
\item $A=E(\overline{S},u_1), B=E(u_{m-1},\overline{S})$ satisfy $x(A),x(B)\geq 1-\eps_\eta$.
\item For any $1\leq i<m-1$, $x(E(u_i,u_{i+1}))\geq 1-\eps_\eta$.
\item $C=\cup_{i=2}^{m-2} E(u_i,\overline{S})$ satisfies $x(C)\leq \eps_\eta$.
\end{itemize}}

We call the sets $A,B,C$ the polygon partition of edges in $\delta(S)$. We say $S$ is left-happy when $A_T$ is odd and  $C_T=0$ and right happy when $B_T$ is odd and $C_T=0$ and happy when $A_T,B_T$ are odd and $C_T=0$.

We abuse notation, and for an (LP) edge $e=(u,v)$ that is not a neighbor of $u_0,v_0$, let $\p(e)$ denote the smallest\footnote{in the sense of the number of vertices that it contains} cut $S'\in \cH$ such that $u,v\in S'$. We say edge $e$ is a \textbf{bottom edge} if $\p(e)$ is a polygon cut and we say it is a \textbf{top edge} if $\p(e)$ is a degree cut.
\end{definition}
Note that when $S$ is a polygon cut $u_1,\dots,u_{m-1}$ will be the atoms $a_1,\dots,a_{m-1}$ that we defined in the previous section, but a reader should understand this definition independent of the polygon definition that we discussed before; in particular, the reader no longer needs to worry about the details of specific cuts $\cC$ that make up a polygon. 
Also, note that since $V\smallsetminus \{u_0,v_0\}$ is the root of the hierarchy, for any edge $e\in E$ that is not incident  to $u_0$ or $v_0$, $\p(e)$ is well-defined; so all those edges are either bottom or top, and edges which are incident to $u_0$ or $v_0$ are neither bottom edges nor top edges.

The following observation is immediate from the above definition.
\begin{observation}\label{obs:childofSD}
	For any polygon cut $S\in\cH$, and any cut $S'\in\cH$ which is a descendant of $S$ let $D=\delta(S')\cap \delta(S)$. If $D\neq \emptyset$, then exactly one of the following is true: $D\subseteq A$ or $D\subseteq B$ or $D\subseteq C$.
\end{observation}

\begin{restatable}[Main Payment Theorem]{theorem}{paymentmain}
\label{thm:payment-main}
For an LP solution $x^0$ and $x$ be $x^0$ restricted to $E$ and a hierarchy $\cH$ for some $\eps_\eta\leq 10^{-10}$ and any $\decrease > 0$,
the maximum entropy distribution $\mu$ with marginals $x$ satisfies the following:
\begin{enumerate}[i)]
\item There is a set of {\em good} edges $E_g\subseteq E\smallsetminus \delta(\{u_0,v_0\})$ such that any bottom edge $e$ is in $E_g$ and  for any (non-root) $S\in \cH$ such that $\p(S)$ is  a degree cut, we have $x(E_g\cap \delta(S))\geq 3/4$. 
\item There is a random vector $s:E_g \to \R$  (as a function of $T\sim\mu$) such that for all $e$, $s_e\geq -x_e \decrease$ (with probability 1), and \label{payment:non-near-min-cuts}
\item If a polygon cut $u$ with polygon partition $A,B,C$ is not left happy, then for any set $F\subseteq E$ with $\p(e)=u$ for all $e\in F$ and $x(F)\geq 1-\eps_\eta/2$, we have
$$ s(A)+s(F)+s^-(C)\geq 0,$$
where $s^-(C)=\sum_{e\in C} \min\{s_e,0\}$.
A similar inequality holds if $u$ is not right happy.
\item 
For every cut $S\in \cH$ such that $\p(S)$ is not a polygon cut, if $\delta(S)_T$ is odd, then $s(\delta(S))\geq 0$. \label{payment:satisfy-non-poly-cuts}
\item 
For a good edge $e\in E_g$, $\E{s_e} \le  - \eps_P \decrease x_e$ (see \cref{eq:epsP} for definition of $\eps_P$) . 
\end{enumerate}
\end{restatable}

The above theorem is the main part of the paper in which we use that $\mu$ is a SR distribution. See \cref{sec:payment} for the proof. We use this theorem to construct a random vector $s$ such that essentially for all cuts $S\in\cH$ in the hierarchy $z/2+s$ is feasible; furthermore for a large fraction of ``good'' edges we have that $\E{s_e}$ is negative and bounded away from $0$. 

As we will see in the this subsection, using part (iii) of the theorem we will be able to show that every leftmost and rightmost cut of any polygon is satisfied.

In the rest of this section we use the above theorem to prove \cref{lem:beforetechnicalthm}. 
We start by explaining how to construct $\cH$.
Given the vector $z=(x+OPT)/2$ run the following procedure on the OPT cycle with  the family of $\eta$-near minimum cuts of $z$ that are crossed on at most one side:

For every connected component ${\cal C}$ of $\eta$ near minimum cuts (w.r.t., $z$) crossed on at most one side, if $|{\cal C}|=1$ then add the unique cut in ${\cal C}$ to the hierarchy. Otherwise, ${\cal C}$ corresponds to a polygon $u$ with atoms $a_0,\dots,a_{m-1}$ (for some $m>3$). Add $a_1,\dots,a_{m-1}$\footnote{Notice that an atom may already correspond to a connected component, in such a case we do not add it in this step.} and $\cup_{i=1}^{m-1} a_i$ to $\cH$. Since every vertex except $u_0,v_0$ has degree 2, they all appear in the hierarchy as singletons. Therefore, every set in the hierarchy is the union of its children. 
Note that since $z(\delta(\{u_0,v_0\}))=2$, the root of the hierarchy is always $V\smallsetminus \{u_0,v_0\}$.


Now, we name every cut  in the hierarchy.
For a cut $S$ if there is a connected component of at least two cuts  with union equal to $S$, then call $S$  a polygon cut with the $A,B,C$ partitioning as defined in \cref{def:abcpolygonpartitioning}.
If $S$ is a cut with exactly two children $X,Y$ in the hierarchy, then also call $S$ a polygon cut\footnote{Think about such set as a {\em degenerate} polygon with atoms $a_1:= X,a_2:= Y,a_0 :=\overline{X\cup Y}$. So, for the rest of this section we call them triangles and in later section we just think of them as polygon cuts.}, $A=E(X,\overline{X}\smallsetminus Y)$, $B=E(Y,\overline{Y}\smallsetminus X)$ and $C=\emptyset$.
Otherwise, call $S$ a degree cut.

\begin{fact}The above procedure produces a valid hierarchy for $\eps_\eta \ge 14\eta$.
\end{fact}
\begin{proof}
First observe that whenever $|{\cal C}|=1$ the unique cut in ${\cal C}$ is a $2\eta$ near min cut (w.r.t, $x$) which is not crossed. For a polygon cut $S$ in the hierarchy, by \cref{lem:NMC}, the set $S$ is a $\eps_\eta$ near min cut w.r.t., $x$. If $S$ is an atom of a polygon, then by \cref{lem:smallatoms} $S$ is a $\eps_\eta$ near min cut.

Now, it remains to show that for a polygon cut $S$ we have a valid ordering $u_1,\dots,u_k$ of cuts in $\cA(S)$. If $S$ is a non-triangle polygon cut, the $u_1,\dots,u_k$ are exactly atoms of the polygon of $S$ and $x(A),x(B)\geq 1-\eps_\eta$ and $x(C)\leq \eps_\eta$ and $x(E(u_i,u_{i+1}))\geq 1-\eps_\eta$ follow by \cref{thm:poly-structure}.
For a triangle cut $S=X\cup Y$    because $S,X,Y$  are $\eps_\eta$-near min cuts (by the previous paragraph), we get $x(A),x(B)\geq 1-\eps_\eta$ as desired, by \cref{lem:shared-edges}. Finally, since $x(\delta(X)),x(\delta(Y))\geq 2$ we have $x(E(X,Y))\geq 1-\eps_\eta$.
\end{proof}

The following observation is immediate:
\begin{observation}
Each cut $S\in \cH$ corresponds to a contiguous interval around OPT cycle. 
For a polygon $u$ (or a triangle) with atoms $a_0,\dots,a_{m-1}$ for $m\geq 3$ we say an OPT edge $e^*$ is {\em interior} to $u$ if $e^*\in E^*(a_i,a_{i+1})$ for some $1\leq i\leq m-2$.
Any OPT edge $e^*$ is interior to at most one polygon.
\end{observation}



\beforetechnical*
\begin{proof}
For $\eps_\eta$ as in \cref{def:eps-eta}, let $E_g, s$ be as defined in \cref{thm:payment-main}, and let {$s_{e_0}=\infty$}.
Also, let $s^*$ be the sum of the $s^*$ vectors from \cref{thm:cutsbothsides} and \cref{thm:crossed-one-side}.
(i) follows (ii) of \cref{thm:payment-main}. 
$\E{s^*_{e^*}}\leq 218\eta\decrease$ follows from \cref{thm:cutsbothsides} and \cref{thm:crossed-one-side} and the fact that every OPT edge is interior to at most one polygon.
 Also, $\E{s_e}\leq -\eps_P \decrease x_e$ for edges $e\in E_g$ follows from (v) of \cref{thm:payment-main}.
 
 Now, we verify (iv): For any (non-root) cut $S\in \cH$ such that $\p(S)$ is not a polygon cut $x(\delta(S)\cap E_g)\geq 3/4$ by (i) of \cref{thm:payment-main}. The only remaining $\eta$-near minimum cuts are sets $S$ which are either atoms or near minimum cuts in the component ${\cal C}$ corresponding to a polygon $u$. So, by \cref{lem:shared-edges}, $x(\delta(S)\cap \delta(u))\leq 1+\eps_\eta$. By (i) of \cref{thm:payment-main} all edges in $\delta(S)\smallsetminus \delta(u)$ are in $E_g$. Therefore, $x(\delta(S)\cap E_g))\geq 1-\eps_\eta\geq 3/4$.
 
 It remains to verify (ii): We consider 4 groups of cuts:
 
{\bf Type 1}: Near minimum cuts $S$ such that $e_0\in \delta(S)$. Then, since $s_{e_0}=\infty$, $s(\delta(S))+s^*(\delta(S))\geq 0$.
 
 {\bf Type 2}: Near minimum cuts $S\in {\cH}$ where $\p(S)$ is not a polygon cut. By (iv) of \cref{thm:payment-main} and that $s^*\geq 0$ the inequality follows.
  
 {\bf Type 3:} Near minimum cuts $S$ crossed on both sides. Then, the inequality follows by \cref{thm:cutsbothsides} and the fact that $s_e\geq -\decrease x_e$ for all $e\in E$.
 
  {\bf Type 4:} Near minimum cuts $S$  that  are crossed on one side (and not in $\cH$) or  $S\in \cH$ and $\p(S)$ is a (non-triangle) polygon cut.
In this case $S$ must be an atom or a $\eta$-near minimum cut (w.r.t., $z$) in some polygon $u \in \cH$. If $S$ is not a leftmost cut/atom or a rightmost cut/atom, then the inequality follows by \cref{thm:crossed-one-side}. Otherwise, say $S$ is a leftmost cut. If $u$ is left-happy then by \cref{thm:crossed-one-side} the inequality is satisfied. Otherwise, for $F=\delta(S)\smallsetminus \delta(u)$, by \cref{lem:shared-edges}, we have $x(F)\geq 1-\eps_\eta/2$. Therefore, by (iii) of \cref{thm:payment-main} we have
$$ s(\delta(S))+s^*(\delta(S))\geq  s(A)+s(F)+s^-(C) \geq 0$$
as desired. 
Note that since $S$ is a leftmost cut, we always have $A\subseteq \delta(S)$. But $C$ may have an unpredictable intersection with $\delta(S)$; in particular, in the worst case only edges of $C$ with negative slack belong to $\delta(S)$. 
A similar argument holds when $S$ is the leftmost atom or a rightmost cut/atom.

	{\bf Type 5:} Near min cut $S$ is the leftmost atom or the rightmost atom of a triangle $u$. This is similar to the previous case except we use \cref{lem:trianglereduction} to argue that the inequality is satisfied when $u$ is left happy.
\end{proof}

\subsection{Hierarchy Notation}
\label{sec:hierarchynotation}
In the rest of the paper we will not work with $z$, OPT edges, or the notion of polygons. So, practically, by \cref{def:hierarchy}, from now on,  a reader can just think of every polygon as a triangle.
In the rest of the paper we adopt the following notation.

 We abuse notation and call any $u\in\hyperlink{tar:AS}{\cA(S)}$ an atom of $S$. 
\begin{definition}[Edge Bundles, Top Edges, and Bottom Edges]\label{def:edgebundletopcut}
 For every degree cut $S$ and every pair of atoms $u,v\in\cA(S)$,  we define a \textbf{top edge bundle} $\bbf=(u,v)$ such that 
	$$\bbf= \{e=(u',v') \in E : \p(e) = S, u'\in u, v'\in v\}.$$
	 Note that in the above definition, $u',v'$ are actual vertices of $G$.
	 
For every polygon cut $S$, we define the \textbf{bottom edge bundle} $\bbf=\{e: \p(e)=S\}$. 
\end{definition}
We will always use bold letters to distinguish top edge bundles from actual LP edges. Also, we abuse notation and write $x_\bbe:=\sum_{f\in\bbe}x_f$ to denote the total fractional value of all edges in this bundle.

In the rest of the paper, unless otherwise specified, we work with edge bundles and sometimes we just call them edges.

 For any $u\in\cH$ with $\p(u)=S$ we write
\begin{align*} 
\delta^\uparrow(u)&:=\delta(u)\cap\delta(S),\\
\delta^\rightarrow(u)&:=\delta(u)\smallsetminus\delta(S).\\
\hypertarget{tar:ErightarrowS}{E^\rightarrow (S)&:= \{e=(u_i, u_j): u_i,u_j\in\cA(S), u_i\neq u_j\}}.
\end{align*}
Also, for a set of edges $A\subseteq \delta(u)$ we write $A^\rightarrow, A^\uparrow$ to denote $A\cap\delta^\rightarrow(u), A\cap\delta^\uparrow(u)$ respectively (when $u$ is clear in context).
Note that $E^\rightarrow(S)\subseteq E(S)$ includes only edges between atoms of $S$ and not all edges between vertices in $S$.  

Finally, for a set of edges $F$ and an edge bundle $\bbe$, we define $F_{-\bbe} = F \smallsetminus \bbe$, and similarly $F_{+\bbe} = F \cup \bbe$.

%% file: probabilistic.tex
\section{Probabilistic statements}
\label{sec:probabilistic}

\input{gurvits.tex}

\input{maxflow.tex}

\subsection{Good Edges}



\begin{definition}[Half Edges]\label{def:halfedges}
We say an edge bundle $\bbe={\bf (u,v)}$ in a degree cut $S\in\cH$, i.e., $\p(\bbe)=S$, is a {\em half edge} if $|x_\bbe-1/2|\leq \eps_{1/2}$, where $\eps_{1/2}$ is defined in \ref{eq:constants}.
\end{definition}

\begin{definition}[Good Edges]\label{def:goodedges}
	We say a top edge bundle $\bbe={\bf(u,v)}$ in a degree cut $S\in\cH$ is (2-2) good, if one of the following holds:
	\begin{enumerate}
		\item $\bbe$ is not a half edge or
		\item $\bbe$ is a half edge and $\P{\delta(u)_T=\delta(v)_T= 2 | u,v\text{ trees}} \geq 3\eps_{1/2}.$
	\end{enumerate}
	We say a top edge $\bbe$ is bad otherwise. 
	We say every bottom edge bundle is good (but generally do not refer to bottom edges as good or bad). We say any edge $e$ that is a neighbor of $u_0$ or $v_0$ is bad.
\end{definition}

In the next subsection we will see that for any top edge bundle $\bbe={\bf (u,v)}$ which is not a half edge, $\P{(\delta(u))_T=(\delta(v))_T=2 | u,v\text{ trees}} = \Omega(1)$. 
The following theorem is the main result of this subsection:
\begin{theorem}\label{thm:badedges}
For $\eps_{1/2}\leq 0.0002$, $\eps_{\eta}\leq \eps_{1/2}^2$, a top edge bundle $\bbe={\bf (u,v)}$ is bad only if the following three conditions hold simultaneously: 
\begin{itemize}
\item $\bbe$ is a half edge, 
\item $x(\delta^\uparrow(u)),x(\delta^\uparrow(v))\leq 1/2+9\eps_{1/2}$,
\item Every other half edge bundle incident to $u$ or $v$ is (2-2) good.
\end{itemize}
\end{theorem}
The proof of this theorem follows from \cref{lem:1/2+eps_1/2sendhigher} and \cref{lem:one-of-two-good} below.

In this subsection, we use repeatedly that for any atom $u$ in a degree cut $S$, $x(\delta(u))\leq 2+\eps_{\eta}$. We also repeatedly use that for a half edge bundle $\bbe={\bf (u,v)}$ in a degree cut, conditioned on $u,v$ trees, $\bbe$ is in or out with probability at least $1/2-\eps_{1/2}-3\eps_\eta>0.49$.

\begin{lemma}\label{lem:2/2goodEvenSum}
Let $\bbe={\bf (u,v)}$ be a good half edge bundle in a degree cut $S\in\cH$. Let $A=\delta(u)_{-\bbe}$ and $B=\delta(v)_{-\bbe}$.
If $\eps_{1/2}\leq 0.001$ and $\eps_{\eta}<\eps_{1/2}/100$, then
$$ \P{A_T+B_T\leq 2 | u,v\text{ trees}}, \P{A_T + B_T\geq 4 | u,v\text{ trees}} \geq 0.4\eps_{1/2}$$
\end{lemma}
\begin{proof}
Throughout the proof all probabilistic statements are with respect to the measure $\mu$ conditioned on $u,v$ trees.
Let $p_{\leq 2}=\P{A_T+B_T\leq 2}$ and similarly define $p_{\geq 4}$.
Observe that whenever $\delta(u)_T=\delta(v)_T	=2$, we must have $A_T+B_T\neq 3$. Since $\bbe$ is 2-2 good, this event happens with probability at least $3\eps_{1/2}$, i.e., 
\begin{equation}\label{eq:<=2+>=4}
p_{\leq 2}+p_{\geq 4} \geq 3\eps_{1/2}
\end{equation}
By \cref{thm:rayleigh_expectconstprob},
using the fact that $p_0=0$, we get $p_{=3} \ge 1/4$. 


 First, we show that $p_{\leq 2} \geq 0.4\eps_{1/2}$. 
 We have
\begin{eqnarray*}
 3+2\eps_{1/2} \geq  \E{A_T+B_T} \geq 4 p_{\geq 4} + 2 p_{= 2} + 3(1-p_{\geq 4}-p_{\leq 2}) = 3 + p_{\geq 4} - p_{=2} -3p_{ =1}.
\end{eqnarray*}
Again, we are using $p_0=0$.
By log-concavity $p_{=2}^2 \geq p_{=3}p_{=1}$, so
 since $p_{=3}\geq 1/4$, $p_{=1}\leq 4p_{=2}^2\leq 4p_{\leq 2}^2$.
Therefore, 
$$p_{\geq 4}-2\eps_{1/2} \leq   p_{=2}+3p_{= 1} = p_{\leq 2} + 2p_{=1}\leq  p_{\leq 2}(1+8p_{\leq 2}).$$ 

Finally, since $\eps_{1/2}<0.001$, plugging this upper bound on $p_{\ge 4}$ into \cref{eq:<=2+>=4} we get $p_{\leq 2}\geq 0.4\eps_{1/2}$. 

Now, we show $p_{\geq 4} \geq 0.4 \eps_{1/2}/2$. 
Assume $p_{\geq 4}<\eps_{1/2}/2$ (otherwise we are done).
Since $p_{=3}\geq 1/4$ by \cref{lem:logconcaveexpecation} with $\gamma\le  (\eps_{1/2}/2)/(1/4) = 2\eps_{1/2}$
$$\E{A_T+B_T | A_T+B_T\geq 4}\cdot p_{\geq 4} \leq \frac{p_{\geq 4}}{1-2\eps_{1/2}}(4+3\eps_{1/2})$$
Therefore,
$$ 3-2\eps_{1/2}-2\eps_\eta\leq \E{A_T+B_T} \leq 2p_{\leq 2} +  \frac{p_{\geq 4}}{1-2\eps_{1/2}}(4+3\eps_{1/2}) + 3(1-p_{\leq 2}-p_{\geq 4})$$
So, $1.01p_{\geq 4}\geq p_{\leq 2}-2.02\eps_{1/2}$ where we used $\eps_{1/2}\leq 0.001$ and $\eps_{\eta}<\eps_{1/2}/100$.  Now, $p_{\geq 4}\geq 0.4\eps_{1/2}$ follows by \cref{eq:<=2+>=4}.
\end{proof}

%

\begin{figure}[htb]\centering
	\begin{tikzpicture}
		\draw [color=blue,line width=1.1pt] (0,0) ellipse   (1.5cm and 1.3cm);
		\node at (-1,-1.2) () {${\color{blue}S}$};
		\node [draw,circle] at (-1,0) (u) {$u$};
		\node [draw,circle] at (0.5,0) (v) {$v$} edge node [below left] {$\bbe$} (u);
		\draw [color=red, line width=1.1pt] (0.5,0) ellipse (0.8 and 1.1);
		\node at (-0.3,-.9) () {${\color{red}W}$};

		\node at (-1,1.5) () {} edge node [above left] {$\delta^\uparrow(u)$} (u);
	\end{tikzpicture}
	\caption{Setting of \cref{lem:1/2+eps_1/2sendhigher}}
	\label{fig:halfkhigher}
\end{figure}
\begin{lemma}\label{lem:1/2+eps_1/2sendhigher}
Let $\bbe={\bf (u,v)}$ be a half edge bundle in a degree cut $S\in\cH$, and suppose $x(\delta^\uparrow(u))\geq 1/2+k\eps_{1/2}$.
If $k\geq 9$, $\eps_{1/2}\leq 0.0002$, and $\eps_\eta \le \eps_{1/2}^2$, then, $\bbe$ is 2-2 good.	
\end{lemma}
\begin{proof}
First, condition $u,v,S$  to be trees.
Let $W=S\smallsetminus \{u\}$. 
Since $S$ is a near mincut,
\begin{eqnarray*}
x(\delta(W))=x(\delta(S))+x(\delta(u))-2x(\delta^\uparrow(u))\leq 2(2+\eps_\eta) - 2(1/2+k\eps_{1/2})=3-2k\eps_{1/2}+2\eps_{\eta}
\end{eqnarray*}
 So, by \cref{lem:treeconditioning}, 
$\P{W\text{ is tree}} \geq 1/2+k\eps_{1/2}-\eps_\eta-\eps_\eta$. Note that the extra $-\eps_{\eta}$ comes from the fact that conditioning $u$ be a tree can decrease marginals of edges in $E(W)$ by at most $\eps_\eta$.

Let $\nu$ be the resulting measure, namely the measure obtained by first conditioning $u,v,S$ to be trees and then $W$ to be a tree. Note that $\nu$ is a strongly Rayleigh distribution on the set of edges in $E(W) \cup E(u,W)\cup E(G/S)$; this is because $\nu$ is a product of 3 SR distributions each supported on one of the aforementioned sets. 

Let $X=\delta^\uparrow(u)_T$ and $Y=\delta(v)_T-1$. Observe that, under $\nu$, $X=Y=1$ iff $\delta(u)_T=\delta(v)_T=2$. Furthermore, $Y\geq 0$ with probability 1, since $v$ is connected to the rest of the graph. So, we just need to lower bound $\PP{\nu}{X=Y=1}.$ First, notice
\begin{equation}\begin{aligned}
 &\EE{\nu}{X}\in [0.5+k\eps_{1/2}-\eps_\eta,1+\eps_\eta]
 \\
 &\EE{\nu}{Y} \in  [0.5+k\eps_{1/2}-4\eps_\eta,1.5-k\eps_{1/2}+3\eps_\eta]
\label{eq:EnuXYeps1/2up}
\end{aligned}\end{equation}
We will give a brief explanation of this: first, note that $\frac{1}{2}+k\eps_{1/2} \le \E{X} \le 1+\eps_\eta$ before conditioning. By conditioning $u,v,S$ to be trees, we can increase $\E{|E(S)|_T}$ by at most $\eps_\eta$, therefore this may decrease $\E{X}$ by at most $\eps_\eta$. Under this measure, $E(S)$ is independent of $X$; therefore conditioning on $W$ to be a tree cannot change $\E{X}$. Second, note that $1 \le \E{Y} \le 1+\eps_\eta$ before conditioning. Now, conditioning on $u,v,S$ to be trees may decrease $\E{Y}$ by at most $2\eps_\eta$ and increase by at most $\eps_\eta$. Conditioning on $W$ to be a tree may increase or decrease $\E{Y}$ by a most $1/2-k\eps_{1/2}+2\eps_\eta$.

Note that using \cref{lem:427gen}, we can immediately argue that $\PP{\nu}{X=Y=1}\geq \Omega(\eps_{1/2})$. We do the following more refined analysis to make sure that this probability is at least $6\eps_{1/2}$ (for $\eps_{1/2}\leq 0.0005$) and $k\geq 9$. Once we prove this, we obtain the lemma:
\begin{align*}
\P{\delta(u)_T=\delta(v)_T=2 \mid u,v \text{ trees}} &\geq \P{S,W\text{ trees} \mid u,v \text{ trees}} \PP{\nu}{X=Y=1} \ge 0.5 \cdot 6 \eps_{1/2}
\end{align*}

{\bf Case 1: $\PP{\nu}{X+Y=2}\geq 48\eps_{1/2}$.}
By \cref{lem:SR>=}, $\PP{\nu}{X\geq 1},\PP{\nu}{Y\geq 1}\geq 1-e^{-0.5}$.
On the other hand, 
by \cref{thm:hoeffding}, $\PP{\nu}{X\leq 1},\PP{\nu}{Y\leq 1}\geq 7/16$. This is because if we have one Bernoulli of value 1, $\PP{\nu}{X\leq 1} \ge (1-\frac{0.5}{n})^n$ is minimized at $n=1$, whereas if we have no Bernoullis of value 1, $\PP{\nu}{X\leq 1} \ge (1-\frac{1.5}{n})^n + 1.5(1-\frac{1.5}{n})^{n-1}$ which is minimized at $n=2$. 
Therefore, by \cref{lem:SRA=nA}, $\PP{\nu}{X=1 | X+Y=2} \geq 0.1269$.
Therefore, we get
$$\PP{\nu}{X=1,Y=1} \geq 48\eps_{1/2} \cdot 0.1269 \geq 6\eps_{1/2}$$ 

{\bf Case 2: $\PP{\nu}{X+Y=2}<48\eps_{1/2} < 0.01$.}  
By \cref{thm:rayleigh_expectconstprob}, $\PP{\nu}{X+Y=1}\geq 0.25$ (if $\EE{\nu}{X+Y} \ge 1.2$ then the assumption of this case obviously fails). So, since $\PP{\nu}{X+Y=2}<0.01$, by log concavity, $\PP{\nu}{X+Y=3} \le 0.01/25$. Furthermore, by \cref{lem:logconcaveexpecation} (with $\gamma=1/25, i=1,k=3$), $\PP{\nu}{X+Y > 2}<0.0005$.

Now, assume that $\PP{\nu}{X \ge 1},\PP{\nu}{Y \ge 1} \ge 0.47$ (we will prove this shortly). Now, applying stochastic dominance, we have
\begin{align*}
\PP{\nu}{X\geq 1| X+Y= 2} &\geq \PP{\nu}{X\geq 1 | X+Y\leq 2}\\
&\geq \PP{\nu}{X\geq 1, X+Y\leq 2}\\
&\geq \PP{\nu}{X\geq 1} - \PP{\nu}{X+Y> 2} \geq \PP{\nu}{X\geq 1}- 0.0005\geq 0.469.
\end{align*}
Similarly, $\PP{\nu}{X\leq 1 | X+Y=2} =\PP{\nu}{Y\geq 1 | X+Y=2} \geq \PP{\nu}{Y\geq  1}-0.0005\geq 0.469$.
Finally since the distribution of $X$ conditioned on $X+Y=2$ is the same as the number of successes in  2 independent Bernoulli trials, with probabilities,  say, $p_1$ and $p_2$, we can minimize $p_1 (1-p_2) + (1-p_1)p_2$ subject to $1-p_1p_2 \ge 0.469$ and $1-(1-p_1)(1-p_2) \ge 0.469$. Solving this yields  $\PP{\nu}{X=1 | X+Y=2}\geq 0.395$.

Lastly, observe that since by \cref{eq:EnuXYeps1/2up} $1.2 \ge \EE{\nu}{X+Y}\geq 1+(2k-1)\eps_{1/2}$, by  \cref{thm:rayleigh_expectconstprob} we can write
$$\PP{\nu}{X+Y=2} \geq (2k-1)\eps_{1/2} e^{-(2k-1)\eps_{1/2}} \geq (2k-2)\eps_{1/2}.$$ 
Therefore,
\begin{align*}
\PP{\nu}{X=Y=1} &= \PP{\nu}{X=1 | X+Y=2}\PP{\nu}{X+Y=2} \ge 0.395 (2k-2)\eps_{1/2}
\end{align*}
To get the RHS to be at least $6\eps_{1/2}$ it suffices that $k\geq 9$.

Now we prove that $\PP{\nu}{X \ge 1} \ge 0.47$; $\PP{\nu}{Y \ge 1} \ge 0.47$ follows similarly.
$$\PP{\nu}{X = 2} \le \PP{\nu}{X+Y \ge 2} \le 0.01 + 0.00042 \le 0.0105$$
Also notice that $\PP{\nu}{X = 1} \ge 0.3$ by \cref{thm:rayleigh_expectconstprob}.
Now, using \cref{lem:logconcaveexpecation} we can write, for $\gamma = 1/25$ and $i=1$,
$$\EE{\nu}{X \mid X \ge 2}\PP{\nu}{X \ge 2} \le 0.0224$$ 
Therefore, since $X$ is integer valued,
$$\PP{\nu}{X \ge 1} \ge \EE{\nu}{X} - \EE{\nu}{X \mid X \ge 2}\PP{\nu}{X \ge 2} \ge \EE{\nu}{X} - 0.0224 \ge 0.47,$$
as desired.
\end{proof}

\begin{figure}\centering
\begin{tikzpicture}[scale=0.8]
\tikzstyle{every node}=[draw,circle]
\foreach \a/\x/\l in {u/-3/U, v/0/V , w/3/W}{
\path  (\x,0) node  (\a) {$\a$};
\path (\a)+(-0.7,0.5) node [color=red,draw=none] (){$\l$};
\foreach \xx in {0,...,2}{
\draw [-] (\a) -- ++(60+\xx*30: 1.5);}
\draw [dashed,color=red,line width=1.2] (\a)+(45:1) arc (45:135:1);
}
\path [-](u) edge node [draw=none,below] {$\bbe$} (v);
\path [-] (w) edge node [draw=none,below] {$\bbf$} (v);
\end{tikzpicture}
\caption{Setting of \cref{lem:one-of-two-good}} 
\label{fig:incident_halfedges}
\end{figure}

\begin{lemma}\label{lem:one-of-two-good}
Let $\bbe={\bf (u,v)},\bbf={\bf (v,w)}$ be two half edge bundles in a degree cut $S\in\cH$. If $\eps_{1/2} < 0.0005$ and $\eps_\eta \le \eps_{1/2}^2$, then one of $\bbe$ or $\bbf$ is good.
\end{lemma}
\begin{proof}
We use the following notation $V=\delta(v)_{-\bbe-\bbf},U=\delta(u)_{-\bbe},W=\delta(w)_{-\bbf}$ (see \cref{fig:incident_halfedges} for an illustration). For a set $A$ of edges and an edge bundle $\bbe$ we write $A_{+\bbe}=A\cup\{\bbe\}$. Furthermore, for a measure $\nu$ we write $\nu_{-\bbe}$ to denote $\nu$ conditioned on $\bbe\notin T$. 

Condition $u,v,w$ to be trees. 
This occurs with probability at least $1-3\epsilon_\eta$.
Let $\nu$ be this measure. By \cref{lem:crossingcorrelation}, without loss of generality, we can  assume 
\begin{equation}\label{eq:Wtgoesup405}
\EE{\nu}{W_T | \bbe\notin T}\leq \EE{\nu}{W_T}+0.405.
\end{equation}
Now, if $\EE{\nu}{V_T | \bbe\notin T}\geq \EE{\nu}{V_T}+0.03$, then we will show $\bbe$ is 2-2 good. First, 
\begin{align*}
	& \EE{\nu_{-\bbe}}{(V_{+\bbf})_T}\in [1.53-\eps_{1/2}-3\eps_\eta, 2+\eps_{\eta}],\\
	&\EE{\nu_{-\bbe}}{U_T}\in [1.5-\eps_{1/2}-3\eps_\eta, 2+\eps_{\eta}],\\
	&\EE{\nu_{-\bbe}}{(V_{+\bbf})_T+U_T} \in [3.03-2\eps_{1/2}-3\eps_\eta, 3.5+2\eps_{1/2}+2\eps_\eta],
\end{align*}
where we may decrease the marginals by $3\eps_\eta$ due to conditioning $u,v,w$ to be trees.

Therefore, by \cref{thm:rayleigh_expectconstprob}, $\PP{\nu_{-\bbe}}{(V_{+\bbf})_T+U_T=4} \geq 0.029$, where we use the fact that $U_T\geq 1$ and $(V_{+\bbf})_T\geq 1$ with probability 1 under $\nu_{-\bbe}$ and apply this and  the remaining calculations to $U_T-1,(V_{+\bbf})_T-1$. In addition, we have
\begin{align*}
&\PP{\nu_{-\bbe}}{U_T\leq 2},\PP{\nu_{-\bbe}}{(V_{+\bbf})_T\leq 2}\geq 0.499 \tag{Markov Inequality}\\
& \PP{\nu_{-\bbe}}{U_T\geq 2},\PP{\nu_{-\bbe}}{(V_{+\bbf})_T\geq 2}\geq   0.39
\tag{\cref{lem:SR>=}}
\end{align*}
It follows by \cref{lem:SRA=nA} applied to $U_T-1$ and $(V_{+\bbf})_T-1$ (with $\eps=0.194$ and $p_m=0.6$) that 
$$\PP{\nu_{-\bbe}}{U_T= 2 | U_T+(V_{+\bbf})_T=4} \geq 0.13,$$
where we use that $U_T \ge 1$, $(V_{+\bbf})_T \ge 1$ with probability 1 under $\nu_{-\bbe}$ because otherwise the tree would be disconnected.

Therefore,
\begin{align*} \P{\delta(u)_T=\delta(v)_T=2 \mid u,v \text{ trees}} &\geq \P{w\text{ is a tree},\bbe\notin T}\PP{\nu_{-\bbe}}{U_T=(V_{+\bbf})_T=2} \\
& \geq (0.49)(0.029)(0.13)\geq 0.0018.
\end{align*}
The lemma follows (i.e., $e$ is 2-2 good) since $0.0018\geq 3\eps_{1/2}$ for $\eps_{1/2}\leq 0.0005$.

%
Otherwise, if $\EE{\nu}{V_T | \bbe\notin T}\leq \EE{\nu}{V_T}+0.03$ then we will show that $\bbf$ is 2-2 good. We have,
\begin{align*}
	& \EE{\nu_{+\bbf}}{(V_{+\bbe})_T},\EE{\nu_{+\bbf}}{W_T} \in  [1-2 \eps_{1/2}-3\eps_\eta,1.5+2\eps_{1/2}+\eps_\eta]\\
&\PP{\nu_{+\bbf}}{(V_{+\bbe})_T\leq 1},\PP{\nu_{+\bbf}}{W_T\leq 1}\geq 0.249 \tag{Markov} \\
&\PP{\nu_{+\bbf}}{(V_{+\bbe})_T\geq 1},\PP{\nu_{+\bbf}}{W_T\geq 1}\geq 0.63 \tag{\cref{lem:SR>=}}
\end{align*}
So, by \cref{lem:SRA=nA} (with $\eps=0.15,p_m=0.7$), we get $\PP{\nu_{+\bbf}}{W_T=1 | (V_{+\bbe})_T+W_T=2}\geq 0.11$. 
On the other hand, 
\begin{align*}
	&\PP{\nu_{+\bbf}}{(V_{+\bbe})_T+W_T=2} \geq \PP{\nu_{+\bbf}}{\bbe\notin T}\PP{\nu_{+\bbf-\bbe}}{(V_{+\bbe})_T+W_T=2} \geq (0.49)(0.0582)\geq 0.0285
\end{align*}
To derive the last inequality, we show  $\PP{\nu_{+\bbf-\bbe}}{(V_{+\bbe})_T+W_T=2}\geq 0.0582$. This is because by negative association and \cref{eq:Wtgoesup405} 
\begin{align*} \EE{\nu_{+\bbf-\bbe}}{(V_{+\bbe})_T+W_T} &= \EE{\nu_{+\bbf-\bbe}}{V_T+W_T}\\
&	
\leq \EE{\nu_{-\bbe}}{V_T+W_T} \leq  \EE{\nu}{W_T}+0.405 + \EE{\nu}{V_T} + 0.03 \leq 2.94;
\end{align*}
So, since $(V_{+\bbe})_T+W_T$ is always at least $1$,   so by \cref{thm:hoeffding}, in the worst case, $\PP{\nu_{-\bbe+\bbf}}{(V_{+\bbe})_T+W_T=2}$ is the probability that the sum of two Bernoullis with success probability $1.94/2$ is 1, which is $0.0582$. 

Therefore, similar to the previous case, 
\begin{align*}\P{\delta(v)_T=\delta(w)_T=2 \mid v,w \text{ trees}}&\geq \P{u\text{ is a tree},f\in T}\PP{\nu_{+\bbf}}{(V_{+\bbe})_T+W_T=2} \\& \quad\quad \cdot \PP{\nu_{+\bbf}}{W_T=1 | (V_{+\bbe})_T+W_T=2}\\
&\geq (0.49) (0.0285)(0.11) \geq 3\eps_{1/2}
\end{align*}
for $\eps_{1/2}\leq 0.0005$ as desired.
\end{proof}

\subsection{2-1-1 and 2-2-2 Good Edges}
\label{sec:211goodedges}

Consider a cut $u \in \cH$, and recall that $x(\delta(u)) \approx 2$. Normally, it is sufficient to have $\delta(u)_T = 2$ when an edge $e \in \delta(u)$ is reduced. In the worst case, the edges of $\delta(u)$ essentially come from two of its descendants $u',v'$, i.e. $x(\delta(u') \cap \delta(u)) \approx 1$ and $x(\delta(v') \cap \delta(u)) \approx 1$. Let $A=\delta(u') \cap \delta(u),B=\delta(v') \cap \delta(u),C=\delta(u)\smallsetminus (A \cup B)$. In such a case, if we condition on reducing an edge in $A$, we may have $A_T$ to be even with probability close to 1, and it will be very expensive to fix the constraint coming from $\delta(u')$, as $(\delta(u') \smallsetminus(\delta(u)))_T$ is 1, i.e. odd, with probability close to 1. Therefore, it is crucial to make sure that when we reduce an edge in $A$ ($B$), we have $A_T$ ($B_T$) is odd with some probability. Since when $\delta(u)$ is even and $A_T$ is odd, $B_T$ will be odd as well (discounting the leftovers $C$, which have negligible expectation), a natural criteria is to ask for $A_T=B_T=1$, hence motivating the upcoming definition of 2-1-1 happy. To get a more high level understanding of how we use these events, see the following two sections of the overview: \hyperlink{ABC-explanation}{dealing with $x_u$ close to 1} and \hyperlink{211-explanation}{dealing with triangles}.   
 
\begin{definition}[$A,B,C$-Degree Partitioning]\label{def:abcdegpartitioning}
\hypertarget{tar:degreepartition}{For $u\in\cH$ and $\eps_{1/1}$ defined in \ref{eq:constants}, we define a partitioning of edges in $\delta(u)$: Let $a,b\subsetneq u$ be {\em minimal} cuts in the hierarchy, i.e., $a,b\in\cH$, such that $a\neq b$  and $x(\delta(a)\cap\delta(u)),x(\delta(b)\cap\delta(u))\geq 1-\eps_{1/1}$. Note that since the hierarchy is laminar, $a,b$ cannot cross. Let $A=\delta(a)\cap\delta(u),B=\delta(b)\cap\delta(u),C=\delta(u)\smallsetminus A\smallsetminus B$. }

If there is no cut $a\subsetneq u$ (in the hierarchy) such that $x(\delta(a)\cap\delta(u))\geq 1-\eps_{1/1}$, we just let $A,B$ be two arbitrary disjoint sets of edges in $\delta(u)$ for which $x(A),x(B)\geq 1-\eps_{1/1}$. As above set $C=\delta(u)\smallsetminus A\smallsetminus B$. Note that this exists WLOG because we may split any edge into an arbitrary number of parallel copies.

If there is just one minimal cut $a\subsetneq u$ (in the hierarchy) with $x(\delta(a)\cap\delta(u))\geq 1-\eps_{1/1}$, i.e., $b$ does not exist in the above definition, then we define $A=\delta(a)\cap\delta(u)$. Let $a'\in\cH$ be the unique child of $u$ such that $a\subseteq a'$, i.e., $a$ is equal to $a'$ or a descendant of $a'$. Then we define $C=\delta(a')\cap\delta(u) \smallsetminus \delta(a)$ and $B=(\delta(u)\smallsetminus A) \smallsetminus C$.
Note that in this case since $x(\delta^\uparrow(a'))\leq 1+\eps_{\eta}$, we have $x(B)\geq 1-\eps_\eta\geq 1-\eps_{1/1}$.
\end{definition}
See \cref{fig:topedgeABCmotivation} for an example.  The following inequalities on $A,B,C$ degree partitioning will be used in this section:
\begin{equation}	
\begin{aligned}
	x(A),x(B) \in [1-\eps_{1,1},1+\eps_\eta],\\
	x(C)\leq 2\eps_{1/1}+\eps_\eta. 
\end{aligned}\label{eq:ABCDegParx}
\end{equation}

In this section we will define a constant $p>0$ which is the minimum  probability that a good edge bundle is happy.
\begin{definition}[2-1-1 Happy/Good]
\label{dfn:211happygood}
Let  $\bbe={\bf (u,v)}$ be a top edge bundle. 
Let $A,B,C\subseteq \delta(u)$ be a Degree Partitioning of edges $\delta(u)$ as defined in \cref{def:abcdegpartitioning}.
We say that $\bbe$ is 2-1-1 happy with respect to $u$ if the event 
$$A_T=1,B_T=1,C_T=0,\delta(v)_T=2,\text{ and $u$ and $v$ are both trees}$$ occurs.

We say $\bbe$ is {\em 2-1-1 good with respect to $u$} if $$\P{\bbe \text{ is 2-1-1 happy wrt $u$}}\geq p.$$
\end{definition}
\begin{remark}\label{rem:poly-and-degree-partition} Note we also use this $A,B,C$ partitioning to help deal with the triangle cut case. In the special case that $u$ is a polygon cut with $A,B,C$-polygon partitioning, let $A',B',C'$ be the degree partitioning of $\delta(u)$. Then, by \cref{def:hierarchy} we have $A' \subseteq A$, $B' \subseteq B$, $C \subseteq C'$. Therefore, if an edge in $\delta(u)$ is reduced and is 2-1-1 happy with respect to $u$, the polygon $u$ is also happy. See \hyperlink{211-explanation}{the overview} for an example.\end{remark}

Many of the lemmas in this section are proved in \cref{app:probabilistic}. In the following, we assume that $\eps_\eta \le \eps_{1/2}^2$ and $12\eps_{1/1}\leq \eps_{1/2}$.

\begin{restatable}{lemma}{lemxesmallhalfeps}\label{lem:x_e<=1/2-eps1}
Let $\bbe={\bf (u,v)}$ be a top edge bundle such that $x_\bbe\leq 1/2-\eps_{1/2}$. If $\eps_{1/2}\leq 0.001$ then, $\bbe$ is 2-1-1 happy with probability at least $0.005\eps_{1/2}^2$.
\end{restatable}

\begin{restatable}{lemma}{xemoreepshalf}\label{lem:x_e>=1/2+eps_1/2}
Let $\bbe={\bf (u,v)}$ be a top edge bundle such that $x_\bbe\geq 1/2+\eps_{1/2}$. If $\eps_{1/2}\leq 0.001$, then, $\bbe$ is 2-1-1 happy with respect to $u$ with probability at least $0.006\eps_{1/2}^2$.
\end{restatable}

Fix $u$ in the hierarchy with degree partitioning $A,B,C$. The above two lemmas show that any edge bundle $\bbe \in \delta(u)$ which is not a half edge bundle is 2-1-1 good, so the difficult case is when the majority of $x(\delta^\rightarrow(u))$ comes from half edge bundles. In \cref{thm:badedges} we showed that $\delta(u)$ can have at most one 2-2 bad edge. Oddly enough, one of the simplest cases of the reduction argument is when there \textit{is} a bad edge in $\delta(u)$. This is because we never reduce bad edges, and therefore we never need to increase edges which are matched to them.\footnote{The main problem with bad edges is that we cannot match them to edges going higher in the matching lemma \ref{lem:matching}. So, in order to prove the matching lemma we need to justify that there are not too many bad edges in any cut. Therefore we cannot simply ``pretend" that one half edge bundle of $\delta(u)$ is bad.}
	So, the main problem is good edges which are not 2-1-1 good. The following key statement, \cref{lem:Ais211good}, shows that these problematic edges are rare in the sense that there is at most one good half edge bundle in $A$ (resp. $B$) which is not 2-1-1 good. 
 
 To prove this we need the following two lemmas. In the first one we show that if $\bbe,\bbf$ are two half edge bundles which almost entirely land in $A$ (or $B$), at least one of them is 2-1-1 good. In the second, we show that if a good half edge bundle does not entirely land in $A$ (or $B$), then it is 2-1-1 good. This is the main tool we use to upper bound the expected increase of good top edges in \cref{sec:payment}.
 
For a set of edges $D$, and an edge bundle $\bbe$, let
$\bbe(D):= \bbe \cap D.$ Note that $\bbe(D)$ is not really an edge bundle. 
 \begin{restatable}{lemma}{lemoneoftwotwooneone}
\label{lem:one-of-two-211}
Let $\bbe={\bf (v,u)}$ and $\bbf={\bf (v,w)}$ be good half top edge bundles and let $A,B,C$ be the \hyperlink{tar:degreepartition}{degree partitioning} of $\delta(v)$ such that $x_{\bbe(B)},x_{\bbf(B)}\leq \eps_{1/2}$.   
Then, one of $\bbe,\bbf$ is 2-1-1 happy with probability at least $0.005\eps_{1/2}^2$.
\end{restatable}

\begin{restatable}{lemma}{lemeABpositivetwooneone}\label{lem:eABpositive211}
Let $\bbe={\bf (u,v)}$ be a good half edge bundle and 	let $A,B,C$ be the \hyperlink{tar:degreepartition}{degree partitioning} of $\delta(u)$ (see \cref{fig:e(A)e(B)large}).  If $\eps_{1/2}\leq 0.001$ and $x_{\bbe(A)},x_{\bbe(B)}\geq \eps_{1/2}$, 
 then
$$ \P{\bbe \text{ 2-1-1 happy w.r.t } u} \geq 0.02\eps_{1/2}^2. $$
\end{restatable}

\begin{lemma}\label{lem:Ais211good}
For a degree cut $S\in\cH$, and $u\in \cA(S)$, let $A,B,C$ be the \hyperlink{tar:degreepartition}{degree partition} of $u$. Then, $A\cap\delta^\rightarrow(u)=:A^\rightarrow$ has fraction at most $1/2+4\eps_{1/2}$ of good edges that are not 2-1-1 good (w.r.t., $u$).
\end{lemma}
\begin{proof}  
	Suppose by way of contradiction that there is a set $D \subseteq A^\rightarrow$ of good edges that are not 2-1-1 good w.r.t. $u$ with $x(D) \ge \frac{1}{2}+4\eps_{1/2}$. By \cref{lem:x_e<=1/2-eps1} and \cref{lem:x_e>=1/2+eps_1/2}, every edge in $D$ is part of a half edge bundle.  
	
	There are at least two half edge bundles $\bbe,\bbf$ such that $x(D \cap \bbe), x(D \cap \bbf) \ge \eps_{1/2}$, as there are at most four half edge bundles in $\delta^\rightarrow(u)$ (and using that for any half edge bundle $\bbe$, $x_\bbe \le \frac{1}{2}+\eps_{1/2}$). 
	Since $D\subseteq A^\rightarrow$, we have 
	$$x(A \cap \bbe),x(A \cap \bbf) \ge \eps_{1/2}.$$ 
	Since $x(A\cap \bbe)\geq \eps_{1/2}$, if $x(B\cap \bbe)\geq \eps_{1/2}$ then, by \cref{lem:eABpositive211} $\bbe$ is 2-1-1 good. But since  every edge in $D$ is not 2-1-1 good w.r.t $u$, we must have $x(B\cap \bbe) <\eps_{1/2}$. The same also holds for $\bbf$. 
	Finally, since $x(B\cap \bbe)<\eps_{1/2}$ and $x(B\cap \bbf)<\eps_{1/2}$ by \cref{lem:one-of-two-211}  at least one of $\bbe,\bbf$ is 2-1-1 good w.r.t $u$. This is a contradiction.
\end{proof}

\paragraph{2-2-2 Good Edges.} While \cref{lem:Ais211good} is sufficient for bounding the increase of top edges, it is not sufficient for bottom edges. Fix a polygon $u$ with partition $A,B,C$ and suppose $\p(u)=S$ is a degree cut (recall that by \cref{rem:poly-and-degree-partition}, the degree partitioning and polygon partitioning of $u$ are essentially the same). Roughly speaking, a bottom edge $g \in E(u)$ is ``matched" to all edges in $\delta(u)$, and needs to increase for edges $f \in A$ when $f$ is reduced and $A_T$ is even, and for edges $f \in B$ when $f$ is reduced and $B_T$ is even. Therefore, $g$ is matched to essentially twice its fraction. If most of the edges in $\delta(u)$ are 2-1-1 good, this is sufficient to bound the expected increase of $g$ because when such an edge is reduced and 2-1-1 happy with respect to $u$, $g$ does not need to increase. 

It turns out that the above lemmas are sufficient to bound the expected increase of $g \in E(u)$ except when $A \cap \delta(S) \approx B \cap \delta(S) \approx 1/2$ and $\bbe \approx A \cap \delta^\rightarrow(u)$ and $\bbf \approx B \cap \delta^\rightarrow(u)$ are both good edge bundles which are not 2-1-1 good. In this extreme case, we employ a new strategy. In \cref{lem:222} below, we prove that the two edge bundles $\bbe,\bbf$ are 2-2 happy \textit{simultaneously} with a constant probability. We call such a pair 2-2-2 good. Later, in \cref{sec:payment}, we use this to ensure that $\bbe$ and $\bbf$ are always reduced simultaneously. The point is that since $\bbe,\bbf$ do not both come from $A$ (or $B$), no cut inside $u$ contains $\bbe$ and $\bbf$. Therefore, $g$ only needs to increase by the maximum of the decrease of $\bbe,\bbf$ (not the sum), effectively saving a factor of 2.

\begin{definition}[2-2-2 Happy/Good]
\label{dfn:222happygood}
	Let $\bbe={\bf (u,v)},\bbf={\bf (v,w)}$ be top half-edge bundles (with $\p(\bbe)=\p(\bbf)$). We say $\bbe,\bbf$ are 2-2-2 happy (with respect to $v$) if $\delta(u)_T=\delta(v)_T=\delta(w)_T=2$ and $u,v,w$ are all trees.
	
	We say $\bbe,\bbf$ are 2-2-2 good with respect to $v$ if $\P{\bbe,\bbf \text{ 2-2-2 happy}}\geq p$.
\end{definition}

%

%


\begin{restatable}{lemma}{lemtwotwotwo}
\label{lem:222}
	Let $\bbe={\bf (u,v)},\bbf={\bf (v,w)}$ be two good top half edge bundles and let $A,B,C$ be \hyperlink{tar:degreepartition}{degree partitioning} of $\delta(v)$ such that $x_{\bbe(B)},x_{\bbf(A)}\leq \eps_{1/2}$. 
If $\bbe,\bbf$ are not 2-1-1 good with respect to $v$, and $\eps_{1/2}\leq 0.0002$, then $\bbe,\bbf$ are 2-2-2 happy with probability at least $0.01$.
\end{restatable}

The following theorem summarizes the above results in a compact form. This is the main result used in the analysis of the increase for bottom edges in \cref{sec:payment}.
\begin{theorem}
	\label{thm:probabilistic}
	Let $v, S \in \cH$ where $\p(v) =S$, and let $A,B,C$ be the \hyperlink{tar:degreepartition}{degree partitioning} of $\delta(v)$. For $p\geq 0.005\eps_{1/2}^2$, with $\eps_{1/2}\leq 0.0002$, $\eps_{1/1}\leq \eps_{1/2}/12$ and $\eps_{\eta}\leq \eps_{1/2}^2$, at least one of the following is true:
	\begin{enumerate}[i)]
		\item $\delta^\rightarrow(v)$ has at least $1/2-\eps_{1/2}$ fraction of bad edges,
		\item $\delta^\rightarrow(v)$ has at least $1/2-\eps_{1/2}-\eps_\eta$ fraction of 2-1-1 good edges with respect to $v$.
		\item There are two (top) half edge bundles $\bbe,\bbf\in\delta^\rightarrow(v)$ such that $x_{\bbe(B)}\leq \eps_{1/2}$, $x_{\bbf(A)}\leq \eps_{1/2}$, and $\bbe, \bbf$
 are 2-2-2 good (with respect to $v$).
	\end{enumerate}
\end{theorem}
\begin{proof}
	Suppose case (i) does not happen. Since every bad edge has fraction at least $1/2-\eps_{1/2}$ this means that $\delta(v)$ has no bad edges.
First, notice by \cref{lem:x_e<=1/2-eps1} and \cref{lem:x_e>=1/2+eps_1/2} any non half-edge in $\delta^\rightarrow(v)$ is 2-1-1 good (with respect to $v$). (Recall we define $\delta^\rightarrow(v) = \delta(v) \smallsetminus \delta(\p(v))$, where $\p(v)$ is the immediate parent of $v$ in the hierarchy).
	If there is only one half edge in $\delta^\rightarrow(v)$, then we have at least fraction $1-\eps_{\eta} - (1/2+\eps_{1/2})$ fraction of 2-1-1 good edges and we are done with case (ii).
	Otherwise,  there are two good half edges $\bbe,\bbf\in \delta^\rightarrow(v)$. 
	
	First, by \cref{lem:eABpositive211} if $x_{\bbe(A)},x_{\bbe(B)}\geq \eps_{1/2}$, then $\bbe$ is 2-1-1 good (w.r.t., $v$) and we are done.
	Similarly, if $x_{\bbf(A)},x_{\bbf(B)}\geq  \eps_{1/2}$, then $\bbf$ is good. So assume none of these happens.
	
	Furthermore by \cref{lem:one-of-two-211} if $x_{\bbe(B)},x_{\bbf(B)}\leq \eps_{1/2}$ (or $x_{\bbe(A)},x_{\bbf(A)}\leq \eps_{1/2}$) then one of $\bbe,\bbf$ is 2-1-1 good.
	
	So, the only remaining case is when $\bbe,\bbf$ are not 2-1-1 good and $x_{\bbe(B)},x_{\bbf(A)}\leq \eps_{1/2}$. But in this case by \cref{lem:222}, $\bbe,\bbf$ are 2-2-2 good; so (iii) holds.
\end{proof}

%% file: gurvits.tex
\subsection{Gurvits' Machinery and Generalizations}
\label{sec:Gurvits}
The following is the main result of this subsection.
\begin{proposition}\label{lem:427gen}
	Given a SR distribution $\mu:2^{[n]}\to\R_+$, let $A_1,\dots,A_m$ be random variables corresponding to the number of elements sampled from $m$ disjoint sets, and let integers  $n_1,\dots,n_m\geq 0$ be such that for any $S\subseteq [m]$,
	\begin{eqnarray*}
		 \P{\sum_{i\in S} A_i \geq \sum_{i\in S} n_i} &\geq & \eps,\\
		 \P{\sum_{i\in S} A_i \leq \sum_{i\in S} n_i} &\geq & \eps,\\
	\end{eqnarray*}
	it follows that,
	$$ \P{\forall i: A_i=n_i} \geq f(\eps)\P{A_1+\dots+A_m=n_1+\dots+n_m},$$
	where $f(\eps)\geq \eps^{2^m}\prod_{k=2}^m \frac1{\max\{n_k, n_1+\dots+n_{k-1}\}+1}$. 
\end{proposition}
We remark that in  applications of the above statement, it is enough to know that for any set $S\subseteq [m]$, $\sum_{i\in S} n_i-1 < \E{\sum_{i\in S} A_i}<\sum_{i\in S} n_i+1$. Because, then by \cref{thm:rayleigh_expectconstprob} we can prove a lower bound on the probability that $\sum_{i\in S} A_i = \sum_{i\in S} n_i$. 

We also remark the above lower bound of $f(\eps)$ is not tight; in particular, we expect the dependency on $m$ should only be exponential (not doubly exponential). We leave it as an open problem to find a tight lower bound on $f(\eps)$.
\begin{proof}	
Let $\cE$ be the event $A_1+\dots+A_m=n_1+\dots+n_m$.
\begin{align*}
	\P{1\leq i\leq m: A_i=n_i} =& \P{\cE}\P{A_m=n_m|\cE}\P{A_{m-1}=n_{m-1} |A_m=n_m,\cE}\\
	&\dots \P{A_2=n_2 | A_3=n_3,\dots,S_{A_m}=n_m,\cE}
\end{align*}	
So, to prove the statement, it is enough to prove that for any $2\leq k\leq n$,
\begin{equation}\label{eq:XAkkp1Am} \P{A_k=n_k | A_{k+1}=n_{k+1},\dots,A_m=n_m,\cE}\geq \eps^{2^{m-k+1}}\frac1{\max\{n_k, n_1+\dots+n_{k-1}\}+1}
\end{equation}
By the following \cref{claim:eps-to-the-2^m},
\begin{eqnarray*}
	\P{A_k\geq n_k | A_{k+1}=n_{k+1},\dots,A_m=n_m,\cE} &\geq & \eps^{2^{m-k+1}},\\
	\P{A_k\leq n_k | A_{k+1}=n_{k+1},\dots,A_m=n_m,\cE} &\geq & \eps^{2^{m-k+1}}.
\end{eqnarray*}
So, \eqref{eq:XAkkp1Am} simply follows by \cref{lem:logconcavity}. Now we prove this claim.
\begin{claim}\label{claim:eps-to-the-2^m} Let $[k]:=\{1,\dots,k\}$.
For any $2\leq k\leq m$, and any set $S\subsetneq [k]$,
\begin{eqnarray*}
	\P{\sum_{i\in S} A_i\geq \sum_{i\in S} n_i | A_{k+1}=n_{k+1},\dots,A_m=n_m,\cE}&\geq& \eps^{2^{m-k+1}},\\
	\P{\sum_{i\in S} A_i\leq \sum_{i\in S} n_i | A_{k+1}=n_{k+1},\dots,A_m=n_m,\cE}&\geq& \eps^{2^{m-k+1}}	
\end{eqnarray*}	
\end{claim}
\begin{proof}
%
We prove by induction. 
First, notice for $k=m$ the statement holds just by lemma's assumption and \cref{lem:updowntruncation}.
Now, suppose the statement holds for $k+1$.
Now, fix a set $S\subsetneq [k]$. 
 Let $\overline{S}=[k]\smallsetminus S$.
Define $A=\sum_{i\in S} A_i$ and $B=\sum_{i\in \overline{S}} A_i$, and similarly define $n_A,n_B$.
 By the induction hypothesis,
 $$ \eps^{2^{m-k}}\leq \P{A\leq n_A | A_{k+2}=n_{k+2},\dots,A_m=n_m,\cE} 
 $$
 The same statement holds for events $A\geq n_A, 
 B\leq n_B, B\geq n_B, A+B\geq n_A+n_B, A+B\leq n_A+n_B$.
Let $\cE_{k+1}$ be the event $A_{k+2}=n_{k+2},\dots,A_m=n_m, \cE$. Note that conditioned on $\cE_{k+1}$,  $A+B=n_A+n_B$  if and only if $A_{k+1}=n_{k+1}$. 
 By \cref{lem:logconcavity}, $\P{A+B=n_A+n_B| \cE_{k+1}}>0$. Therefore,
 by \cref{lem:updowntruncation},
$$ \P{A\geq n_A | A+B=n_A+n_B, \cE_{k+1}}, \P{A\leq n_A | A+B=n_A+n_B, \cE_{k+1}} \geq (\eps^{2^{m-k}})^2=\eps^{2^{m-k+1}}$$
as desired. 
\end{proof}
This finishes the proof of \cref{lem:427gen}
\end{proof}

\begin{lemma}\label{lem:logconcavity}
Let $\mu:2^{[n]}\to\R_{\geq 0}$ be a $d$-homogeneous SR distribution. If for an integer $0\leq k\leq d$, $\PP{S\sim\mu}{|S|\geq k}\geq \eps$ and $\PP{\mu}{|S|\leq k}\geq \eps$. Then, 
\begin{align*}\P{|S|=k}&\geq \min\{\frac{\eps}{k+1},\frac{\eps}{d-k+1}\},\\
\P{|S|=k}&\geq \min\left\{p_m, \eps\left(1-\left(\frac{\eps}{p_m}\right)^{1/\max\{k,d-k\}}\right)\right\}.
\end{align*}
where $p_m \leq \max_{0\leq i\leq d}\P{|S|=i}$ is a lower bound on the mode of $|S|$.
\end{lemma}
\begin{proof}
Since $\mu$ is SR, the sequence $s_0, s_1,\dots,s_d$ where $s_i=\P{|S|=i}$ is log-concave and unimodal. So, either the mode is in the interval $[0,k]$ or in $[k,d]$. We assume the former and prove the lemma; the latter can be proven similarly.
First, observe that since $s_k\geq s_{k+1}\geq \dots\geq s_d$, we get $s_k\geq \eps/(d-k+1)$. In the rest of the proof, we show that $s_k\geq \eps(1-(\eps/p_m)^{1/k})$ or $s_k \ge p_m$.

Suppose $s_i$ is the mode. 
It follows that there is $i\leq j\leq k-1$ such that $\frac{s_j}{s_{j+1}}\geq \left(\frac{s_i}{s_k}\right)^{1/(k-i)}$.  So, by \cref{lem:logconcaveexpecation}, 
$$\eps\leq  s_k+\dots+s_d\leq \frac{s_k}{1-\left(\frac{s_k}{s_i}\right)^{1/(k-i)}}$$ 
If $s_k\geq p_m$ or $s_k\geq \eps$ then we are done. Otherwise, 
$$ s_k\geq \eps \left(1-(s_k/p_m)^{1/(k-i)}\right)\geq \eps\left(1- \left(\eps/p_m\right)^{1/k}\right)$$
where we used $s_i\geq p_m$ and $s_k\leq \eps$.
\end{proof}

\begin{lemma}\label{lem:updowntruncation}
Given a strongly Rayleigh distribution $\mu:2^{[n]}\to \R_{\geq 0}$, let $A,B$ be two (nonnegative) random variables corresponding to the number of elements sampled from two {\em disjoint} sets such that  $\P{A+B=n}>0$ where $n=n_A+n_B$. Then,
\begin{eqnarray}
	\P{A\geq n_A | A+B=n}=\P{B\leq n_B | A+B=n} & \geq & \P{A\geq n_A}\P{B\leq n_B},  \\
	\P{A\leq n_A|A+B=n}=\P{B\geq n_B | A+B=n}  &\geq &  \P{A\leq n_A}\P{B\geq n_B}.
\end{eqnarray} 
\end{lemma}
\begin{proof}
We prove the second statement. The first one can be proven similarly.
First, notice
\begin{align*}
	&\P{A\leq n_A , A+B \geq n} + \P{B\geq n_B , A+B < n}\\
	=&\P{B\geq n_B, A\leq n_A, A+B\geq n} + \P{A\leq n_A, B\geq n_B, A+B< n}\\
	=& \P{B\geq n_B, A\leq n_A} \geq \P{B\geq n_B}\P{A\leq n_A}=:\alpha,
\end{align*}
where the last inequality follows by negative association.
Say $q=\P{A+B\geq n}$.
From above, either $\P{A\leq n_A, A+B \geq n}\geq \alpha q$ or $\P{B\geq n_B, A+B<n}\geq \alpha (1-q)$. In the former case, we get $\P{A\leq n_A | A+B\geq n}\geq \alpha$ and in the latter we get $\P{B\geq n_B | A+B<n}\geq \alpha$.
Now the lemma follows by the stochastic dominance property
\begin{eqnarray*} 
\P{A\leq n_A | A+B=n} &\geq& \P{A\leq n_A | A+B \geq n} \\
\P{B\geq n_B | A+B=n} &\geq& \P{B\geq n_B | A+B < n}	
\end{eqnarray*}
Note that in the special case that $A+B<n$ never happens, the lemma holds trivially.
\end{proof}

Combining the previous two lemmas, we get
%

\begin{corollary}\label{lem:SRA=nA}
Let $\mu:2^{[n]}\to\R_{\geq 0}$ be a SR distribution. Let $A,B$ be two random variables corresponding to the number of elements sampled from two disjoint sets of elements such that $A \ge k_A$ with probability 1 and $B \ge k_B$ with probability 1. If $\P{A\geq n_A},\P{B\geq n_B}\geq \eps_1$ and $\P{A\leq n_A},\P{B\leq n_B}\geq \eps_2$, then, letting $n'_A = n_A-k_A, n'_B = n_B-k_B$,
\begin{align*}
	&\P{A=n_A | A+B=n_A+n_B}\geq \eps \min\{\frac1{n'_A+1},\frac{1}{n'_B+1}\},\\
&\P{A=n_A | A+B=n_A+n_B}\geq \min\left\{p_m, \eps(1-(\eps /p_m)^{1/\max\{n'_A,n'_B\}})\right\}
\end{align*}
where $\eps=\eps_1\eps_2$ and $p_m\leq\max_{k_A\leq k\leq n_A+n_B-k_B} \P{A=k | A+B=n_A+n_B}$ is a lower bound on the mode of $A$.

In the special case that $n_A=1,n_B=1$, $k_A=0,k_B=0$,  if   $\P{A=1 | A+B=2} \le \eps$,  $p_m \ge   1-2\eps$. If $\eps\leq 1/3$, 
$$\P{A=1 | A+B=2}\geq  \max\left\{\eps/2, \eps\left(1-\frac{\eps}{1-2\eps}\right)\right\}.$$
\end{corollary}
To get the first statement, we construct a new SR distribution from $\mu$ as follows. First, we symmetrize $g_\mu$ by setting all $x_a \in A$ to $x$ and all $x_b \in B$ to $y$; call the resulting polynomial $q_\mu$. Then, notice $q'_\mu = q_\mu/(x^{k_A}x^{k_B})$ is real stable. Therefore, we can apply the above corollary to a distribution with generating polynomial $q'_\mu$.\footnote{To be precise, we apply the above corollary to the polarization of $q'_\mu$, where $x,y$ are polarized by a disjoint set of variables of size equal to their maximum degree.}

To get the second statement, notice that since the distribution of $A$ is unimodal, $$\min\{\P{A=0}, \P{A=2}\} \le \eps$$


%


%% file: maxflow.tex
\subsection{Max Flow}
This proposition and the max flow event are crucially used in the analysis of the bottom-bottom case in the payment theorem (\cref{thm:payment-main}). See \cref{ex:bottombottom} and the preceding discussion for more high-level intuition. The main consequences of this section are \cref{lem:bottompaymentprob} and \cref{lem:bottompayment-polyprob}. 
\begin{proposition}\label{lem:maxflow}
Let $\mu:2^{E} \to\R_{\geq 0}$ be a homogeneous SR distribution. For any $330\eps <\zeta < 0.002$ and disjoint sets $A,B\subseteq E$ such that  $\E{A_T},\E{B_T} \in [1-\eps,1+\eps]$ (where $T\sim\mu$) there is an event $\cE_{A,B}(T)$ such that $\P{\cE_{A,B}(T)}\geq 0.0246\zeta^2(1-\zeta/2.1-\eps)$  
and 
it satisfies the following three properties.
\begin{enumerate}[i)]
	\item $\P{A_T=B_T=1 | \cE_{A,B}(T)}=1$,
	\item $\sum_{e\in A} |\P{e} - \P{e| \cE_{A,B}(T)}| \leq \zeta$, and
	\item $\sum_{e\in B} |\P{e} - \P{e | \cE_{A,B}(T)}|\leq \zeta$.
\end{enumerate}	
\end{proposition}
In other words, under event $\cE _{A,B}$ which has a constant probability, $A_T=B_T=1$ and the marginals of all edges in $A,B$ are preserved up to total variation distance $\zeta$. 
We also remark that above statement holds for a much larger value of $\zeta$ at the expense of a smaller lower bound on $\P{\cE_{A,B}(T)}$.

Before, proving the above statement we prove the following lemma.
\begin{lemma}\label{lem:A'B'ABalpha}
Let $\mu:2^E \to\R_{\geq 0}$ be a homogeneous SR distribution. Let $A,B\subseteq E$ be two disjoint sets such that $ \E{A_T},\E{B_T} \in [1-\eps,1+\eps]$ (where $T\sim\mu$),  $A'\subset A$ and $B'\subseteq B$  and $\E{A'_T\cup B'_T}\geq 1+\alpha$ for some $\alpha>100\eps$. If $\alpha < 0.001$, we have
$$ \P{{A'_T}={B'_T}=A_T=B_T=1}\geq 0.11 \alpha^3.$$ 
\end{lemma}
\begin{proof}
	First, condition on $(A\smallsetminus A')_T=(B\smallsetminus B')_T=0$. This happens with probability at least $\alpha-2\eps \ge 0.98 \alpha$ because $\E{A_T}+\E{B_T}\leq 2+2\eps$ and $\E{A'_T}+\E{B'_T}\geq 1+\alpha$. Call this measure $\nu$. 
	It follows by negative association that  
	\begin{equation}\label{eq:ABpTflow} \EE{\nu}{A'_T},\EE{\nu}{B'_T}\in [\alpha-\eps, 2+3\eps - \alpha].	
 \end{equation}
	\begin{itemize}
	\item {\bf Case 1: $\EE{\nu}{A'_T+B'_T} > 1.5$.} 
Since $\EE{\nu}{A'_T+B'_T}\leq 2+2\eps$, by \cref{thm:rayleigh_expectconstprob}, 
$\PP{\nu}{A'_T+B'_T=2} \geq  0.25$.
Furthermore,
\begin{align*}
&\PP{\nu}{A'_T\geq 1},\PP{\nu}{B'_T\geq 1} \geq 1-e^{-(\alpha-\eps)}\geq 0.98\alpha \tag{\cref{lem:SR>=}, $\alpha<0.001$}\\
& \PP{\nu}{A'_T\leq 1},\PP{\nu}{B'_T\leq 1} \geq \alpha/2-1.5\eps	\tag{Markov's Inequality}
\end{align*}
Therefore, by \cref{lem:SRA=nA} and using $\alpha\leq 0.001$, $\P{A'_T=1 | A'_T+B'_T=2} \geq 0.45\alpha^2$.
It follows that 
$$\P{A_T=B_T=A'_T=B'_T=1} \geq (0.98\alpha)\PP{\nu}{A'_T=B'_T=1}\geq (0.98\alpha)0.25(0.45\alpha^2) \ge 0.11\alpha^3.$$

	\item {\bf Case 2: $\E{A'_T+B'_T}\leq 1.5$.}
Since $\EE{\nu}{A'_T+B'_T}\geq 1+\alpha$, by \cref{thm:rayleigh_expectconstprob},  $\P{A'_T+B'_T=2} \ge \alpha e^{-\alpha}\geq 0.99\alpha$. But now $\E{A'_T},\E{B'_T}\leq 1.5$ and therefore by Markov's Inequality, 
$$\PP{\nu}{A'_T\leq 1},\PP{\nu}{B'_T\leq 1}\geq 0.25.$$ 
On the other hand, by \cref{lem:SR>=} (similar to case 1) $\PP{\nu}{A'_T\geq 1},\PP{\nu}{B'_T\geq 1} \geq 1-e^{-\alpha+\eps}\geq 0.98\alpha$.
It follows by \cref{lem:SRA=nA} that 
$\P{A'_T=1 | A'_T+B'_T=2} \geq 0.2\alpha$.  Therefore,
$$ \P{A_T=B_T=A'_T=B'_T=1} \geq (0.98\alpha)\PP{\nu}{A'_T=B'_T=1}\geq (0.98\alpha)(0.2\alpha)(0.99\alpha)\geq 0.11\alpha^3$$
as desired.
	\end{itemize}
\end{proof}

It is worth noting that $\alpha^3$ dependency is necessary in the above example. For an explicit Strongly Rayleigh distribution consider the following product distribution:
	$$ (\alpha x_1 + (1-\alpha)y_2) (\alpha y_1 + (1-\alpha) z_2)(\alpha z_1 + (1-\alpha)x_2),$$
	and let $A=\{x_1,x_2\}$, $B'=B=\{y_1,y_2\}$, and $A'=\{x_1\}$.
	Observe that 
	$$\P{A_T=B_T=A'_T=B'_T=1} = \P{x_1=1,y_1=1,z_1=1}=\alpha^3.$$

\begin{proof}[Proof of \cref{lem:maxflow}]
	To prove the lemma, we construct an instance of the max-flow, min-cut problem. Consider the following graph with vertex set $\{s,A,B,t\}$. For any $e\in A, f\in B$ connect $e$ to $f$ with a directed edge of capacity $y_{e,f}=\P{e,f\in T| A_T=B_T=1}$. 
	For any $e \in E$, let $x_e := \P{e \in T}$. Connect $s$ to $e\in A$ with an arc of capacity $\beta x_e$ and similarly connect $f\in B$ to $t$ with arc of capacity $\beta x_f$, where $\beta$ is a parameter that we choose later. We claim that the min-cut of this graph is at least $\beta(1-\eps-\zeta/2.1)$. Assuming this, we can prove the lemma as follows: let $\bf{z}$ be the maximum flow, where $z_{e,f}$ is the flow on the edge from $e$ to $f$. We define the event $\cE_{A,B}(T) = \cE(T)$ to be the union of events $z_{e,f}$. More precisely, conditioned on $A_T=B_T=1$ the events $e,f\in T | A_T=B_T=1$ are disjoint for different pairs $e\in A, f\in B$, so we know that we  have a specific $e,f$ in the tree $T$ with probability $y_{e,f}$. And, of course, $\sum_{e\in A,f\in B} y_{e,f}=1$. So, for $e\in A, f\in B$ we include a $z_{e,f}$ measure of trees, $T$, such that $A_T=B_T=1, e,f\in T$.
		First, observe that  
		\begin{equation}
		\label{cEdefn}	
		\P{\cE}=\sum_{e\in A,f\in B} z_{e,f} \P{A_T=B_T=1}\geq \beta(1-\zeta/2.1-\eps)\P{A_T=B_T=1}.
		\end{equation}
		Part (i) of the proposition follows from the definition of $\cE$. Now, we check part (ii): Say  $z=\sum_{e\in A,f\in B} z_{e,f}$, and the flow into $e$ is $z_e$. Then,
		$$ \sum_{e\in A} |x_e-\P{e\in T | \cE}|=\sum_{e\in A} \left|x_e - \sum_f \frac{z_{e,f}}{z}\right| = \sum_{e\in A} |x_e -\frac{z_e}{z}|  $$
		Note that both $x$ and $z_e/z$ define a probability distribution on edges in $A$; so the RHS is just the total variation distance between these two distributions. 
		We can write
		\begin{eqnarray*}\sum_{e\in A} |x_e-\P{e\in T | \cE}|&=&2\sum_{e\in A: z_e/z>x_e} \left(\frac{z_e}{z} -x_e \right)\\
		&\leq& 2\sum_{e\in A: z_e/z > x_e} \left(\frac{\beta x_e}{\beta(1-\zeta/2.1-\eps)} - x_e\right) \\
		&\leq& 2 \cdot \sum_e x_e \frac{\zeta/2.1+\eps}{1-\zeta/2.1-\eps}\leq 2\frac{(1+\eps)(\zeta/2.1+ \eps)}{1-\zeta/2.1- \eps} \le  \zeta.	
		\end{eqnarray*}
		The first inequality uses that the max-flow is at least $\beta(1-\zeta/2.1-\epsilon)$ and that the incoming flow of $e$ is at most $\beta x_e$, and the last inequality follows by $\zeta<0.003$ and $\eps<\zeta/330$.
		(iii) follows by the same argument.

	It remains to lower-bound the max-flow or equivalently the min-cut.
	Consider an $s,t$-cut $S,\overline{S}$, i.e., assume $s\in S$ and $t\notin S$. Define $S_A=A\cap S$, $S_B=B\cap S$, and similarly $\oS_A = A \cap \bar{S}$, $\oS_B = B \cap \overline{S}$. We write
	\begin{eqnarray*}
		\text{cap}(S,\oS) &=& \beta x(\oS_A) + \beta x(S_B) + \sum_{e\in S_A, f\in \oS_B} y_{e,f}\\
		&= & \beta x(\oS_A\cup S_B) + \P{(S_A)_T=(\oS_B)_T=1 | A_T=B_T=1}
	\end{eqnarray*}
	If $x(S_B)\geq x(S_A)-\zeta/2.1$, then 
	$$\text{cap}(S,\oS) \geq \beta x(\oS_A\cup S_B)\geq \beta(x(\oS_A\cup S_A)-\zeta/2.1) \geq \beta (1-\eps-\zeta/2.1),$$
    and we are done.
	Otherwise, say $x(S_B) +\gamma= x(S_A)$, for some $\gamma>\zeta/2.1$. So, 
	
    $$x(\oS_B) + x(S_A) = x(\oS_B) + x(S_B)+\gamma \geq 1-\eps+\gamma$$
	So, by \cref{lem:A'B'ABalpha} with ($\alpha = \gamma-\eps > \zeta/2.1 - \eps > 100\eps$) 
	$$\P{(S_A)_T=(\oS_B)_T=1 | A_T=B_T=1} \geq \frac{\P{(S_A)_T=(\oS_B)_T=A_T=B_T=1}}{ \P{A_T=B_T=1}}\geq \frac{0.11(\gamma-\eps)^3}{\P{A_T=B_T=1}}.$$
	It follows that 
	\begin{eqnarray*}\text{cap}(S,\oS) &\geq& \beta x(\oS_A\cup S_B) +\frac{0.11(\gamma-\eps)^3}{\P{A_T=B_T=1}}\\
	&\geq &	 \beta (x(\oS_A \cup S_A) - \gamma)+\frac{0.11(\gamma-\eps)^3}{\P{A_T=B_T=1}}\\
	&\geq & \beta(1-\eps-\gamma)+\frac{0.11(\gamma-\eps)^3}{\P{A_T=B_T=1}}
	\end{eqnarray*}

	To prove the lemma we just need to choose $\beta$ such that RHS is at least $\beta(1-\eps-\zeta/2.1)$. Or equivalently,
	$$ \frac{0.11(\gamma-\eps)^3}{\P{A_T=B_T=1}} \geq \beta(\gamma-\zeta/2.1).$$
	In other words, it is enough to choose $\beta \leq \frac{0.11(\gamma-\eps)^3}{\P{A_T=B_T=1}(\gamma-\zeta/2.1)}$. Since $\gamma > \zeta/2.1$ and $\zeta>330\eps$, we have $\gamma-\eps \geq 0.473\zeta$. Therefore, we can set $\beta=\frac{0.11(0.473\zeta)^2}{\P{A_T=B_T=1}}$. Finally, this plus \eqref{cEdefn} gives
	$$ \P{\cE} \geq (1-\zeta/2.1-\eps)\beta \P{A_T=B_T=1} = 0.11(0.473\zeta)^2(1-\zeta/2.1-\eps) \ge 0.0246\zeta^2(1-\zeta/2.1-\eps)$$
	as desired.
%
%
\end{proof}

\begin{definition}[Max-flow Event]\label{def:max-flow-event}
\hypertarget{tar:max-flow-event}{For a polygon cut $S\in\cH$ with polygon partition  $A,B,C$, let $\nu$ be the max-entropy distribution conditioned on $S$ is a tree and $C_T=0$. By \cref{lem:treeconditioning}, we can write $\nu: \nu_{S} \times \nu_{G/S}$, where $\nu_S$ is supported on trees in $E(S)$ and $\nu_{G/S}$ on trees in $E(G/S)$.
	For a sample $(T_S,T_{G/S})\sim \nu_S\times \nu_{G/S}$, we say $\cE_S$ occurs if  $\cE_{A,B}(T_{G/S})$ occurs, where $\cE_{A,B}(.)$ is the event defined in \cref{lem:maxflow} for sets $A,B$ and $\zeta=\eps_M:=\frac1{4000}$ and $\eps=2\eps_\eta$. }
\end{definition}

\begin{corollary}\label{cor:poly-reduction}
For a polygon cut $S\in\cH$ with polygon partition $A,B,C$, we have, 
\begin{enumerate}[i)]
\item $\P{{\cal E}_S} \ge 0.0245\eps_M^2.$
\item For any set $F\subseteq \delta(S)$ 	conditioned on ${\cal E}_S$ marginals of edges in $F$ are preserved up to $\eps_M+\eps_\eta$ in total variation distance.
\item For any $F\subseteq E(S)\cup \delta(S)$ where either $F \cap A=\emptyset$ or $F\cap B=\emptyset$, there is some $q\in x(F)\pm (\eps_M + 2\eps_\eta)$ such that the law of $F_T | \cE_S$ is the same as a $BS(q)$.
\end{enumerate}
\end{corollary}
\begin{proof}
Condition $S$ to be a tree and $C_T=0$ and let $\nu$ be the resulting measure. 
It follows that  
$$\P{{\cal E}_S}  = \PP{\nu}{\cE_S} \P{C_T=0,S\text{ tree}}\geq 0.0246 \eps_M^2(1-\eps_M/2.1-\eps)\P{C_T=0,S\text{ tree}} \geq 0.0245\eps_M^2,$$
using $\epsilon = 2\epsilon_\eta$ and $\eps_M = 1/4000$, which proves (i). 

Now, we prove (ii). By \cref{lem:maxflow}, the marginals of edges in $\delta(S)$ are preserved up to a total variation distance of $\eps_M$, so
$$\EE{\nu}{(F\cap \delta(S))_T | \cE_{A,B}(T_{G/S})} = \EE{\nu}{(F\cap \delta(S))_T}\pm \eps_M.$$
Since $x(C)\leq \eps_\eta$ and $x(\delta(S)) \le 2+\eps_\eta$, by negative association, 
$$x(F\cap \delta(S))-\eps_\eta/2\leq \EE{\nu}{(F\cap \delta(S))_T} \leq x(F\cap \delta(S)) + \eps_\eta.$$
This proves (ii).
Also observe that since conditioned on $\cE_S$, we choose at most one edge of $F\cap \delta(S)$, $(F\cap \delta(S))_T$ is a $BS(q_{G/S})$ for some 
 $q_{G/S}=x(F \cap \delta(S))\pm(\eps_M+\eps_\eta)$.

On the other hand, observe that conditioned on $\cE_S$, $S$ is a tree, so 
$$x(F\cap E(S))\leq \E{(F\cap E(S))_T | \cE_S} \leq x(F\cap E(S))+\eps_\eta/2.$$
Since the distribution of $(F\cap E(S))_T$ under $\nu | \cE_S$ is SR,  there is a random variable $BS(q_S)=(F\cap E(S))_T$  where $x(F\cap E(S))\leq q_S\leq x(F\cap E(S))+\eps_\eta/2$.

Finally, $F_T|\cE_S$ is exactly $BS(q_S)+BS(q_{G/S}) = BS(q)$ for $q=x(F)\pm(\eps_M+2\eps_\eta)$.
\end{proof}




Normally, conditioning on $\delta(S)_T$ for a polygon $S \in \cH$ may dramatically change the distribution of any random variable $\delta(u)_T$ for any $u$ which is an ancestor of $S$ and for which $\delta(u) \cap \delta(S) \not= \emptyset$. For example, it may essentially determine the parity of $\delta(u)_T$. 
On the other hand, the following two corollaries show that after conditioning on $\cE_S$ the probability $\delta(u)$ is even remains a (large) constant. So in some sense, conditioning on the max-flow event $\cE_S$ decouples the random variables $\delta(S)_T$ and $\delta(u)_T$.

\begin{corollary}	\label{lem:bottompaymentprob}
For $u\in \cH$ and a polygon cut $S\in \cH$ that is an ancestor of $u$,
	 $$\P{\delta(u)_T\text{ odd} | \cE_S}\leq  0.5678.$$
\end{corollary}
\begin{proof}
First, notice by \cref{obs:childofSD}, $\delta(u)\cap\delta(S)$ is either a subset of $A$, $B$, or $C$. Therefore, by (iii) of \cref{cor:poly-reduction} we can write $\delta(u)_T | \cE_S$ as a $BS(q)$ for $q\in 2\pm [0.001]$ (where we use that $\eps_M+3\eps_\eta<0.001$). Furthermore,  since $\delta(u)_T\neq 0$ with probability $1$, we can write this as a $1+BS(q-1)$. 
Therefore, by \cref{cor:bernoullisumeven},
$$\P{\delta(u)_T \text{ odd} | \cE_S} = \P{BS(q-1)\text{ even}}\leq \frac12(1+e^{-2(q-1)})\leq \frac12(1+e^{-1.999})\leq 0.5678$$
as desired.
\end{proof}

\begin{corollary}\label{lem:bottompayment-polyprob}
For a polygon cut $u\in \cH$  and a polygon cut $S\in \cH$ that is an ancestor of $u$,
	 $$\P{u \text{ not left happy} | \cE_S}\leq  0.56797.$$
	 and the same follows for right happy. 
\end{corollary}
\begin{proof}
Let $A,B,C$ be the polygon partition of $u$. Recall that for $u$ to be left-happy, we need $C_T=0$ and $A_T$  odd.
Similar to the previous statement, we can write $A_T | \cE_S$ as a $BS(q_A)$ for $q_A\in 1\pm [0.00026]$ (where we used that $\eps_M=1/4000$ and $\eps_\eta\leq \eps_M/300$).
 Therefore, by \cref{cor:bernoullisumeven},
$$\P{A_T \text{ even} | \cE_S} \leq \frac12(1+e^{-2q_A})\leq  \frac12(1+e^{-1.99948})\leq 0.56771$$
Finally, $\E{C_T | \cE_S}\leq x(C_T)+\eps_M+2\eps_\eta \leq 0.00026$. 
Now using the union bound, 
$$\P{u \text{ not left happy} \mid \cE_S} \le 0.56771 + 0.00026 \le 0.56797$$
as desired.
\end{proof}

%% file: matching.tex

\section{Matching}
\label{sec:matching}



The main result of this section is to construct a matching that we use in order to decide which edges will have positive slack to compensate for the negative slack of edges going higher. 
Refer to \cref{ex:simple} for a high-level motivation to construct a matching. . 

\begin{definition} [$\eps_F$ fractional edge]
\hypertarget{tar:epsF}{For $z\geq 0$ we say that $z$ is $\eps_{F}$-fractional if $\eps_F \le z\leq 1-\eps_{F}$.}
\end{definition}
\noindent The following lemma is the main result of this section
\begin{lemma}[Matching Lemma]
\label{lem:matching}
For any $S \in \cH$, $\eps_{F}\leq 1/10, \eps_B\geq 21\eps_{1/2}, \alpha\geq 2\eps_\eta$, $\eps_{1/2}\leq 0.0002$, there is a matching from good edges (see \cref{def:goodedges}) in $E^\rightarrow(S)$ to edges in $\delta(S)$ where every good edge bundle $\bbe={\bf(u,v)}$ (where $u,v\in\cA(S)$) is matched to a fraction $m_{\bbe,u}$ of edges in $\delta^\uparrow(u)$ and a fraction $m_{\bbe,v}$ of $\delta^\uparrow(v)$, and:
\begin{eqnarray}
\label{eq:uvbadematch}m_{\bbe,u} F_u  + m_{\bbe,v} F_v &\le & x_\bbe (1 + \alpha)\\
\label{eq:matchedamount}\sum_{e \in \delta^\rightarrow (u)} m_{\bbe,u} &=& x(\delta^\uparrow(u)) Z_u,
\end{eqnarray}
where for every atom $u\in \cA(S)$, define
$$\hypertarget{tar:Fu}{F_u=1-\eps_B\I {x(\delta^\uparrow(u))\text{ is $\eps_F$ fractional}},} \quad
\hypertarget{tar:Zu}{Z_u:=\left(1 + \I{|\cA(S)|\geq 4, x(\delta^\uparrow(u))\leq \eps_F}\right).}$$
\end{lemma}

Roughly speaking, the intention of the above lemma is to match good edges in $E^\rightarrow(S)$ to a similar fraction of edges that go higher (such that an edge bundle $e$ adjacent to atoms $u,v$ is only matched to edges in $\delta^\uparrow(u),\delta^\uparrow(v)$). Since we never ``reduce'' bad edges in the proof of payment theorem (\cref{thm:payment-main}), we don't use them in the matching. That inherently can cause a problem, as there could not be ``enough'' good edges in $E^\rightarrow(S)$ to saturate the edges going higher in the matching. The parameter $F_u$ help us in this regard; in particular, it allows us to match some of the (good) edges in $E^\rightarrow(S)$ to more than their fraction in $\delta(S)$.
 
Next, we motivate the parameter $Z_u$. If  $x(\delta^\uparrow(u))\approx 0$, when those edges are reduced the conditional probability that $\delta(u)_T$ is even could be very close to 0. 
The parameter $Z_u$ lets us match twice as many edges to $\delta^\uparrow(u)$; so there will be only half a burden to fix the parity of $\delta(u)_T$.
See the discussion in \hyperlink{xu-close-to-zero}{overview section} for more details. 



Throughout this section we adopt the following notation: For a cut $S\in \cH$ and a set $W\subseteq \cA(S)$, we write 
\begin{align*}
E(W,S\smallsetminus W)&:=\cup_{u\in W, v\in \cA(S)\smallsetminus W} E(u,v),\\
\delta^\uparrow(W)&:=\cup_{u\in W} \delta^\uparrow(u)=\delta(W)\cap \delta(S), \\
\delta^\rightarrow(W) &:= \cup_{u\in W} \delta^\rightarrow(u).
\end{align*}
Note that in $\delta^\rightarrow(W)\not\subseteq \delta(W)$ since it includes edge bundles between atoms in $W$. 

Before proving the main lemma we record the following facts.
\begin{lemma}
\label{fact:anothercutineq}
For any $S\in \cH$ 
and $W \subsetneq \cA(S)$ (recall $\cA(S)$ is the set of $u \in \cH$ with $\p(u) = S$), we have 
$$ x(\delta^\rightarrow(W))  
	\geq \frac12 \sum_{u\in W} x(\delta(a)) - \eps/2 \ge |W| - \eps/2.$$
\end{lemma}
\begin{proof}
We have
$$ x(\delta^\rightarrow (W))  
= \frac12 \left(\sum_{u \in W} (x(\delta (u)) +   x(	E(W, S\smallsetminus W)) -  x(\delta ^\uparrow (W))\right).$$
Since $x(\delta (S\smallsetminus W)) \ge 2$ and $x(\delta (S)) \le 2 + \eps$, we have:
$$\text{(a) }  x(	E(W, S\smallsetminus W)) + x(\delta^\uparrow(S\smallsetminus W)))\ge 2 \quad\text{ and }\quad\text{(b) }  x(\delta^\uparrow(W)) +  x(	\delta^\uparrow( S\smallsetminus W)) \le 2 + \eps.$$
Subtracting (b) from (a), we get
$$x(	E(W, S\smallsetminus W)) - x(\delta ^\uparrow (W)) \ge -\eps,$$
which after substituting into the above equation,
completes the proof of the first inequality in the lemma statement. The second inequality follows from the fact that $\delta (u) \ge 2$ for each atom $u$.
\end{proof}

\begin{lemma}\label{lem:three-nodes-good}
For $S \in\cH$, if $|\cA(S)|=3$ then there are no bad edges in \hyperlink{tar:ErightarrowS}{$E^\rightarrow(S)$}.
\end{lemma}
\begin{proof}
Suppose  $\cA(S)=\{u,v,w\}$ and
$\bbe=(u,v)$  is a bad edge bundle. Then $|x_\bbe - \frac{1}{2}| \le \epsilon_{1/2}$. In addition, by \cref{thm:badedges}, $x(\delta^\uparrow(u)),x(\delta^\uparrow(v)) \le 1/2+9\epsilon_{1/2}$. Therefore, 
$$x_{\bf(u,w)} = x(\delta(u)) - x_{\bbe} -  x(\delta^\uparrow(u))\geq 1-10\epsilon_{1/2}.$$
Similarly, $x_{\bf(v,w)}\geq 1-10\eps_{1/2}$.
Finally, since $x(\delta (S)) \ge 2$, and $x(\delta^\uparrow(u)),x(\delta^\uparrow(v))\leq 1/2+9\eps_{1/2}$, we must have $x(\delta^\uparrow(w)) \ge 1-18\epsilon_{1/2}$.
But, this contradicts the assumption that $w\in \cH$ must satisfy $x(\delta(w))\leq 2+\epsilon_\eta$.
\end{proof}

\begin{proof}[Proof of \cref{lem:matching}]
	We will prove this by setting up a max-flow min-cut problem. 
	Construct a graph with vertex set $\{s,X,Y,t\}$, where $s,t$ are the source and sink. We identify $X$ with  the set of good edge bundles in $E^\rightarrow(S)$ and $Y$ with the set of atoms in $\cA(S)$. For every edge bundle $\bbe \in X$, add an arc from $s$ to $\bbe$ of capacity $c(s,\bbe):= (1+\alpha) x_\bbe$.  For every $u\in \cA(S)$, 
		there is an arc $(u,t)$ with capacity
		$$ c(u,t) = x(\delta^\uparrow(u))F_u Z_u.$$
		
Finally, connect $\bbe=(u,v)\in X$ to nodes $u$ and $v\in Y$ with a directed edge of infinite capacity, i.e., $c(\bbe,u)=c(\bbe,v)=\infty$.
We will show below that there is a flow saturating $t$, i.e. there is a flow of value 
$$c(t):=\sum_{u \in \cA(S)} c(u,t) = \sum_{u\in \cA(S)} x(\delta^\uparrow(u))F_uZ_u.   $$

Suppose that in the corresponding max-flow, there is a flow of value $f_{\bbe,u}$ on the edge $(\bbe,u)$.
Define $$m_{\bbe,u} := \frac{f_{\bbe,u}}{F_u}.$$
Then \eqref{eq:uvbadematch} follows from the fact that the flow leaving $\bbe$ is at most the capacity of the edge from $s$ to $\bbe$, and \eqref{eq:matchedamount} follows by conservation of flow on the node $u$ (after cancelling out $F_u$ from both sides).

We have left to show that for any $s$-$t$ cut $A,\overline{A}$ where $s\in A, t\in \overline{A}$ that the capacity of this cut is at least $c(t)$.

\begin{claim} If $A=\{s\}$, then capacity of $(A,\overline{A})$ is at least $c(t)$.	
\end{claim}
\begin{proof}
First, note that 
\begin{align}
	c(t) &= \sum_{u \in \cA(S)} x(\delta^\uparrow(u))F_uZ_u \le \sum_{u \in \cA(S)} x(\delta^\uparrow(u))Z_u\nonumber \\ &\le \I{|\cA(S)| \ge 4} \cdot |\{u \in \cA(S): x(\delta^\uparrow(u)) \le \eps_F\}| \cdot \eps_F + x(\delta(S))\nonumber \\
	&\le 2+\eps_\eta + \eps_F\I{|\cA(S)| \ge 4}|\cA(S)| \label{eq:bound-ct}
\end{align}
because $F_u \le 1$ and $Z_u = 1 + \I{|\cA(S)| \ge 4,x(\delta^\uparrow(u)) \le \eps_F}$. 

Second, note that
\begin{align*}x(E^\rightarrow (S))  = \frac12 \sum_{u \in \cA(S)} (x(\delta (u)) - x(\delta ^\uparrow (u))) 
\ge \frac{2 |\cA(S)| - (2 + \eps_\eta)}{2}
= |\cA(S)| - 1 - \eps_\eta/2.
\end{align*}
Therefore, if there are $k$ bad edges in $E^\rightarrow(S)$, then
\begin{align}
	x_G \ge |\cA(S)| - 1 - \eps_\eta/2 - k(\frac{1}{2}+\eps_{1/2}) \label{eq:sum-inside}
\end{align}

\textbf{Case 1: $|\cA(S)|=3$.} Then $Z_u=1$ for all $u\in\cA(S)$ and by \cref{lem:three-nodes-good} all edges are good. 
So, by \cref{eq:sum-inside},
$x(E ^\rightarrow (S))\ge 2- \eps_\eta/2$. 
Thus, for $\alpha\geq 2\eps_\eta$ we have
$$c(s) = (1+ \alpha) x_G \ge (2-\epsilon_\eta/2)(1+\alpha) \ge 2+\epsilon_\eta \underset{\cref{eq:bound-ct}}{\ge} c(t)$$
as desired. 

\textbf{Case 2: $|\cA(S)|\geq 5$.} By \cref{thm:badedges} there is at most one bad half edge adjacent to every vertex. Therefore there are at most  $|\cA(S)|/2$ bad edges, so by \cref{eq:sum-inside}, 
$$(1+\alpha)x_G \ge (1+\alpha)\left(|\cA(S)| - 1 - \eps_\eta/2 - \frac{1}{2}|\cA(S)|(\frac{1}{2}+\eps_{1/2})\right) \ge 2+\epsilon_\eta + \eps_F |\cA(S)| \underset{\cref{eq:bound-ct}}{\ge} c(t)$$
where the second to last inequality holds, using $\alpha \ge  2 \eps_\eta$, $|\cA(S)|\ge 5$, $\eps_{1/2} \le 0.01$, and $\eps_F \le 0.1$.

\textbf{Case 3: $|\cA(S)|=4$, and we have 0 or 1 bad edges.} Then by \cref{eq:sum-inside}, $x_G\geq 2.5-\eps_\eta /2 - \eps_{1/2}$, so by \cref{eq:bound-ct}, $(1+\alpha)x_G \geq 2+\eps_\eta + 4\eps_F \ge c(t)$ for $\eps_F \le 0.1$, $\alpha \ge 2\eps_\eta, \eps_{1/2} \le 0.01$. 

\textbf{Case 4: $|\cA(S)|=4$, and there are 2 bad edges.}
Then they form a perfect matching inside $S$ and  for each
$u\in\cA(S)$, $x(\delta^\uparrow(u))\le 1/2 + 9\eps_{1/2}$ (see \cref{thm:badedges}).

Therefore it must also be the case that $x(\delta^\uparrow(u))\ge  \eps_F$ for each $u\in\cA(S)$. If not,  there would have to be a node $u' \in \cA(S)$ such that
$x(\delta^\uparrow(u')) \ge (2- \eps_F)/3 > 1/2 + 9\eps_{1/2}$,
which is a contradiction to $u'$ having an incident bad edge.
Thus, for each $u \in \cA(S)$, $x(\delta^\uparrow(u))$ is $\eps_F$-fractional, i.e., $F_u=1-\eps_B$ and $Z_u=1$ implying that 
 $c(t) \le (2 + \eps_\eta)(1- \eps_B)$.
 Therefore, by \cref{eq:sum-inside},
 $$c(s) = (1+ \alpha) x_G \ge (1+ \alpha) (2-2\eps_{1/2} - \eps_\eta/2,)$$ 
 and the rightmost quantity is at least $c(t)$ for $\eps_B \ge 2 \eps_{1/2}$ and $\alpha \ge 2 \eps_\eta$.
\end{proof}

From now on, we assume that the min s-t cut $A\neq \{s\}$. In the following we will prove that for any set of \textit{atoms} $W \subsetneq S$, we have:
\begin{equation}\label{eq:goalTdeltarigthup}
 c(s,\delta^\rightarrow(W))  = (1+\alpha) x_G(\delta^\rightarrow(W)) \geq c(\delta^\uparrow(W),t) 
\end{equation}
where for a set $F$ of edges we write $x_G(F)$ to denote the total fractional value of good edges in $F$.

Let $A_X=A\cap X, A_Y=A\cap Y$ and so on. Assuming the above inequality, let us prove the lemma: First, for the set of edges $A_X$ chosen from $X$, let $Q$ be the set of endpoints of all edge bundles in $A_X$ (in $\cA(S)$). 

Observe that we must choose all atoms in $Q$ inside $A_Y$ due to the infinite capacity arcs, i.e., $Q\subseteq A_Y$. 
Let $W =  S \smallsetminus Q$. Note that $W\neq S$. Then:
\begin{eqnarray*}
 c(A,\overline{A}) &= & c(A_Y,t) + c(s,\overline{A}_X)\\
 &\geq  & c(\delta^\uparrow(Q),t) + c(s,\delta^\rightarrow(W))  \\
 &=& c(\delta^\uparrow(S),t)-c(\delta^\uparrow(W)) + c(s,\delta^\rightarrow(W)) \geq c(\delta^\uparrow(S),t),
\end{eqnarray*}
where the last inequality follows by \eqref{eq:goalTdeltarigthup}.

Finally, we prove \eqref{eq:goalTdeltarigthup}. Suppose atoms in $W$ are adjacent to $k$ bad edges. Then
\begin{align}
 x_G(\delta^\rightarrow(W))	 &=  x(\delta^\rightarrow(W))  - x_B(\delta^\rightarrow(W))\notag\\
\intertext{which by \cref{fact:anothercutineq} and the fact that each bad edge has fraction at most $1/2+\eps_{1/2}$, is}
&\geq  |W| - \eps_\eta/2 - k(1/2+\eps_{1/2}). \label{eq:xGEdeltalower}
\end{align}
 
 To upper bound $ c(\delta^\uparrow(W),t)$, we observe that for any $u\in\cA(S)$, 
 $$c(u,t)\leq \begin{cases} x(\delta^\uparrow(u))Z_u\leq 1/5  & \text{if $x(\delta^\uparrow(u))< \eps_F$}\\
  (1/2+9\eps_{1/2})(1-\eps_B) & \text{if 
 $x(\delta^\uparrow(u))>\eps_F$ and $u$ incident to bad edge}	\\
 1 + \eps_\eta & \text{otherwise, using \cref{lem:shared-edges}.}
 \end{cases}
 $$
Therefore, we can write,
 $$ c(\delta^\uparrow(W),t) \leq k(1/2+9\eps_{1/2})(1-\eps_B) + (|W|-k)(1+\eps_{\eta}).$$
 Now, to prove \eqref{eq:goalTdeltarigthup}, using \eqref{eq:xGEdeltalower}, it is enough to choose $\alpha$ and $\eps_B$ such that,
 $$
 (1+\alpha) \left(|W| - \eps_{\eta}/2 - k(1/2+\eps_{1/2})\right) \geq k(1/2+9\eps_{1/2})(1-\eps_B) + (|W|-k)(1+\eps_{\eta}),
 $$
 or equivalently,
 $$ |W|(\alpha-\eps_{\eta}) \geq k (\alpha/2 + 10\eps_{1/2} + \alpha \eps_{1/2} -\eps_B/2 - 9\eps_B \eps_{1/2} -\eps_{\eta}) + \frac{\eps_{\eta}}{2}(1+\alpha)$$
Since  every atom is adjacent to at most one bad edge, $k\leq |W|$
and $|W|\ge 1$, the inequality follows using
$\eps_B \ge  21 \eps_{1/2}$ and $\alpha > 2 \eps_\eta$ and $\eps_{1/2}\leq 0.0002$ and $\eps_\eta \leq \eps_{1/2}^2$.
\end{proof}

%

%% file: payment.tex

\section{Reduction and payment}\label{sec:payment}


In this section we prove \cref{thm:payment-main}. 






In  \cref{sec:probabilistic} we defined a number of happy events, such as 2-1-1 happy or 2-2-2 happy and showed that each of these events occurs with probability at least $p$. In this section, we will subsample these events to define a corresponding decrease  event that occurs with probability {\em exactly}\footnote{Suppose that under the distribution $\mu$ on spanning trees, some event ${\cal D}'$  has probability $q \ge p$ and we seek to define an event ${\cal D} \subseteq {\cal D}'$ that has probability {\em exactly} $p$.
To this end, one can copy every tree $T$ in the support of $\mu$, exactly $\lfloor \frac{kq}{p}\rfloor$ times for some integer $k>0$ and whenever we sample $T$ we choose a copy uniformly at random. So, to get a probability exactly $p$ for an event, we say this event occurs if for a ``feasible'' tree $T$ one of the first $k$ copies are sampled. Now, as $k\to\infty$ the probability that $\cD$ occurs converges to $p$. Now, for a number of decreasing events, $\cD_1,\cD_2,\dots,$ that occur with probabilities $q_1,q_2,\dots$ (respectively), we just need to let $k$ be the least common multiple of $p/q_1, p/q_2,\dots$ and follow the above procedure. Another method is to choose an independent Bernoulli with success probability $p/q$ for any such event ${\cal D}$.} $p$.

\paragraph{Reduction Events.} 
\begin{itemize}
\item \textbf{Bottom edges.} \hypertarget{reduction-events}{For each polygon cut}  $S\in\cH$, let $\cR_S$ be the indicator of a uniformly random subset of measure $p$ of the \hyperlink{tar:max-flow-event}{max flow event} ${\cal E}_S$. Note that when $\cR_S=1$ then in particular we know that the polygon $S$ is happy.\label{def:RS}
\item \textbf{Top edges.} \label{def:Reu} For a top edge bundle $\bbe=(u,v)$ define
$$\cH_{\bbe,u} = \begin{cases}
 1 & 	\text{if $\bbe$ is 2-1-1 happy and good w.r.t. } u\\
  1 & \text{if $\bbe$ is 2-2 happy and good, but not 2-1-1 good with respect to $u$} \\
  0 & \text{otherwise.}
  \end{cases}
  $$ 
  and let $\cH_{\bbe,v}$ be defined similarly. Since $p$ is a lower bound on the probability a good edge is happy, we may now let $\cR_{\bbe,u}$ and $\cR_{\bbe,v}$ be indicators of subsets of measure $p$ of $\cH_{\bbe,u}$ and $\cH_{\bbe,v}$ respectively (note $\cR_{\bbe,u}$ and $\cR_{\bbe,v}$ may overlap). 
In this way every top edge bundle $\bbe=(u,v)$ is associated with indicators $\cR_{\bbe,u}$ and $\cR_{\bbe,v}$. In the special case that $u$ is in case 3 (and not case 1 or 2) of \cref{thm:probabilistic}, fix two half edge bundles $\bbe,\bbf$ that are neighbors of $u$ which satisfy the conditions of case 3. For these edges, by \cref{thm:probabilistic}, $\cH_{\bbe,u} \cap \cH_{\bbf,u}$ has measure at least $p$. This is because $\cH_{\bbe,u} \cap \cH_{\bbf,u}$ happens if and only if $\bbe,\bbf$ are 2-2-2 happy with respect to $u$. Here, we choose $\cR_{\bbe,u},\cR_{\bbf,u}$ to be the same subset of measure $p$ of $\cH_{\bbe,u} \cap \cH_{\bbf,u}$. 
\end{itemize}

Define $r:E\to\R_{\geq 0}$ as follows: For any (non-bundle) edge $e$,
 $$r_e = \begin{cases} 
 \beta x_e \cR_{S} & \text{if $\p(e)=S$ for a polygon cut $S\in\cH$} \\
 \frac{1}{2}\tau x_e (\cR_{\bbf,u} + \cR_{\bbf,v}) & \text{if $e\in \bbf$ for a top edge bundle $\bbf=(u,v)$},
\end{cases}$$ 
for $\beta$, the parameter of \cref{thm:payment-main} and $\tau$ as defined in \ref{eq:constants}. 

\paragraph{Increase Events} 
Let ${\bf E}$ be the set of edge bundles, i.e., top/bottom edge bundles. 
Now, we define the increase vector $I: {\bf E} \to \R_{\geq 0}$ as follows:
\begin{itemize}
\item \textbf{Bottom edges.} For each polygon $S \in \cH$ (and corresponding bottom edge bundle) with polygon partition $A,B,C$, let 
$r(A):= \sum_{f \in A} r_f$, $r(B):= \sum_{f \in B} r_f$, and  $r(C):= \sum_{f \in C} r_f$.
Then set
\begin{align}
I_S:= (1+\eps_\eta)
\Big(&
\max \{r(A)\cdot \I{S \text{ not left happy}}, r(B) \cdot\I{S \text{ not right happy}} \}\nonumber \\
&+ r(C) \I{S\text{ not happy}}\Big).
\label{eq:IeBottom}	
\end{align}
\item \textbf{Top edges.} 
For every degree cut $S \in \cH$, invoke \cref{lem:matching} with \begin{align}\label{eq:matching-params} \alpha = 2\eps_\eta, \eps_B = 21\eps_{1/2}, \eps_F = 1/10\tag{Matching parameters}\end{align} and let $m_{\bbe,u}$ be the resulting matching for every $u \in \cA(S)$. For each  top edge bundle $\bbe= (u,v)$, let
\begin{equation}
\label{eq:Ieu}	
I_{\bbe,u} := \sum_{g \in \delta^\uparrow(u)} r_g \cdot\frac{ m_{\bbe,u}}{ \sum_{\bbf\in\delta^\rightarrow(u)} m_{\bbf,u}}\I{u \text{ is odd}},\end{equation}
and define $I_{\bbe,v}$ analogously. Let $I_\bbe = I_{\bbe,u} + I_{\bbe,v}$.
\end{itemize}

The following theorem is the main technical result of this section.
\begin{theorem}\label{thm:expI}
For any good top edge bundle $\bbe$, 	
$\E{I_\bbe} \le (1-\frac{\eps_{1/1}}{6})p\tau x_\bbe$,
and for any bottom edge bundle $S$,
$\E{I_S} \le 0.99994 \beta p$.
\end{theorem}
Using this theorem, we can prove the desired theorem: 
\paymentmain*
\begin{proof}[Proof of \cref{thm:payment-main}]
First, we set the constants: 
\begin{align}\label{eq:constants}
\eps_{1/2}=0.0002, \eps_{1/1}=\frac{\eps_{1/2}}{12}, p=0.005\eps_{1/2}^2, \hyperlink{tar:max-flow-event}{\eps_M}=0.00025, \tau=0.571\beta \tag{Global constants} 
\end{align}
Define $E_g$ to be the set of bottom edges together
with any edge $e$ which is part of a good top edge bundle. Now, we verify (i): We show for any $S\in \cH$ such that $\p(S)$ is a degree cut, $x(E_g\cap \delta(S))\geq 3/4$. 
First, by \cref{thm:badedges}, if $x(\delta^\uparrow(S))\geq 1/2+9\eps_{1/2}$ then all edges in $\delta^\rightarrow(S)$ are good, so the claim follows because by \cref{lem:shared-edges}, $x(\delta^\rightarrow(S))\geq 1-\eps_\eta\geq 3/4$. Otherwise, $x(\delta^\uparrow(S))\leq 1/2+9\eps_{1/2}$. Then, by \cref{thm:badedges} there is at most one bad edge in $\delta^\rightarrow(S)$. Therefore, there is a fraction at least $x(\delta^\rightarrow(S))-(1/2+\eps_{1/2})\geq 3/4$ of good edges in $\delta^\rightarrow(S)$.

For any edge $e\in E'$ define
\begin{equation}
	s_e = -r_e + \begin{cases}I_\bbf \frac{x_e}{x_{\bbf}} & \text{ if $e\in \bbf$ for a top edge bundle $\bbf$,}\\
	I_S x_e & \text{ if $\p(e)=S$ for a polygon cut $S\in\cH$.}
 \end{cases}
\end{equation}

Now, we verify (ii): First, we observe that $s_e=0$ (with probability 1) if $e$ is part of a bad edge bundle since  we defined reduction events only for good edges and $m_{\bbe,u}$ is non-zero only for good edge bundles.
 Since $r_e \le \beta x_e$ for bottom edges and $r_e \le \tau x_e$ for top edges, and $\tau \le \beta$, it follows that $s_e \ge -x_e \beta$ with probability 1.

Now, we verify (iii): 
Suppose a polygon cut $u$ is not \hyperlink{tar:hierarchy}{left-happy}. Since $u$ is not happy we must have $\cR_u=0$ and $r_e=0$ for any $e\in F$.
Therefore,
\begin{align*} s(A) + s(F)+ s^-(C) &= s(A) + I_S x(F) + s^-(C) \\
&\geq -r(A) + (1+\eps_\eta)(r(A)+r(C))(1-\eps_\eta/2) -r(C)\geq 0. 	
\end{align*}
where we used that $x(F)\geq 1-\eps_\eta/2$.

Now, we verify (iv): Let $S\in \cH$, where $\p(S)$ is a degree cut. If $S$ is odd, then $r_e=0$ for all edges $e\in \delta^\rightarrow(S)$; so  by \cref{eq:Ieu} 
\begin{align*}
	s(\delta(S)) &\geq -\sum_{g\in\delta^\uparrow(S)} r_g + \sum_{\bbe\in \delta^\rightarrow(S)} I_{\bbe,S}\\
	&=-\sum_{g\in\delta^\uparrow(S)} r_g+ \sum_{\bbe\in\delta^\rightarrow(S)}\sum_{g\in\delta^\uparrow(S)} r_g \frac{m_{\bbe,S}}{\sum_{\bbf\in\delta^\rightarrow(S)} m_{\bbf,S}}=0.
\end{align*}

Finally, we verify (v): Here, we use \cref{thm:expI}. For a good top edge $e$ that is part of a top edge bundle $\bbf$ we have
$$ \E{s_e} = -\E{r_e} + \E{I_\bbf} \frac{x_e}{x_\bbf} \leq -\tau p x_e + (1-\frac{\eps_{1/1}}{6})p\tau x_e = -\frac{\eps_{1/1}}{6} p\tau x_e.$$
On the other hand, for a bottom edge $e$ with $\p(e)=S$, then
$$ \E{s_e} = -\E{r_e} + \E{I_S} x_e \leq  -\beta p x_e + 0.99994p\beta x_e \leq -0.00006p\beta x_e.$$
 Finally, we can let 
\begin{equation}\label{eq:epsP}
\eps_P:=\frac{\eps_{1/1}}{6}p\frac{\tau}{\decrease} = \frac{\eps_{1/2}}{72} 0.005\eps_{1/2}^2 0.571 \geq  0.000039 \eps_{1/2}^3 \geq 3.12 \cdot 10^{-16}
\end{equation}
as desired.
\end{proof}

\begin{table}\centering
\begin{tabular}{|c|c|c|c|}
	\hline Name & Value & Set In & Explanation\\
	\hline $\eps_{1/2}$ & 0.0002 & \ref{eq:constants} & Half edge threshold, \cref{def:halfedges}\\
	\hline $\eps_{1/1}$ & $\frac{\eps_{1/2}}{12}$ & \ref{eq:constants} & $A,B,C$ partitioning threshold, \cref{def:abcdegpartitioning} \\
	\hline $p$ & $0.005\eps_{1/2}^2$ & \ref{eq:constants} & Min prob. of happiness for a (2-*) good edge \\
	\hline $\eps_M$ & $0.00025$ & \ref{eq:constants} & Marginal errors due to max flow, \cref{def:max-flow-event} \\
	\hline $\tau$ & $0.571\beta$ & \ref{eq:constants} & \hyperlink{reduction-events}{Top edge decrease} \\ 
	\hline $\eps_P$ & $\frac{\eps_{1/1}}{6}p\frac{\tau}{\decrease}$ & \eqref{eq:epsP} & Expected decrease constant, \cref{thm:payment-main}\\
	\hline $\alpha$ & $2\eps_\eta$ & \ref{eq:matching-params} & Parameter of \cref{lem:matching} \\ 
	\hline $\eps_B$ & $21\eps_{1/2}$ & \ref{eq:matching-params} & Parameter of \cref{lem:matching} \\
	\hline $\eps_F$ & $1/10$ & \ref{eq:matching-params} & Parameter of \cref{lem:matching} \\
	\hline $\eps_\eta$ & $14\eta$ & \eqref{def:eps-eta} & \cref{def:hierarchy} \\
	\hline $\eta$ & $\frac{1}{1308}\eps_P$ & \eqref{eq:whatiseta} & Near min cut constant \\
	\hline $\decrease$ & $\eta/4.1$ & \eqref{def:decrease-param} & Slack shift constant e.g. \cref{thm:payment-main,thm:crossed-one-side,thm:cutsbothsides}\\
	\hline 
\end{tabular}
\caption{A table of all constants used in the paper.}\label{table:constants}
\end{table}


In the rest of this section we prove \cref{thm:expI}.
Throughout the  proof, we will repeatedly use the following facts proved in \cref{sec:probabilistic}:
If a top edge $e = (u,v)$ that is part of a bundle $\bbf$ is reduced (equivalently $\cH_{\bbf,u}=1$ or $\cH_{\bbf,v}=1$), then $u$ and $v$ are trees, which means that tree sampling inside $u$ and $v$ is independent of the reduction of $e$. 

Note however, that conditioning on a near-min-cut or atom to be a tree increases marginals inside and reduces marginals outside as specified by \cref{lem:treeconditioning}. Since for any $S\in \cH$, $x(\delta(S)) \le 2+\eps_\eta$, the overall change is $\pm \eps_\eta/2$. 
 

The proof of \cref{thm:expI} simply follows from  \cref{cor:topIncrease} and \cref{cor:botincrease} that we will prove in the following two sections.


\subsection{Increase for Good Top Edges}\label{sub:degcutincrease}
The following lemma is the main result of this subsection.
\begin{restatable}[Top Edge Increase]{lemma}{topincrease}
\label{cor:topIncrease}
Let $S\in\cH$ be a degree cut and $\bbe=(u,v)$ a good edge bundle with $\p(\bbe)=S$.
If $\eps_{1/2}\leq 0.0002$, $\eps_{1/1}\leq \eps_{1/2}/12$ and $\eps_\eta\leq \frac{\eps_{1/1}}{100}$, $\hyperlink{tar:epsF}{\eps_F}=1/10$ then 
$$\E{I_{\bbe,u}}+ \E{I_{\bbe,v}} \le p\tau x_\bbe \left(1 -\frac{\eps_{1/1}}{6}\right).$$
\end{restatable} 

We will use the following technical lemma to prove the above lemma.
\begin{lemma}
\label{lem:topedge}
Let $S\in\cH$ be a degree cut with an atom $u\in\cA(S)$. If $x(\delta ^\uparrow (u)) > \eps_F$, $\eps_{1/2}\leq 0.0002$, $\eps_{1/1}\leq \eps_{1/2}/12$, $\eps_\eta \le \frac{\eps_{1/1}}{100}$, then we have \begin{align}
\label{eq:topred}
	\sum_{\substack{ g\in \delta ^\uparrow (u),\\ g\in \bbf=(u',v') \text{ good top}}} &\frac{1}{2} \tau x_g \cdot  (\P{\delta(u)_T\text{ odd}|\cR_{\bbf,u'}} + \P{\delta(u)_T\text{ odd} | \cR_{\bbf,v'}})\\
&\quad	+\sum_{g\in \delta^\uparrow(u), \p(g)=S'\text{ polygon}} \beta x_g \cdot \P{\delta(u)_T \text{ odd} | \cR_{S'}}
\le \tau(1-\frac{\eps_{1/1}}{5}) x(\delta^\uparrow(u)) \hyperlink{tar:Fu}{F_u},\nonumber
\end{align}
where recall we set $ F_u := 1-\eps_B \I{x(\delta^\uparrow(u))\text{ is $\eps_F$ fractional}}$ in \cref{lem:matching}, where $\eps_B := 21 \eps_{1/2}$ and $\eps_F = 1/10$ as in \ref{eq:matching-params}.
\end{lemma}

\begin{proof}[Proof of \cref{cor:topIncrease}]

By linearity of expectation and using \cref{eq:Ieu}:
\begin{align}
\label{eq:Dec-eu}
	\E{I_{\bbe,u}} &= \frac{ m_{\bbe,u}}{ \sum_{\bbf \in \delta^\rightarrow(u)} m_{\bbf,u}} \E{\sum_{g \in \delta^\uparrow(u)} r_g \cdot \I{u \text{ is odd}}}\notag \\
	&= \frac{ m_{\bbe,u}}{ \sum_{\bbf \in \delta^\rightarrow(u)} m_{\bbf,u}} \Big(\sum_{\substack{g\in \delta ^\uparrow (u): \\ g\in \bbf=(u',v') \text{ good top}}  } \frac{1}{2}\tau x_g (\P{\cR_{\bbf,u'},\delta(u)_T\text{ odd}} + \P{\cR_{\bbf,v'},\delta(u)_T\text{ odd}}) \\
	&\quad +\sum_{g\in \delta^\uparrow(u): \p(g)=S'\text { polygon}} \beta x_g \P{\cR_{S'},\delta(u)_T \text{ odd}}\Big)\nonumber 
\end{align} 
A similar equation holds for $\E{I_{e,v}}$.
 
The case where $x(\delta^\uparrow(u))\leq \eps_F$ or $x(\delta^\uparrow(v))\leq \eps_F$ is dealt with in \cref{lem:xdeltau<=epsF}. So, consider the case where $x(\delta ^\uparrow(u)), x(\delta ^\uparrow(v)) > \eps_F$. Now recall that from \eqref{eq:matchedamount}, 
\begin{equation}\label{eq:summeu}
	\sum_{\bbf\in\delta^\rightarrow(u)} m_{\bbf,u} = \hyperlink{tar:Zu}Z_u x(\delta^\uparrow(u))
\end{equation}
 where $Z_u = 1 + \I{|S|\geq 4, x(\delta^\uparrow(u))\leq \eps_F}$. In this case, $Z_u=Z_v=1$. 

Using $\P{\cR_{\bbf,u'},\delta(u)_T\text{ odd}}=p\P{\delta(u)_T\text{ odd}|\cR_{\bbf,u'}}$, and plugging  \eqref{eq:topred}  into \eqref{eq:Dec-eu} for $u$ and $v$, we get (and using \cref{eq:summeu}):
\begin{align}
\label{eq:DegreeCutPayment}
\E{I_{\bbe,u}}+ \E{I_{\bbe,v}} &\le p \tau(1-\frac{\eps_{1/1}}{5}) \left(x(\delta^\uparrow(u)) \hyperlink{tar:Fu}{F_u}\frac{m_{\bbe,u}}{x(\delta^\uparrow(u))} +  x(\delta^\uparrow(v)) \hyperlink{tar:Fu}{F_v} \frac{m_{\bbe,v}}{x(\delta^\uparrow(v))}\right)\\
&=  p \tau (1-\frac{\eps_{1/1}}{5}) (\hyperlink{tar:Fu}{F_u}m_{\bbe,u}
+\hyperlink{tar:Fu}{F_v} m_{\bbe,v})\notag\\
& \le p\tau(1-\frac{\eps_{1/1}}{5}) (1 + 2\eps_\eta) x_\bbe < p\tau x_\bbe (1- \frac{\eps_{1/1}}{6}).\notag
\end{align}
where on the final line we used \eqref{eq:uvbadematch} and $\eps_{\eta} < \frac{\eps_{1/1}}{100}$.
\end{proof}




\begin{proof}[Proof of \cref{lem:topedge}]
Suppose that $S_i \in\cH$ are the ancestors of $S$ in the hierarchy (in order) such $S_1=S$ and for each $i$, $S_{i+1}=\p(S_i)$.
Let
$$\delta^{\ge i}:= \delta (u) \cap \delta (S_i) \quad \quad \text{ and }\quad \quad  \delta^i := \delta(u) \cap \delta ^\rightarrow (S_i).$$
Each group of edges $\delta^i$ is either entirely top edges or entirely bottom edges. 
First note that if $g\in \delta^i$ and $g$ is a bottom edge, i.e., $S_{i+1}$ is a polygon cut, then by \cref{lem:bottompaymentprob},  
$$\P{\delta (u)_T\text{ odd} | \cR_{S_{i+1}}} = \P{\delta(u)_T\text{ odd} | \cE_{S_{i+1}}} \le 0.5678$$
(see \cref{def:max-flow-event} and \cref{def:RS} for definition of $\cE_{S_{i+1}},\cR_{i+1}$) where in the equality we used that $\cR_{S_{i+1}}$ is a uniformly random event chosen in $\cE_{S_{i+1}}$.
Therefore, to prove \cref{eq:topred} it is enough to show
\begin{align}
\sum_{\substack{ g\in \delta ^\uparrow_\text{good}(u):\\ g\in \bbf=(u',v') \text{ top},\\ }} &\frac{1}{2} \tau x_g  (\P{\delta(u)_T\text{ odd}|\cR_{\bbf,u'}} + \P{\delta(u)_T\text{ odd} | \cR_{\bbf,v'}}) \notag\\
&\quad 
\leq \tau\left((1-\frac{\eps_{1/1}}{5})  \hyperlink{tar:Fu}{F_u}\left( x(\delta^\uparrow_\text{good}(u)) +x(\delta^\uparrow _\text{bad}(u))\right) + 0.0014 x(\delta^\uparrow_\beta(u))\right)\label{payineq}	
\end{align}
where we write $\delta_\beta(u),\delta_\text{good}(u), \delta_\text{bad}(u)$ to denote the set of bottom edges, good top edges, and bad (top) edges in $\delta(u)$ respectively and we used that 
$$\tau(1-\frac{\eps_{1/1}}{5})(1-\eps_B)-0.5678\beta \geq 0.0014\tau$$
since $\tau=0.571\beta$, $\eps_{1/1}\leq \frac{\eps_{1/2}}{12}$, $\eps_{1/2}\leq 0.0002$, and $\eps_B = 21\eps_{1/2}$ as defined in \ref{eq:matching-params}.

Since  $h(\bbf):= \frac{1}{2}  (\P{\delta(u)_T\text{ odd}|\cR_{\bbf,u'}} + \P{\delta(u)_T\text{ odd} | \cR_{\bbf,v'}})\le 1$
 and  $(1-\frac{\eps_{1/1}}{5})F_u$ is nearly 1, in each of the following cases
\begin{equation}\label{eq:toptopgoalbottom}
	x(\delta^\uparrow_\beta(u)) \ge \begin{cases}
 	0.003 & \text{ when $F_u = 1$}\\
 	\frac{4}{5} x(\delta^\uparrow(u)) & \text{ when $F_u = 1- \eps_B$}
 \end{cases}\quad \text{ or }\quad x(\delta^\uparrow_\text{bad}(u)) \ge 0.006 \quad\text{when $F_u \ge 1- \eps_B$},
\end{equation}
\eqref{payineq} holds. To see this, just plug in $\eps_{1/1} \leq \frac{\eps_{1/2}}{12}$,  $\eps_{1/2}\leq 0.0002$,   $\eps_B = 21\eps_{1/2}$,  $\eps_\eta \leq 10 ^{-10}$,  $x(\delta^\uparrow(u))\le 1 + \eps_\eta$ and any inequality from \eqref{eq:toptopgoalbottom} into \eqref{payineq}, using the upper bound $h(\bbf) = 1$.


Alternatively, for $\delta_\text{top}(u)=\delta_\text{good}(u)\cup\delta_\text{bad}(u)$ be the set of top edges in $\delta(u)$, if we can show the existence of a set $D\subseteq \delta^\uparrow_\text{top}(u)$ such that 
\begin{equation}	\label{eq:toptopgoalgoodtop}
 x(D)\cdot \min_{\substack{g\in D:\\ g\in\bbf=(u',v') \text{ good}}}1-\frac{\P{\delta(u)_T\text{ odd}|\cR_{\bbf,u'}} + \P{\delta(u)_T\text{ odd} | \cR_{\bbf,v'}}}{2}\geq  \left(\frac{\eps_{1/1}}{5} + 1-\hyperlink{tar:Fu}{F_u}\right)x(\delta^\uparrow_\text{top}(u)),
\end{equation}
then, again, \eqref{payineq} holds.

In the rest of the proof, we will consider a number of cases and show that in each of them, either one of the inequalities in \eqref{eq:toptopgoalbottom} or the inequality in \eqref{eq:toptopgoalgoodtop} for some set $D$ is true, which will imply the lemma.




\begin{figure}[htb]\centering
\begin{tikzpicture}
	\node [inner sep=2,draw,circle] at (0,0) (u) {\footnotesize$u$};
	\node [fill=red,opacity=0.2,circle,inner sep=10] (0,0)   () {};
	\node [color=red] at (-0,-0.75) () {\footnotesize $S=S_1$};
	\draw [line width=1pt,color=red] (45:1) arc (45:135:1) (45:1.5) arc (45:135:1.5);
	\draw [line width=1pt,color=blue] (50:3) arc (50:130:3);
	\draw [line width=4pt,color=blue,->] (50:3) -- +(0,2.1);
	\draw [line width=4pt,color=green,->] (45:2) -- +(0,3);
	\draw [line width=1pt,color=orange] (50:3.8) arc (50:130:3.8);
	\draw [line width=4pt,color=orange,->] (50:3.8) -- +(0,1.5);
	\node [xshift=45,color=orange] at (50:3.8) () {\footnotesize$x(\delta^{\geq \ell})\geq 2\eps_\eta +\eps_{1/1}$};
	\node [color=green,xshift=-8] at (135:2) () {\footnotesize$S_j$};
	\node [color=blue,xshift=-8] at (130:3) () {\footnotesize$S_k$};
	\node [color=orange,xshift=-8] at (130:3.8) () {\footnotesize$S_\ell$};
	\node [color=blue,xshift=40] at (50:3) () {\footnotesize $x(\delta^{\geq k})\geq 2\eps_\eta + \frac{\eps_F}{2}$};
	\node [color=green,xshift=40] at (45:2) () {\footnotesize $x(\delta^{\geq j})\geq 1-\eps_{1/1}$};
	\draw [color=gray,line width=1.5pt,->] (0,.6) -- node [right=3] {\footnotesize $\delta^1$} +(0.5,0);
	\draw [color=gray,line width=1.5pt,->]  (0,1.2) -- node [right=3] {\footnotesize $\delta^2$} +(0.5,0); 
	\draw [color=gray,line width=1.5pt,->] (0,4.2) -- +(0.5,0);
	\draw [color=gray,line width=1.5pt,->]  (0,2.2) -- node [right=3] {\footnotesize$\delta^j$} +(0.5,0);
	\draw [color=gray,line width=1.5pt,->]  (0,2.6) -- +(0.5,0);
	\draw [color=green,line width=1pt]  (45:2) arc (45:135:2);
	\draw [color=gray,line width=4pt] (u) -- +(0,4.23);
	\draw [dotted,line width=1.1pt] (140:2.5) -- (137:3);
	\node [color=red,xshift=-5] at (135:1) () {\footnotesize $S_2$};
	\node [color=red,xshift=-5] at (135:1.5) () {\footnotesize $S_3$};
\end{tikzpicture}	
\end{figure}

 First, let
\begin{align*}
j&=\max\{i:x(\delta^{\ge i}) \ge  1- \eps_{1/1}\}\\
k&=\max\{i:x(\delta^{\ge i}) \ge 2\eps_{\eta} + \eps_F/2\},\\
\ell &=\max\{i:x(\delta^{\ge i}) \ge  2\eps_\eta +\eps_{1/1}\}
\end{align*}
Just note $j\leq k\leq \ell$.
Note that levels $\ell$ and $k$  exist since $x(\delta^\uparrow (u)) \ge \eps_F$, whereas level $j$ may not exist (if $x(\delta^\uparrow (u)) <  1 - \eps_{1/1}$). We consider three cases:

\paragraph{Case 1: $ x(\delta^\uparrow (u)) \ge 1- \eps_{1/1}$:}
Then $j$ exists and $S_j$ has a valid $A,B,C$ degree partitioning (\cref{def:abcdegpartitioning}) where $A = \delta(v) \cap \delta (S_j)$ such that either $u=v$ or $v$ is a descendant of $u$ in $\cH$. Note that, $x(\delta(u) \cap \delta(S_j)) \ge 1-\eps_{1/1}$, and by \cref{def:abcdegpartitioning}, $B \cap \delta(u) = \emptyset$. In addition, in this case, $x(\delta^\uparrow(u))$ is not $\eps_F$ fractional (see \cref{lem:matching}), so $\hyperlink{tar:Fu}{F_u}=1$. 
\begin{description}
\item [Case 1a:  $x(\delta^{j}) \ge 3/4$.]
If $\delta^j$ are bottom edges then \eqref{eq:toptopgoalbottom} holds. 
So, suppose that $\delta^j$ is a set of top edges.  
By \cref{lem:Ais211good}, at most $1/2+4\eps_{1/2}$ fraction of edges in $A\cap \delta^j$ are good but not 2-1-1 good (w.r.t., $u$). 
So, the rest of the edges in $A\cap\delta^j$ are either bad or 2-1-1 good. Since 
$$x(A\cap\delta^j)\geq 3/4-x(C)\ge  3/4-2\eps_{1/1}-\eps_\eta,$$
$\delta^j$ either has a mass  of $\frac12(1/4 - 2 \eps_{1/1} - \eps_\eta- 4 \eps_{1/2} ) > 1/8 - 3 \eps_{1/2}$ of  bad edges or of 2-1-1 good edges.\footnote{We are using the fact that $\eps_{1/1} =\eps_{1/2}/12 $ and that $\eps_\eta$ is tiny by comparison to these.}
The former case implies that \eqref{eq:toptopgoalbottom} holds. In  the latter case, by \cref{claim:211} for any 2-1-1 good edge $g\in \delta^j$ with $g\in\bbf=(u',v')$ we have  $\P{\delta(u)_T\text{ odd} | \cR_{\bbf,u'}} \leq 2\eps_\eta + \eps_{1/1}$; so  \eqref{eq:toptopgoalgoodtop}  holds for $D$ defined as the set of 2-1-1 good edges in $\delta^j$.

\item [Case 1b: $x(\delta^{j}) < 3/4$.] 
If $x(\delta^\uparrow_\beta(u))\geq 0.003$, then  \eqref{eq:toptopgoalbottom} holds. 
Otherwise, we apply \cref{claim:top} with $\eps = \eps_{1/1}$ to all good top edge bundles $\bbf \in D=\delta^{\ge j +1} \smallsetminus \delta^{\ge \ell+1}$ and we get that 
$$\frac12(\P{\delta(u)_T\text{ odd}|\cR_{\bbf,u'}} + \P{\delta(u)_T\text{ odd} | \cR_{\bbf,v'}})\leq 1- \eps_{1/1}+\eps_{1/1}^2.$$
Since $x(D)\geq 1-\eps_{1/1} - 3/4  - 2 \eps_\eta -\eps_{1/1}-0.003>  0.24 $,  \eqref{eq:toptopgoalgoodtop} holds. 
\end{description}


\paragraph{Case 2: $ 1-\eps_F < x(\delta^\uparrow (u)) < 1- \eps_{1/1}$. } Again we have $\hyperlink{tar:Fu}{F_u}=1$.
So we can either show that $x(\delta^\uparrow_\beta(u))\geq 0.003$ or take $D$ to be the top edges in  $\delta^\uparrow (u) \smallsetminus \delta^{\ge \ell+1} $ and use \cref{claim:top} with $\eps = \eps_{1/1}$. This will enable us  to show that \eqref{eq:toptopgoalgoodtop} holds as in 
 the previous case.

\paragraph{Case 3: $ \eps_F < x(\delta^\uparrow (u)) < 1- \eps_F$: }
In this case $\hyperlink{tar:Fu}{F_u}=1-\eps_B$.
If at least  $4/5$ of the edges in $ \delta^\uparrow(u)$ are bottom edges, then we are done by \eqref{eq:toptopgoalbottom}.

Otherwise, let $u'=\p(u)$. For any top edge $e\in\delta^\uparrow(u)$ where $e\in\bbf=(u'',v'')$ we have
$$ \P{\delta(u)_T\text{ odd} | \cR_{\bbf,u''}} \leq \P{u'\text{ tree}|\cR_{\bbf,u''}} \P{\delta(u)_T\text{ odd} | u'\text{ tree},\cR_{\bbf,u''}} + \P{u'\text{ not tree}| \cR_{\bbf,u''}}
$$
Using that $u'\subseteq u''$ is a tree under $|\cR_{\bbf,u''}$ with probability at least $1-\eps_\eta/2$, and applying \cref{claim:top} (to $u$ and $u'$) with $\eps = \eps_F$ 
we have $\P{\delta(u)_T\text{ odd}|u'\text{ tree}, \cR_{\bbf,u''}}\leq 1-\eps_F+\eps_F^2$ we get
$$\P{\delta(u)_T\text{ odd} | \cR_{\bbf,u''}} \leq 1-\eps_F+\eps_F^2 + \eps_\eta/2.$$
Now, let $D$ be all top edges in $\delta^\uparrow(u)$. Then, we apply \cref{eq:toptopgoalgoodtop} to this set of mass at least $x(\delta^\uparrow(u))/5$, and we are done, using that $(\eps_F-2\eps_F^2)/5 \ge (\frac{\eps_{1/1}}{5} + \eps_B)$ which holds for $\eps_F\geq 1/10$, $\eps_B=21\eps_{1/2}$, and $\eps_{1/2}\leq 0.0002$.
\end{proof}

\begin{claim}
\label{claim:211}
For $u\in\cH$ and a top edge $e \in \bbf=(u',v')$ for some $u'\in\cH$ that is an ancestor of $u$, if
 $x(\delta  (u) \cap \delta(u')) \ge 1-\eps_{1/1}$ and $\bbf$ is 2-1-1 good, then 
	$$\P{\delta (u)_T\text{ odd} | \cR_{\bbf,u'}} \le 2\eps_\eta + \eps_{1/1}.$$
\end{claim}
\begin{proof}
Let $A,B,C$ be the \hyperlink{tar:degreepartition}{degree partitioning} of $\delta(u')$.
By the assumption of the claim, without loss of generality, assume $A\subseteq \delta(u)\cap\delta(u')$. Furthermore, by definition, $B \cap \delta(u) = \emptyset$. 
This means that if $\cR_{\bbf,u'}=1$ then  $u'$ is a tree and $A_T = 1= (\delta  (u) \cap \delta(u'))_T$ (also using $C_T=0$ and $B \cap \delta(u) = \emptyset$). 
Therefore,
	$$\P{\delta (u)_T\text{ odd} | \cR_{\bbf,u'}} = \P{(\delta (u)\smallsetminus \delta(u'))_T \text{ even} |  \cR_{\bbf,u'}}.$$
To upper bound the RHS first observe that
$$\E{(\delta (u)\smallsetminus \delta(u'))_T| \cR_{\bbf,u'}} \leq \eps_\eta/2 + x(\delta (u)\smallsetminus \delta(u')) \leq \eps_\eta/2 + x(\delta(u)) - x(A) < 1+2\eps_\eta+\eps_{1/1}. $$
Under the conditional measure $|\cR_{\bbf,u'}$,  $u'$ is a tree ,  so $u$ must be connected inside $u'$, i.e.,  $(\delta (u)\smallsetminus \delta(u'))_T\geq 1$ with probability 1.  Therefore, 
$$\P{(\delta (u)\smallsetminus \delta(u'))_T \text{ even} |  \cR_{\bbf,u'}} \leq \P{(\delta (u)\smallsetminus \delta(u'))_T-1 \neq 0 |  \cR_{\bbf,u'}} \leq 2\eps_{\eta}+\eps_{1/1}$$
as desired.
\end{proof}

\begin{claim}
\label{claim:top}
For $u,u'\in \cH$ such that $u'$ is an ancestor of $u$. Let $\nu=\nu_{u'}\times\nu_{G/u'}$ be the measure resulting from conditioning $u'$ to be a tree. 
if $x(\delta (u) \cap \delta(u')) \in [ \eps, 1 -\eps]$, then
	\begin{equation}
	\label{eq:uandSprime}
	 \PP{\nu}{\delta (u)\text{ odd} |  (\delta(u)\cap \delta(u'))_T} \leq 1-\eps+\max\{2\eps_\eta,\eps^2\}.
	\end{equation}
	In other words, for any integer $k \ge 0$, we have $\PP{\nu}{\delta (u)\text{ odd} |  (\delta(u)\cap \delta(u'))_T = k} \le 1-\eps+\max\{2\eps_\eta,\eps^2\}$. 
\end{claim}
\begin{proof}
Let $D=\delta(u)\smallsetminus \delta(u')$. By assumption, $u'$ is a tree, so $D_T\geq 1$ with probability 1. Therefore, since we have no control over the parity of $(\delta(u)\cap\delta(u'))_T$
\begin{align*}
	\PP{\nu}{\delta (u)_T\text{ even} |(\delta(u)\cap\delta(u'))_T} \geq \min\{\P{D_T-1\text{ odd}|u'\text{ tree}}, \P{D_T-1=0|u'\text{ tree}}\}
\end{align*}
where we removed the conditioning by taking the worst case over $(\delta(u) \cap \delta(u'))_T$ even, $(\delta(u) \cap \delta(u'))_T$ odd. First, observe by the assumption of the claim and that $x(\delta^\uparrow(u))\leq 2+\eps_\eta$ we have
$$\E{D_T-1 | u'\text{ tree}} \in [\eps, 1 - \eps+2\eps_\eta].$$
Furthermore, since we have a SR distribution on $G[u']$, $D_T-1$ is a \hyperlink{tar:BS}{Bernoulli sum} random variable.
Therefore, 
$$\P{D_T-1=0|u'\text{ tree}}\geq \eps-2\eps_\eta$$ 
and by \cref{cor:bernoullisumeven} 
$$\P{D_T-1\text{ odd}|u'\text{ tree}} \geq 1-1/2(1+e^{-2\eps})\geq \eps-\eps^2$$
as desired. 
%
%
%
\end{proof}

\begin{lemma}\label{lem:xdeltau<=epsF}
Let $S\in\cH$ be a degree cut and $\bbe=(u,v)$ a good edge bundle with $\p(\bbe)=S$. If $x(\delta ^\uparrow(u)) < \eps_F$, $\eps_{1/2}\leq 0.0002$, $\eps_{1/1}\leq \eps_{1/2}/10$, then, 
$$\E{I_{\bbe,u}}+ \E{I_{\bbe,v}} \le p\tau x_\bbe \left(1 -\frac{\eps_{1/1}}{6}\right)$$ 
\end{lemma}
\begin{proof}
First notice, by \cref{lem:bottompaymentprob} for any bottom edge $g\in\delta^\uparrow(u)$ with $\p(g)=S'$, we have
$$\P{\delta(u)_T\text{ odd} | \cR_{S'}} = \P{\delta(u)_T\text{ odd} | \cE_{S'}} \leq 0.5678,$$
using $0.5678\beta \leq \tau$ and $F_u = 1$ (as $x(\delta^\uparrow(u)) \le \eps_F$) we can write,
\begin{align}\label{eq:Ieu3exp}
\E{I_{\bbe,u}} \leq	\sum_{h \in \delta ^\uparrow (u)} x_h p \tau \hyperlink{tar:Fu}{F_u}\cdot \frac{ m_{\bbe,u}}{ Z_u x (\delta ^\uparrow (u))}.
\end{align}
Secondly, if $x(\delta^\uparrow(v))\geq \eps_F$,  applying  \eqref{eq:topred} and \eqref{eq:Dec-eu} to $I_{\bbe,v}$ and using $Z_v \ge 1$ we get 
\begin{align}\label{eq:Ievexp3}
\E{I_{\bbe,v}} &\le \frac{m_{\bbe,v}}{\sum_{\bbf\in\delta^\rightarrow(v)} m_{\bbf,v}}
p\tau(1-\frac{\eps_{1/1}}{5})x(\delta^\uparrow(v)) \hyperlink{tar:Fu}{F_v} = m_{\bbe,v}p\tau\left(1-\frac{\eps_{1/1}}{5}\right)\hyperlink{tar:Fu}{F_v}
\end{align}
\textbf{Case 1: $|\cA(S)|=3$, where $\cA(S)=\{u,v,w\}$.} Let $\bbf = (u,w)$,  $\bbg = (v,w)$ (and of course $\bbe = (u,v)$). We will use the following facts below:  
\begin{figure}[htb]
\centering	
	\begin{tikzpicture}
		\node[draw,circle] at (0,0) (u) {$u$};
		\node[draw,circle] at (1,-1) (v) {$v$} edge node [right] {$\bbe$} (u);
		\node[draw,circle] at (-1,-1) (w) {$w$} edge node [left] {$\bbf$} (u) edge node [below] {$\bbg$} (v);
		\draw [dotted,color=red,line width=1pt] (0,-0.75) ellipse (1.75 and 1.25);
		\draw (u) -- node [left] {$\delta^\uparrow(u)$} +(0,1.5)  
		(w) -- node [left] {$\delta^\uparrow(w)$} +(-0.5,2)
		(v) -- node [right] {$\delta^\uparrow(v)$} +(0.5,2);
		\node [color=red] at (-1.75,-1.5) (){$S$};
	\end{tikzpicture}
\end{figure}
\begin{align*}
&x_\bbe + x_\bbf \ge 2 - \eps_F \tag{$x(\delta (u)) \ge 2$ and $x(\delta^\uparrow(u))\leq \eps_F$}\\
&x(\delta ^\uparrow (v)) + x(\delta ^\uparrow (w)) \ge 2- \eps_F \tag{$x(\delta (S))\geq 2$}\\
&x_\bbf, x(\delta^\uparrow(w))\leq 1+\eps_\eta,\tag{\cref{lem:shared-edges}}
\end{align*}
so we have, 
\begin{equation}\label{eq:edeltupvlarge}x_\bbe, x(\delta ^\uparrow (v)) \ge 1- \eps_F - \eps_\eta. 	
\end{equation}

Now we bound $\E{ I_{\bbe,u}} + \E{ I_{\bbe,v}}$. By \cref{eq:Ieu3exp} and \cref{eq:Ievexp3} (which we may apply to $\E{I_{\bbe,v}}$ since $x(\delta^\uparrow(v))\geq \eps_F$),
\begin{align}
\E{I_{\bbe,u}}+\E{I_{\bbe,v}}&\leq \sum_{h \in \delta ^\uparrow (u)} x_h p \tau \hyperlink{tar:Fu}{F_u}\cdot \frac{ m_{\bbe,u}}{ Z_u x (\delta ^\uparrow (u))} +p\tau\left(1-\frac{\eps_{1/1}}{5}\right)\hyperlink{tar:Fu}{F_v} m_{\bbe,v}
\notag\\
&= p\tau \hyperlink{tar:Fu}{F_u} m_{\bbe,u}
+p\tau\left(1-\frac{\eps_{1/1}}{5}\right)\hyperlink{tar:Fu}{F_v} m_{\bbe,v} \tag{$Z_u=1$ as $|\cA(S)|=3$}\\
& = p\tau (\hyperlink{tar:Fu}{F_u}m_{\bbe,u}+\hyperlink{tar:Fu}{F_v} m_{\bbe,v}) - \frac{\eps_{1/1}}{5}p\tau \hyperlink{tar:Fu}{F_v}m_{\bbe,v} \notag\\
& \leq p\tau (1+2\eps_\eta)x_{\bbe} - \frac{\eps_{1/1}}{5}p\tau \hyperlink{tar:Fu}{F_v}m_{\bbe,v}\label{IeuIevS3}
\end{align}
where the final inequality follows from \eqref{eq:uvbadematch}.
To complete the proof, we lower bound $m_{\bbe,v}$.

Using \eqref{eq:matchedamount} for $v$ and $w$, we can write,
\begin{align*}
x(\delta^\uparrow(v))+x(\delta^\uparrow(w)) &= m_{\bbe,v} + m_{\bbg,v} + m_{\bbf,w}+m_{\bbg,w}\\ &\leq m_{\bbe,v} +    \frac{(1+2\eps_\eta)}{(1-\eps_B)} (x_\bbf + x_{\bbg}) \tag{using \eqref{eq:uvbadematch}}\\
&= m_{\bbe,v} + \frac{(1+2\eps_\eta)}{(1-\eps_B)} \left(\sum_{a\in\cA(S)} \frac{x(\delta(a))}{2} - \frac{x(\delta(S))}{2} - x_\bbe\right)\\
&\leq m_{\bbe,v} +  \frac{(1+2\eps_\eta)}{(1-\eps_B)}  (2+3\eps_\eta-x_\bbe) 
\end{align*}
and using the fact that $x(\delta^\uparrow(v))+x(\delta^\uparrow(w))\geq 2-\eps_F$, we get $$ m_{\bbe,v}\geq x_\bbe -\eps_F-4\eps_B  \geq (1- 1.2\eps_F)x_\bbe,$$ 
 where the second inequality follows from  \eqref{eq:edeltupvlarge}
 and $\eps_B = 21 \eps_{1/2}$ and $\eps_\eta <\eps_{1/2}^2$ and $\eps_F\geq 1/10$.
Plugging this back into \eqref{IeuIevS3} and using $\hyperlink{tar:Fu}{F_v}\geq 1-\eps_B= 1-21\eps_{1/2}$ we get
$$ \E{I_{\bbe,u}}+\E{I_{\bbe,v}}\leq p\tau x_\bbe \left(1+2\eps_\eta - \frac{\eps_{1/1}}{5} (1-1.2\eps_F)(1-21\eps_{1/2})\right) \leq p\tau x_\bbe (1-\frac{\eps_{1/1}}{6}) $$
as desired. In the last inequality we used $\eps_F\leq 1/10$ and $\eps_{1/2}\leq  0.0002$.

{\bf Case 2: $|S|	\geq 4$.} 
In this case,  $Z_u = 2$. 
Therefore, by \cref{eq:Ieu3exp}
$$ \E{I_{\bbe,u}} \leq \sum_{e\in\delta^\uparrow(u)} x_e p\tau \hyperlink{tar:Fu}{F_u} \frac{m_{\bbe,u}}{Z_u x(\delta^\uparrow(u))} = \frac12 p\tau \hyperlink{tar:Fu}{F_u} m_{\bbe,u}.$$
If $x(\delta^\uparrow(v))<\eps_F$, we get the same inequality for $I_{\bbe,v}$. Then, 
$$\E{I_{\bbe,u}} + \E{I_{\bbe,v}} \le \frac12 p\tau (F_um_{\bbe,u} + F_vm_{\bbe,v}) \underset{\eqref{eq:uvbadematch}}{\le} \frac12 p\tau x_e(1+2\eps_\eta),$$
which is clearly sufficient for the lemma statement.

Otherwise, $x(\delta^\uparrow(v))\geq \eps_F$ in which case by \eqref{eq:Ievexp3} we get $\E{I_{\bbe,v}}\leq m_{\bbe,v} p\tau \hyperlink{tar:Fu}{F_v}(1-\eps_{1/1}/5)$. We conclude the lemma similar to the previous case.
%
\end{proof}

	
%

\subsection{Increase for Bottom Edges}\label{sub:polycutincrease}

The following lemma is the main result of this subsection.
\begin{lemma}[Bottom Edge Increase]\label{cor:botincrease}
If $\eps_{1/2}\leq 0.0002$, $\eps_{\eta} \leq \eps_{1/2}^2$, for any polygon cut $S\in\cH$,
$$\E{I_{S}} \le 0.99994 \beta p.$$	
\end{lemma}
\begin{proof}
For a set of edges $D \subseteq \delta(S)$ define the random variable.
\begin{align}I_{S}(D):=(1+\eps_\eta) (\max\{&r(A\cap D)\I{S\text{ not left happy}} , r(B\cap D)\I{S\text{ not right happy}}\}\nonumber \\
&\quad + r(C\cap D)\I{S\text{ not happy}}).\label{def:ISD}
\end{align}
Note that by definition $I_S(\delta(S))=I_S$ and for any two disjoint sets $D_1,D_2$, $I_S(D_1\cup D_2)\leq  I_S(D_1)+I_S(D_2)$.
Also, define $I_S^\uparrow = I_S(\delta^\uparrow(S))$ and $I_S^\rightarrow = I_S(\delta^\rightarrow(S))$.

First, we upper bound $\E{I^\uparrow_S}$.
Let $f\in\delta^\uparrow(S)$ and suppose that $f$ with $\p(f)=S'$ is a bottom edge. Say we have $f\in A^\uparrow(S)$ ($f \in B^\uparrow(S)$ is similar). 
We write,
\begin{align*}\E{I_S(f)}&= (1+\eps_\eta) \beta x_f \P{\cR_{S'}}\P{S \text{ not left happy } | \cR_{S'}}\\
& \le 0.568 x_f p\beta \le x_f p \tau 	
\end{align*}
where in the inequality we used \cref{lem:bottompayment-polyprob} and that 
$$\P{S\text{ not left happy} | \cR_S} = \P{S\text{ not left happy} | \cE_S}$$ 
since $\cR_S$ is a uniformly random subset of $\cE_S$. If $f \in C^\uparrow(S)$, we use the trivial guarantee $\E{I_S(f)}\le (1+\eps_\eta) x_f p \beta$. 

On the other hand, if  $f$ is a top edge, then we use the trivial bound
\begin{equation}\label{eq:trivialbottomupS}\E{I_S(f)} \le (1+\eps_\eta) \tau p x_f.	
\end{equation}
Therefore, 
\begin{equation}
\label{eq:ISup}
\E{I_S^\uparrow} \le (1+\eps_\eta) \tau p x(\delta^\uparrow(S))  + (1+\eps_\eta)\eps_\eta p \beta \le (1+\eps_\eta) (0.571)\beta p x(\delta^\uparrow(S)) + 2\eps_\eta p \beta 
\end{equation}
since $x(C) \le \eps_\eta$.

Now, we consider three cases: 

\textbf{Case 1:  $\hat{S}=\p(S)$ is a degree cut.}
Combining \eqref{eq:ISup} and \cref{lem:BT} below, we get
\begin{align*}\E{I_S} 
	&\leq (1+\eps_\eta) p(0.571)\beta (7/4+6\eps_{1/2}+\eps_\eta) + 2\eps_\eta p \beta \leq 0.99994\beta p
\end{align*}
using $\eps_{1/2}\leq 0.0002$ and $\eps_\eta\leq \eps_{1/2}^2$.

\textbf{Case 2: $\hat{S}=\p(S)$ is a polygon cut with ordering $u_1,\dots,u_k$ of $\cA(\hat{S})$, $S=u_1$ or $S=u_k$}
Then, by \cref{lem:leftmostISright} below, 
\begin{align*}
\E{I_S}\leq (1+\eps_\eta)\beta p (0.571x(\delta^\uparrow(S)) + 0.31) + 2\eps_\eta p \beta \leq 0.89 \beta p 
\end{align*}
where we used $x(\delta^\uparrow(S))\leq 1+\eps_\eta$.

\textbf{Case 3: $\hat{S}=\p(S)$ is a polygon cut with ordering $u_1,\dots,u_k$ of $\cA(\hat{S})$, $S\neq u_1, u_k$}
Then, by \cref{lem:middleISright} below
\begin{align*}
\E{I_S}\leq (1+\eps_\eta)\beta p (0.571x(\delta^\uparrow(S)) + 0.85) + 2\eps_\eta p \beta \leq 0.86 \beta p
\end{align*}
where we use that $x(\delta^\uparrow(S))\leq \eps_\eta$ since we have a \hyperlink{tar:hierarchy}{hierarchy}.
This concludes the proof.	
\end{proof}

\subsubsection{Case 1: $\hat{S}$ is a degree cut}
\label{sec:BT}
\begin{lemma}
\label{lem:BT}
Let $S\in \cH$  be a polygon cut with parent $\hat S$ which is a degree cut. Then
$$	\E{I^\rightarrow_S} \le  (1+\eps_\eta)p\tau (x(\delta^\rightarrow(S))-(1/4-6\eps_{1/2})).$$
\end{lemma}
\begin{proof}
 Let $A,B,C$ be the polygon partition of $S$.
 We will show that for a constant fraction of the edges in $\delta ^\rightarrow (S)$, we can improve over the trivial bound in \eqref{eq:trivialbottomupS}.
To this end, consider the cases given by \cref{thm:probabilistic}.

{\bf Case 1: There is a bad half edge $\bbe$ in $\delta ^\rightarrow(S)$.}
Since bad edges never decrease, no corresponding increase occurs, so by the trivial bound \cref{eq:trivialbottomupS}
$$\E{I_{S}^\rightarrow} \leq  (1+\eps_\eta) p\tau(x(\delta^\rightarrow(S))-(1/2-\eps_{1/2})) .$$
This concludes the proof.

{\bf Case 2: There is a set of  2-1-1 good edges  (w.r.t., $S$) $D\subseteq  \delta^\rightarrow (S)$, such that $x_{D} \ge 1/2 - \eps_{1/2}-\eps_\eta$.} 
For any (top) edge $e\in \bbf=(S,u)$ such that $e \in D$, 
if $\cR_{\bbf,S}$, then 
 $S$ is happy, that is $A_T=B_T=1, C_T=0$ by \cref{rem:poly-and-degree-partition}.

Therefore,
\begin{align*}
\E{I_{S}(D)} &\le \sum_{e\in D: e\in \bbf=(S,u)} \frac{1+\eps_\eta}{2}\tau x_e \P{S\text{ not happy} | \cR_{\bbf,u}}\P{\cR_{\bbf,u}} \\
&\leq \frac{1+\eps_\eta}{2} p \tau x(D).
\end{align*}
Using the trivial inequality \cref{eq:trivialbottomupS} for edges in $\delta^\rightarrow(S)\smallsetminus D$ we get
$$ \E{I^\rightarrow_S} \leq (1+\eps_\eta) p\tau (\frac{x(D)}{2}+x(\delta^\rightarrow(S))-x(D)) \leq (1+\eps_\eta)p\tau (x(\delta^\rightarrow(S))-(1/4-\eps_{1/2}))$$
as desired. In the last inequality we used $x(D)\geq 1/2-\eps_{1/2}-\eps_\eta$.

{\bf Case 3: Cases 1 and 2 do not hold.} Therefore, by \cref{thm:probabilistic} there are least two 2-2-2 good top half edge bundles. In this case, $S$ has chosen a fixed pair of 2-2-2 good edges $\bbe= (S,v)$, $\bbf = (S,w)$ in $\delta ^\rightarrow(S)$ (as defined in \hyperlink{reduction-events}{the reduction events}) such that $x_{\bbe(B)},x_{\bbf(A)}\leq \eps_{1/2}$ and $\cR_{\bbe,S}=\cR_{\bbf,S}$ with probability 1. 
(Recall that $\bbe(A)=\bbe\cap A$.) Let $D=\bbe(A) \cup \bbf(B)$. 
In this case, $\bbe$ and $\bbf$ are reduced simultaneously by $\tau$ when they are 2-2-2 happy (w.r.t., $S$), i.e., when $\cR_{\bbe,S}=\cR_{\bbf,S}=1$. In such a case we have $\delta(S)_T = \delta(v)_T = \delta(w)_T = 2$. 
Therefore,
\begin{align*}\E{I_S(D)} &\leq  (1+\eps_\eta) \E{\max\{r(A\cap D), r(B\cap D)\}} \\
&\leq (1+\eps_\eta) \frac{\tau}{2} \max\{x_{\bbe(A)},x_{\bbf(B)}\} (\P{\cR_{\bbe,S} \land \cR_{\bbf,S}} + \P{\cR_{\bbe,v}} + \P{\cR_{\bbf,w}}) \\
&\leq (1+\eps_\eta) \tau \frac{3p}{2} x(D)\left(\frac12+3\eps_{1/2}\right) = (1+\eps_\eta) \tau p x(D)\left(\frac34+4.5\eps_{1/2}\right)
\end{align*}
where we used that $1/2-2\eps_{1/2}-x(C)\leq x_{\bbe(A)},x_{\bbf(B)} \leq 1/2+\eps_{1/2}$ and that $x(C)\leq \eps_{\eta}$.
Using the trivial inequality \cref{eq:trivialbottomupS} for edges in $\delta^\rightarrow(S)\smallsetminus D$ we get
\begin{align*} \E{I^\rightarrow_S} &\leq (1+\eps_\eta) p\tau (x(D)(3/4+4.5\eps_{1/2})+x(\delta^\rightarrow(S))-x(D)) \\
&	\leq (1+\eps_\eta)p\tau (x(\delta^\rightarrow(S))-(1/4-6\eps_{1/2}))
\end{align*}
where we used $x(D)\geq 1-4\eps_{1/2}-\eps_\eta$.

%
%
%
%
%
\end{proof}

\subsubsection{Case 2: $S$ and its parent $\hat{S}$ are both polygon cuts}
\label{sec:BB}



%
In this subsection we prove two lemmas: \cref{lem:leftmostISright}, which bounds $\E{I_S^\rightarrow}$ when $S$ is the leftmost or rightmost atom of $\hat S$, and \cref{lem:middleISright}, which bounds this quantity when $S$ is not leftmost or rightmost.
\begin{lemma}\label{lem:leftmostISright}
Let $S\in\cH$ be a polygon cut with $\p(S)=\hat{S}$ also a polygon cut. Let $u_1,\dots,u_k$ be the ordering of cuts in $\cA(\hat{S})$ (as defined in \cref{def:hierarchy}). If $\eps_M\leq 0.001$, $\eps_\eta\leq \eps_M^2$, $S=u_1$ or $S=u_k$, then 
$$ \E{I_S^\rightarrow} \leq 0.31 \beta p.  $$
\end{lemma}
\begin{proof}
Let $S$  be the leftmost atom of $\hat{S}$
and let $A,B,C$ be the polygon partition of $\delta(S)$. First, note 
\begin{equation}
\label{eq:Dec-eBTu2}
\E{I_S^\rightarrow} \le  (1+\eps_\eta) \left( \E{\max(r(A^\rightarrow), r(B^\rightarrow) ) \cdot \I{S\text{ not  happy}}} + \E{ r(C^\rightarrow)\I{S\text{ not  happy}} }
\right).
\end{equation}
where recall that $A^\rightarrow=A\cap \delta^\rightarrow(S)$. 
WLOG assume $x(A^\rightarrow)\geq x(B^\rightarrow)$. Then, 

\begin{align*}
&\E{\max\{r(A^\rightarrow), r(B^\rightarrow)\}\I{S\text{ not happy}}} = 	
  \beta p x(A^\rightarrow) \cdot \P{S\text{ not happy} |\cR_{\hat{S}}}\\
 \intertext{By \cref{lem:uhappy} we have}
  x(A^\rightarrow) & \cdot \P{S\text{ not happy} |\cR_{\hat{S}}}
 	 \le x(A^\rightarrow)   \left(1-((1- x(A^\rightarrow))^2 + (x(A^\rightarrow))^2 -2\eps_M 
	- 17\eps_\eta)\right)  \\
	&\leq \left(2x(A^\rightarrow)^2-2x(A^\rightarrow)^3 +2\eps_M x(A^\rightarrow)+17\eps_\eta x(A^\rightarrow)\right)\\
	&\leq (8/27+2\eps_M+17\eps_\eta),
\end{align*}
where in the final inequality we used that the function $x\mapsto x^2(1-x)$ is maximized at $x=2/3$, and using $\eps_M\leq 0.00 1,\eps_\eta<\eps_M^2$.

Plugging this back into \eqref{eq:Dec-eBTu2}, and using $x(C) \le 
\eps_\eta$, we get
 $$\E{I_{S}^\rightarrow} \le  (1+ \eps_\eta)\beta p(\frac{8}{27}+2\eps_M+18\eps_\eta)\leq 0.31 \beta p, $$
 where the last inequality follows since $\eps_M\leq 0.001$ and $\eps_\eta <\eps_M^2$. 
\end{proof}


\begin{lemma} \label{lem:uhappy} 
Let $S\in\cH$ be a polygon cut with $\p(S)=\hat{S}$ also a polygon cut.  Let $u_1,\dots,u_k$ be the ordering of cuts in $\cA(\hat{S})$. If $S=u_1$, (or $S=u_k$) then
 $$ \P{S \text{ happy}| \cR_{\hat{S}}}
 \ge  (1- x(A^\rightarrow))^2 + (x(A^\rightarrow))^2 -2\eps_M 
	- 17\eps_\eta.$$	
\end{lemma}
\begin{proof}
Let $A,B,C, \hat{A},\hat{B},\hat{C}$ be the polygon partition of $S, \hat{S}$ respectively. Observe that since $S=u_1$, we have $\hat{A}= E(u_1,\overline{\hat{S}})=A^\uparrow\cup B^\uparrow\cup C^\uparrow$ and $\hat{B},\hat{C} \cap (A\cup B\cup C)=\emptyset$. 
Conditioned on $\cR_{\hat{S}}$, $\hat{S}$ is a tree,  and marginals of all edges in $\hat{A}$ is changed by a total variation distance at most $\eps'_M:=\eps_M+2\eps_\eta$ from $x$ (see \cref{cor:poly-reduction}) and they are independent of edges inside $\hat S$. The tree conditioning increases marginals inside by at most $\eps_\eta/2$.  Since after the changes just described
$$\E{C_T} \le x_C+ \eps_\eta + \eps'_M \le 4\eps_\eta +\eps_M,$$
it follows that $\P{C_T=0 | \cR_{\hat{S}}}\geq 1- 4\eps_\eta -\eps_M$.
So, 
\begin{equation}\label{eq:prob-110-before-nu}\P{S\text{ happy} \mid \cR_{\hat S}} \geq (1-4\eps_\eta - \eps_M)\P{A_T=B_T=1 | C_T=0, \cR_{\hat S}}.	
\end{equation}

Let $\nu$ be the conditional measure $C_T=0,~\cR_{\hat S}$. We see that
\begin{align*}
&\PP{\nu}{A_T=B_T=1} 
= \PP{\nu}{A_T^\uparrow = 1, B_T^\uparrow = 0, A_T^\rightarrow = 0, B_T^\rightarrow = 1} + \PP{\nu}{A_T^\uparrow = 0, B_T^\uparrow = 1, A_T^\rightarrow = 1, B_T^\rightarrow = 0} 
&\intertext{so using  independence of $(\delta^\uparrow(S))_T$ and $(\delta^\rightarrow(S))_T$.}
&\quad = \PP{\nu}{A_T^\uparrow = 1, B_T^\uparrow = 0}\PP{\nu}{A_T^\rightarrow = 0, B_T^\rightarrow = 1} + \PP{\nu}{A_T^\uparrow = 0, B_T^\uparrow = 1}\PP{\nu}{A_T^\rightarrow = 1, B_T^\rightarrow = 0}\\
	&\quad \geq (x(A^\uparrow)-\eps'_M)\PP{\nu}{A_T^\rightarrow = 0, B_T^\rightarrow = 1} + (x(B^\uparrow)-\eps'_M)\PP{\nu}{A_T^\rightarrow = 1, B_T^\rightarrow = 0}.
\end{align*}
In the final inequality, we used the fact that conditioned on $\cR_{\hat S}$, $\hat A =(A^\uparrow \cup B^\uparrow \cup C^\uparrow)_T = 1$ and marginals in $A^\uparrow$ and $B^\uparrow$ are approximately preserved.
Now, we lower bound $\PP{\nu}{A^\rightarrow_T=1,B^\rightarrow_T=0}$. Let $\eps_A,\eps_B$ be such that 
$$ \EE{\nu}{A^\rightarrow_T}=\PP{\nu}{A^\rightarrow_T=1,B^\rightarrow_T=0}+\eps_A,\hspace{0.5cm} \EE{\nu}{B^\rightarrow_T}=\PP{\nu}{A^\rightarrow_T=0,B^\rightarrow_T=1}+\eps_B$$
First notice that $\PP{\nu}{A^\rightarrow_T + B^\rightarrow_T \ge 1} = 1$, and so $\PP{\nu}{A^\rightarrow_T + B^\rightarrow_T \ge 2} \le \EE{\nu}{A^\rightarrow_T + B^\rightarrow_T}-1$. So,
\begin{align*} \eps_A+\eps_B &= \EE{\nu}{A^\rightarrow_T+B^\rightarrow_T} - \PP{\nu}{A^\rightarrow_T+B^\rightarrow_T=1} = \EE{\nu}{A^\rightarrow_T+B^\rightarrow_T} - (1-\PP{\nu}{A^\rightarrow_T + B^\rightarrow_T \ge 2}) \\ &\le 2(\EE{\nu}{A^\rightarrow_T+B^\rightarrow_T}-1) \le 5\eps_\eta.
\end{align*}
To see the last inequality, first, by \cref{def:hierarchy}, $x(\delta^\uparrow(S)) \ge 1-\eps_\eta$. Since $x(\delta(S)) \le 2+\eps_\eta$, we get that $x(\delta^\rightarrow(S)) \le 1+2\eps_\eta$. Therefore, $$\EE{\nu}{A^\rightarrow_T+B^\rightarrow_T} \le \E{\delta^\rightarrow(S) \mid \cR_{\hat{S}}} \le x(\delta^\rightarrow(S)) + \eps_\eta/2 \le 1+2.5\eps_\eta.$$

Therefore, 
\begin{align*}
\PP{\nu}{A_T=B_T=1}	&\geq (x(A^\uparrow)-\eps'_M) (\EE{\nu}{B^\rightarrow_T}-\eps_B) + (x(B^\uparrow)-\eps'_M) (\EE{\nu}{A^\rightarrow_T}-\eps_A)\\
&\geq (x(A^\uparrow)-\eps'_M)  (x(B^\rightarrow)-5\eps_\eta) + (x(B^\uparrow)-\eps'_M)(x(A^\rightarrow)-5\eps_\eta)
\end{align*}
where the second inequality uses that the  tree conditioning and $C^\rightarrow_T=0$ can only increase the marginals of edges in $A^\rightarrow$ and $B^\rightarrow$.
Simplify the above using  $x(A^\uparrow)  +x(A^\rightarrow) \ge 1-\eps_\eta$, and similarly for $B$,  
\begin{align*}
	&\PP{\nu}{A_T=B_T=1} \\
	&\ge (1- x(A^\rightarrow) - \eps_\eta-\eps'_M) (x(B^\rightarrow)-5\eps_\eta) + (1- x(B^\rightarrow) - \eps_\eta -\eps'_M)(x(A^\rightarrow)-5\eps_\eta) \\ 
	\intertext{ and since $x(A^\rightarrow) + x(B^\rightarrow) \ge 1- 2\eps_\eta$ (because
$x(A^\uparrow) + x(B^\uparrow) \le 1+ \eps_\eta$ and $x_C \le \eps_\eta$), this is}
	&\ge (1- x(A^\rightarrow) - \eps_\eta-\eps'_M) (1 - x(A^\rightarrow)-7\eps_\eta) + (x(A^\rightarrow) - 3\eps_\eta -\eps'_M)(x(A^\rightarrow)-5\eps_\eta) \\
	&\ge   (1- x(A^\rightarrow))^2 + (x(A^\rightarrow))^2 -\eps'_M 
	- 8\eps_\eta.
	\end{align*}
Plugging this into \cref{eq:prob-110-before-nu}, we obtain 
\begin{align*}
\P{A_T=B_T=1,C_T=0 \mid \cR_{\hat S}} &\geq (1-2\eps_\eta - \eps'_M)\P{A_T=B_T=1 | C_T=0, \cR_{\hat S}} \\
& \ge (1-2\eps_\eta - \eps'_M)((1- x(A^\rightarrow))^2 + (x(A^\rightarrow))^2 -\eps'_M - 8\eps_\eta) \\
& \ge (1- x(A^\rightarrow))^2 + (x(A^\rightarrow))^2 -2\eps'_M - 10\eps_\eta,
\end{align*}
which noting $\eps'_M=\eps_M+2\eps_\eta$ completes the proof of the lemma. 
\end{proof}

 \begin{lemma}\label{lem:middleISright}
Let $S\in\cH$ be a polygon cut with $\p(S)=\hat{S}$ also a polygon cut with $u_1,\dots,u_k$ be the ordering of cuts in $\cA(\hat{S})$. If $S\neq u_1,u_k$, then 
$$ \E{I_S^\rightarrow} \leq  0.85 \beta p.$$
\end{lemma}

\begin{proof}
Let $S=u_i$ for some $2\leq i\leq k-1$.
 Let $A,B,C$ be the polygon partitioning of  $\delta(u_i)$ and $\hat{A},\hat{B},\hat{C}$ be the polygon partition of $\hat{S}$.
 Since $u_i$ is in the \hyperlink{tar:hierarchy}{hierarchy} $A^\uparrow\cup B^\uparrow\cup C^\uparrow\subseteq \hat{C}$.
 So, conditioned on $\cR_{\hat{S}}$, $A^\uparrow_T= B^\uparrow_T= C^\uparrow_T=0$.
 
Once again, let $\nu$ be the conditional measure $C_T=0,~ \cR_{\hat{S}}$. Similar to the previous case, we will lower-bound
\begin{align}
\P{S\text{ happy} |\cR_{\hat{S}}} &\geq (1-2\eps_\eta) \P{A^\rightarrow_T=1,B^\rightarrow_T=1,| C_T=0,  \cR_{\hat{S}}}\nonumber\\
&=(1-2\eps_\eta) \PP{\nu}{A^\rightarrow_T=1 | A^\rightarrow_T+B^\rightarrow_T=2}\PP{\nu}{A^\rightarrow_T+B^\rightarrow_T=2 }
\label{eq:bbhappyGreduced}
\end{align}
where we used $\E{C^\rightarrow_T | \cR_{\hat{S}}} \leq 2\eps_\eta$ in the first inequality. 
So, it remains to lower-bound each of the two terms in the RHS. 

We start with the first one. Since $x(A)\in [1-\eps_\eta,1+\eps_\eta]$ and $x(A^\uparrow)\leq \eps_\eta$ we have
$$ \EE{\nu}{A^\rightarrow_T}\in [1-2\eps_\eta,1+3\eps_\eta]. 
$$
The same bounds hold for $\EE{\nu}{x(B^\rightarrow)}$. 

Therefore,
\begin{align*}
&\PP{\nu}{A^\rightarrow_T \ge 1}, \PP{\nu}{B^\rightarrow_T\geq 1} \geq 1-e^{-1+2\eps_\eta}
 \tag{\cref{lem:SR>=}} \\
&\PP{\nu}{A^\rightarrow_T\leq 1},\PP{\nu}{B^\rightarrow_T\leq 1} \geq 0.495 \tag{Markov}
\end{align*}


Therefore, by \cref{lem:SRA=nA} (with $\eps=0.495(1-e^{-1+2\eps_\eta})\geq 0.31$) we have
$$\PP{\nu}{A^\rightarrow_T =1 \mid A^\rightarrow_T + B^\rightarrow_T = 2} \ge 0.155.$$ 
By \cref{lem:treeoneedge}, 
$\PP{\nu}{E(u_{i-1},u_i)_T=1}\geq 1-4\eps_\eta$. Similarly, $\PP{\nu}{E(u_{i},u_{i+1})_T=1}\geq 1-4\eps_\eta$. And, 
$$\PP{\nu}{\delta^\rightarrow(u_i)_T- E(u_{i-1},u_i)_T -E(u_i,u_{i+1})_T=0} \geq 1-4\eps_\eta$$
So, by a union bound all of these events happen simultaneously and we get $\PP{\nu}{\delta^\rightarrow(u_i)_T=2}\geq 1-12\eps_\eta$.
Therefore,
$$ \PP{\nu}{(A^\rightarrow)_T=(B^\rightarrow)_T=1} \geq 0.155 (1-12\eps_\eta) \ge 0.153.$$
Plugging this back into \eqref{eq:bbhappyGreduced}, we get
$$ \P{S\text{ happy} | \cR_{\hat{S}}}\geq 0.153(1-2\eps_\eta) \ge 0.152.$$
Plugging this in \eqref{eq:Dec-eBTu2}
we get
\begin{align*} \E{I_S^\rightarrow}&\leq (1+\eps_\eta)\beta p \P{S\text{ not happy}|\cR_{\hat{S}}}(\max\{x(A^\rightarrow),x(B^\rightarrow)\} + x(C^\rightarrow)) \\
	&\leq (1+\eps_\eta) \beta p (1-0.152) (1+\eps_\eta+\eps_\eta) \leq 0.85 \beta p
\end{align*}
as desired.

\end{proof}
%

%% file: probabilistic-app.tex
\section{Proofs from \cref{sec:probabilistic}}
\label{app:probabilistic}

In all of the following lemmas, we assume that $\eps_\eta \le \eps_{1/2}^2$ and $12\eps_{1/1}\leq \eps_{1/2}$.

\begin{figure}[htb]\centering
\begin{tikzpicture}
\node [draw,circle] at (0,0) (u) {$u$};
\node [draw,circle] at (2,0) (v) {$v$} edge node [below] {$\bbe$} (u);
\draw (v) -- +(60:1) (v) -- + (90:1) (v) -- +(120:1);
\draw [color=red,dashed,line width=1.1pt](2.5,0.5) arc (50:130:0.75);
\node [color=red] at (2.8,0.5) () {$V$};
\end{tikzpicture}
\caption{Setting of \cref{lem:x_e<=1/2-eps1}}
\label{fig:x_e<=1/2-eps1}
\end{figure}

\lemxesmallhalfeps*
\begin{proof}
Let $A,B,C$ be the \hyperlink{tar:degreepartition}{degree partitioning} of  $\delta(u)$.
Let $V:= \delta(v)_{-\bbe}$ (see \cref{fig:x_e<=1/2-eps1}).
Condition $u,v$ be trees, $\bbe$ and $C$ to 0, let $\nu$ be the resulting measure. This happens with probability at least  $0.5$ and increases marginals in $A_{-\bbe},B_{-\bbe}, V$ by at most $x_\bbe+2\eps_{1/1}+\eps_{\eta}\leq x_\bbe+2.1\eps_{1/1}$ and by tree conditioning decreases marginals by at most $2\eps_{\eta}$.  
After conditioning, we have
\begin{align*}
&\EE{\nu}{A_T} \in x(A)-x_{\bbe(A)}+[-2\eps_\eta, x_\bbe+2.1\eps_{1/1}] 
 \subset [0.5, 1.5], \text{ similarly }\EE{\nu}{B_T}\subset [0.5,1.5]\\
&\EE{\nu}{V_T} \in  x(\delta(v))-x_\bbe+ [-2\eps_\eta, x_\bbe+2.1\eps_{1/1}]\subset [1.5,2.01] 
\\
&\EE{\nu}{B_T+V_T} \in x(B)+x(\delta(v))-x_\bbe - x_{\bbe(B)} + [-2\eps_\eta, x_\bbe+2.1\eps_{1/1}] 
\subset [2+1.8\eps_{1/2}, 3.01],\\ 
&\EE{\nu}{A_T+B_T}\in x(A)+x(B)-x_{e(A)}-x_{\bbe(B)} + [-2\eps_\eta, x_\bbe+2.1\eps_{1/1}]\subset 
[1.5, 2.01],\\
& \EE{\nu}{A_T+B_T+V_T} \in x(A)+x(B)+x(\delta(v))-x_\bbe-x_{e(A)}-x_{\bbe(B)}+ [-2\eps_\eta, x_\bbe+2.1\eps_{1/1}] \\ 
&\quad \quad \quad \quad \subset 
[3+1.75\eps_{1/2},4.01].  
\end{align*}
where we used $\eps_{1/2}\leq 0.001$ and $12\eps_{1/1}<\eps_{1/2}$ and $x_{e(A)},x_{\bbe(B)},x_{e(A)}+x_{\bbe(B)}\leq x_\bbe\leq 1/2-\eps_{1/2}$.
It immediately follows from \cref{lem:427gen} that $\PP{\nu}{A_T=B_T=1,V_T=2}$ is at least a constant. In the rest of the proof, we do a more refined analysis. 
Using $A_T+B_T\geq 1, V_T\geq 1$,
\begin{align*}
&\PP{\nu}{A_T+B_T+V_T=4}\geq 	(1.75\eps_{1/2})e^{-1.75\eps_{1/2}} \geq 1.7\eps_{1/2}, \tag{\cref{thm:rayleigh_expectconstprob}} \\
&\PP{\nu}{A_T+B_T\geq 2},\PP{\nu}{V_T\geq 2} \geq 0.39, \tag{\cref{lem:SR>=}} \\
&\PP{\nu}{A_T+B_T\leq 2},\PP{\nu}{V_T\leq 2} \geq 0.5, \tag{Markov, $A_T+B_T\geq 1, V_T\geq 1$ under $\nu$} \\
&\PP{\nu}{A_T\leq 1}\geq 0.25, \PP{\nu}{B_T+V_T\leq 3} \geq 0.33. \tag{Markov's Inequality and $V_T\geq 1$ under $\nu$}\\
&\PP{\nu}{A_T\geq 1}\geq 0.39, \PP{\nu}{B_T+V_T\geq 3}\geq 1.75\eps_{1/2}, \tag{\cref{lem:SR>=}}
\end{align*}

It follows by \cref{lem:SRA=nA} (with $\eps=0.195, p_m\geq 1-2\eps\geq 0.6$) that 
$$\PP{\nu}{V_T=2 | A_T+B_T+V_T=4} \geq 0.13.$$
Note that since $V_T\geq 1, A_T+B_T\geq 1$ with probability 1, we apply \cref{lem:SRA=nA} to $V_T-1, A_T+B_T-1$.

Furthermore, by \cref{lem:updowntruncation}, $\PP{\nu}{A_T\geq 1 | A_T+B_T+V_T=4} \geq 0.128$, $\PP{\nu}{A_T\leq 1 | A_T+B_T+V_T=4} \geq 0.43\eps_{1/2}$. The same holds for $B_T$. Therefore, by \cref{lem:SRA=nA} (with $\eps=0.055\eps_{1/2}$), using that $\eps_{1/2}<0.001$,
$$\PP{\nu}{A_T=1 | A_T+B_T=2, V_T=2} \geq 0.05\eps_{1/2}.$$
Putting these together we have
\begin{align*} \P{e\text{ 2-1-1 happy}} &\geq 0.5 \PP{\nu}{A_T=B_T=1,V_T=2} \\
&= 0.5\PP{\nu}{A_T+B_T+V_T=4}\PP{\nu}{V_T=2|A_T+B_T+V_T=4}\\
&\quad \quad \cdot \PP{\nu}{A_T=1| V_T=2,A_T+B_T=2}\\
&\geq 0.5 (1.7\eps_{1/2})(0.13)(0.05\eps_{1/2}) \geq  0.005\eps_{1/2}^2
\end{align*}
as desired.
\end{proof}

\xemoreepshalf*
\begin{proof}
Let $A,B,C$ be the \hyperlink{tar:degreepartition}{degree partitioning} of the edges in $\delta(u)$, $V=\delta_{-\bbe}(v)$.
Condition $u, v$ be trees, $C_T=0$ and $u \cup v$ to be a tree (in order). This happens with probability at least $\frac12+\eps_{1/2}-3\eps_\eta-2\eps_{1/1}\geq 0.5$. Let $\nu$ be the resulting measure restricted to edges in $A,B,V$. 
Note that $\nu$ on edges in $A,B,V$ is SR.
This is because $\nu$ is a product of two strongly Rayleigh distribution on the following two disjoint set of edges (i) the edges between $u,v$ and (ii) the edges in $A_{-\bbe},B_{-\bbe},V$.

Furthermore, observe that under $\nu$, every set of edges in $A_{-\bbe},B_{-\bbe},V$ increases by at most $2\eps_{1/1}+\eps_\eta<0.2\eps_{1/2}$ (using $12\eps_{1/1}\leq \eps_{1/2}$), and decreases by at most $1-x_\bbe+2\eps_{\eta}$.  Therefore,
\begin{align*}
&\EE{\nu}{A_T} \in x(A)+[-(1-x_\bbe)-2\eps_{\eta},1-x_\bbe+0.2\eps_{1/2}]\subset   [0.5,1.5], \text{ similarly, }\EE{\nu}{B_T} \in [0.5,1.5]\\
& \EE{\nu}{V_T} \in x(\delta(v))-x_\bbe+ [-(1-x_\bbe)-2\eps_{\eta},0.2\eps_{1/2}] \subset [ 0.995 , 1.5].\\
&\EE{\nu}{A_T+B_T} \in x(A)+x(B)+1-x_{e(A)}-x_{\bbe(B)}+ [-(1-x_\bbe)-2\eps_{\eta},0.2\eps_{1/2}] \subset [1.995, 2.5],  \\
&\EE{\nu}{B_T+V_T}\in x(B)+x(\delta(v))-x_\bbe+ [-(1-x_\bbe)-2\eps_{\eta}, 1-x_\bbe+0.2\eps_{1/2}]\subset [1.99, 3-1.75\eps_{1/2}].\\
&\EE{\nu}{A_T+B_T+V_T} \in x(A)+x(B)+x(\delta(v))+1-x_\bbe-x_{e(A)}-x_{\bbe(B)}+[-(1-x_\bbe)-2\eps_{\eta},0.2\eps_{1/2}] \\
&\quad \quad \quad \quad \subset [2.99, 4-1.75\eps_{1/2}].
\end{align*}
where in the upper bound on $\EE{\nu}{A_T}$, $\EE{\nu}{B_T}$, $\EE{\nu}{B_T+V_T}$ we used that the marginals of edges in the bundle $\bbe$ can only increase by $1-x_\bbe$ (in total) when conditioning $u\cup v$ to be a tree.
So, 
\begin{align*} 
&\PP{\nu}{A_T+B_T+V_T=3} \geq \eps_{1/2}, \tag{By \cref{thm:hoeffding}}\\
&\PP{\nu}{A_T+B_T\geq 2}\geq 0.63, \PP{\nu}{V_T\geq 1} \geq 0.63\tag{\cref{lem:SR>=}, $A_T+B_T\geq 1$}\\
&\PP{\nu}{A_T+B_T\leq 2}\geq 0.25, \PP{\nu}{V_T\leq 1}\geq 0.25, \tag{Markov Inequality, $A_T+B_T\geq 1$}\\
& \PP{\nu}{A_T\geq 1} \geq 0.39, \PP{\nu}{B_T+V_T\geq 2}\geq 0.59 \tag{\cref{lem:SR>=} } \\
&\PP{\nu}{A_T\leq 1} \geq 0.25, \PP{\nu}{B_T+V_T\leq 2}\geq 1.75\eps_{1/2}, \tag{Markov, In worst case $\P{B_T+V_T<2}=0$}
\end{align*}
It follows by \cref{lem:SRA=nA} (with $\eps=0.157, p_m=0.68$) that
$$ \PP{\nu}{A_T+B_T=2 | A_T+B_T+V_T=3} \geq 0.12.$$
Note that since $A_T+B_T\geq 1$ with probability $1$, we apply \cref{lem:SRA=nA} to $A_T+B_T-1, V_T$. 

Furthermore, by \cref{lem:updowntruncation},
$\PP{\nu}{A_T\geq 1 | A_T+B_T+V_T=3} \geq 0.68\eps_{1/2}$ and $\PP{\nu}{A_T\leq 1 | A_T+B_T+V_T=3}\geq 0.147$.
By symmetry, the same holds for $B_T$.
Therefore, by \cref{lem:SRA=nA},
$$ \PP{\nu}{A_T=1 | A_T+B_T=2, V_T=1}\geq 0.09\eps_{1/2}.$$
where we used $\eps_{1/2}<0.001$.

Finally,
$$ \P{\bbe\text{ 2-1-1 happy}}\geq (0.09\eps_{1/2})0.12(\eps_{1/2})0.5 \geq 0.005\eps_{1/2}^2,$$
as desired.
\end{proof}

\begin{figure}[htb]\centering
\begin{tikzpicture}
		\node [draw=none,color=purple,inner sep=0] at (0.1,0) (A) {{ $A$}};
		\node [draw=none,color=blue,inner sep=0] at (0.8,0) (B) {{ $B$}};
		\node [draw=none] at (-0.3,-.5) () {$u$};
		\node [draw,circle] at (2.75,-0.25) (v) {{ $v$}};
		\node [draw,circle,inner sep=13] at (0.5,0) (u) {};
		\draw   (u) --  +(60:1.5) (u) -- +(90:1.5) (u) -- +(120:1.5);
		\draw  (v) --  +(60:1.25) (v) -- +(90:1.25) (v) -- +(120:1.25);
		\draw [color=red,dashed,line width=1.1pt] (3.4,0.3) arc (50:130:1);
		\node [color=red] at (3.6,0.3) () {$V$};
		\draw [color=purple] (v) edge [bend left=60] node [below] {$\bbe(A)$}(A); 
		\draw [line width=1.1pt, dotted, color=blue] (v) edge [bend left=30] node [above] {$\bbe(B)$}(B); 		
	\end{tikzpicture}
	\caption{Setting of \cref{lem:<=1211good}}
	\label{fig:e(B)small}
\end{figure}

\begin{lemma}\label{lem:<=1211good}	
For a good half top edge bundle $\bbe={\bf (u,v)}$, let $A,B,C$ be the \hyperlink{tar:degreepartition}{degree partitioning} of $\delta(u)$, and  let $V=\delta(v)_{-\bbe}$ (see \cref{fig:e(B)small}). If $\eps_{1/2}\leq 0.001$, $x_{\bbe(B)}\leq \eps_{1/2}$, and $\P{(A_{-\bbe})_T + V_T\leq 1}\geq 5\eps_{1/2}$  then $\bbe$ is 2-1-1 good,
$$ \P{\bbe \text{ 2-1-1 happy w.r.t. $u$}} \geq 0.005\eps_{1/2}^2$$
\end{lemma}
\begin{proof}
The proof is similar to \cref{lem:x_e>=1/2+eps_1/2}. We condition $u,v$ to be trees, $C_T=0$, $u\cup v$  to be a tree.
Let $\nu$ be the resulting SR measure on edges in $A,B,V$.
 The main difference is since $x_\bbe\not\geq 1/2+\eps_{1/2}$ we use the lemma's assumptions to lower bound $\PP{\nu}{A_T+B_T+V_T=3},\PP{\nu}{A_T+V_T\leq 2},\PP{\nu}{B_T+V_T\leq 2}$.

First, since $\bbe$ is 2-2 good, by \cref{lem:2/2goodEvenSum} and negative association,
\begin{align*}
\PP{\nu}{(\delta(u)_{-\bbe})_T+V_T\leq 2} \geq \P{(\delta(u)_{-\bbe})_T+V_T\leq 2} - \P{C_T=0} \geq 0.4\eps_{1/2} - 2\eps_{1/1}-\eps_\eta\geq 0.22\eps_{1/2},	
\end{align*}
where we used $\eps_{1/1}\le \eps_{1/2}/12$.
Letting $p_i = \P{(\delta(u)_{-\bbe})_T+V_T = i}$, we therefore have $p_{\le 2} \ge 0.22 \eps_{1/2}$. In addition, by
\cref{thm:rayleigh_expectconstprob}, $p_3 \ge 1/4$.
If $p_2 < 0.2 \eps_{1/2}$, then from
$p_2/p_3 \le 0.8 \eps_{1/2}$, we could use log-concavity to derive a contradiction to $p_{\le 2} \ge 0.22 \eps_{1/2}$ (analogously to what's done in the proof of \cref{lem:logconcaveexpecation}).
Therefore, we must have
$$\PP{\nu}{A_T+B_T+V_T=3}=\PP{\nu}{(\delta(u)_{-\bbe})_T+V_T=2}\geq 0.2\eps_{1/2}.$$

Next, notice since $\P{u,v,u\cup v\text{ trees}, C_T=0}\geq 0.49$, by the lemma's assumption, $\PP{\nu}{\bbe(B)}\leq 2.01\eps_{1/2}$. Therefore, 
$$\EE{\nu}{B_T+V_T}\leq x(V)+x(B)+1.01\eps_{1/2}+2\eps_{1/1}+\eps_\eta \leq 2.51.$$
So, by Markov, $\PP{\nu}{B_T+V_T\leq 2}\geq 0.15$. 
Finally, by negative association, 
$$\PP{\nu}{A_T+V_T\leq 2} \geq \PP{\nu}{(A_{-\bbe})_T+V_T\leq 1} \geq \P{(A_{-\bbe})_T+V_T\leq 1} - \P{C_T=0} \geq 4.8\eps_{1/2}$$
where we used the lemma's assumption. 

Now, following the same line of arguments as in \cref{lem:x_e>=1/2+eps_1/2}, we have \\$\PP{\nu}{A_T+B_T=2 | A_T+B_T+V_T=3} \geq 0.12$.
Also, $\PP{\nu}{A_T\geq 1 | A-T+B_T+V_T=3} \geq 3.02$, which implies $\PP{\nu}{A_T=1 | A_T+B_T=2,V_T=1} \geq 0.42\eps$. This implies
$$
	\P{\bbe \text{ 2-1-1 happy}} \geq (0.42\eps_{1/2})0.12(0.2\eps_{1/2})0.498\geq 0.005\eps_{1/2}^2
$$
as desired.
\end{proof}

\lemoneoftwotwooneone*
\begin{proof}
Let $U=\delta(u)_{-\bbe}$.
By \cref{lem:crossingcorrelation}, we can assume, without loss of generality, that
\begin{equation}
\label{eq:5-26-1}
	 \E{U_T | \bbf\notin T,u,v,w\text{ tree}}\leq x(U_T)+0.405 +3\eps_{\eta}.
\end{equation}
On the other hand,  
\begin{align*} \E{(A_{-\bbe-\bbf})_T} &\geq \E{(A_{-\bbe-\bbf})_T|\bbf\notin T, u,v,w\text{ tree}} \P{\bbf\notin T, u,v,w,\text{ tree}}\\
&	\geq \E{(A_{-\bbe-\bbf})_T|\bbf\notin T, u,v,w\text{ tree}} 0.49
\end{align*}

So, 
\begin{equation}
\label{eq:5-26-2} 	
\E{(A_{-\bbe-\bbf})_T | \bbf\notin T, u,v,w,\text{ tree}} \leq \frac{1}{0.49} x(A_{-\bbe-\bbf}) \leq \frac{1}{0.49} (4\eps_{1/2}+\eps_{\eta}) \leq 8.2\eps_{1/2}.
\end{equation}

Combining \eqref{eq:5-26-1} and \eqref{eq:5-26-2}, we get
$\E{U_T+(A_{-\bbe}) | \bbf\notin T,u,v,w \text{ tree}}\leq 1.91$ where we used $\eps_{1/2}\leq 0.001$.
Therefore, using \cref{thm:rayleigh_expectconstprob}, we get
$$\P{U_T + (A_{-\bbe})_T\leq 1} \geq 0.49 \P{U_T + (A_{-\bbe})_T\leq 1 | \bbf\notin T,u,v,w\text{ tree}}\geq 0.01,$$
Since $\eps_{1/2}\leq 0.001$, by \cref{lem:<=1211good}, $\bbe$ is 2-1-1 good.
\end{proof}

\begin{figure*}[htb]\centering
\begin{tikzpicture}
		\node [draw=none,color=purple,inner sep=0] at (0.1,0) (A) {{ $A$}};
		\node [draw=none,color=blue,inner sep=0] at (0.8,0) (B) {{ $B$}};
		\node [draw,circle] at (2.75,-0.25) (v) {{ $v$}};
		\node [draw,circle,inner sep=13] at (0.5,0) (u) {};
		\draw   (u) --  +(60:1.5) (u) -- +(90:1.5) (u) -- +(120:1.5);
		\draw  (v) --  +(60:1.25) (v) -- +(90:1.25) (v) -- +(120:1.25);
		\draw [color=red,dashed,line width=1.1pt] (3.4,0.3) arc (50:130:1) (1.2,0.75) arc (50:130:1);
		\node [color=red] at (3.6,0.3) () {$Y$} ;
		\node [color=red] at (-0.3,0.75) () {$X$} ;
		\draw [color=purple] (v) edge [bend left=60] node [below] {$\bbe(A)$}(A); 
		\draw [color=blue] (v) edge [bend left=30] node [above] {$\bbe(B)$}(B); 		
	\end{tikzpicture}
	\caption{Setting of \cref{lem:eABpositive211}.}
	\label{fig:e(A)e(B)large}
\end{figure*}

\lemeABpositivetwooneone*
\begin{proof}
Condition $C_T$ to be zero, $u,v$ and $u\cup v$ be trees. This happens with probability at least  $0.49$.
Let $\nu$ be the resulting measure.
Let $X=A_{-\bbe}\cup B_{-\bbe}, Y=\delta(v)_{-\bbe}$ 
Since $\bbe$ is 2-2 good by \cref{lem:2/2goodEvenSum} and stochastic dominance, 
$$\PP{\nu}{X_T + Y_T\leq 2}\geq \P{(\delta(u)_{-\bbe})_T+Y_T\leq 2} - \P{C_T=0} \geq 0.4\eps_{1/2} - 2\eps_{1/1}-\eps_\eta\geq 0.22\eps_{1/2},$$
where we used $\eps_{1/1}<12\eps_{1/2}$.
It follows by log-concavity of $X_T+Y_T$ that $\PP{\nu}{X_T+Y_T=2}\geq 0.2\eps_{1/2}$. 
Now, 
\begin{align*}
&\EE{\nu}{X_T}, \EE{\nu}{Y_T}\in [1-3\eps_{1/1}, 1.5+\eps_{1/2}+2\eps_{1/1}+3\eps_{\eta}]\subset [0.995,1.51]
\end{align*}
So,
\begin{align*}
&\PP{\nu}{X_T\geq 1}, \PP{\nu}{Y_T\geq 1}\geq 0.63, \tag{\cref{lem:SR>=}} \\
&\PP{\nu}{X_T\leq 1}, \PP{\nu}{Y_T\leq 1}\geq 0.245.\tag{Markov}
\end{align*}
Therefore, by \cref{lem:SRA=nA} $\PP{\nu}{X_T= 1 | X_T+Y_T=2}\geq 0.119$.
$$ \PP{\nu}{X_T=Y_T=1} \geq (0.2\eps_{1/2}) 0.119 \geq 0.023\eps_{1/2},$$
Let $\cE$ be the event $\{X_T=Y_T=1 | \nu\}$.
Note that in $\nu$ we always choose exactly 1 edge from the $\bbe$ bundle and that is independent of edges in $X,Y$, in particular the above event. Therefore, we can  correct the parity of $A,B$ by choosing from $e_A$ or $e_B$.
It follows that
$$ \P{\bbe \text{ 2-1-1 happy w.r.t $u$}} \geq \PP{\nu}{\cE}(1.99\eps_{1/2}) 0.49 \geq 0.02\eps_{1/2}^2,$$
where we used that $\EE{\nu}{\bbe(A)_T}  \geq 1.99\eps_{1/2}$, and the same fact for $\bbe(B)_T$. To see why this latter fact is true, observe that conditioned on $u,v$ trees, we always sample at most one edge between $u,v$. Therefore, since under $\nu$ we choose exactly one edge between $u,v$, the probability of choosing from $e(A)$ (and similarly choosing from $\bbe(B)$) is at least 
$$\frac{\E{\bbe(A)_T | u,v\text{ trees}, C_T=0}}{\P{\bbe | u,v\text{ trees}, C_T=0}} \geq \frac{x_{\bbe(A)}-2\eps_\eta}{x_{\bbe}+3\eps_{1/1}}\geq \frac{\eps_{1/2}-2\eps_{\eta}}{1/2+1.3\eps_{1/2}}\geq 1.99\eps_{1/2}$$
as desired.
\end{proof}

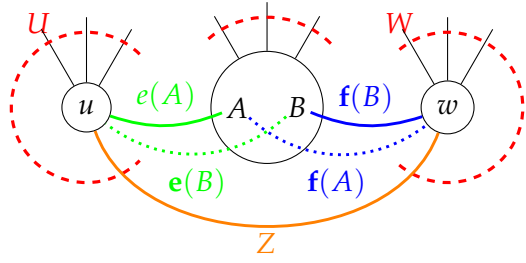
\begin{figure}[htb]\centering
\begin{tikzpicture}[scale=0.8]
\tikzstyle{every node}=[draw,circle]
\foreach \a/\x/\l in {u/-3/U, w/3/W}{
\path  (\x,0) node  (\a) {$\a$};
\path (\a)+(-0.8,1.4) node [color=red,draw=none] (){$\l$};
\foreach \xx in {0,...,2}{
\draw [-] (\a) -- ++(60+\xx*30: 1.5);}
}
\draw [dashed,color=red,line width=1.2] (u)+(45:1.25) arc (45:315:1.25);
\draw [dashed,color=red,line width=1.2] (w)+(-130:1.25) arc (-130:130:1.25);

\node [inner sep=15] at (0,0) (v) {};
\foreach \xx in {0,...,2}{
\draw [-] (v) -- ++(60+\xx*30: 1.75);}
\draw [dashed,color=red,line width=1.2] (v)+(45:1.5) arc (45:135:1.5);

\node [draw=none,inner sep=0] at (-.5,0) (A) {$A$};
\node [draw=none,inner sep=0] at (.5,0) (B) {$B$};

\path [-] (A) edge [bend left=20,color=green,line width=1.1pt] node [draw=none,above=-7pt] {$e(A)$} (u)
(B) edge [bend left=40,color=green, line width=1.1pt,dotted] node [draw=none,below=-7pt] {$\bbe(B)$} (u)
(A) edge [bend right=40,color=blue, line width=1.1pt, dotted] node [draw=none, below=-7pt] {$\bbf(A)$} (w)
(B) edge [bend right=20,color=blue, line width=1.1pt] node [draw=none,above=-7pt] {$\bbf(B)$} (w)
;
\path (u) edge [color=orange,bend right=70,line width=1.1pt] node [draw=none,below=-5pt] {$Z$} (w);
\end{tikzpicture}
\caption{Setting of \cref{lem:222}. We assume that the dotted green/blue edges are at most $\eps_{1/2}$. Note that edges of $C$ are not shown.} 
\label{fig:eABpositive211}
\end{figure}
\lemtwotwotwo*
\begin{proof}
%
	First, observe that by \cref{lem:<=1211good} if $\P{U_T+(A_{-\bbe})_T\leq 1}\geq 0.25\eps$, where $\eps\geq 20\eps_{1/2}$ is a constant that we fix later, then $\bbe$ is 2-1-1 good, which is a contradiction. 
	So, assume, $\P{U_T+(A_{-\bbe})_T\geq 2}\geq 1-0.25\eps.$
	Furthermore, let $q=\P{U_T+(A_{-\bbe})_T\geq 3}$. Since $x(U)+x(A_{-\bbe})\leq 2+3\eps_{1/2}+2\eps_{1/1}+3\eps_\eta\leq 2+3.2\eps_{1/2}$ (where we used $x_{\bbe(A)}\geq x_\bbe - x_{\bbe(B)} - x_C\geq  1/2-2\eps_{1/2}-2\eps_{1/1}-\eps_{\eta}$ and where we used $12\eps_{1/1}\leq \eps_{1/2}$), 
	$$ 2(1-q-0.25\eps) +3q \leq 2+3.2\eps_{1/2}.$$
	This implies that $q\leq 0.5\eps + 3.2\eps_{1/2}\leq 0.75\eps$ (for $\eps\geq  13\eps_{1/2})$.
	Therefore, 
	\begin{equation}\label{eq:211-bad}
		\P{U_T+(A_{-\bbe})_T=2},\P{W_T+(B_{-\bbf})_T=2}\geq 1-\eps
	\end{equation}
	where the second inequality follows by a similar argument.
	\begin{claim} 	Let $Z=\delta(u)\cap\delta(w)$. If $\eps<1/15$, then either
		$\E{Z | u,v,w\text{ tree}}\leq 3\eps$ or $\E{Z | u,v,w\text{ tree}}\geq (1-3\eps)$.
	\end{claim}
\begin{proof}
For the whole proof we work with $\mu$ conditioned on $u,v,w$ are trees.
	Let $z=\E{Z}$.
	Let $D=U\cup W\cup A_{-\bbe}\cup B_{-\bbf}\smallsetminus Z$. Note that $D_T+2Z_T=U_T\cup W_T\cup (A_{-\bbe})_T\cup (B_{-\bbf})_T$.
	By \cref{eq:211-bad} and a union bound $\P{D_T+2Z_T=4}\geq 1-2\eps-3\eps_\eta$. Therefore,
	\begin{align*}
		2.1\eps\geq 2\eps+3\eps_\eta  \geq \P{D_T+2Z_T\neq 4} \geq \P{D_T=3} \geq \sqrt{\P{D_T=2}\P{D_T=4}}
	\end{align*}
	where the last inequality follows by log-concavity.
	On the other hand,  
	\begin{align*} &z=\P{Z=1} \leq \P{D_T=2,Z=1}+\P{D_T+2Z_T\neq 4} \leq \P{D_T=2} + 2.1\eps,\\	
&	1-z=\P{Z=0} \leq \P{D_T=4,Z=0}+\P{D_T+2Z_T\neq 4} \leq \P{D_T=4} + 2.1 \eps
	\end{align*}
Putting everything together,
	$$(2.1\eps)^2 \geq (z-2.1\eps)(1-z-2.1\eps)=z(1-z)-2.1\eps + 2.1\eps^2.$$
	Therefore, using $\eps\leq 1/15$, we get that either $z\leq 3\eps$ or $z\geq 1-3\eps$.
\end{proof}

So, for the rest of proof we assume $\E{Z_T|u,v,w \text{ trees}}<3\eps$. A similar proof shows $\bbe,\bbf$ are 2-2-2 good when  $\E{Z_T|u,v,w\text{ trees}}>1-3\eps$.
We run the following conditionings in order: $u,v,w$ trees, $Z_T=0$, $C_T=0$, $\bbe(B),\bbf\notin T$, $\bbe(A)\in T$. Note that $\bbe(A)\in T$ is equivalent to $u\cup v$ be a tree. Call this event $\cE$ (i.e., the event that all things we conditioned on happen).
First, notice 
\begin{equation}
\label{eq:cElb}\P{\cE}\geq (1-3\eps_\eta) (1-3\eps-2\eps_{1/1}-\eps_\eta-\eps_{1/2}-(1/2+\eps_{1/2}))  (1/2-3\eps_{1/2})\geq 0.22\geq 1/5
\end{equation}

Moreover, since all of these conditionings correspond to upward/downward events,  $\mu | \cE$ is strongly Rayleigh.
The main statement we will show is that $$\P{\bbe,\bbf\text{ 2-2-2 happy}|\cE}\ge \P{U_T=(A_{-\bbe})_T=1, (B_{-\bbf})_T=0,W_T=2|\cE}= \Omega(1).$$
The main insight of the proof is that \cref{eq:211-bad} holds (up to a larger constant of $\eps$), even after conditioning $\cE, B_{-\bbf}=0,A_{-\bbe}=1$; so, we can bound the preceding event by just a union bound. The main non-trivial statement is to argue that the expectations of $B_{-\bbf}$ and $A_{-\bbe}$ do not change so much under $\cE$.

Combining \eqref{eq:211-bad} and \eqref{eq:cElb},
\begin{equation}\label{eq:211-badcE} \P{U_T+(A_{-\bbe})_T=2 | \cE},\P{W_T+(B_{-\bbf})_T=2 | \cE}\geq 1-5\eps.	
\end{equation}
We claim that 
\begin{equation}
\label{eq:BTcondcE}
	\E{B_T | \cE}=\E{(B_{-\bbf})_T|\cE} \leq x(B_{-\bbf}) + 3\eps_\eta+ 3\eps_{1/1}+ \eps_{1/2} +35 \eps \leq 0.66 
\end{equation}
using $\eps_{1/2}<0.0002$ and $\eps= 20\eps_{1/2}$.
To see this, observe that after each conditioning in $\cE$ either all marginals increase or all decrease. Furthermore, the events $C_T=0, Z_T=0, \bbe(B)_T=0$ can increase marginals by at most $3\eps_\eta+3\eps_{1/1}+\eps_{1/2}$; the only other event that can increase $B_{-\bbf}$ is $\bbf\notin T$. Now 
we know $\P{(B_{-\bbf})_T+W_T=2 |\cE}\geq 1-5\eps$ before and after conditioning $\bbf\notin T$. Therefore, by Corollary 2.19, $2-10\eps \leq \E{(B_{-\bbf})_T+W_T}\leq 2+25\eps$. 
But if $\E{(B_{-\bbf})_T}$ increased by more than $35\eps$, then either before conditioning $\bbf\notin T$,  $\E{(B_{-\bbf})+W_T}<2-10\eps$ or afterwards it is more than $2+25\eps$, which is a contradiction, and completes the proof of \eqref{eq:BTcondcE}.
A similar argument shows that $\E{(A_{-\bbe})_T|\cE}\leq 0.66$.

We also claim that
$$\E{(A_{-e})_T | \cE} \geq x(A_{-e}) - 3\eps_{\eta} -35\eps \ge 0.33.$$
As above, everything conditioned on in $\cE$ increases $\E{(A_{-e})_T }$ except for possibly $\bbe(A)\in T$.  As above, we know that $\P{U_T+(A_{-e})_T=2 |\cE}\geq 1-5\eps$ before and after $\bbe(A)\notin T$. So again applying Corollary 2.19, we see that it can't decrease by more than $35 \eps$.

It follows that
$$ 0.33\leq \E{(A_{-\bbe})_T |\cE}\leq \E{(A_{-\bbe})_T | \cE,(B_{-\bbf})_T=0} \leq 0.66 +0.66\leq 1.32.  $$
So, by \cref{thm:rayleigh_expectconstprob} and \cref{thm:hoeffding}, $\P{(A_{-\bbe})_T=1 | \cE, (B_{-f})_T=0}\geq 0.33e^{-.33}\geq 0.237$.


Therefore, by \cref{thm:rayleigh_expectconstprob}
$$ \P{\cE,(A_{-\bbe})_T=1,(B_{-\bbf})_T=0} \geq (0.22)(0.39)(0.23)\geq  0.019.$$
Therefore, by  \eqref{eq:211-badcE}
$$ \P{U_T=1 | \cE,(A_{-\bbe})_T=1,(B_{-\bbf})_T=0}, \P{W_T=2 | \cE,(A_{-\bbe})_T=1,(B_{-\bbf})_T=0}\geq 1-5\eps/0.019$$
Finally, by union bound 
$$ \P{U_T=1, W_T=2 | \cE,(A_{-\bbe})_T=1,(B_{-\bbf})_T=0}\geq 1-\eps/0.009$$
Using $\eps=20\eps_{1/2}$ and $\eps_{1/2}\leq 0.0002$ this means both of the above events happens, so  $\bbe,\bbf$ are 2-2-2-happy with probability $0.019(1-\eps/0.009) > 0.01$ as desired.
\end{proof}